\documentclass[12pt]{article}
\usepackage{amsmath,amssymb,amsthm,color}
\usepackage{graphicx}
\usepackage{cite}
\usepackage{caption}
\usepackage{epsfig}
\usepackage{epstopdf}
\renewcommand{\baselinestretch}{2}
\begin{document}
%
\title{Geometric and electronic properties of graphene-related systems: Chemical bondings}
\author{
\small Ngoc Thanh Thuy Tran$^{a}$, Shih-Yang Lin$^{a,*}$, Chiun-Yan Lin$^{a}$, Ming-Fa Lin$^{a,*}$\\
\small $^a$Department of Physics, National Cheng Kung University, Tainan 701, Taiwan \\
 }
\renewcommand{\baselinestretch}{1}
\maketitle

\renewcommand{\baselinestretch}{1.4}
\begin{abstract}

This work presents a systematic review of the feature-rich essential properties in graphene-related systems using the first-principles method. The geometric and electronic properties are greatly diversified by the number of layers, the stacking configurations, the sliding-created configuration transformation, the rippled structures, and the distinct adatom adsorptions. The top-site adsorptions can induce the significantly buckled structures, especially for hydrogen and fluorine adatoms. The electronic structures consist of the carbon-, adatom- and (carbon, adatom)-dominated energy bands. There exist the linear, parabolic, partially flat, sombrero-shaped and oscillatory band, accompanied with various kinds of critical points. The semi-metallic or semiconducting behaviors of graphene systems are dramatically changed by the multi- or single-orbital chemical bondings between carbons and adatoms. Graphene oxides and hydrogenated graphenes possess the tunable energy gaps. Fluorinated graphenes might be semiconductors or hole-doped metals, while other halogenated systems belong to the latter. Alkali- and Al-doped graphenes exhibit the high-density free electrons in the preserved Dirac cones. The ferromagnetic spin configuration is revealed in hydrogenated and halogenated graphenes under certain distributions. Specifically, Bi nano-structures are formed by the interactions between monolayer graphene and buffer layer. Structure- and adatom-enriched essential properties are compared with the measured results, and potential applications are also discussed.
\vskip 1.0 truecm
\par\noindent

\noindent \textit{Keywords}: graphene oxide; chemical bonding; $\pi$ bonding; energy gap; charge density
\vskip1.0 truecm

\par\noindent  * Corresponding authors. {~ Tel:~ +886-6-2757575-65272 (M.F. Lin)}\\~{{\it E-mail address}: mflin@mail.ncku.edu.tw (M.F. Lin); sylin.1985@gmail.com (S.Y. Lin)}
\end{abstract}
\pagebreak
\renewcommand{\baselinestretch}{2}
\newpage

{\bf 1. Introduction}
\vskip 0.3 truecm

The condensed-matter systems purely made up of carbon atoms include diamond,\cite{kamo1983diamond} graphites,\cite{tuinstra1970raman,huang1997collective} graphenes,\cite{novoselov2004electric,novoselov2005two} graphene nanoribbons,\cite{son2006half,li2008chemically,cai2010atomically} carbon nanotubes,\cite{iijima1991helical,iijima1993single,de1995carbon} carbon toroids,\cite{haddon1997electronic,lin1998persistent} C$_{60}$-related fullerenes,\cite{mintmire1992fullerene,liu1998fullerene} and carbon onions.\cite{kratschmer1990solid,chen2001new} Such systems, with various dimensions, are formed by the very active bondings related to the four atomic orbitals; furthermore, they are suitable for studying the unique phenomena. All of them possess the $sp^2$-bonding surfaces except the $sp^3$ bonding in diamond. Specifically, the few- and multi-layer graphenes have been manufactured using various methods since the first discovery in 2004 by the mechanical exfoliation.\cite{novoselov2004electric} Graphene exhibits a plenty of remarkable essential properties arising from the hexagonal symmetry and the nanoscale size, such as an anomalous quantum Hall effect,\cite{zhang2005experimental,zhang2011experimental} an exceedingly high mobility of charge carriers,\cite{bolotin2008ultrahigh,orlita2008approaching} a high transparency,\cite{nair2008fine,chen2013transparent} the largest Young's modulus of materials ever tested,\cite{lee2008measurement,lee2012estimation} the diverse optical selection rules,\cite{ho2010magneto,chung2011exploration} the rich magnetic quantization,\cite{guinea2006electronic,lin2015magneto} the unusual plasmon modes,\cite{ho2006coulomb,ju2011graphene} and the semiconducting and semi-metallic behaviors.\cite{castro2007biased,bao2011stacking,ho2009semimetallic} To create the novel phenomena and extend the potential applications, the electronic properties can be easily modulated by the layer number,\cite{wu2009synthesis,hao2010probing} stacking configuration,\cite{hao2010probing,bao2011stacking} mechanical strain,\cite{guinea2010energy,wong2012strain} sliding,\cite{son2011electronic,tran2015configuration} doping,\cite{giovannetti2008doping,dai2009gas} electric field,\cite{lu2006influence,l2007electronic,yan2007electric,castro2007biased} magnetic field,\cite{lai2008magnetoelectronic,goerbig2011electronic} and temperature.\cite{bolotin2008temperature,wu2011plasma} This review work is focused on the layer number-, stacking-, shift-, ripple- and doping-dependent geometric structures and electronic properties by using the first-principles calculations. The structure-enriched pristine graphenes, graphene oxides, hydrogenated graphenes, and the halogen-, alkali-, Al- and Bi-doped graphenes are taken into consideration. The physical and chemical pictures are proposed to comprehend the newly created phenomena, especially for the multi-orbital chemical bondings. The orbital hybridizations among the same/distinct atoms, which account for the main features of electronic structures, are explored in detail, e.g., the effects of chemical bondings on the Dirac-cone structures, energy gaps, semiconducting, semi-metallic and metallic behaviors, adatom-dependent energy bands, free electrons and holes, spatial charge distributions, specific adatom nanosclusters, and special structures in density of states (DOS). Spin configuration is another focus on hydrogenated and halogenated graphenes. A detailed comparison with the other theoretical studies and the experimental measurements is made. The high potentials in various materials applications are also discussed.

The stacking symmetry plays a critical role in the essential properties of layered graphenes. The pristine graphenes, with the stacking order\cite{lee2008growth,yankowitz2014electric,coletti2013revealing,ohta2007interlayer,norimatsu2010selective,rong1993electronic,xu2012electronic,pong2007observation,kazemi2013stacking} or the turbostratic structure,\cite{ferrari2007raman,li2008carbon,lenski2011raman} have been observed by the various experimental methods. In general, the stacking configurations are presented in the forms of AAA, ABA, ABC and AAB. These typical systems exhibit rather different characteristics in electronic properties,\cite{lu2006low,koshino2010interlayer,mak2010electronic,lui2011observation,mohammadi2014effects,do2015configuration} magnetic quantization,\cite{yuan2011landau,zhang2012hund,lin2015magneto,do2015configuration} optical excitations,\cite{lu2006influence,koshino2013stacking,do2015rich,chiu2016influence} transport properties,\cite{yuan2010electronic,ghosh2012monte,ma2012stacking,khodkov2012electrical} and phonon spectra.\cite{yan2008phonon,yan2009electron,lui2013tunable,cong2015magnetic} The electronic structures are extensively investigated by the first-principles calculations\cite{aoki2007dependence,zhang2010band,avramov2011ab,menezes2014ab} and the tight-binding model.\cite{avetisyan2010stacking,yuan2011landau,do2015configuration,lin2015magneto} These studies show that the interlayer atomic interactions of carbon $2p_z$ orbitals dominate the low- and middle-energy electronic structure. The main energy bands, the linear, parabolic, partially flat, sombrero-shaped and oscillatory ones, depend on the stacking configurations. The stacking-induced important differences in energy dispersions can diversify the quantization phenomena, selection rules, Hall conductances and Raman spectra. For example, the magnetoelectronic properties of the higher-symmetry AAA stacking have the monolayer-like quantized Landau levels (LLs) with the well-behaved modes,\cite{lin2015magneto} while those of the lower-symmetry AAB stacking present the splitting LLs and the frequently anti-crossing behavior.\cite{do2015configuration} On the experimental side, scanning tunneling microscopy (STM),\cite{lee2008growth,rong1993electronic} angle-resolved photoemission spectroscopy (ARPES),\cite{coletti2013revealing,ohta2007interlayer,siegel2013charge,bostwick2007quasiparticle,ohta2006controlling} and scanning tunneling spectroscopy (STS) are,\cite{luican2011single,li2010observation,cherkez2015van,lauffer2008atomic,yankowitz2013local,que2015stacking,pierucci2015evidence} respectively, utilized to verify geometric structures, band structures, and density of states (DOS). Up to now, the experimental measurements have identified the interlayer distance of $3.55$ {\AA} ($3.35$ {\AA}) in AA (AB) stacking,\cite{lee2008growth,rong1993electronic} the Dirac-cone structure in monolayer system,\cite{ohta2007interlayer,siegel2013charge,bostwick2007quasiparticle} the parabolic dispersions in bilayer AB stacking,\cite{ohta2007interlayer,ohta2006controlling} the linear and parabolic bands in tri-layer ABA stacking,\cite{coletti2013revealing,ohta2007interlayer} the weakly dispersive, sombrero-shaped and linear bands in tri-layer ABC stacking,\cite{coletti2013revealing} and a lot of stacking- and layer-enriched van Hove singularities in DOS.\cite{luican2011single,li2010observation,cherkez2015van,lauffer2008atomic,yankowitz2013local,que2015stacking,pierucci2015evidence}

How to create the dramatic transitions of essential properties in pristine graphenes is one of the mainstream topics. The critical factors are the obvious changes in geometric structures, such as, the electrostatic manipulation of STM on stacking symmetry\cite{xu2012electronic,feng2013superlubric,liu2012observation,xu2013graphene} and the successful synthesis of graphene ripple using chemical vapor deposition (CVD).\cite{meng2013strain,yan2013strain,bai2014creating} The sliding bilayer graphene and monolayer rippled graphene are suitable for a model study. The former has been studied in the electronic,\cite{son2011electronic,tran2015configuration} magnetic,\cite{huang2014feature} optical,\cite{chiu2016influence,chen2014shift} transport\cite{koshino2013electronic} and phonon\cite{choi2013anomalous} properties. The transformation between two specific stacking configurations will induce the continuous variation in the interlayer atomic interactions of $2p_z$ orbitals and thus leads to the diverse/unique physical phenomena. For example, two isotropic Dirac-cone structures are transformed into two pair of parabolic bands during AA$\to$AB.\cite{son2011electronic,tran2015configuration} Furthermore, the sliding bilayer graphene exhibits three kinds of LLs, the well-behaved, perturbed and undefined LLs, and three magneto-optical selection rules of $\Delta\,n=0$, $\pm\,1$ \& $\pm\,2\,$ (n quantum number for each LL).\cite{huang2014feature,lin2016optical} As for rippled graphenes, the charge distribution of carbon four orbitals becomes highly non-uniform on a curved surface. The $sp^2$ bonding is dramatically changed into the $sp^3$ bondings in the increase of curvature. The theoretical calculations show that the curvature effects, the misorientation of $2p_z$ orbitals and hybridization of ($2s,2p_x,2p_y,2p_z$) orbitals, dominate the essential properties, e.g., the seriously distorted Dirac cone and the semiconductor-semimetal transition.\cite{wehling2008midgap,lin2015feature} The similar effects are clearly revealed in 1D carbon nanotubes, e.g., the largely reduced energy gaps in zigzag nanotubes\cite{hamada1992new,kane1997size,l2002electronic} and the metallic ones with very small radii.\cite{blase1994hybridization,gulseren2002systematic} However, the distinct dimensions and boundary conditions will account for the main differences between rippled graphenes and carbon nanotubes. The position-dependent STS measurements on graphene ripple are made to explore the relation between curvatures and electronic structures,\cite{meng2013strain,yan2013strain,bai2014creating} e.g., the V-shaped spectrum and many special structures in the differential conductance. The curvature-induced charge distribution provides an available chemical environment for potential materials applications.

The chemical modification on graphene surface is emerging as a useful tool to engineer new electronic properties. The adatom adsorption
is a very efficient method in generating the gap-tunable semiconductors and the 2D metals with high free carrier density. All the pristine graphenes exhibit the gapless or narrow-gap behaviors in the absence and presence of external fields,\cite{lu2006influence,l2007electronic,yan2007electric,castro2007biased,lai2008magnetoelectronic,goerbig2011electronic} clearly illustrating the high barriers in gap engineering. The significant gap opening in the electronic properties is an important issue for the future nanoelectronic and optical devices. Graphene oxides (GOs) are a potential candidate, since their energy gaps could be easily modulated by the distribution and concentration of oxygen adatom. GOs have been successfully synthesized by the distinct chemical processes, especially for the four methods developed by Brodie,\cite{brodie1859atomic} Staudenmaier,\cite{staudenmaier1898verfahren} Hummers,\cite{hummers1958preparation} and Tang.\cite{tang2012bottom} Three kinds of chemical bonds, C-O, O-O and C-C bonds, are formed in GOs, as indicated from the theoretical studies.\cite{tran2016chemical,tran2016pi,lian2013big,ito2008semiconducting,mkhoyan2009atomic} Apparently, the atomic interactions of carbon $2p_z$ orbitals are not the critical factors in determining the essential properties. The comprehensive orbital hybridizations, which are responsible for the gap opening and the rich electronic properties, are worthy of a systematic investigation. The first-principles calculations show that the atom- and orbital-related hybridizations could be analyzed by using the atom-dominated energy bands, the spatial charge distributions, and the orbital-projected DOS.\cite{tran2016chemical,tran2016pi} Such hybridizations are very useful in understanding the relation between the forms of chemical bondings and the concentration-dependent energy gaps, and the oxidization-induced drastic changes of band structures. For example, the various chemical bondings can destroy the of Dirac cone and generate the O-, C- and (C,O)-dominated energy bands with many van Hove singularities. Energy gaps in the absence of Dirac cone are confirmed by the ARPES\cite{schulte2013bandgap} and STS\cite{pandey2008scanning} measurements. GOs present the diverse chemical environments suitable for materials engineering.

Graphane, a 2D hydrocarbon compound with the extended covalent C-H bonds, is first proposed by Sofo et al.\cite{sofo2007graphane} In general, hydrocarbon systems could be readily oxidized to generate carbon dioxides and water with a considerable energy release, so that they are good fuel materials. Graphane is predicted to be a wide-gap semiconductor of $E_g\sim\,3.5$ eV. Energy gaps are easily modulated by changing the concentration and distribution of H adatoms;\cite{chandrachud2010systematic,yang2010two,gao2011band,huang2016configuration} therefore, hydrogenated graphenes are very suitable for essential science and technical applications. The chemical adsorption of H adatoms on graphene surface is successfully produced by the hydrogen plasma under electrical discharge,\cite{elias2009control,jaiswal2011controlled} electron irradiation of adsorbates (H$_2$O and NH$_3$),\cite{jones2010formation} and plasma-enhanced CVD.\cite{wang2010toward} The hydrogen concentration could be estimated from the intensity ratio of D and G bands in Raman spectroscopy.\cite{elias2009control,jaiswal2011controlled} Moreover, the hydrogen adatom-induced energy gap is verified from the ARPES\cite{balog2010bandgap,grassi2011scaling} and STS measurements.\cite{guisinger2009exposure,castellanos2012reversible} This clearly illustrates that hydrogen adatoms have strong chemical bondings with carbon atoms, even if the former only possess one 1s orbital. They are located at the top sites, as revealed from the theoretical predictions\cite{sofo2007graphane,chandrachud2010systematic,yang2010two,gao2011band,huang2016configuration} and the experimental identifications.\cite{balog2009atomic} These optimal adsorption positions can create an obvious buckling of a hexagonal structure corresponding to the various bond lengths and angles. The curvature effects will lead to the drastic changes of carbon sp$^2$ bondings and thus the extra orbital hybridizations. In addition to the rich geometric and electronic properties, the theoretical calculations present the non-magnetic and magnetic properties, strongly depending on the hydrogen distribution.\cite{boukhvalov2008hydrogen} Hydrogenated graphenes might exhibit the splitting of spin-up and spin-down energy bands. Such electronic states in transport measurements are expected to produce the spin-polarized currents, indicating potential applications in spintronic devices.\cite{hussain2012polylithiated,hussain2012strain,antipina2012high,gao2011band,nechaev2011high}

Halogen atoms have very strong electron affinities, so that the significant chemical bondings with carbon atoms can create the dramatic changes of essential properties. They are ideal adsorbates on graphene surface for studying the diversified phenomena. Whether five kinds of halogen adatoms possess the similar features or present the important differences deserves a systematic investigation. The fluorinated graphenes have been successfully synthesized by the fluorination methods of direct gas,\cite{cheng2010reversible,mazanek2015tuning} plasma,\cite{wang2014fluorination,sherpa2014local} photochemistry\cite{gong2013photochemical,lee2012selective} and hydrothermal reaction.\cite{samanta2013highly,wang2012synthesis} A plenty of experimental\cite{nair2010fluorographene,jeon2011fluorographene} and theoretical\cite{medeiros2010dft,sahin2012chlorine} studies show that such systems belong to the unusual semiconductors or the $p$-type metals, depending on the fluorination conditions. The large energy gaps due to the high fluorination are clearly evidenced in the experimental measurements of electrical resistances,\cite{nair2010fluorographene,cheng2010reversible} optical transmissions,\cite{nair2010fluorographene,jeon2011fluorographene} and photoluminscence spectra.\cite{jeon2011fluorographene} Furthermore, the $p$-type doping is identified from ARPES measurements on the redshift of the Fermi level.\cite{walter2011highly} The tunable electronic properties clearly indicate the potential applications in various areas, e.g., lithium batteries,\cite{sun2014solvothermally} supercapacitors,\cite{zhao2014fluorinated} biological scaffold materials,\cite{wang2012fluorinated} printing technologies,\cite{nebogatikova2016fluorinated} molecule detectors,\cite{katkov2015backside} and biosensors.\cite{urbanova2016fluorinated} Specifically, the theoretical calculations predict the ferromagnetic behavior,\cite{liu2012electronic} while the measured magnetic moment presents the temperature-dependent paramagnetism.\cite{nair2012spin} This difference might be due to the non-uniform adatom clusters formed in rippled graphene. There also exist the experimental syntheses on the chlorinated,\cite{li2011photochemical,wu2011controlled} brominated,\cite{jankovsky2014towards} and iodinated\cite{yao2012catalyst} graphenes. From few theoretical calculations,\cite{medeiros2010dft,sahin2012chlorine} the (Cl,Br,I,At)-absorbed graphenes are predicted to be in sharp contrast with the fluorinated systems, including the geometric, electronic and magnetic properties. The main differences might lie in the critical orbital hybridizations of the halogen-C bonds. The former have many free holes for any adatom distribution and concentration, as confirmed by the upshift of the graphitic G-band phonon.\cite{wu2011controlled}

Alkali atoms can create active chemical environments to form the critical interactions with other atoms or molecules, since each of them possesses a s-state electron in the outmost orbital. The significant interactions between graphene-related systems and alkali-metal atoms have long been intensively investigated, such as, graphite intercalation compounds,\cite{dresselhaus1981intercalation} and alkali-doped carbon nanotubes\cite{lee1997conductivity,rao1997evidence}, graphenes\cite{chen2008charged,ohta2006controlling}, and fullerenes.\cite{hebard1991potassium,haddon1992electronic} The intercalation of alkali atoms within graphitic layers and their adsorptions on hexagonal surfaces are very efficient n-type chemical dopings. The alkali-induced free electrons are rather suitable for studying the diverse quasiparticle phenomena in the different dimensional systems.\cite{bostwick2007quasiparticle} Potassium,\cite{chen2008charged,ohta2006controlling,bostwick2007quasiparticle,bianchi2010electron} lithium,\cite{yoo2008large} sodium,\cite{papagno2011large} rubidium,\cite{watcharinyanon2011rb} and caesium\cite{watcharinyanon2011rb} adatoms have been successfully utilized to chemically dope pristine graphene systems. The tunable conduction electron density is directly confirmed by ARPES.\cite{ohta2006controlling,bostwick2007quasiparticle,virojanadara2010epitaxial,sugawara2011fabrication,papagno2011large} The alkali-adsorbed graphenes present an obvious red shift in the Dirac-cone structure, as revealed from the experimental and theoretical investigations.\cite{chan2008first,jin2010crossover,praveen2015adsorption,lin2016feature} The critical orbital hybridizations in alkali-carbon interactions could be understood from the detailed analyses on the first-principles calculations. They will determine where the carbon- and alkali-dominated energy bands come to exist.\cite{lin2016feature} For example, the occupied conduction electrons are related to the former or both, depending on the free  carrier density. The creation of 2D free carrier density is a quite important topic, while it is difficult to get an accurate value from energy bands within the first Brillouin zone. The alkali-adsorbed graphene nanoribbons will be introduced to see how many electrons contributed by alkali adtoms become free conduction carriers, mainly owing to a linear relation between the Fermi momentum and the 1D carrier density. A systematic study predicts that each alkali adatom contributes the outermost s-state orbital to built the Fermi sea through the critical interactions. Specially, the free-carrier density is linearly proportional to the adatom concentration, but independent of the edge structure, the ribbon width, and the kind and distribution of alkali atoms. In the first-principles calculations, the Bader analysis is frequently used to judge the charge transfer between two different atoms. Whether this method is suitable in calculating the free-carrier density will be discussed. The very high free-carrier density generated by alkali adatoms in graphene-related systems is very useful in developing the next-generation nanoelectronic devices.\cite{obradovic2006analysis,wang2008room}

Al atoms possess three valence electrons in their outer shells, creating the active chemical environments in the strong interactions with other elements, such as carbon, hydrogen, silicon, and some heavy metals. The theoretical studies show that the most stable adsorption of Al adatoms on graphene surface corresponds to the hollow site.\cite{chan2008first,ao2010high} The Al-doped graphenes are predicted to present the n-type doping,\cite{chan2008first,lin2016feature} as revealed in the alkali-doped systems. These two systems might have the comparable DOS at the Fermi level and the free electron density.\cite{lin2016feature} The Al-substituted systems might serve as potential hydrogen storage materials at room temperature, in which the functional capacity is largely enhanced by the Al atoms due to the strong Al-C bonds.\cite{ao2009doped} Carbon monoxide could be detected by the drastic changes in electrical conductivity before and after molecule adsorption,\cite{ao2008enhancement} mainly originating from the more valence electrons of Al. On the experimental side, Al-related molecules exhibit the significant chemical reactions, being available in the high-performance rechargeable aluminium-ion battery.\cite{lin2015ultrafast} Specifically, layered graphene systems are frequently grown on various substrates, such asthe widely used silicon carbide, Au and Ag metals. Controlling the adatom nano-structures by the critical interactions between adsorbates and substrates is very important in the fabrication of low-dimensional materials with unique electronic properties. Recently, the corrugated substrate and buffer graphene layer have been observed by STM.\cite{chen2015long,chen2015tailoring} A large-scale hexagonal array of Bi atoms is clearly revealed at room temperature, while such adatoms are aggregated in triangular and rectangular nanoclusters as a meta-stable structure during the annealing process. The high structural stability and tunable adatom distributions is very useful in developing the next-generation nanoelectronic devices.

In this work, the essential properties are investigated by the first-principle density functional theory using the Vienna ab initio simulation package.\cite{kresse1996efficient} The exchange-correlation energy due to the electron-electron interactions is calculated from the Perdew-Burke-Ernzerhof functional under the generalized gradient approximation.\cite{perdew1996generalized} The projector-augmented wave pseudopotentials are employed to evaluate the electron-ion interactions.\cite{blochl1994projector} The wave functions are built from the plane waves with a maximum energy cutoff of $500$ eV. The spin configurations are taken into account for the adatom-adsorbed graphene systems. The vacuum distance along the z-axis is set to be $15$ {\AA} for avoiding the interaction between two neighboring cells. Furthermore, the van der Waals force is employed in the calculations using the semiemprical DFT-D2 correction of Grimme\cite{grimme2006semiempirical} to correctly describe the interactions between two graphene layers. The first Brillouin zone is sampled in a Gamma scheme along the two-dimensional periodic direction by $12\times\,12\times\,1$ k points for structure relaxations, and by $100\times\,100\times\,1$ for further calculations on electronic properties. The convergence criterion for one full relaxation is mainly determined by setting the Hellmann-Feynman forces smaller than $0.01$ eV/{\AA} and the total energy difference of $\Delta\,E_t < 10^{-5}$ eV. For the pristine structure-dependent graphenes and the adatom-doped graphenes, the calculated results include the ground state energy, binding energy, interlayer distance, bond length, bond angle, position and height of adatom, adatom nanostructure, band structure, adatom-induced energy gap and free carriers, charge distribution, spin configuration; total and orbital-projected DOSs. They are compared with the ARPES and STS measurements. Specifically, the multi-orbital chemical bondings are identified from the atom-dominated energy bands, the spatial distributions for charge and charge difference, and the orbital-projected DOSs. Such bondings play a critical role in the essential properties; that is, they are responsible for the diverse electronic properties of the graphene-related systems. This point of view could be further extended to any condensed-matter systems. In addition, the tight-binding model is utilized to understand part of the theoretical predictions.

This review article broadly covers the fields related to few-layer graphene systems with the forms of AAA, ABA, ABC, AAB, sliding, and rippling configurations, and the dopings due to various adatoms. Furthermore, it investigates a lot of factors affecting the geometric and electronic properties, e.g., the hexagonal symmetry, mirror/inversion symmetry, rotational symmetry, stacking configuration, curvature, boundary condition, adatom distribution and concentration, charge transfer, orbital hybridization, spin arrangement, and buffer layer/substrate. The optimal geometric structures focus on the interlayer distance, bond length, bond angle, critical curvature, position and height of adatom, and nanostructure of adatom. The main features of energy bands are revealed in the contribution of atom, energy dispersion, existence of Dirac cone, isotropy or anisotropy, symmetry or asymmetry about $E_F$, energy gap, critical point, constant-energy surface, state degeneracy, and spin-induced splitting. The 2D critical points can exhibit the special structures in DOS, including the V-shaped form, shoulders, asymmetric peaks, and symmetric peaks. In addition to the essential properties, materials functionalities and potential applications are also discussed. In chapter 2, the stacking-enriched electronic properties are studied for four kinds of tri-layer graphenes, in which the similarities and differences are explored detailedly. The discussions on the consistence with the ARPES and STS measurements and the further examinations are made. The dramatic transformations of electronic structures could be observed by sliding in bilayer graphene and rippling in monolayer graphene, as clearly indicated in chapter 3. The effects arising from the interlayer atomic interactions and the non-uniform charge distribution on a curved surface deserve a systematic investigation, especially for the semiconductor-semimetal transition in rippled graphens. To see strong $sp^3$ hybridizations and boundary conditions, carbon nanotubes are shown in chapter 3.2. Chapters 4, 5, 6, 7 and 8, respectively, correspond to O-, H-, halogen-, alkali-, and Al- and Bi-doped graphenes. The atom-dominated energy bands, spatial distributions for total charges and charge differences, magnetic configurations, and total and orbital-projected DOSs are analyzed in detail. The complex orbital hybridizations are obtained for various adatoms. Five types of $\pi$ bondings are proposed to account for the concentration-dependent energy gaps in semiconducting graphene oxides. The significant orbital hybridizations in C-H and C-C bonds induce the middle, narrow and zero gaps, being sensitive to the changes of distribution and concentration. The spin-split energy bands are examined for the H-, Cl- and Br-doped graphenes. The strong affinity of halogen adatoms can create free holes, while the F-doped systems might become semiconductors at sufficiently high concentration. The metal-semiconductor transition is associated with the complex multi-orbital hybridizations. On the other hand, the alkali-doped graphenes have high free electron densities, being insensitive to the distinct adatoms. The relation between the $\pi$ bonding and the adatom-carbon hybridization is explored to comprehend why the red-shift Dirac-cone structures could be observed in alkali- and Al-doped graphenes. The alkali-doped graphene nanoribbons are introduced to accurately evaluate the free carrier densities. Specifically, for Bi-doped graphene, the six-layered substrate, corrugated buffer layer, and slightly deformed monolayer graphene are all simulated. The transformation from the most stable hexagonal array to the meta-stable nanocluster is determined by the interactions between buffer layer and monolayer graphene. Finally, chapter 9 contains concluding remarks and future outlook.

\vskip 0.6 truecm
\par\noindent
{\bf 2. Few-layer graphenes}
\vskip 0.3 truecm

Up to now, four kinds of typical stacking configurations, AAA,\cite{lee2008growth} ABA,\cite{yankowitz2014electric,coletti2013revealing,ohta2007interlayer} ABC,\cite{yankowitz2014electric,coletti2013revealing,norimatsu2010selective} and AAB,\cite{rong1993electronic,xu2012electronic,pong2007observation,kazemi2013stacking} have been successfully synthesized by the experimental methods. The AAA stacking is revealed in graphite intercalation compounds, whereas the ABA and ABC stackings exist in natural graphite.\cite{dresselhaus1981intercalation} The AA-stacked graphene is grown on diamond (111) surface using a high-density dc plasma in hydrogen-methane mixtures.\cite{lee2008growth} Both ABA and ABC stacking are frequently observed by mechanical exfoliation of kish graphite,\cite{hass2008growth,coraux2008structural} CVD,\cite{ismach2010direct,park2010growth,reina2008large,chae2009} chemical and electrochemical reduction of graphite oxide,\cite{zhao2011few,stankovich2007synthesis,rao2016study} arc discharge,\cite{wu2009synthesis,qin2016growth,wu2010efficient} an electrochemical method that uses oxalic acid (C$_2$H$_2$O$_4$) as the electrolyte,\cite{mahanandia2014electrochemical} and the flame method.\cite{li2011flame} Specifically, the AAB stacking could be produced by rotating the top graphite layer against the sublayers,\cite{rong1993electronic} the electrostatic-manipulation STM,\cite{xu2012electronic} mechanical cleavage,\cite{pong2007observation} liquid phase exfoliation of natural graphite,\cite{kazemi2013stacking} and chemical growth on Ru (0001) surface.\cite{que2015stacking} The STM technology is utilized to tune the stacking configuration continuously.\cite{xu2012electronic}

Few-layer graphenes can exhibit feature-rich essential properties, especially for the high-symmetry stacking systems. The tri-layer AAA-, ABA-, ABC- and AAB-stacked graphenes are suitable for a systematic study. Two neighboring layers in the  AA stacking, as shown in Fig. 1, have the same (x,y) projections, while those in the AB stacking shift to each other by the C-C bond length along the armchair direction. Each graphene layer is a rather stable plane because of the quite strong $\sigma$ bondings formed by (2s,2p$_x$,2p$_y$) orbitals. The interplane attractive forces come from the weak Van der Waals interaction associated with the 2p$_z$ orbitals. The first-principles calculations show that the ground state energies are, respectively, $-$55.832866 eV, $-$55.857749 eV, $-$55.862386 eV, and $-$55.864039 eV for the AAA, ABA, ABC and AAB stackings. Only slight energy differences among them indicate the near-future promise in the experimental synthesis. Also, the inter-layer distance (3.52-3.53 \AA) in the AA stacking is larger than that (3.26-3.27 \AA) in the AB stacking.

The electronic properties of planar graphenes are mainly determined by the 2$p_z$-orbital bondings, the honeycomb symmetry, the stacking configurations and the number of layers. The tri-layer AAA, ABA, ABC and AAB stackings are predicted to have the unusual energy bands, respectively: (1) the linearly intersecting bands (the almost isotropic Dirac-cone structures), (2) the parabolic bands and the linear bands, (3) the weakly dispersive bands, the sombrero-shaped bands, and the linear bands; (4) the oscillatory bands, the sombrero-shaped bands, and the parabolic bands. Such energy dispersions can create the special van Hove singularities in density of states (DOS). The former three systems are semimetals with valence and conduction band overlaps, while the last one is a narrow-gap semiconductor of $E_g\sim 10$ meV. The diverse electronic structures can be verified by ARPES. Furthermore, the differences in the DOSs are useful in distinguishing the stacking configurations by the STS measurements.

{\bf 2.1 AAA stacking }
\vskip 0.3 truecm

The interlayer atomic interactions in few-layer graphene can induce free electrons and holes with the same carrier density. Monolayer graphene has one pair of linear energy bands intersecting at the Fermi level ($E_F$) because of the honeycomb symmetry. The isotropic Dirac cone only creates a vanishing DOS at $E_F$; therefore, the single-layer graphene is a zero-gap semiconductor. The low-lying Dirac-cone structures are preserved in the AA-stacked few-layer graphenes. The tri-layer AAA stacking exhibits three pairs of vertical Dirac-cone structures, as indicated in Fig. 2(a). Three Dirac points are just located at the K point (3D energy bands in Fig. 3(a)), but their energies are $E^{c,v}=0.36$ eV, 0 and $-0.36$ eV, respectively (inset in Fig. 2(a)). The separated energy spacings are mainly determined by the vertical interlayer hopping integral, according to the analytic calculations of the tight-binding model \cite{l2007electronic}. Specifically, the second Dirac-cone structure is independent of the 2$p_z$ orbitals on the middle plane. Apparently, free electrons and holes are, respectively, distributed in the conduction band of the first cone and the valence band of the third cone. The free carrier density is $\sigma_e\,=\sigma_h\,=1.56$x$10^{13}$ e/cm$^2$, as evaluated  from $k_F^2/\pi$ ($k_F$ the Fermi momentum). It is deduced that a N-layer AA stacking has free electrons or holes in the N/2 ((N-1)/2) isotropic Dirac cones for an even-layer (odd-layer) system, i.e., a few- or multi-layer AA-stacked graphene is a semimetal with the sufficiently high free carrier density.\cite{l2007electronic,lobato2011multiple} Furthermore, the Dirac-cone structures remain isotropic in the AA-stacked graphite even if the Dirac points strongly depend on the wavevector along $\hat k_z$.\cite{chiu2010absorption}

The main features of energy bands, dispersion relation, isotropy, symmetry about $E_F$, and orbital dominance, are sensitive to the changes in state energies. The low-lying linear bands dominated by the 2$p_z$ orbitals change into the parabolic bands with the increasing state energies, and so does the variation from the isotropic to the anisotropic behavior. The parabolic dispersions have the band-edge states at the M point, in which they belong to the saddle points in the energy-wave-vector space. This clearly indicates that many $\pi$- and $\pi^\ast$-electronic states are accumulated near the M point and will make important contributions to the essential physical properties, e.g., the $\pi$-electronic optical\cite{chiu2010absorption,chiu2011excitation} and Coulomb excitations.\cite{shyu2000loss,ho2006coulomb} As for the saddle-point energies, they are distinct for the valence and conduction bands, mainly owing to the interlayer hopping integrals \cite{l2007electronic}. That is to say, the interlayer atomic interactions might result in the asymmetry of valence and conduction bands about $E_F$. On the other hand, the 2$p_x$ and 2$p_y$ orbitals can generate deeper or higher energy bands, e.g., two valence and two conduction bands ($\sigma$ and $\sigma^*$ bands) initiated from the $\Gamma$ point at $E^v=-3$ eV and $E^c\approx $3.4-3.7 eV, respectively.

The special structures in DOS directly reflect the diverse energy dispersions. DOS of the tri-layer AAA stacking, shown in Fig. 4(a), exhibits the linear energy dependence near $E_F$ (inset), the logarithmically divergent peaks at middle energy, and the shoulder structures at deeper or higher energy. These structures, respectively, originate from the isotropic linear bands, the parabolic bands near the saddle points, and the initial band-edge states of the parabolic bands. A finite DOS at $E_F$ clearly illustrates the semi-metallic behavior; furthermore, it is the largest one compared with those of other stacking configurations. The large energy difference between valence and conduction peaks obviously reveals the band asymmetry, when state energies are measured from $E_F$. Moreover, the peak and shoulder structures, respectively, correspond to the 2$p_z$- and (2$p_x$,2$p_y$)-dominated energy bands, as indicated from the orbital-projected DOSs. The highly accumulated $\pi\,$- and $\pi^\ast\,$-electronic states, with three symmetric peaks, are, respectively, presented in $-$2.8 eV $\le\,E\le\,-$1.9 eV and 1.2 eV$\le\,E\le\,$2.2 eV. It is also noticed that the prominent peaks at the middle energy are the common characteristics of the graphene-related systems with the s$p^2$ bondings, such as, graphites,\cite{shyu2000loss,ooi2006density} graphene nanoribbons,\cite{lin2000opticalJ} carbon nanotubes,\cite{lin2000optical,lin1994plasmons,lin1996collective} and fullerenes.\cite{gensterblum1991high,bolognesi2012collective} Such peaks can induce the pronounced absorption peaks \cite{lin2000opticalJ,lin2000optical} and the strong $\pi$ plasmons.\cite{shyu2000loss,lin1994plasmons,lin1996collective,gensterblum1991high,bolognesi2012collective}

{\bf 2.2 ABA stacking }
\vskip 0.3 truecm

The low-energy essential properties are drastically changed by the stacking configuration. The tri-layer ABA stacking has one pair of linear bands and two pairs of parabolic bands, i.e., it exhibits the monolayer- and bilayer-like electronic structure. The former, with the Dirac-cone structure (Fig. 3(b)), only comes from the 2$p_z$ orbitals on the first and third layers, and the atomic interactions of two next-neighboring layers \cite{lin2015magneto} are responsible for a very small spacing in separated Dirac points ($\sim$13 meV in the inset of Fig. 2(b); Fig. 3(b)). The isotropic cone structures are seriously distorted in the latter. The first and second pairs of parabolic bands are, respectively, initiated at ($E^c=0.005$ eV, $E^v=-0.007$ eV) and ($E^c=0.59$ eV, $E^v=-0.54$ eV), in which the former has an observable spacing of $\sim\,12$ meV near the Fermi level (Fig. 3(b)). These band-edge state energies are closely related to the neighboring-layer atomic interactions. The saddle-point energies are affected by the distinct interlayer atomic interactions (a comparison of Figs. 2(b) and 2(a)), while such points remain at the M point. Moreover, the (2$p_x$,2$p_y$)-dominated energy bands are insensitive to the change of the stacking configurations. As for AB-stacked multi-layer graphenes, all energy bands belong to bilayer-like parabolic dispersions \cite{lin2015magneto,lu2006influence} except the only linear Dirac-cone structure in an odd-layer system. In addition, this parity effect has been confirmed by measuring the electric double-layer capacitance associated with the layer-number-dependent DOS.\cite{goto2013parity}

The ABA stacking is rather different from the AAA stacking in the low-energy DOS. Apparently, the former has an asymmetric DOS about E=0 (Fig. 4(b)), owing to the interlayer atomic interactions. DOS is very small at the Fermi level, indicating the low free carrier density. A merged shoulder structure, which arises from the band-edge states of the first pair of valence and conduction bands, is revealed below the Fermi level (the black arrow). Two obvious shoulders at E=0.6 eV and $-$0.54 eV are induced by the second pair of parabolic bands. Moreover, there exist  three  very strong $\pi$ and $ \pi^\ast$ symmetric peaks, in which their energies depend on the intralayer and  interlayer atomic interactions.\cite{lin2015magneto,lu2006influence}

{\bf 2.3 ABC stacking }
\vskip 0.3 truecm

The stacking configuration can diversify the energy dispersions of electronic structures. The tri-layer ABC stacking exhibits three pairs of unusual low-lying energy bands, as indicated in the inset of Fig. 2(c). A pair of partially flat bands, which is initiated from the Fermi level, is mostly contributed by 2$p_z$ orbitals situated at the outmost layers. The surface states could survive in the ABC-staked graphenes,\cite{linyp2015magneto} while they are absent in the other stacking systems or 3D rhombohedral graphite.\cite{ho2016evolution} They will result in the unique quantization phenomena, e.g., the special LL energy spectra\cite{lin2014energy,lin2014stacking,koshino2011landau} and Hall effects.\cite{koshino2011landau} The second pair has the sombrero-shaped energy dispersions, in which two constant-energy valleys occur at $E^v=-0.4$ eV and $E^c=0.38$ eV (Fig. 3(c)). The finite-radius loops are effectively treated as the 1D parabolic bands and could induce the new structures in DOS (inset in Fig. 4(c)). Furthermore, the non-monotonous energy dispersions account for the abnormal LL energy spectra and the frequent LL anti-crossings \cite{lin2014energy,lin2014stacking}. The third pair of linear bands is the momolayer-like Dirac-cone structures (Fig. 3(c)), and this pair is revealed in any ABC-stacked systems. With the increase of layer number, there exist parabolic bands and more sombrero-shaped bands.

The low-energy DOS of the ABC-stacked graphene presents two kinds of new structures, as shown in the inset of Fig. 4(c). A delta-function-like symmetric peak occurs at the Fermi level as a result of the partially flat bands of surface states. Two asymmetric peaks, which are similar to those induced by the 1D parabolic bands, are revealed at E=$-$0.4 eV and 0.38 eV. These structures further illustrate that the tri-layer ABC stacking does not exhibit the bilayer-like behavior associated with two pairs of parabolic bands. This system cannot be regarded as the superposition of monolayer and bilayer graphenes; that is, it possesses the unique intrinsic properties among tri-layer systems.

{\bf 2.4 AAB stacking }
\vskip 0.3 truecm

The $z=0$ mirror  symmetry is absent in the tri-layer AAB stacking. The interlayer atomic interactions are more complicated in the lower-symmetry system \cite{do2015configuration}, and so do the essential electronic properties. There is a small direct gap of $E_g\sim\,8$ meV near the K point, as indicated in the inset of Fig. 2(d). The tri-layer AAB stacking is a narrow-gap semiconductor, being in sharp contrast to the other semi-metallic stacking systems. The first pair is the oscillatory valence and conduction bands nearest to $E_F$. Three constant-energy loops could survive simultaneously (Fig. 3(d)), leading to the novel and abnormal magnetic quantization, such as, the splitting LL energy spectra and the multi-anti-crossing behavior \cite{do2015configuration}. The second and third pairs, respectively, belong to the sombrero-shaped and parabolic bands, with the band-edge states at ($E^v=-0.28$ eV, $E^c=0.28$ eV) and ($E^v=-0.46$ eV, $E^c=0.46$ eV). The similar energy bands have also been obtained from the tight-binding model, in which nine distinct interlayer interactions of $2p_z$ orbitals are used in the calculations.\cite{do2015configuration,do2015rich}

The tri-layer AAB stacking exhibits more special structures in the low-energy DOS, compared with the other stackings. A dip structure at the Fermi level is clearly shown in the inset of Fig. 4(d). It is a combination of energy gap and broadening factor ($\sim\,$10 meV in the calculations). A pair of asymmetric peaks at $E=-0.05$ eV and 0.05 eV, respectively, comes from the constant-energy loops of the oscillatory valence and conduction bands, in which their energies are associated with the oscillation widths measured from the Fermi level. Another two similar peaks at larger energies reflect the band-edge states of the sombrero-shaped bands. Moreover, two shoulder structures are induced by the deeper/higher parabolic bands. According to the feature-rich DOS, the AAB stacking represents a unique tri-layer system, as observed in the ABC stacking.

There are certain important differences among the tri-layer AAA, ABA, ABC and AAB stackings. The former three systems are semi-metals, while the last one is a narrow-gap semiconductor. The AAA stacking, with the rigid-shift Dirac-cone structures, has the largest band overlap or the highest free carrier density. The ABA stacking exhibits the monolayer- and bilayer-like band structures. However, the latter two systems present the unique band structures, e.g., the surface-related flat bands, the sombrero-shaped bands and the oscillatory bands. They both cannot be regarded as the superposition of monolayer and bilayer systems. The great differences are also revealed in the energy, form, intensity and number of special structures in DOS. The stacking-enriched electronic structures and DOSs are directly reflected in the other essential physical properties. For example, the magnetic quantization is diversified by the stacking configuration,\cite{lin2015magneto,do2015configuration} and so do the main features of optical spectra.\cite{lin2015magneto,do2015rich}

ARPES is the most powerful experimental technique to identify the wave-vector-dependent electronic structures. The experimental measurements on graphene-related systems could be used to investigate the feature-rich band structures under the different dimensions,\cite{ruffieux2012electronic,sugawara2006fermi,gruneis2008electron,siegel2013charge,bostwick2007quasiparticle,ohta2007interlayer} layer numbers,\cite{ohta2007interlayer,coletti2013revealing,kim2013coexisting,sutter2009electronic} stacking configurations,\cite{ohta2006controlling} substrates,\cite{siegel2013charge,ohta2007interlayer,coletti2013revealing} and adatom/molecule adsorptions.\cite{zhou2008metal,gruneis2008tunable,papagno2011large} The 1D parabolic energy bands in graphene nanoribbons have been directly observed.\cite{ruffieux2012electronic,sugawara2006fermi} The AB-stacked graphite presents the 3D $\pi$ bands,\cite{gruneis2008electron} in which they have the bilayer- and monolayer-like energy dispersions at $k_z=0$ and zone boundary of $k_z=1$ (K and H points in the 3D first Brillouin zone). As to few-layer graphenes, the verified electronic structures include the Dirac-cone structure in single-layer system,\cite{siegel2013charge,bostwick2007quasiparticle,ohta2007interlayer} two pairs of parabolic bands in bilayer AB stacking,\cite{ohta2007interlayer,ohta2006controlling} the coexisting linear and parabolic dispersions in symmetry-broken bilayer system,\cite{kim2013coexisting} the linear and parabolic bands in tri-layer ABA stacking,\cite{ohta2007interlayer,coletti2013revealing} and the partially flat, sombrero-shaped and linear bands in tri-layer ABC stacking.\cite{coletti2013revealing} Specially, the substrate effects might induce the dramatic changes in electronic structures, e.g., a sufficiently large energy spacing between $\pi$ and $\pi^*$ bands in bilayer AB stacking,\cite{ohta2007interlayer,coletti2013revealing} and the oscillatory bands in few-layer ABC stackings.\cite{ohta2007interlayer} Up to now, the ARPES measurements on the unusual energy bands in tri-layer AAA and AAB stackings are absent. The further examinations are useful in providing the tight-binding model parameters or the complicated interlayer atomic interactions. In addition, the deeper-energy $\sigma$ bands are identified to be initiated at $\sim\,-$3.5 eV for monolayer and bilayer graphene grown on substrate,\cite{mazzola2013kinks,jung2016sublattice} being roughly consistent with $-$3 eV in the theoretical predictions (Figs. 2(a)-2(d)).

The STS measurements, in which the tunneling conductance (dI/dV) is proportional to DOS, can serve as efficient methods to examine the special structures in DOS. They have been successfully utilized to verify the diverse electronic properties in graphene nanoribbons,\cite{huang2012spatially,sode2015electronic,chen2013tuning} carbon nanotubes,\cite{wilder1998electronic,odom1998atomic} graphite,\cite{klusek1999investigations,li2009scanning} few-layer graphenes,\cite{luican2011single,li2010observation,cherkez2015van,lauffer2008atomic,yankowitz2013local,que2015stacking,pierucci2015evidence} and adatom-adsorbed graphenes.\cite{chen2015long,chen2015tailoring,gyamfi2011fe} The geometry-dependent energy gaps and the asymmetric peaks of 1D parabolic bands have been verified for graphene nanoribbons\cite{huang2012spatially,sode2015electronic,chen2013tuning} and carbon nanotubes.\cite{wilder1998electronic,odom1998atomic} The measured DOS of AB-stacked graphite exhibits the splitting $\pi$ and $\pi^\ast$ peaks,\cite{klusek1999investigations} and it is finite near the Fermi level because of the semi-metallic behavior.\cite{li2009scanning} The experimental measurements on few-layer graphene show that the low-lying DOS characteristics are revealed as a linear E-dependence  vanishing at the Dirac point in monolayer system,\cite{li2009scanning} the asymmetry-created peak structures in bilayer graphene,\cite{luican2011single,li2010observation,cherkez2015van} an electric-field-induced gap in bilayer AB stacking and tri-layer ABC stacking,\cite{lauffer2008atomic,yankowitz2013local} a pronounced peak at $E=0$ characteristic of partial flat bands in tri-layer and penta-layer ABC stacking,\cite{que2015stacking,pierucci2015evidence} and a dip structure at $E_F=0$ accompanied with a pair of asymmetric peaks in tri-layer AAB stacking,\cite{que2015stacking} On the other hand, two critical features, a V-shaped spectrum with a finite value near $E_F$ in tri-layer AAA stacking and a merged shoulder structure below $E_F$ in tri-layer ABA stacking, require further experimental verifications.

\vskip 0.6 truecm
\par\noindent
{\bf 3. Structure-enriched graphenes}
\vskip 0.3 truecm

The various stacking configurations, with high and low symmetries, can enrich and diversify the essential properties. The sliding bilayer graphene, which presents the transformation between the highly symmetric stackings, is an ideal system for studying the electronic topological transitions. Stacking boundaries including relative shifts between neighboring graphene layers have been grown by the CVD method.\cite{alden2013strain} The sliding of graphene flakes on graphene substrate is initiated by the STM tip to overcome the Van der Waals interactions.\cite{feng2013superlubric} Micrometer-size graphite flakes will slide spontaneously after stirred by a STM tip.\cite{liu2012observation} Specially, the electrostatic-manipulation STM is performed on a highly oriented pyrolytic graphite surface and can induce a continuous and large-scaled movement of the top layer. \cite{xu2012electronic} The sliding bilayer graphene is expected to be achieved by this method.\cite{xu2013graphene}

The first-principle calculations are used to comprehend the electronic topological transitions in the sliding bilayer graphene. The ground state energy, interlayer distance, band structure and DOS are very sensitive to the change in stacking configuration, mainly owing to the diverse interlayer atomic interactions of $2p_z$ orbitals. The dramatic transitions occur between Dirac-cone structures and parabolic bands, accompanied with the creation of arc-shaped stateless regions, variation of free carrier density, distorted energy dispersions, extra low-energy critical points, and splitting of middle-energy states. A lot of van Hove singularities derived from the saddle points are revealed in DOS The stacking-enriched energy bands and DOS are useful in the identification of the sub-angstrom misalignment. The theoretical predictions agree with the tight-binding model calculations,\cite{huang2014feature,chen2014shift} but require further experimental examinations.

The drastic changes in electronic properties could also be driven by the curved\cite{guinea2008midgap,guinea2008gauge,guinea2009gauge,isacsson2008electronic,wehling2008midgap,boukhvalov2009enhancement,guinea2010energy,atanasov2010tuning,vozmediano2010gauge,iannuzzi2011comparative,santos2012magnetism,zhang2012electron,mohammed2012tunnel,seyed2013computational,imam2014first,zwierzycki2014transport,koskinen2014graphene,monteverde2015under,pudlak2015cooperative,lin2015feature,wu2016tuning} structures or the uniaxial\cite{pereira2009tight,choi2010effects,wong2012strain} deformations. The strong sp$^2$-bondings in the low-dimensional carbon-related systems provide the flexible surfaces\cite{l2002electronic,yoosefian2015pd,josa2015tailoring} suitable for the creation of various deformations. The curved systems, with distinct structures, are successfully synthesized in experimental laboratories, such as, rippled graphenes,\cite{meng2013strain,yan2013strain,bai2014creating,capasso2014graphene,deng2016wrinkled,lee2016multilayer,rakic2016large} graphene bubbles,\cite{de2008periodically,levy2010strain,ramirez2016interference} carbon nanotubes,\cite{iijima1991helical} carbon nanoscrolls,\cite{savoskin2007carbon,xie2009controlled,berman2015macroscale} and folded/curved graphene nanoribbons.\cite{kosynkin2009longitudinal,liu2012folded} Such systems present the diverse chemical bondings on the curved surface, thus leading to the unusual essential properties.\cite{guinea2008gauge,choi2010effects,monteverde2015under,pudlak2015cooperative} Specifically, the 2D graphene ripples might induce unique phenomena in surface science; material physics and chemistry. The corrugated environments are promising for potential applications in electronic devices\cite{yan2012high}, toxic sensors,\cite{chi2009adsorption,ao2008enhancement} and energy storages.\cite{tozzini2011reversible,ning2012high,tozzini2013prospects}

The first-principles calculations are focused on how the the various corrugated structures can create the feature-rich electronic properties in rippled graphenes.\cite{wehling2008midgap,imam2014first,lin2015feature} The misorientation of $2p_z$ orbitals and hybridization of (2s,2$p_x$,2$p_y$,2$p_z$) are induced by a non-uniform charge distribution on a curved surface, and there exist similarities and differences between 2D graphene ripples and 1D carbon nanotubes.\cite{kane1997size,l2002electronic,mintmire1992fullerene,hamada1992new,saito1992electronic,saito1992electronicprb} The corrugated direction, curvature and period play a critical role in charge distributions, bond length, band structure, and DOS. Armchair ripples are zero-gap semiconductors with a complete Dirac-cone structure, while zigzag ripples present the anisotropic energy spectra, the semi-metallic behaviors due to the destroyed Dirac cone, and the newly created van Hove singularities. Both systems have the splitting middle-energy electronic states under the reduced rotation symmetry. There are some curvature-induced prominent peaks in the low-energy DOS of zigzag ripples. On the experimental side, graphene ripples are successfully grown on Rh(111)\cite{meng2013strain,yan2013strain} or Cu\cite{bai2014creating} substrates via the CVD process. The position-dependent STS measurements\cite{meng2013strain,yan2013strain,bai2014creating} present two types of energy spectra in the dI/dV-V diagram, a V shape and some low-lying peaks with a specific dependence on the effective quantum number. These are consistent with the theoretical predictions dominated by the corrugated structures and the unusual energy dispersions.\cite{lin2015feature} Also noticed that the effective-field model\cite{guinea2008midgap,guinea2009gauge,bai2014creating,pudlak2015cooperative} is proposed to investigate the peculiar properties of the highly corrugated graphenes. The effective magnetic field associated with the extremely rigid jump structures will induce the quasi-LLs.\cite{levy2010strain}

{\bf 3.1 Sliding bilayer graphene }
\vskip 0.3 truecm

A relative shift between two graphene layers occurs along the armchair direction in the range of $0\le\delta_a\le 12/8$ (in units of the C-C bond length b = 1.424 $\mbox\AA$) and then along the zigzag direction for $0\le\delta_z\le 4/8$ (in units of $\sqrt 3\,b$), leading to the transformation of high-symmetry stacking configurations: AA ($\delta_a$=0) $\rightarrow$ AB ($\delta_a$=1) $\rightarrow$ $\mbox{AA}^\prime$ ($\delta_a$=12/8; $\delta_z$=0) $\rightarrow$ AA ($\delta_z$=4/8), as shown in Fig. 5(a). All C atoms in the AA and $\mbox{AA}^\prime$ stackings possess the same chemical environment, while the latter present the different x-y projections. The AA and AB stackings have the largest and shortest interlayer distances (3.52 $\mbox\AA$ and 3.26 $\mbox\AA$ in Fig. 5(b)), respectively, corresponding to the highest and lowest ground state energies.\cite{bhattacharyya2016lifshitz} The shorter the interlayer distance is, the stronger the attractive van der Waals interactions among the 2$p_z$ orbitals are. The predicted distances are consistent with the measured results (3.55 $\mbox\AA$ and 3.35 $\mbox\AA$).\cite{lee2008growth,rong1993electronic}

The structure transformation leads to the dramatic changes in band structure. The inversion symmetry in bilayer graphene is unbroken during the variation of shift; that is, the low-energy electronic structure remains doubly degenerate for the K and K$^\prime$ valleys. However, the six-fold rotational symmetry is changed into the two-fold one except for AA and AB stackings. The AA stacking, as shown in Fig. 6(a), exhibits two pairs of vertical Dirac cones centered at the K point. With an increase in $\delta_a$ (1/8 in Fig. 6(b)), the electronic states in the valence band of the upper cone strongly hybridize with those of the conduction cone of the lower cone, so that an arc-shaped stateless region along $\hat k_x$ is created near $E_F$. This clearly indicates that the carrier density of free electrons and holes is reduced. The low-energy states around the K point are transferred to its neighbor regions with the saddle points, according to the conservation of electronic states. There is a pair of saddle points above or below $E_F$, in which their splitting energies depend on the relative shift (Fig. 9(a)). The arc-shaped region grows rapidly, and the touching Dirac points start to separate at $\delta_a\,=$4/8 (Fig. 6(c)). And then in the range of $5/8\le\,\delta_a\,<1$, the splitting saddle points merge together, and the distorted energy dispersions gradually become parabolic ones. Specially, when the destroyed Dirac-cone structure is magnetically quantized, it will induce the undefined LLs without a specific quantum mode.\cite{huang2014feature}

The AB stacking of $\delta_a\,=1$ is characterized by two pair of parabolic bands, with a weak valence and conduction band overlap (Fig. 6(d)). In the further increase of $\delta_a$, the first pair near $E_F$ is seriously distorted along $\hat k_y$ and -$\hat k_y$ simultaneously, as indicated in Fig. 6(e) for $\delta_a\,=$11/8. Specifically, two Dirac points in the growing arc-shape stateless are revealed at distinct energies, and two neighboring conduction (valence) bands strongly hybridize with each other. Finally, two pairs of the isotropic cone structures are formed in the $\mbox{AA}^\prime$ stacking (Fig. 6(f)), in which two Dirac points are situated at different wave vectors. The two cone axes are tilted for the conduction and valence bands, being in sharp contrast to the unique non-tilted axis in the AA stacking. The magnetic quantization of the tilted cone structures will create the well-behaved LLs with the different localization centers and thus the new magneto-optical rules.\cite{huang2014feature}

The non-vertical Dirac-cone structure is transformed into the vertical one for the configuration variation from $\mbox{AA}^\prime$ to AA. When the relative shift occurs along the zigzag direction, two neighboring conduction (valence) bands form strong hybrids, as shown in Fig. 6(g) at $\delta_z=$1/8. In the further shift, the valence band of the upper cone hybridizes with the conduction band of the lower cone and the Dirac-cone structures are reformed at $\delta_z=$3/8 (Fig. 6(h)). The created stateless region along $\hat k_y$, with two Dirac points at distinct energies, first expands, then declines, and finally disappears in the AA stacking ($\delta_z=$4/8). Outside this region, there exist the highly distorted energy bands with the saddle points.

For the sliding bilayer graphene, the 6-fold rotational symmetry becomes the 2-fold one in various low-symmetry systems. The 2D band structures, as shown in the insets of Figs. 7(a)-7(h) along the high-symmetry points, exhibit the anisotropic and non-monotonous energy dispersions except for AA and AB stackings (Figs. 7(a) and 7(d)). In general, the K point is not a high-symmetry critical point anymore. These clearly illustrate that the highly distorted energy dispersions are too complicated to be characterized by the perturbation expansion, i.e., the effective low-energy Hamiltonian model might be not suitable for studying the essential physical properties. The generalized tight-binding model has been developed to explore the electronic properties in external fields (discussed later). As to the middle-energy electronic states, the saddle points are located at the M and M$^\prime$ points. These two points are split by the reduced rotational symmetry, in which the energies of van Hove singularities depend on the stacking configuration. The splitting saddle points can induce more absorption peaks. In addition, the deeper/higher $\sigma$ and $\sigma^*$ bands ($|E^{c,v}|>$ 3 eV) are hardly affected by the relative shift; that is, the changes in the interlayer atomic interactions only affect the electronic structures of 2$p_z$ orbitals. The main features of the stacking-enriched band structures, the anisotropic and distorted energy dispersions, the created arc-shaped stateless region, the tilted Dirac-cone structures and the splitting middle-energy $\pi$ and $\pi^\ast$ bands, could be further examined by the APRES measurements.\cite{ruffieux2012electronic,sugawara2006fermi,gruneis2008electron,siegel2013charge,bostwick2007quasiparticle,ohta2007interlayer,coletti2013revealing,kim2013coexisting,sutter2009electronic,ohta2006controlling,zhou2008metal,gruneis2008tunable,papagno2011large}

The main features of DOS, the height, form, energy and number of special structures, strongly depend on the change in stacking symmetry, as indicated in Figs. 8(a)-8(h). Among all stacking configurations, the AA stacking has the highest DOS at $E_F$=0 (inset in Fig. 8(a)), mainly owing to the largest overlap of conduction and valence bands. This system is expected to have the highest electrical conductivity. However, the $\delta_a=$4/8 stacking presents very low DOS in the almost vanishing band overlap (Fig. 8(c)). The forms of special structures are determined by the energy dispersions near the band-edge states. The plateau and cusp in the AA and AA$^\prime$ stackings come from the overlap of two pairs of isotropic Dirac-cone structures (Figs. 8(a) and 8(e)). The parabolic bands near the K point in the AB stacking only present the shoulder structures (Fig. 8(d)). Specifically, the symmetric peak structures in the other stackings are induced by the saddle points around the arc-shaped stateless regions. As for the rather strong $\pi$ and $\pi^\ast$ peaks, they are greatly enriched by the splitting M and M$^\prime$ points. The low- and middle-energy DOS peaks are predicted to exhibit the rich absorption peaks.\cite{chen2014shift}

The number and energy of van Hove singularities deserves a closer examination. The diverse and non-monotonous dependence on the relative shift is clearly indicated in Figs. 9(a) and 9(b). Two low-energy peaks, which correspond to the created saddle points above and below $E_F$ (Fig. 6), are revealed in most of stacking systems (Fig. 9(a)). The energy spacing in each pair of saddle points is observable only when the drastic changes occur between the vertical Dirac-cone structure and the parabolic bands. There are three or four distinguishable peaks near $\delta_a=$4/8. On the other hand, it is relatively easy to observe the splitting $\pi$ and $\pi^\ast$ peaks under the reduced rotational symmetry (Fig. 9(b)). The middle-energy DOSs have 6-8 strong peaks except that the bilayer systems close to AA and AB stackings exhibit four single peaks. The STS measurements \cite{huang2012spatially,sode2015electronic,chen2013tuning,wilder1998electronic,odom1998atomic,klusek1999investigations,li2009scanning,li2010observation,cherkez2015van,lauffer2008atomic,yankowitz2013local,que2015stacking,pierucci2015evidence,chen2015long,chen2015tailoring,gyamfi2011fe} on the shift-dependent van Hove singularities are very useful in the identification of sub-angstrom misalignment between two graphene layers.

The transformation of band structure is induced by the relative shift between two high-symmetry stacking configurations,\cite{bhattacharyya2016lifshitz,son2011electronic,lee2015extreme} and so do the other essential physical properties. The first-principles calculations show that the novel optical phonon splittings come from the difference in their frequency renormalizations due to the interlayer couplings.\cite{choi2013anomalous} The dependence of phonon frequency on stacking configuration has a strong effect on the polarized Raman scattering intensity. On the other hand, the generalized tight-binding model is developed to understand the essential electronic properties under external fields, e.g., the magneto-electronic and optical properties.\cite{huang2014feature} The stacking symmetry, the complicated interlayer atomic interactions, and the effects due to a perpendicular magnetic field ($B_0\hat z$) are taken into consideration simultaneously. Three kinds of LLs, the well-defined, perturbed and undefined LLs, are predicted to exist in the sliding graphene systems. Among them, the third kind is never observed in experimental measurements. The $B_z$-related LL energy spectra exhibit the monotonous and abnormal dependences, and the crossing and anti-crossing behaviors. Moreover, they diversify the magneto-optical selection rule, including the well-known rules of ${\Delta\,n=\pm\,1}$ in AA and AB stackings, and the new rules of ${\Delta\,n=0}$ \& ${\pm\,2}$ in AA$^\prime$ stacking. Specifically, the undefined LLs can generate many absorption peaks in the absence of selection rule. In short, the experimental measurements on band structures, van Hove singularities, phonon spectra, magneto-electronic properties and optical spectra can identify various misalignments in sliding bilayer graphene.

{\bf 3.2 Graphene ripples}
\vskip 0.3 truecm

The geometric structure of graphene ripple will diversify the chemical bondings on a curved surface and determine the essential properties. It is characterized by the corrugated direction, period and height. Two types of typical structures are armchair and zigzag ripples, as shown in Figs. 10(a) and 10(b), respectively. Their periodical lengths (l's) are proportional to the number ($N_\lambda$) of zigzag and armchair lines along the y-axis. Armchair and zigzag ripples, respectively, correspond to armchair and zigzag nanotubes, while the main differences of geometric structures lie in the dimensions and periodical boundary conditions (open and closed ones). The ratio of squared height to width, $C_r=h^2/lb$ ($b$ the C-C length; Fig. 10(c))) is taken to describe the curvature of ripple. The rippled structure is modeled as the sinusoidal form in the initial calculations. The C-C bonds would be broken at the crests and troughs for a very large $C_r$. The critical curvature grows with the increase of period, mainly owing to the reduced mechanical bending force.\cite{chang2014geometric} For example, the highest $C_r$ is, respectively, $0.68$ and $1.56$ for $N_\lambda=6$ and $12$ armchair ripples ($0.27$ and $1.33$ for $N_\lambda=5$ and $11$ zigzag ones). The C-C bond lengths depend on the positions in each ripple, in which the largest change is revealed at the crests and troughs compared with a planar structure, e.g., $1.2 \%$ for $N_\lambda=6$ armchair ripple and $7.9 \%$ for $N_\lambda=5$ zigzag one. Apparently, they are strongly affected by the corrugated direction. Armchair and zigzag ripples, respectively, exhibit two and three different nearest C-C bond lengths. This will induce strong effects on the low-lying energy bands, as identified from carbon nanotubes.\cite{kane1997size,l2002electronic} The curved surface brings about the misorientation of $2p_z$ orbitals and the significant hybridizations of ($2s$,$2p_x$,$2p_y$,$2p_z$) orbitals; that is, the curvature effects can create the new $\pi$ and $\sigma$ bondings. Concerning the rotation symmetry, it is only two-fold for a rippled structure along a specific direction, but six-fold for a planar structure. The above-mentioned changes in bond lengths, chemical bondings and rotation symmetry are expected to greatly enrich the electronic properties.

Band structure is dominated by the longitudinal structure, curvature and period. All armchair ripples have a pair of linearly intersecting energy band at the Fermi momentum ($k_F$), being the same with monolayer graphene. The Dirac point, as shown in Fig. 11(a), is revealed along the $\Gamma\,$Y direction (Fig. 10(d)). It has an obvious red shift, compared with that of a planar structure ($[k_x=0,k_y=2/3$]; the black solid curve for $C_r=0$). The original Dirac point is directly obtained from the zone-folding condition, in which the hexagonal Brillouin zone is mapped into the rectangular one. The curvature effects do not destroy the Dirac-cone structures, as clearly indicated in Figs. 11(b) \& 11(c). However, they can create the red-shift Fermi momentum, the reduced Fermi velocity, and the slight anisotropy. These variations grow with an increasing curvature, but decrease in the increment of period (Fig. 11(a)). Concerning the middle-energy states, several pairs of parabolic bands are initiated near the Y point. The local minimum (maximum) of the conduction band (valence band) along Y$\Gamma\,$ is the local maximum (minimum) along the perpendicular direction. Such saddle points are sampled from the electronic states near the M and M$^\prime$ points in graphene under the reduced rotation symmetry. The splittings of parabolic bands are easily observed for various curvatures and periods (a comparison between $C_r=0$ \& $C_r\neq\,0$). That is to say, the two-fold rotation symmetry destroys the double degeneracy of the M and M$^\prime$ points and induces the significant energy splittings. The similar splittings are revealed in zigzag ripples (Fig. 12(e)).

The low-lying energy bands are drastically changed by the corrugated structures. Zigzag ripples exhibit three kinds of energy dispersions: linear, partially flat and parabolic ones, corresponding to the low, middle, and high curvatures, respectively. For the $N_\lambda=5$ zigzag ripple, the linear bands intersecting with $E_F=0$ survive at low curvature, as shown in Fig. 12(a) for $C_r=0.05$. A larger $C_r$ accounts for the distorted Dirac-cone structure and a partial flat band at $E^c\sim\,0.4$ eV (blue curve in Fig. 12(b) for $C_r=0.1$). These two features are clearly displayed in the energy-wave-vector space (Fig. 12(h)). The latter is mostly contributed by the specific carbon atoms at the ripple crests and troughs, as denoted by the radii of gray circles. At high curvature, two low-lying parabolic bands, which are dominated by the crest- and trough-related carbon atoms, come into existence (Figs. 12(c) and 12(d) for $C_r=0.19$ and $0.27$). Such bands are accompanied with the highly distorted Dirac cone, as revealed in Fig. 12(i) for $C_r=0.27$. They are very close to $E_F=0$ and even cross it. This will induce a finite DOS at the Fermi level or the semi-metallic property (Fig. 14(b)). Their band-edge states belong to the saddle points or the local extreme points. Also, the number and energy of critical points are easily tuned by the increasing period. For example, the $N_\lambda=9$ zigzag ripple has five critical points asymmetric about $E_F=0$ under the critical curvature ($C_r=1.17$ in Fig. 12(g) with blue arrows). Apparently, the dramatic transformation between two distinct band structures could be achieved by changing the curvature and period in zigzag ripples. Up to now, the experimental measurements on the electronic structures of rippled graphenes are absent. The predicted results, the Dirac-cone structures in armchair ripples, the diverse energy dispersions in zigzag ripples, and the splitting middle-energy bands deserve a closer examination using the ARPES measurements.

A detailed comparison between graphene ripples and carbon nanotubes is very useful in understanding the effects due to the periodic boundary condition and curvature. These two systems present certain important similarities and differences. The energy bands of 2D graphene ripples along the $\Gamma\,$Y direction correspond to those of 1D carbon nanotube along the axial direction. An armchair nanotube has a pair of linear bands intersecting at $k_F$, e.g., a ($3,3$) nanotube in Fig. 11(a) (the heavy dashed curve; notation for carbon nanotubes in Refs. [283,284]).
The Fermi-momentum state is sampled from the K point of graphene in the presence of a closed periodic boundary condition.\cite{mintmire1992fullerene,hamada1992new,saito1992electronic,kane1997size} An obvious red shift of $k_F$ mainly comes from the misorientation of $2p_z$ orbitals on a cylindrical surface,\cite{hamada1992new} or the reduced $\pi$ bondings and the created $\sigma$ bondings of $2p_z$ orbitals (the different hopping integrals for the nearest neighbors).\cite{l2002electronic} Both armchair nanotubes and ripple possess linearly intersecting bands, while they, respectively, belong to 1D conductors and 2D semiconductors, being determined by the finite and vanishing DOSs at the Fermi level (Fig. 14(a)). The metallic properties of the former keep the same even under the strong hybridizations of ($2s$,$2p_x$,$2p_y$,$2p_z$) orbitals.\cite{liu2002properties} In general, a ($m,0$) zigzag nanotube, with $m\neq\,3I$ ($I$ an integer) ($m=3I$) is a middle-gap (narrow-gap) semiconductor, when the periodic boundary condition and the curvature effects are taken into consideration.\cite{kane1997size,l2002electronic} But for a very small nanotube, with $m\le\,6$,\cite{blase1994hybridization,gulseren2002systematic} the serious $sp^3$ orbital hybridizations can create the metallic behavior, e.g., a gapless ($5,0$) nanotube (the heavy dashed curve in Fig. 12(e)). Specifically, the middle-energy states of carbon nanotubes are doubly degenerate because of the cylindrical symmetry.

The complicated chemical bondings, which account for the rich electronic properties, are clearly evidenced by the non-uniform spatial charge distributions on a curved surface. For a planar graphene, the parallel $2p_z$ orbitals and the planar ($2s,2p_x,2p_y$) orbitals can form the $\pi$ bondings and the $\sigma$ bondings, respectively, as shown for $V_{pp\pi}$ and ($V_{pp\sigma}$, $V_{ps\sigma}$, $V_{ss\sigma}$) in Fig. 13(a). The orbital hybridizations could be characterized by the carrier density ($\rho$) and the difference of carrier density ($\Delta\rho$). The latter is obtained by subtracting the carrier density of an isolated carbon (hollow circles in Figs. 13(a)-13(c)) from that of a graphene (graphene ripple). Carbon atoms contribute four valence electrons to create various chemical bondings, leading to the largely reduced charge density near the atomic site (region (I)). However, the strong $\sigma$ bondings of ($2s,2p_x,2p_y$) orbitals induce the increased charge density between two atoms (dumbbell shapes in region (II)). There are more charge densities normal to the surface (region (III)) because of the hybridized $2p_z$ orbitals. These charge distributions strongly depend on the atomic positions of a graphene ripple (Figs. 13(b) and 13(c)), quite different from the uniform ones of a monolayer graphene (Fig. 13(a)). That is, the non-parallel $2p_z$ orbitals can create the extra $\sigma$ bondings ($V_{pp\sigma.}$). Moreover, at the crests and troughs, it is easy to observe the strong hybridizations of four orbitals (region (IV)). Specifically, armchair and zigzag ripple, respectively, possess two and three distinct charge distributions along three nearest neighbors (a detailed examination made for various directions). Both systems exhibit the misorientation and hybridization of atomic four orbitals; furthermore, zigzag ripples present the curvature effects in a stronger manner. This is the main cause for the shift, distortion and destruction of the Dirac-cone band structure. The similar curvature effects are also revealed in very small armchair and zigzag nanotubes (Figs. 13(b) \& 13(c)), leading to the metallic behavior (Figs. 11(a) \& 12(e)).

The multi-orbital hybridizations on a curved surface provide a suitable chemical environment for materials engineering and potential applications. The non-uniform charge distributions originate from carbon four orbitals, especially for those at the crest and trough (region (IV) in Fig. 13). Such orbitals might have the strong bondings with gas molecules, e.g., carbon monoxide (CO), formaldehyde (H$_{2}$CO), and hydrogen. The high sensitivity to CO and H$_{2}$CO indicates a promising material for toxic gas sensor applications, as done for aluminum-doped graphenes.\cite{chi2009adsorption,ao2008enhancement} These characteristics could also be utilized for hydrogen storages,\cite{tozzini2011reversible,ning2012high,tozzini2013prospects} where the curvature inversion leads to the efficient and fast hydrogen release.\cite{tozzini2011reversible} Moreover, the high-curvature graphene ripples exhibit the semimetallic behavior. The electrical conductivity is expected to be largely enhanced by the high free carrier density near $E_F$. This is useful in the further applications of nanoelectronic devices.\cite{yan2012high}

The main differences between armchair and zigzag ripples are further reflected in the low-energy DOS. The former have a vanishing DOS at $E_F=0$ and a V-shaped energy spectrum, as shown in Fig. 14(a) for various ripple structures. This clearly illustrates the semiconducting behavior. The linear DOS grows with the increasing curvature, a result of the reduced Fermi velocity. There are several symmetric $\pi$ and $\pi^\ast$ peaks at the higher energy, mainly owing to the reduced rotation symmetry and the curvature effects (Fig. 14(c)). On the other hand, the latter exhibit the special peak and shoulder structures at low energy under a sufficiently large curvature, as observed in Fig. 14(b). The valence and conduction structures are highly asymmetric about $E_F$, in which most of them occur at $E\ge\,0$. Their number is largely enhanced by the increment of curvature. A special peak is revealed near $E_F$, indicating the semi-metallic behavior. In addition, an armchair nanotube possesses a finite DOS in the plateau form near $E_F$, while a very small zigzag nanotube presents several prominent peaks there. All carbon nanotubes present the prominent asymmetric peaks arising from the 1D parabolic subbands. Such peaks have been verified by the STS measurements.\cite{wilder1998electronic,odom1998atomic}

The orbital-projected DOS could be used to further examine the various chemical bondings and verify the curvature effects.
At low curvature, the low- and middle-energy special structures in DOS only depend on the bondings of $2p_z$ orbitals
(not shown). But for the high-curvature armchair and zigzag ripples, they are dominated by $2p_z$ and $2p_x$ orbitals, as clearly indicated in Figs. 14(c) and 14(d). Their contributions are comparable to each other even near $E_F$; that is, the feature-rich energy bands are mainly determined by these two orbitals. Apparently, there exist the strong hybridizations of ($2p_z,2p_x$) orbitals in the nearest-neighbor carbon atoms. Three kinds of orbital interactions, ($2p_z,2p_z$), ($2p_z,2p_x$) and ($2p_x,2p_x$), can create the $\pi$ and $\sigma$ bondings simultaneously.\cite{lin2015feature} The complex chemical bondings, which rely on the arrangement of orbitals an surface curvature, can account for a lot of special structures. In addition, the ($2p_x,2p_y,2p_z$) orbitals also make the important contributions to the $\sigma$ peaks, e.g., those near $E=-3$ eV, an evidence of the multi-orbital hybridizations. The peak structures related to $2s$ orbitals only occur at the deeper energy (not shown).

The main features of the calculated DOSs are partially in agreement with the position-dependent STS measurements.\cite{meng2013strain} The measured dI/dV-V spectra have a prominent peak near $E_F$ and some low-energy peaks with a linear relation between the peak energies and the square root of the effective quantum number, or they are in the featureless V-shaped form. The former corresponds to the semi-metallic zigzag ripples at high curvatures, as shown in the inset of Fig. 14(b). Furthermore, the latter are related to the low-curvature zigzag ripples with the weak orbital hybridizations. On the other hand, the predicted results, the V-shaped DOS in armchair ripples (Fig. 14(a)), the curvature-dependent Fermi velocities, and the curvature- and hybridization-induced $\pi$ and $\pi^*$ peaks need to be further verified by the experimental measurements.

The experimental measurements by STM can provide the spatially atomic distributions at the local nano-structures. They have been successfully utilized to resolve the unique geometric structures of the graphene-related systems, such as, graphene,\cite{de2008periodically,meng2013strain,bai2014creating} graphene compounds,\cite{balog2010bandgap,pandey2008scanning} graphite,\cite{vcervenka2009room,kondo2012atomic} graphene nanoribbons\cite{ruffieux2012electronic,tao2011spatially} and carbon nanotubes.\cite{wilder1998electronic,odom1998atomic} The atomic-scaled observations clearly show the rippled and buckled structures of graphene island,\cite{meng2013strain,bai2014creating,de2008periodically} the adatom distributions on graphene surface,\cite{balog2010bandgap,pandey2008scanning} the 2D networks of local defects,\cite{vcervenka2009room} the pyridinic-nitrogen and graphitic-N structures,\cite{kondo2012atomic} the nanoscale width of achiral armchair nanoribbon,\cite{ruffieux2012electronic} and the chiral arrangements of the hexagons on the planar edges\cite{tao2011spatially} and a cylindrical surface.\cite{wilder1998electronic,odom1998atomic} The corrugation direction, the curvature and period of graphene ripple and the relative shift in sliding bilayer graphene are worthy of a closer experimental examination. Also, the measurements are useful in understanding the orbital interactions due to the curvature effects and the interlayer distances.

\vskip 0.6 truecm
\par\noindent
{\bf 4. Graphene oxides}

\vskip 0.3 truecm

The electronic properties of adatom-doped graphenes is one of important topics in physics, chemistry, and materials. They are greatly diversified by various adatom adsorptions. The adatom-doped graphenes have attracted a lot of theoretical \cite{praveen2015adsorption,tran2016chemical,tran2016pi,nakada2011migration,brar2011gate} and experimental \cite{zanella2008electronic,gao2010first,dai2010adsorption} researches. Furthermore, such systems might have high potentials in near-future applications as a result of the tunable electronic properties, e.g., supercapacitors,\cite{fan2013nitrogen,karthika2013phosphorus,xue2015multiscale,chen2011high,gao2011direct} optoelectronics,\cite{loh2010graphene,wu2009organic,su2009composites} energy storage,\cite{porro2015memristive,zhang2013synthesis,wang2013situ} and sensors.\cite{wang2010nitrogen,sheng2012electrochemical,veerapandian2012synthesis,robinson2008reduced} According to the modified electronic properties, there are two kinds of adatom-doped graphenes, namely semiconducting and metallic systems. The experimental measurements have identified the adatom-dependent semiconducting graphenes, such as the O-,\cite{nourbakhsh2010bandgap,ito2008semiconducting,tran2016pi} H-,\cite{huang2016configuration,gao2011band}, Si-\cite{azadeh2011tunable,zhang2016opening} S-,\cite{denis2010band,denis2013concentration}, P-\cite{denis2010band,denis2013concentration} and BN-doped graphenes.\cite{shinde2011direct} The two typical semiconducting systems, graphene oxides and hydrogenated graphenes, are chosen for a detailed discussion. The optimal adatom positions, the atom-dominated energy bands, the spin configuration, the spatial charge distributions, and the orbital-projected DOS are obtained from the first-principles calculations. They can provide the full informations in the complex chemical bondings associated with the C-adatom, C-C and adatom-adatom bonds. The various orbital hybridizations are responsible for the adatom concentration- and configuration-enriched electronic properties, such as the opening of energy gaps, the destruction or recovery of the Dirac-cone structure, the formation of atom- and orbital-dependent energy bands, and a lot of van Hove singularities in DOS. On the other hand, the alkali- \cite{praveen2015adsorption,denis2011chemical} and halogen-doped \cite{xu2015electronic,chu2012charge} graphenes exhibit the metallic behaviors, as verified from the experimental measurements.\cite{virojanadara2010epitaxial,gruneis2008tunable,papagno2011large} The former and the latter, respectively, have free electrons and holes in the conduction and valence Dirac cones. The alkali- and Al-doped graphenes are predicted to have the almost same Dirac-cone structure or free carrier density. Such 2D systems might possess the high free carrier density, compared with the other layered systems.\cite{jin2014high,xu2016quasi} The Dirac-cone structure is almost preserved in the alkali-doped graphenes, while it is distorted in the halogen-doped graphenes. The outermost s orbital of alkali adatom only creates an conduction band. However, more outer orbitals in halogen adatom can induce certain valence energy bands. The main differences between these two metallic systems will be identified to come from the critical orbital hybridizations. Specifically, the fluorinated graphenes would change into the gap-opened semiconductors for very high adatom concentrations, since the dense top-site absorptions can fully terminate the $\pi$ bondings on graphene surface. Part of theoretical predictions have been confirmed by the experimental measurements.

One of the effective methods to obtain large-scale graphene is chemical exfoliation in which graphite is exfoliated through oxidation and then is reduced to graphene monolayers subsequently.\cite{gomez2007electronic,tung2009high,eda2008large} This method was first discorvered by Schafhaeutl in 1840.\cite{schafhaeutl1840lxxxvi} Since the first successful synthesis of monolayer graphene in 2004,\cite{novoselov2004electric} GOs have been extensively investigated. Although graphene has potential applications, its zero-gap band structure will create high barriers in the further development of graphene-based nano electronic devices. The band-gap engineering becomes a critical issue in graphene-related systems. Oxygen adsorption on graphene layers can induce middle-energy band gaps which can be tuned by the concentration and distribution of adatom. GOs are considered as promising materials in a wide variety of applications, especially in energy storage and enviromental science, such as supercapacitors,\cite{xue2015multiscale,chen2011high,gao2011direct} memristor devices,\cite{porro2015memristive} sensors,\cite{veerapandian2012synthesis,robinson2008reduced} and water purification.\cite{li2015graphene}

There are three main methods to produce GOs from graphite, namely Brodie, Staudenmaier, and Hummers. The synthesis of GOs was first reported by Brodie in 1859 from the detailed investigations on the graphitic structure by using KClO$_3$ as the oxidant to oxidize graphite in HNO$_3$.\cite{brodie1859atomic} Later in 1898, Staudenmaier improved Brodie's method by adding H$_2$SO$_4$ to increase the acidity of the mixture.\cite{staudenmaier1898verfahren} In 1958, Hummers developed an alternative method in which a mixture of NaNO$_3$, H$_2$SO$_4$ and KMnO$_4$ is used in synthesis process.\cite{hummers1958preparation} The third method is the most widely used because of its shorter reaction time and no ClO$_2$ emission.\cite{poh2012graphenes,bai2011functional} Recently, other methods have modified Hummers method by adding H$_3$PO$_4$ combined with H$_2$SO$_4$ in the absence of NaNO$_3$, and increasing the amount of KMnO$_4$.\cite{marcano2010improved} GOs have also been manufactured using bottom-up method (Tang-Lau method) since its process is simpler and more environmentally friendly compared to traditionally top-down methods.\cite{tang2012bottom} Different O-concentrations can be synthesized by controlling the amount of oxidant compound (KMnO$_4$), \cite{C5RA02099A} or the oxidation time.

First-principles calculations are a suitable method to study the oxygen-enriched essential properties, especially for the relation between the opened gap and the multi-orbital chemical bondings. Geometric and electronic properties are mainly determined by the competition or cooperation of the critical orbital hybridizations in C-C, C-O, and O-O bonds. They are very sensitive to variations in the number of graphene layer, stacking configuration, oxygen concentration and distribution. The rich electronic structures exhibit the destruction or distortion of the Dirac cone, tunable band gaps, C-, O- and (C,O)-dominated energy dispersions, and many critical points. The total DOS has a lot of special structures related to the parabolic, partially flat and composite energy bands. The atom-dorminated energy bands, the orbital-projected DOS, and the spatial charge distributions can be used to comprehend the critical orbital hybridizations in C-C, C-O and O-O bonds, being responsible for the diverse essential properties. There exists five types of $\pi$ bondings during the variation from 50\% to vanishing adsorptions, namely the complete termination, the partial suppression, the 1D bonding, the deformed planar bonding, and the well-behaved ones. Such bondings account for the finite and zero gaps, corresponding to the O-concentrations of $\ge$25\% and $\le$3\%, respectively. In addition, GOs do not have the ferromagnetic or anti-ferromagnetic spin configurations, as identified from various calculations.

\vskip 0.3 truecm
{\bf 4.1 Single-side adsorption }
\vskip 0.3 truecm

Numerous theoretical and experimental researches have been carried out to explore the geometric structure of GOs; furthermore, the effects due to concentration and distribution of the oxygen-containing groups are investigated in detail. Oxygen atoms are adsorbed at the bridge site so that the epoxy C-O-C functional group is formed, as shown in Fig. 15. The bridge site is the most stable site compared to the hollow and top positions, consistent with previous studies.\cite{nakada2011migration,saxena2011investigation,erickson2010determination} The unit cells are classified as zigzag (Z), armchair (A), and chiral (C) configurations, in which carbon atoms are arranged along these edge structures, respectively. In comparison with the two latter, the first one has a lower total ground state energy, demonstrating higher stability. The zigzag-edge configurations will be chosen as model systems. The optimized C-C bond lengths of GOs change in the range from 1.42 $\mbox\AA$ to 1.53 $\mbox\AA$ when the O-concentration gradually grows from zero to 50\%, as illustrated in Table 1. However, the C-O bond lengths decrease from $1.47$ {\AA} to $1.41$ {\AA}. The binding energy of each oxygen adatom, the reduced energy due to the oxidization, is characterized as $E_{b}$ = $(E_{sys} - E_{gra} - nE_{O})/n$, where $E_{sys}$, $E_{gra}$, and $E_{O}$ are the total energies of the GOs system, the graphene sheet, and the isolated O atom, respectively. The slightly buckled structures are revealed in the decrease of O-concentration. In general, the stabilities are enhanced by the higher O-concentration and distribution symmetry, in agreement with other theoretical predictions.\cite{huang2012oxygen} The rich geometric structures can diversify electronic properties.

The functional group and the concentration and distribution of oxygen could be verified using experimental measurements. Spectroscopic techniques, such as solid-state nuclear magnetic resonance (NMR) \cite{cai2008synthesis,hontoria1995study,stankovich2007synthesis} and X-ray photoelectron spectroscopy (XPS) \cite{gao2009new} can provide essential insights into the types of oxygenated functional groups in GOs and their concentrations. Multi-dimensional NMR spectra are available in the identification of the major functional groups in GOs, namely the thermodynamic stable epoxy (C-O-C) and hydroxyl (C-OH).\cite{lerf1998structure,casabianca2010nmr} XPS is powerful in examining how many atoms are doped into graphene and the types of adatoms,\cite{gao2009new} e.g., the estimated concentrations for oxidized carbons and graphitic carbons.\cite{marcano2010improved} Furthermore, microscopic techniques, which include transmission electron microscopy (TEM), STM, and scanning transmission electron microscope (STEM), have been utilized to determine the distribution of GOs. The high-resolution TEM image of a suspended GOs sheet reveals holes, graphitic regions, and oxygen-containing groups with approximate area percentages of 2, 16, and 82 \%, respectively.\cite{erickson2010determination} A nanoscale periodic arrangement of O atoms is confirmed using STM.\cite{pandey2008scanning} Morever, the high-resolution annular dark field imaging in STEM measurement has verified the O distribution of monolayer GO.\cite{mkhoyan2009atomic}

The 2D energy bands along high symmetric points, as shown in Fig. 16, are determined by the C-O, O-O and C-C bonds; the O-concentration and distribution. In sharp contrast to pristine graphene, the isotropic Dirac-cone structure near the K point is destroyed by oxidation at high O-concentrations (Figs. 16(a)-16(e)), mainly owing to the strong C-O bonds. The multi-hybridizations between orbitals of C and O atoms lead to the termination of the complete $\pi$ bondings formed by parallel $2p_z$ orbitals of C atoms. A middle energy gap is created by the O-dominated energy bands near $E_F$ (blue circles), as obtained  from other researches.\cite{ito2008semiconducting,huang2012oxygen} Such bands originate from the significant O-O  bonds, and they exhibit the partially flat band or parabolic bands along different directions. With the decrease of O-concentration from $50$ \% to $0$, the O-dominated bands become narrower and contribute at deeper energy, while the $\pi$ and $\pi^\ast$ bands are gradually recovered (Figs. 16(f)-16(j)). Also, energy gap declines from $3.54$ eV to $0$. As mentioned above, the strong C-O bonds induce the (C,O)-related energy bands with very weak dispersions at $-$2 eV$\,\le\,E^v\le\,-$4 eV (blue and red circles revealed simultaneously). On the other hand, the planar C-C bond possesses the strongest $\sigma$ bondings formed by the (2$p_x$, 2$p_y$, 2s) orbitals and thus creates the deeper $\sigma$ bands with $E^v\le\,-$3.5 eV. Previous theoretical studies stated that energy gap declines monotonically with the decrease of O-concentration.\cite{nourbakhsh2010bandgap,ito2008semiconducting} However, other studies demonstrated that the dependence of energy gap on O-concentration is non-monotonous.\cite{huang2012oxygen,tran2016pi} Particularly, energy gap also depends on the O-distributions (distinct relative positions and edge configurations in Table 1).\cite{lian2013big} The critical low-energy electronic properties can be further understood from the 3D energy bands shown in the insets of Fig. 16.

Energy gaps of GOs could be divided into three categories. For the high O-concentration of $\ge$ 25\%, the Dirac cone is seriously distorted as a result of the strong O-O and C-O bonds (Figs. 16(a)-16(e)). These systems always have a finite gap, independent of the O-distribution. $E_g$'s decline quickly and fluctuate widely in the concentration range of ~25-3\% (Figs. 16(f)-16(h)). Such small or zero gaps correspond to the reformed Dirac cone with the significant distortion. Energy gaps become zero thoroughly for the low concentration of $\le$3\% as a result of the fully recovered Dirac cones without energy spacing (Figs. 16(i)-16(j)). Most of theoretical researches about GOs are focused on how the band gap is modulated with the variation of O-concentration,\cite{lian2013big,ito2008semiconducting} however, the termination and reformation of $\pi$ bonds are not clearly illustrated. The band-decomposed charge density distributions provide useful information about the relation between the $\pi$-bondings and energy gap. There exists five types of $\pi$ bondings as O-concentration decreases from the full to vanishing adsorptions \cite{tran2016pi}: the complete termination (50\% (Z) in Fig. 17.(b)), the partial suppression (33.3\% (A) in Fig. 17(c)), the 1D bonding (4.2\% (Z) in Fig. 17(d)), the deformed planar bonding (system 1\% (Z) in Fig. 17(e)), and the well-behaved one (pristine graphene in Fig. 17(a)). The suppression or extension degree of $\pi$ bondings can account for the strong dependence of $E_g$ on O-concentration. In addition, the electrical conductivity of GOs has been investigated to be greatly enhanced by chemical reduction.\cite{du2004novel,szabo2005composite,stankovich2006graphene} Therefore, GOs are expected to have high potentials in the near-future nanoelectronic devices.\cite{xue2015multiscale,chen2011high,gao2011direct,veerapandian2012synthesis,robinson2008reduced}

The charge density (Figs. 18(a)-18(d)) and the charge density difference (Figs. 18(g)-18(j)) can provide very useful information about the chemical bondings and thus comprehend the dramatic changes of energy bands. The latter is created by subtracting the charge density of graphene and O atoms from that of GOs system. $\rho$ illustrates the chemical bonding as well as the charge transfer. As O atoms are adsorbed on graphene, charges are transfered from C to O atoms (red region) $\approx$ 1e using Bader analysis. The strong C-O bonds induce the terminated $\pi$-bonding (black rectangles in Figs. 18(b)-18(d)) accounting for the absence of Dirac cone and the opening of band gap. The orbitals of O atoms have strong hybridizations with those of passivated C, as seen from the green region enclosed by the dashed black rectangles (Figs. 18(h))-18(j). This region lies between O and passivated C atoms and bents toward O. Such strong C-O bonds lead to the termination of the complete $\pi$ bondings between parallel $2p_z$ orbitals of C atoms and thus the destruction of the Dirac-cone structure. To explain the opening of band gap, Lian et al. \cite{lian2013big} proposed 2$p_z-$2$p_{x,y}$ orbital hybridizations of C and O atoms. Afterwards, Tran et al. \cite{tran2016chemical} has investigated orbital hybridizations of 2p$_{x,z}-$2$p_{x,z}$ or 2p$_{x,y,z}-$2$p_{x,y,z}$ between passivated C and O atoms, obviously including the 2$p_z-$2$p_z$ hybridizations. Between two non-passivated C atoms of GOs, $\Delta \rho$ shows a strong $\sigma$ bonding (pink squares), which becomes a bit weaker after the formation of C-O bonds (e.g., 50\% (Z),  50\% (A) and 33.3\% (A) systems). This demonstrates that not only 2$p_z$ but also 2$p_{x,y}$ orbitals of passivated C atoms hybridize with orbitals of O atoms. As to the O-O bonds, the orbital hybridizations depend on the adsorption positions of O atoms. For instance, the orbitals with high charge density enclosed in the dashed pink rectangle ( 50\% (Z); Fig. 18(h)) are lengthened along $\hat{y}$, clearly illustrating the 2$p_y$-2$p_y$ hybridization (further supported by the projected DOS in Fig. 19(a)). In case of 50\% (A) system, these lengthened orbitals are equally projected on $\hat{x}$ and $\hat{y}$, indicating the 2$p_{x,y}$-2$p_{x,y}$ orbital hybridizations (sees Fig. 19(b)). These are also viewed in xy-plane projections. The O-O bond belongs to a weak $\sigma$ bonding at larger distance, being responsible for the O-dominated energy bands near E$_F$ (Fig. 16(a)). The similar orbital hybridizations in the O-O bonds are also revealed in chiral distributions (not shown).

The form, number and energy of the special structures in DOS are very sensitive to the chemical bondings and O-concentration. Furthermore, the projected DOS can comprehend the orbital hybridizations in O-O, C-O and C-C bonds. The main characteristics of band structures are directly reflected in the orbital-projected DOS, which is responsible for the orbital contributions and hybridizations in chemical bonds (Figs. 19(a)-19(j)). The lower-energy DOS is dramatically altered after oxygen adsorption. For pristine graphene, the $\pi$ and $\pi^*$ peaks due to 2$p_z-$2$p_z$ bondings between C atoms will dominate DOS within the range of $|E|\le\,$ 2.5 eV (solid curve in Fig. 14(a)). \cite{neto2009electronic}. Furthermore, the vanishing DOS at $E_F$ indicates the zero-gap semiconducting behavior. However, for high O-concentration (Figs. 19(a)-19(e)), these two prominent peaks are absent because of the strong C-O bond. Instead, there exist an energy gap and several O-dominated special structures, a result of the O-O bonds in a wide range of $E\sim\,-$2.5 eV to $E_F$. Such structures consist of the symmetric peaks (triangles), the shoulder structures (arrows), or the asymmetric peaks (circles), corresponding to the saddle points of parabolic bands, the local minimum/maximum states, or the composite band of partial flat and parabolic dispersions, respectively. They mainly originate from the 2$p_y$-2$p_y$ hybridization (Fig. 19(a)) or 2$p_{x,y}$-2$p_{x,y}$ hybridization (Figs. 19(b) and 19(c)). Specifically, the symmetric distribution of O adatoms leads to the same contribution of 2$p_x$ and 2$p_y$ orbitals in 50\% (A) configuration. Energy gaps are sensitive to the change in O-O bond strength, as indicated from distinct gaps among various distributions and concentrations. It is revealed that the range of O-dominated structures becomes narrower and shifts away from $E_F$ as the O-concentration declines (Figs. 19(e)-19(j)), clearly illustrating the competition between O-O and C-C bonds. For low O-concentration (from 3\% (Z); Figs. 19(i)-19(j)), the $\pi$ and $\pi^*$ peaks are reformed due to the complete $\pi$-bondings. As to the middle-energy DOS ($-$2 eV$\,\le\,E^v\le\,-$4 eV), it grows quickly and exhibits prominent symmetric peaks due to the strong C-O bonds, including the delta-function-like peaks. Such special peaks reflect the almost flat bands along any directions. They indicate the strong hybridizations among the (2$p_x$,2$p_z$) orbitals of O and those of passivated C atoms.

Electronic properties of GOs are enriched by the number of graphene layers and stacking configuration. Energy dispersions of O-adsorbed few-layer graphenes are shown in Fig. 20 for the O-concentration of 50\% in zigzag distribution. For O-adsorbed AA stacking, there is one anisotropic Dirac-cone structure with $E_F$ located at the conduction band, being in great sharp contrast with two isotropic Dirac-cone structures of the pristine system.\cite{tran2015configuration} Some free electrons appear in the conduction Dirac one, indicating the same free holes in the partial flat band due to O-O bonds (slightly penetrates the Fermi level). The band-structure differences between the O-adsorbed AA and AB stackings (Figs. 20(a) and 20(b)) are small and only lie in the crossing and anti-crossing behaviors (green ellipses). As the number of graphene layer increases, there are more distorted Dirac cones and saddle points. Specifically, the energy dispersions of ABA and ABC stackings are almost identical, since these two systems correspond to the same structure of AB bilayer graphene with monolayer GO on the top, as investigated in the previous study.\cite{tran2016chemical} However, they are different from the AAA stacking in the distorted Dirac-cone structures as well as the crossing and anti-crossing bands (Figs. 20(c) and 20(d)). There also exist the O-dominated energy dispersions near $E_F$ and the (C,O)-related bands at middle energy, as revealed in monolayer GO. All of the O-adsorbed few-layer graphenes exhibits the semi-metalic behavior with a finite DOS at $E_F$, different from the semiconducting monolayer systems.

{\bf 4.2 Double-side adsorption }

So far we have discussed the essential properties of single-side adsorbed GOs. They are dramatically changed by the double-side adsorption and O-concentration. Various O-configurations with different coverages are illustrated in Fig. 21. Binding energy $E_b$ in Table 2 indicates that the configuration stability is enhanced by the higher O-distribution symmetry and concentration. Furthermore, the double-side adsorbed systems are more stable compared to the single-side ones (Table 1). This is consistent with another DFT calculation which shows that the oxygen-containing groups prefer to reside on both sides of graphene.\cite{boukhvalov2008modeling} For the high O-concentration (50\% (Z) in Fig. 21(a)), the O-O bonds dominate the lower-energy electronic properties (blue circles), and the Dirac-cone structure arising from the $\pi$ bondings is absent. Similarly to the single-side adsorption, when the O-concentration decreases, the competition between the weakened O-O bonds and the gradually recovered $\pi$ bondings (from non-passivated C atoms) will greatly reduce the band gap (33.3\% (A), 25\% (Z), 11.1\% (Z), and 4\% (Z) in Figs. 21(b)-21(e), respectively). The double-side adsorbed bilayer graphene is regarded as the superposition of two monolayer GOs. The band structures of AA (Fig. 21(f)) and AB systems (Fig. 21(g)) are almost the same, revealing the semiconducting behavior. As for tri-layer systems, the double-side adsorbed AAA, ABA and ABC stackings have the almost identical energy bands, since they correspond to the same sandwich structure with monolayer graphene at the middle and two  monolayer GOs on its top and bottom. There exists a semi-metallic Dirac-cone structure coming from the C atoms of the middle layer without passivation (Fig. 21(h)). The critical orbital hybridizations in O-O, C-O and C-C bonds of both-side adsorbed systems remain similar to the single-side adsorption ones (see the charge density in Figs. 18(e)-18(f) and charge density difference in Figs. 18(k)-18(l)). The terminated $\pi$ bondings and the reformed $\pi$ bondings are clearly illustrated in the black rectangles of Figs. 18(e) and 18(f). The former and the latter are responsible for the (C,O)-dominated energy bands and the distorted Dirac-done structure, respectively. The strong covalent $\sigma$ bondings with high charge density exist between two non-passivated C atoms and become weaker when the C atoms are bonded with O atoms (white squares in Figs. 18(k) and 18(l)).


The adsorption-induced dramatic changes in band structures could be directly examined by ARPES. For example, such measurements have identified the Dirac-cone structure of graphene grown on SiC,\cite{ohta2008morphology} and observed the opening of band gap for graphene on Ir(111) through oxidation \cite{schulte2013bandgap}. As expected, the feature-rich energy bands of GOs, including the absence and presence of the distorted Dirac-cone structures, the concentration- and distribution-dependent band gap, the O-dominated bands near $E_F$, and the (C,O)-dominated bands at middle energy can be examined by ARPES. In addition to ARPES, the unusual electronic structures can be verified using the photoluminescense spectroscopy,\cite{liang2015band} and internal photoemission spectroscopy,\cite{xu2012direct} such as, the verifications on band gap, Dirac point energy, etc. The comparisons between theoretical predictions and experimental measurements can comprehend the roles of chemical bondings on the electronic properties of GOs and the effects due to the O-concentration and distribution.

The STS measurements on the special structures of DOS can provide the critical informations coming from the chemical adsorption. For example, the band gap of exfoliated oxidized graphene sheets has been directly confirmed by STS.\cite{pandey2008scanning} The main features in electronic properties, including the concentration- and distribution-dependent energy gaps, the O-dominated prominent structures near $E_F$, the high $\pi$- and $\pi^*$-peaks, and the strong (C,O)-related peaks at middle energy, can be further investigated with STS. The STS measurements on the asymmetric and symmetric peaks at distinct energy ranges can identify the O-related band widths, and the specific chemical bondings in GOs.

The electrical conductivity is an important quantity to characterize the electronic properties of GOs sheets. When O atoms are adsorbed on graphene, the $\pi$ and sp$^2$ bondings in pristine graphene are changed into the complex hybridizations of (2$p_x$,2$p_y$,2$p_z$) orbitals in O-O, in O-C and C-C bonds. The high-concentration GOs are typically considered as an electrical insulator due to the destruction of $\pi$ bondings. The electrical conductivity of GOs is gradually enhanced as the O-concentration decreases \cite{becerril2008evaluation}. In order to improve the electrical conductivity of GOs, numerous methods of reducing O-concentration are applied. GO films that have been reduced with hydroiodic acid show a higher electrical conductivity compared to those reduced by hydrobromic acid and hydrazine hydrate.\cite{mohan2015characterisation} Moreover, the electrical conductivity grows quickly with the increase of temperature and reduction time.\cite{mattevi2009evolution,jung2008tunable}

By appropriately tuning the O-concentration and distribution, and the number of graphene layers, GOs are dramatically changed among the insulators, semiconductors and semi-metals. The tunable and controllable electronic properties make GOs become high potential candidates in materials applications, especially for electronic devices. The experimental measurements show that the reduced GOs could serve as the conducting channels in the field-effect transistors (FETs).\cite{trung2014flexible,truong2014reduced,joung2010high} Furthermore, such FETs are deduced to have potential applications in chemical and biological sensors. The great advantages is that the fabrication of large-scale reduced GO films is fast, facile, and substrate independent, compared to graphene films.\cite{he2010centimeter} Moreover, by adjusting the O-concentration or extra adsorbing molecules, the conductance of GOs is drastically changed, clearly indicating a good promise in electrical sensors.\cite{eda2010chemically,basu2012recent,su2014electrical} On the other hand, monolayer GOs have high optical transmittance in the visible spectrum region due to the atomic thickness.\cite{loh2010graphene} The conducting and transparent properties of reduced GOs can be applied in transparent conductors, \cite {zheng2014graphene,nekahi2014transparent} as well as in optoelectronic devices, including photovoltaic \cite{murray2011graphene,yin2010organic,saha2014solution} and light-emitting \cite{saha2014solution,lu2012novel} devices.

\vskip 0.3 truecm

{\bf 5. Hydrogenated graphenes}

Nonstop at the success of graphene fabrication, a Manchester team lead by Nobel laureates A. Geim and K. Novoselov has used hydrogen atoms to modify a highly conductive graphene into a new 2D material-graphane.\cite{elias2009control} The discovery of graphane has opened the flood gates to further chemical modifications of graphene. The addition of hydrogen atoms to carbon atoms in graphene can generate a new system without damaging the distinctive one-atom-thick construction itself. By gradually binding hydrogen to graphene, they are able to drive the process of transforming a semiconducting material into an insulating one. As well as being an insulator that could prove useful in creating graphene-based electronic devices,\cite{gharekhanlou2010bipolar,lu2016ferromagnetism,zhou2014graphene} graphane is also suitable for the application of hydrogen storage.\cite{hussain2014enriching,nechaev2011solid}

Graphane, the fully hydrogenated analogue of graphene, was first theoretically proposed by Sofo et al.,\cite{sofo2007graphane} in which its electronic properties could be controlled by adatom adsorption and desorption. However, it is not easy to be synthesized. The main issue is that hydrogen molecules must first be separated into atoms. This process usually requires high temperature, while it would alter or damage the crystallographic structure of graphene. In 2009, A. Geim 's team has worked out a way to fabricate graphane by passing hydrogen gas through an electrical discharge. \cite{elias2009control} By this method, H atoms are created, and then drift towards a sample of graphene and bond with C atoms. In 2010, Jones et al. have synthesized graphane and partially hydrogenated graphene on both sides and single side by electron irradiation of adsorbates on graphene (H$_2$O and NH$_3$).\cite{jones2010formation} Not long afterward, Wang et al. reported a new method to prepare high-quality and monolayer graphane by plasma-enhanced CVD.\cite{wang2010toward} The great advantage of this method is that it can produce a large-area graphane film with very short deposition time (<5 min) and low growth temperature ($\sim$920 K) compared to the CVD method ($\sim$1270 K).

The essential properties of hydrogenated graphenes with various concentrations and distributions are explored by the first-principles calculations in detail. The geometric structures strongly depend on the significant H-C bonds and the curvature-induced hybridizations in C-C bonds, and so do the electronic properties. The planar structure is changed into the buckled one, leading to the sensitive changes in bond lengths, bond angles and carbon heights. There exist middle, narrow and zero energy gaps even at high adatom concentrations. Furthermore, whether ferromagnetic spin configurations could survive is mainly determined by the adatom distributions. The band structures display the rich features, including the destruction or recovery of the Dirac-cone structure, critical points, weakly dispersive bands, and (C,H)-related partially flat bands. The orbital-projected DOSs are clearly marked by the low-energy prominent peaks, delta-function-like peaks, discontinuous shoulders, and logarithmically divergent peaks. By the detailed analyses, the strong competitions in the critical chemical bondings of sp$^3$s and sp$^2$ are responsible for the diverse properties. In general, hydrogenated graphenes exhibit tunable band gaps as well as ferromagnetism, and they are potential candidates for hydrogen storage applications.

\vskip 0.3 truecm
{\bf 5.1 Single-side adsorption }
\vskip 0.3 truecm

In sharp contrast to oxygens, hydrogens are preferably adsorbed at the top or bottom of carbon-related structures. With the decreasing H-concentration, the geometric structures can be classified into three types, namely zigzag (Z), armchair (A), and chiral (C), based on the arrangement of C atoms between two H atoms (Fig. 22). There are two kinds of well-behaved adsorption distributions, in which hydrogen adatoms are distributed at the same sublattice (Figs. 22(a)-22(f)) or the two different sublattices simultaneously (Figs. 22(g)-22(j)). The latter is predicted to be more stable than the former because of the lower binding energy (Table 3). The half-hydrogenation (Fig. 22(a)), the case which hydrogen atoms on one side of graphane are removed, is referred to as "graphone".\cite{zhou2009ferromagnetism} The main characteristics of geometric structures, including the C-C bond lengths, C-H bond lengths, H-C-C angles, and heights of passivated carbon atoms are strongly dependent on the concentration and distribution of hydrogen atoms, as shown in Table 3. With hydrogen adsorptions, the C-C bond lengths in graphene become non-uniform. The nearest and next-nearest C-C bond lengths are in the ranges of 1.46$-$1.49 $\mbox\AA$ and 1.35$-$1.41 $\mbox\AA$, respectively. As to the C-H bond lengths, they are slightly influenced by H-concentration with the range of about 1.12$-$1.16 $\mbox\AA$. The C-H bond is formed by the sp$^3$s hybridization of four orbitals (2s,2p$_x$,2p$_y$,2p$_z$) of carbon and 1s orbital of hydrogen (evidences discussed later in Figs. 26 \& 27). The pristine planar C-C bond lengths are mainly determined by the $\sigma$ bonding due to the (2s,2p$_x$,2p$_y$) orbitals. These orbitals partially participate in the hybridizations between C and H atoms. Consequently, the weakened $\sigma$ bonding on the plane has an effect to lengthen the nearest C-C bond lengths and thus shorten the second-nearest ones. The H-C-C angles present a significant change from 103.26$^o$ to 107.57$^o$. Apparently, the passivated carbon atoms deviate from the graphene plane. Their heights slightly varry from 0.34 $\mbox\AA$ to 0.38 $\mbox\AA$, leading to the buckled graphene structure. The calculated parameters are in between those of graphene and diamond. The dramatic changes of geometric structures are due to the transformation from sp$^2$ to sp$^3$s hybridizations.

Recently, single-side hydrogenated graphenes have attracted a great deal of attention since they could serve as magnetic materials with ferromagnetic configurations.\cite{zhou2009ferromagnetism,boukhvalov2010stable,lu2016ferromagnetism} The total ground state energy is effectively reduced by the ordered spin arrangement, e.g., $\sim$ 0.8 eV for graphone. The adatom distributions which can induce the local magnetic momenta are investigated in detail. There are spin- and non-spin-polarized energy bands, being determined by the adatom distributions. When hydrogen atoms are adsorbed at the same sublattice, electronic structures exhibit the spin-dependent energy bands, as clearly shown in Fig. 23. The spin splitting obviously appears in the range of ${|E^{c,v}|\le\,2}$ eV, and it is almost negligible at larger state energies. For the spin-up and spin-down states, the energy bands nearest to $E_F$ are fully occupied and unoccupied, respectively, indicating the preferred magnetism. The pair number of spin-split energy bands is determined by hydrogen atoms in a unit cell, e.g., four, two, and  one pairs of splitting energy bands (Fig. 23(a); Figs. 23(b), 23(c) \& 23(e); Figs. 23(d) \& 23(f)), respectively, corresponding to the same adatom numbers (Fig. 22(a); Figs. 22(b), 22(c) \& 22(e); Figs. 22(d) \& 22(f)). All the energy bands are strongly anisotropic; furthermore, valence bands are highly asymmetric to conduction bands about $E_F$. The band-edge states, the extreme or saddle points in the energy-wave-vector space, are situated at the highly symmetric points and the other wave vectors between them. The linear isotropic Dirac cone of graphene is destroyed in most of distributions, mainly owing to the absence of the hexagonal symmetry. However, the deformed cone structure (Figs. 23(d) \& 23(e)) could survive at the higher-symmetry distributions (Figs. 22(d) \& 22(e)). In general, the low-lying energy bands have the weakly dispersive, linear and parabolic dispersions. They determine direct or indirect gaps of ${E_g<1}$ eV, being consistent with the previous calculations.\cite{lu2016ferromagnetism} Specifically, the first kind of energy bands mainly come from the second- and fourth-nearest carbon atoms. This is associated with the strong C-H bonds. Moreover, H atoms make important contributions to the energy bands in the deeper energy range of $-3 \le E^v \le -8$ eV, as revealed in the (C,H)-related partially flat bands (the coexistent red ad blue circles).

The main differences between the non-polarized and polarized spin configurations lie in the low-lying energy bands, including state degeneracy and band gap (Figs. 24 \& 23). The  spin  magnetism is absent under the uniform adatom distribution at the two sublattices (Figs. 22(g)-22(j)). There are no spin-split energy bands separately crossing $E_F$, while they are replaced by the pairs of weakly dispersive energy bands. Energy gap might be largely reduced by the spin splitting, e.g., $E_g$=3.56 eV and 0.86 eV for the former and latter cases at 25\%, respectively (Table 3). As to lower H-concentrations, the low-lying energy bands, which are weakly dependent on hydrogens and passivated carbon atoms, are built from the reformed $\pi$ bondings. Specifically, the predicted polarization-dependent energy bands could be verified from the spin-resolved ARPES.\cite{usachov2015observation,gierz2010giant} The experimental measurements of magneto-optical Kerr effect are available in studying the spin-split absorption spectra.\cite{ellis2013magneto,chen2016time} Moreover, the ferromagnetic configurations are expected to create the spin-polarized currents in transport measurements, indicating potential applications in spintronic devices.\cite{friedman2015hydrogenated,han2014graphene}

The spatial spin densities could provide more information about electronic and magnetic properties. The single-side hydrogenated graphenes exhibit the ferromagnetic configuration (Fig. 25), being different from the anti-ferromagnetic one in zigzag graphene nanoribbons (Fig. 46).\cite{son2006energy,chang2014configuration} The spin-up magnetic moments almost dominate the spatial arrangement, directly reflecting the fully occupied states in weakly dispersive energy bands below $E_F$ (Fig. 23). They only appear at the same sublattice without hydrogen adsorptions, in which the second and fourth neighboring carbon atoms make the main contributions. This clearly indicates that spin momenta are seriously suppressed by the strong orbital hybridizations in C-H bonds; that is, the magnetic structure is mainly determined by the competition between the orbital- and spin-dependent interactions. The typical magnetic moments in a unit cell are about 1-4 ${\mu_B}$ (Bohr magneton), and they depend on H-concentration and -distribution (Table 3). Specifically, graphone can present a large magnetic moment of 4$\mu_{B}$, in which each non-passivated carbon atom contributes 1 $\mu_{B}$. In addition, the similar ferromagnetic configurations are also observed in double-side adsorptions (Figs. 30(a) \& 30(b)). The hydrogen-created magnetic properties on graphene surface could be examined using the spin-polarized STM.\cite{serrate2010imaging,wulfhekel2007spin}

The above-mentioned characteristics are closely related to the strong C-C and C-H chemical bonds. The significant orbital hybridizations could be fully comprehended by analyzing the spatial charge density. The 1s orbitals of H atoms strongly bond the 2p$_z$ orbitals of passivated carbon atoms (dashed rectangles in Figs. 26(a)-26(d)), being further revealed in the drastic density variations associated with the 2$p_z$ orbitals (Figs. 26(g)-26(j); $\Delta\rho$ related to the second-nearest carbons indicated by the black arrows). Furthermore, they only partially hybridize with (2p$_x$,2p$_y$,2s) orbitals. This could be identified from the lower charge densities in the nearest C-C bond, compared with the others (black solid rectangles in Figs. 26(b)-26(d); another evidences in DOS). As a result, there exist the sp$^3$s orbital hybridizations in C-H bonds. All the $\pi$ bondings are absent in graphone system (Figs. 26(a) \& 26(g)), accounting for the destruction of Dirac-cone structure (Fig. 23(a)). With the decrease of H-concentration, the $\pi$ bondings due to the pairs of non-passivated carbon atoms gradually appear (pink rectangles in Figs. 26(b)-26(d); Figs. 26(h)-26(j)). They could reform the $\pi$ and $\pi^\ast$ bands at lower H-concentrations (Fig. 23(f); Figs. 24(b)-24(d)), and even the Dirac-cone structure (Figs. 23(d) \& 23(e)). However, the $\sigma$ bondings do not present very serious changes, so that their energy bands could remain at the almost same energy range. The critical sp$^3$s bondings also induce the similar charge distributions in the double-side adsorptions, as indicated from Figs. 26(e), 26(f), 26(k) \& 26(l).

The magnetic configurations and the competition of sp$^3$s and sp$^2$ bondings are clearly evidenced in the orbital-projected DOSs as many special structures. There are two groups of low-lying strong peaks almost symmetric about the Fermi level, as shown in Fig. 27 (two arrows). For the ferromagnetic systems, they are, respectively, associated with the weakly dispersive spin-up and spin-down energy bands at $E<0$ and $E>0$ (Figs. 27(a)-27(f); Fig. 23). However, two distinct spin states make the same contribution to the non-magnetic systems (Figs. 27(g)-27(j); Fig. 24). This critical difference could be detailedly examined using the spin-polarized STS.\cite{corbetta2012magnetic,berbil2007spin} The graphone system presents a shoulder structure at ${E=-2.1}$ eV and a lot of merged special structures at ${E<-5}$ eV, arising from the $\sigma$ and various orbitals, respectively (blue curve and other curves in Fig. 27(a)). This indicates the full termination of $\pi$ bonding, the dominant sp$^3$ bonding, and the partial contribution of $\sigma$ orbitals in the H-induced hybridization. Concerning the former, the dominance is gradually changed from $\sigma$ orbitals into 2p$_z$ ones with the decrease of H-concentration (red curves). Furthermore, the shoulder structure might become the linear $E$-dependence at certain distributions (Figs. 27(d) \& 27(e)), corresponding to the slightly deformed Dirac cones (Figs. 23(d) \& 23(e)). Most of 2$p_z$ orbitals form the non-well-behaved $\pi$ bonding, although the C-H bonds could survive at deeper energies.

{\bf 5.2 Double-side adsorption }
\vskip 0.3 truecm

Graphane could exist in three types of configurations, namely, chair, boat and twist boat. The first configuration, as shown in Fig. 28(a), is identified to be the most stable system.\cite{sofo2007graphane,wen2011graphane,samarakoon2009chair} As illustrated in Table 4, the double-side adsorptions possess the lower binding energies, compared with the single-side ones. This means that the the double-side adsorbed systems are more stable, in agreement with the previous study.\cite{yi2015stability} Both of them exhibit the similar buckled structures, with the significant changes in the C-C bond lengths, C-H bond lengths, heights of passivated carbons and H-C-C angles (Tables 5 \& 4). Among all the hydrogenated systems, graphane has a uniform geometric structure, including the longest C-C bond length (1.53 $\mbox\AA$), the shortest C-H bond length (1.12 $\mbox\AA$), and the lowest height (0.22 $\mbox\AA$). This originates from the full hydrogenation. The attractive C-H interactions, which coexist on both sides, compete with the $\sigma$ bondings to create the stable structure. Such competition is responsible for the main geometric properties. With the variation of concentration and distribution, the height obviously grows and even reaches 0.57 $\mbox\AA$ at 6.3\% (Z), being related to the highly non-uniform adsorption in a unit cell.

On the experimental side, the hydrogenation of layered graphenes is confirmed by Raman\cite{eng2013highly,luo2009thickness,wojtaszek2011road} and micro-Raman spectroscopies.\cite{byun2011nanoscale} Furthermore, STM images have revealed the hydrogen adsorbate structures on graphene surface, including the top-site positions and the distinct configurations: ortho-dimers, para-dimers, and various extended dimer structures and monomers.\cite{balog2009atomic} Moreover, the H-concentrations could be obtained from Fourier transform infrared spectroscopy combined with CHNO combustion analysis.\cite{pumera2013graphane} The predicted geometric properties of hydrogenated graphenes, the C-H bond lengths, the C-C bond lengths, the heights of passivated carbons and the H-C-C angles (Tables 4 \& 5), deserve further experimental measurements. Such examinations are very useful in identifying the competition between the sp$^3$s and sp$^2$ hybridizations.

Graphane presents a non-magnetic characteristic, in which the spin configuration is fully suppressed by the significant C-H bonds. Band structure, as indicated in Fig. 29(a), belongs to a large direct-gap semiconductor of $E_g=3.57$ eV. This gap is associated with the $\sigma$ and $\pi^\ast$ bands at the $\Gamma$ point. The Dirac-cone structure and the $\pi$ bands disappear, indicating the thorough termination of the $\pi$ bondings. The H-dominated energy bands are absent near $E_F$, while the O-dominated ones could survice and determine $E_g$ (Figs. 16(a)-16(d)). The 1s orbital of each H atom has strong hybridizations with carbon orbitals, so that only the (C,H)-co-dominated bands are revealed at deeper energy, as observed in any hydrogenated graphenes (Figs. 23 \& 24). With the decrease of H-concentration in the zigzag distribution, one or two pairs of weakly dispersive energy bands come to exist (Figs. 29(b)-29(e)), accompanied with  the gradual recovery of the other $\pi$ and $\pi^\ast$ bands. Energy gaps decline and fluctuate, as observed in Table 4. On the other hand, the armchair-distribution systems present zero gap, since there exist Dirac-cone structures near the $\Gamma$ point (Fig. 29(f)). Furthermore, energy gaps are very narrow or vanishing under the chiral distributions (Figs. 29(g) \& 29(h)), mainly owing to the deformed Dirac-cone structure and the almost flat bands near $E_F$.

The double-side hydrogenated graphenes, as shown in Figs. 30(a) \& 30(b), exhibit the spin-polarized energy bands under the same passivated sublattice (Figs. 28(g) \& 28(h)). The similar magneto-electronic properties are revealed in the single-side adsorptions (Figs. 23(a)-23(f)). The splitting spin-up and spin-down energy bands become pairs of valence and conduction bands, so that the ferromagnetic configurations in hydrogenated graphenes belong to semiconductors. However, the ferromagnetic ordering, with the metallic free carriers, could exist in the halogenated graphenes (Figs. 33-36 in Chap. 6.1) and alkali-adsorbed graphene nanoribbons (Fig. 46(d) in Chap. 7.2). Such weakly dispersive bands, which mainly come from the second and fourth neighboring carbon atoms, are clearly evidenced in the spatial spin densities. Two pairs of splitting energy bands and the magnetic moment of 1 $\mu_{B}$ correspond to two H adatoms.

Graphane quite differs from the other hydrogenated graphenes in electronic properties, and so do the special structures in the orbital-projected DOS (Figs. 31 \& 27). The former does not exhibit two groups of low-lying prominent peaks centered at $E_F$ (Fig. 31(a)), since the pairs of weakly dispersive bands are absent (Fig. 29(a)). Only two shoulder structures, which originate from (2p$_x$,2p$_y$) and 2p$_z$ orbitals of carbon atoms, are, respectively, located at $-$1.75 eV and 1.75 eV. That is, energy gap is determined by the $\sigma$ and $\pi^\ast$ bands. Specifically, at deeper energies of ${E<-3}$ eV, many structures due to (2s,2p$_x$,2p$_y$,2p$_z$,1s) orbitals appear simultaneously, clearly indicating the sp$^3$s hybridizations in C-H bonds. Such chemical bondings exist in all the hydrogenated graphenes (Figs. 31 \& 27). However, DOSs dramatically change with various distributions and concentrations (Figs. 31(b)-31(h)), especially for the recovery of 2p$_z$-orbital structures (red curves due to $\pi$ bands near $E_F$) and the existence of spin-split or spin-degenerate strong peaks across $E_F$ (arrows).

Up to now, the hydrogen-induced bandgap opening is confirmed by ARPES\cite{balog2010bandgap,grassi2011scaling} and STS\cite{guisinger2009exposure,castellanos2012reversible} measurements. The rich features of energy bands, the concentration- and distribution-dependent energy gaps, the strong anisotropy, three kinds of energy dispersions, the occupied spin-up electronic states, and the (C,H)-co-dominated energy bands at deeper energies, deserve further experimental examinations. ARPES\cite{balog2010bandgap,grassi2011scaling} and spin-resolved ARPES\cite{usachov2015observation,gierz2010giant} measurements could verify the important differences between graphane and graphone, including the spin-degenerate or spin-split bands and the large or small gaps. Moreover,  STS\cite{guisinger2009exposure,castellanos2012reversible} and spin-polarized STS\cite{corbetta2012magnetic,berbil2007spin} are available in the identification of  the unusual electronic and magnetic properties. Whether two groups of low-lying DOS peaks coexist in the experimental measurements determines the non-magnetic or ferromagnetic configurations.


As mentioned in many articles, hydrogenated graphene materials have potential applications in various fields. The experimental studies show that the process of hydrogenation is reversible and produces high hydrogen density and low mass of graphane. These create outstanding candidates for energy storage systems\cite{hussain2012polylithiated,hussain2012strain,antipina2012high} and open a bright future for hydrogen-powered vehicles.\cite{nechaev2011high} Furthermore, the band gap is engineered under the reversible hydrogenation, i.e., the degree of hydrogenation is available in the modulation of conductive properties. Through this process, graphene-graphane mixed structures offer greater possibilities for the manipulation of semiconducting materials, and they can be potentially applied in the fields of transistor,\cite{gharekhanlou2010bipolar} superconductor,\cite{lu2016ferromagnetism,zhou2014graphene} photonics,\cite{zhou2012hydrogenated} and others applications.\cite{zhou2014graphene} For instance, the bipolar junction transistor based on graphane, in which profiles of carriers and intrinsic parameters are calculated and discussed.\cite{gharekhanlou2010bipolar} Moreover, partially hydrogenated graphene and graphene can be used as a sensing platform to detect Trinitrotoluene (the main component in explosives) in seawater.\cite{seah2014towards} It is expected that future studies on functionalized hydrogenated graphenes will achieve many interesting applications.

\vskip 0.6 truecm
\par\noindent

{\bf 6. Halogenated graphenes}
\vskip 0.3 truecm

The fluorinated graphene has become a rising star of graphene-based systems due to its high stability and interesting properties, e.g., a high room-temperature resistance (>10 G$\Omega$),\cite{nair2010fluorographene,cheng2010reversible} optical transparency,\cite{nair2010fluorographene,jeon2011fluorographene,robinson2010properties} tunable band gaps\cite{liu2012electronic,medeiros2010dft,nair2010fluorographene,jeon2011fluorographene} and magnetic properties.\cite{liu2012electronic,nair2012spin} A typical method of preparing fluorinated graphite is called fluorination, such as the fluoinations using direct gas, \cite{cheng2010reversible,mazanek2015tuning} hydrothermal reaction,\cite{samanta2013highly,wang2012synthesis} plasma,\cite{wang2014fluorination,sherpa2014local} and photochemistry.\cite{gong2013photochemical,lee2012selective} Both F-concentrations and C-F bonds usually depend on the fluorination conditions covering the temperature, pressure, and time of  treatment. The fluorinated graphite could be further reduced to monolayer system by means of the exfoliation methods, mainly including liquid-phase exfoliation,\cite{zhang2013two} modified Hummer's exfoliation,\cite{romero2013fluorinated} thermal exfoliation,\cite{dubois2014thermal} and solvothermal exfoliation.\cite{sun2014solvothermally} The fluorinated graphene could also be synthesized by the direct chemical fluorination of monolayer graphene, e.g., by heating the mixture of graphene sheet and XeF$_2$.\cite{jeon2011fluorographene,robinson2010properties} Moreover, the chlorinated,\cite{li2011photochemical,wu2011controlled} brominated\cite{jankovsky2014towards} and iodinated\cite{yao2012catalyst} graphenes have been produced by the similar methods. However, the experimental synthesis on the astatine-doped graphenes is absent up to now. This might be due to weak At-C bodings or the small binding energies.

The previous theoretical calculations show that the full fluorination of graphene leads to the destruction of the Dirac-cone structure and thus a direct gap of ${E_g\sim\,3.1}$ eV.\cite{medeiros2010dft,liu2012electronic} The fluorination-induced large gap is confirmed by the experimental measurements of the transport\cite{nair2010fluorographene,cheng2010reversible} and optical\cite{jeon2011fluorographene} measurements. Moreover, the partial fluorination can create the diverse electronic and magnetic properties, being sensitive to the concentrations and distributions.\cite{medeiros2010dft,liu2012electronic} The fluorinated graphenes exhibit the metallic or semiconducting behaviors,\cite{medeiros2010dft,liu2012electronic} and the ferromagnetism or non-magnetism.\cite{liu2012electronic} The p-type dopings are evidenced in the experimental measurements of the G-band phonon$^{ref}$ and Fermi level.\cite{walter2011highly} The experimental measurements on the induced magnetic moment indicate the paramagnetic behavior with the specific temperature- and field-dependence, being attributed to the non-uniform adatom clusters.\cite{nair2012spin} On the other hand, there are only few theoretical studies on the Cl-, Br-, and I-doped graphenes.\cite{medeiros2010dft,sahin2012chlorine} Their electronic properties are predicted to strongly depend on the degree of halogenation. Obviously, it is worthy of making a systematic investigation on five kinds of halogenated graphenes, in which the critical mechanisms responsible for the diversified properties and the important differences among them are explored in detail.

The geometric, electronic and magnetic properties are investigated for all the halogen-adsorbed graphenes. The dependence on the concentration and arrangement of halogen adatoms (X = F, Cl, Br, I, and At) is discussed extensively. Absorption energies, bond lengths, buckled or planar structures, carbon- or adatom-dominated energy bands, band-edge states, free hole density, charge distribution, spin configuration and DOS are included in the calculations. Apparently, there exist the diversified essential properties, covering the opening of band gap or the distortion of the Dirac-cone structure, the metallic behaviors due to free holes, the creation of the adatom-dominated or (adatom,C)-co-dominated energy bands, the degeneracy or splitting of the spin-related energy bands, the multi- or single-orbital hybridizations in halogen-C bonds; ferromagnetism and non-magnetism. They are further reflected in a lot of special structures of DOS. Such properties are strongly affected by the distinct kinds of adatoms, in which the critical importances between fluorinated and other halogenated graphenes are investigated in detail.

{\bf 6.1 Halogenation-diversified essential properties}
\vskip 0.3 truecm



The essential properties of halogen-doped graphenes are investigated for the distinct adatoms, concentrations and distributions. The optimal adatom position at the top site, as shown in Fig. 32, is consistent with previous calculations.\cite{medeiros2010dft,xu2015electronic} Obviously, the X-C bond length strongly depends on the atomic number, in which it varies from 1.57 {\AA} to 3.72 {\AA} as F$\rightarrow$At at the same concentration 3.1\% (Table 5). F adatom is much lower than the other halogen adatoms, indicating the serious orbital hybridizations in the F-C bonds. Fluorination results in the buckled graphene layer, in which the heights of the passivated carbons are sensitive to the adatom concentrations. Moreover,  the strength of X-C bonding is directly reflected in the C-C bond lengths and the binding energies. For fluorinated graphenes, the nearest C-C bond lengths are greatly enhanced at high concentration, and their binding energies are the lowest ones among the halogenated systems. The stability of halogenated graphenes presents in the order of F $>$ Cl $>$ Br $>$ I $>$ At, which agrees well with other studies.\cite{medeiros2010dft,zbovril2010graphene} In addition, the configuration parameters are also affected by the adatom distribution, e.g., those of single- and double-side fluorinated graphenes at 50\%.

It should be noticed that the Cl-, Br-, I- and At-absorbed graphenes possess the meta-stable configurations with the much lower adatom heights. This configuration presents a smaller binding energy compared to the most stable one, e.g., the chlorinated systems (Table 5).\cite{medeiros2010dft,klintenberg2010theoretical} The enhanced ground state energy mainly results from the weakened sp$^2$ bondings in the longer C-C bonds. The very strong orbital hybridizations in Cl-C bonds can destroy the Dirac-cone structure and create an observable energy gap (${E_g=1.41}$ eV at 100\%). But for the optimal configurations, all the above-mentioned halogenated systems exhibit the metallic behavior with free holes even at the saturated adsorption. The $p$-type doping has been confirmed by the experimental measurements.\cite{chu2012charge,wu2011controlled,subrahmanyam2009comparative} The drastic changes in concentration-dependent energy bands will be explored in detail. In short, the fluorinated graphenes sharply contrast to the other halogenated ones in geometric configurations, and so do the essential electronic properties.

Electronic and magnetic properties are dramatically altered by graphene halogenations, such as the Dirac-cone structure, the Fermi level, the free carrier density, the energy gap, the adatom-dominated bands, and the spin configurations. For the sufficiently low concentrations, halogen adatoms will donate free holes by the very strong affinity. The Dirac-cone structure is seriously or slightly deformed,  relying on the strength of orbital hybridization in halogen-C bonds, as shown for halogenated graphenes in Figs. 33(a)-33(f). The Fermi level is situated at the valence Dirac cone, so that free holes exist between the valence Dirac point and the Fermi level. The free carrier density is lower than that in alkali-doped graphene (Fig. 2 in Chap. 7). The linearly intersecting bands (Fig. 33(a)) become parabolic bands with two separated Dirac points for the F-adsorbed system (Fig. 33(b)), while the other halogenated graphenes only exhibit the slight distortions (Figs. 33(c)-33(f)). This clearly illustrates the more complicated orbital hybridizations in F-C bonds, being attributed to the lower height of adatom. Such hybridizations induce the thorough changes of all energy bands. Apparently, the quasi-rigid blue shifts of the pristine energy bands are absent. Moreover, the halogen-C bond strength can determine whether the (halogen,C)-co-dominated or halogen-dominated energy bands survive. The rather strong F-C bonds cause the (F,C)-co-dominated valence bands to be located in $-$3.5 eV${\le\,E^v\le\,-2.5}$ eV. However, the Cl-, Br, I- and At-dominated bands are very similar to one another. They are characterized by the almost flat bands in $-$1 eV${\le\,E^v\le\,-2.5}$ eV. The weak mixing with the C-dominated $\pi$ bands directly reflects the bonding strength. The significant splitting of the spin-up and spin-down energy bands is revealed near $E_F$ (red and blue curves), in which the largest energy spacing reaches $\sim$ 0.5 eV. The former has more occupied electronic states (red curves), leading to the ferromagnetic configuration.

The band structure of fluorinated graphenes strongly depend on the F-concentration and distribution, as shown in Figs. 34(a)-34(f). The fully fluorinated graphene has the most stable chair configuration (Fig. 32(a)).\cite{liu2012electronic,leenaerts2010first} The Dirac-cone structure (the $\pi$ bonding) thoroughly vanishes under the overall F-C bondings (Fig. 34(a)). A direct energy gap of 3.12 eV appears at the $\Gamma$ point, in agreement with the previous theoretical calculations\cite{liu2012electronic,leenaerts2010first,medeiros2010dft,robinson2010properties} and experimental measurements.\cite{nair2010fluorographene} It is related to the highest (F,C)-co-dominated valence bands and the lowest C-dominated conduction bands.  All valence bands are co-dominated by  F and C. They are built from their 2p$_x$+2p$_y$ orbitals, as indicated from DOS in Fig. 40(a). Furthermore,  energy dispersions (band widths) are sufficiently strong (wide) in $-$4 eV${\le\,E^v\le\,-1.56}$ eV, but become weak at deeper energies. These clearly illustrate the strong orbital interactions in F-F, C-C and F-C bonds. The former two can create energy bands, leading to the hybridized valence bands by the last one. With the decrease of fluorination, the semiconducting or metallic behaviors are mainly determined by the C-dominated energy bands (Figs. 34(b)-34(f)); that is, the $\pi$-electronic structure (the 2p$_z$-orbital bonding) is gradually recovered. The F-F bonding quickly declines, while the strength of F-C bond almost keeps the same, as implied from the weakly dispersive (F,C)-co-dominated bands in $-$4 eV${\le\,E^v\le\,-2.5}$ eV even at lower concentrations (Figs. 34(f) and 33(b)). In addition, hole doping and ferromagnetism might coexist under certain concentrations and distributions (Figs. 34(c) and 34(e)).

As for the other halogenated graphenes, they possess the metallic energy bands under various adatom adsorptions, as shown in Figs. 35(a)-35(f). Free carriers can occupy the Dirac-cone structure, as well as the halogen-created energy bands with wide band-widths, at sufficiently high concentrations (Figs. 35(a), (b), (e); (f)). This clearly indicates the coexistence of the $\pi$ bonding in carbon atoms and the significant atomic interactions among halogen adatoms. The latter are largely  reduced in the decrease of concentration, so that halogen-dependent energy dispersions become very weak, and they are below or near the Fermi level (Figs. 35(e), 35(d); 33(c)-33(f)). That is, free carriers only exist in the Dirac cone at lower concentrations. The drastic changes in the halogen-induced band widths are also revealed in the concentration-dependent DOS (Fig. 40).

The halogenated graphenes might present the ferromagnetic spin configuration under the same adsorption sublattice, as observed in hydrogenated systems (Chap.5). The fluorinated systems, as shown in Figs. 34(c) \& 34(e),  have the spin-split energy bands across the Fermi level simultaneously; that is, they exhibit the metallic ferromagnetisms. Such bands are dominated by carbon atoms (2p$_z$ orbitals), but not F adatoms. This is clearly identified from the spatial spin density near carbon atoms (Figs. 36(a) \& 36(b), and the spin-split DOSs of C-2p$_z$ orbitals asymmetric about the Fermi level (Figs. 39(c) \& 39(d)). The occupied/unoccupied carrier densities in the spin-up and spin-down bands differ from each other (the dashed red and solid blue curves), in which their difference will determine the strength of the net magnetic moment. 1.94 $\mu_B$ and 0.22 $\mu_B$ are, respectively, revealed at the single-side 50\% (Z) and 16.7\% fluorinations (Table 5). On the other hand, the  metallic ferromagnetisms in the other halogenated graphenes are closely related to the adatom-dominated spin-split energy bands below and near the Fermi level (Figs. 33(c), 33(d), 35(e); 35(h)). This is consistent with the spin density accumulated around adatoms (Figs. 36(c)-36(f)), and the asymmetric and large low-energy DOS of adatom orbitals (Figs. 40(c), 40(d), 40(e); 40(h)). In short, there are certain important differences between halogenated and hydrogenated graphenes in the spin-dependent properties, covering the metallic or semiconducting behaviors, the uniform or non-uniform magnetic-moment contributions from each non-passivated carbon atoms in a unit cell, and the spin density concentrated near adatoms or carbons.

The multi- or single-orbital hybridizations in halogen-C bonds, which dominate the essential properties, could be comprehended from the spatial charge distributions ($\rho$ and ${\Delta\rho}$ in Figs. 37 \& 38). The F-adsorbed graphenes, as shown in Figs. 37(b)-37(g), exhibit the strong covalent F-C bonds (dashed black rectangles), the F-F bonds (purple rectangles), the weakened $\sigma$ C-C bonds (black rectangles), and the deformed $\pi$ bonds (pink rectangles) compared to the pristine graphene (Fig. 37(a)). Apparently, these can create the (F,C)-co-dominated energy bands, the F-induced energy bands, the more complicated $\pi$ and $\sigma$ bands, and the seriously distorted Dirac cone (Figs. 34 \& 39). There exist very complicated charge density differences in three kinds of bonds. The obvious spatial distribution variations on ${yz}$ and ${xz}$ planes clearly illustrate the multi-orbital hybridizations of (2p$_x$,2p$_y$,2p$_z$) in F-C bonds (Figs. 37(e)). Two fluorine adatoms present the significant orbital hybridizations at sufficiently short distance (Figs. 37(e) \& 37(f)) The sp$^3$ bonding is evidenced in a buckled graphene structure, as indicated from the deformations of the $\sigma$ bonding in the nearest C atom (black arrows) and the $\pi$ bonding in the next-nearest one (pink arrows). Electrons are effectively transferred from C to F atoms, leading to the p-type doping with the deformed $\pi$ bonding at low concentration (Fig. 37(g)). On the contrary, the other halogenated graphenes do not have rather strong X-C bonds and thus almost keep the same in the planar $\sigma$ bonding, as shown for chlorinated systems in Figs. 38(a)-38(f). Halogen and carbon are bound to each other by their $p_z$ orbitals. This critical single-orbital hybridization cannot destroy the $\pi$ bonding and the Dirac cone (Figs. 35 \& 40). This is responsible for the $p$-type doping; that is, all of them belong to metals. For high concentrations, halogen adatoms possess the (p$_x$,p$_y$)-orbital hybridizations, even creating conduction bands with many free carriers. In short, the important differences between fluorinated graphenes and other halogenated systems in the essential properties originate from the orbital hybridizations of chemical bonds.

The halogenated graphenes, as shown in Figs. 39 and 40, present the diverse structures in DOSs as a result of the adatom-dependent chemical bondings. Under the full fluorination (Fig. 39(a)), DOS is vanishing within a 3.12 eV energy range centered at the Fermi level, illustrating the destruction of Dirac cone by the strong F-C interactions. (2p$_x$,2p$_y$) orbitals of fluorine and carbon atoms dominate the valence-state DOS at ${E<-1.56}$ eV (pink and blue curves). The effective energy range of the former ($\sim$3.5 eV) is even wider than that of the latter ($\sim$2.5 eV). Furthermore, the shoulder and peak structures due to them appear simultaneously. These further indicate the coexistence of the F-F, C-C and F-C bondings. With the decrease of concentration (Figs. 39(b)-39(f)), there are more special structures in DOS, corresponding to the reduced energy widths of F-dependent valence bands, the creation of carbon $\pi$ bands (red curves), and the partial contribution of F-2p$_z$ orbitals (cyan curves; a lot of low structures). Specifically, F-(2p$_x$,2p$_y$) orbitals can make obvious contribution even for the dilute fluorination, but are revealed at narrow ranges, e.g., certain sharp peaks in Figs. 39(e) and 39(f). Such structures arise from the weakly dispersive (F,C)-co-dominated energy bands (Figs. 34(f) and 33(b)). The enhancement of $\pi$-band width becomes the critical factor in determining the magnitude of energy gap and the metallic behavior. For the sufficiently low concentration, a finite DOS near $E=0$ is combined with one dip structure (arrows in Figs. 39(d) \& 39(e)) or a zero plateau (Fig. 39(f)) at the right-hand side, an evidence of the distorted valence Dirac cone (Figs. 34(e), 34(f); 33(b)). The above-mentioned features are mainly due to the complicated orbital hybridizations of (2p$_x$,2p$_y$,2p$_z$) in F-F, C-C and F-C bonds. It is also noticed that the metallic ferromagnetism is clearly revealed in the occupied/unoccupied spin-dependent DOSs of C-2p$_z$ orbitals near the Fermi level (the solid and dashed red curves in Figs. 39(c) \& 39(d)), in the agreement with the carbon-dominated magnetism (Figs. 36(a) \& 36(b)).

The other halogenated graphenes sharply contrast with fluorinated systems in certain important characteristics of DOS. For any adatom adsorptions, the former have an obvious $\sigma$ shoulder about 3 eV below the Fermi level (blue curves in Figs. 40(a)-40(h)). The $\sigma$ valence bands keep the same after halogenation, further indicating the negligible hybridizations among C-(2p$_x$,2p$_y$) orbitals and adatom ones. All systems present a dip structure near $E=0$ (arrows), in which the blue shift can reach $\sim$0.7 eV. This reflects the distorted Dirac cone formed by the $\pi$ bonding of C-2p$_z$ orbitals. Furthermore, there are obvious special structures due to the $\pi$ bands (red curves). Under the sufficiently high halogenation, the (3p$_x$,3p$_y$) and 3p$_z$ orbitals of Cl adatoms can, respectively, form the conduction and valence bands, and the valence bands (pink and cyan curves in Figs. 40(a) and 40(b)), as revealed in brominated graphenes (Figs. 40(f) and 40(g)). Their band widths decline quickly in the decrease of concentration, being characterized by the sharp peak structures (Figs. 40(c), 40(d), 40(e) \& 40(h)). The (3p$_x$,3p$_y$) or (4p$_x$,4p$_y$)-dominated weakly dispersive energy bands can greatly enhance DOS near the Fermi level. Specifically, the special structures related to the (2p$_z$,3p$_z$,4p$_z$) orbitals appear at the same energies. Their hybridizations are the significant interactions between halogen and carbon atoms, according to the atom- and orbital-dependent features in DOSs.  Moreover, the spin-up and spin-down DOSs  might be very asymmetric about the Fermi level, in which they come from the adatom p$_z$ orbitals (the solid and dashed cyan curves in Figs. 40(c), 40(e); 40(h)), or p$_x$+p$_y$ orbitals (the solid and dashed pink curves in Fig. 40(d)). This corresponds to the adatom-dominated ferromagnetic metals (Figs. 36(c)-36(f)).

{\bf 6.2 Comparisons, measurements and applications}
\vskip 0.3 truecm

Halogenation, hydrogenation and oxidization are quite different from one another in geometric and electronic properties, being closely related to the orbital-dependent chemical bondings. Halogen and hydrogen have the optimal top-site positions, and oxygen is situated at the bridge site. F-C, H-C and O-C bonds present the multi-orbital hybridizations, while the other X-C bonds only possess the critical $p_z$-orbital interactions. The former can create the buckled graphene structure, the destruction or serious distortion of $\pi$ bonding, the obvious change of $\sigma$ bonding, and the (adatom,C)-co-dominated energy bands. Halogenated graphenes and graphene oxides possess the tunable energy gaps, and fluorinated graphenes might be semiconductors or metals. As to the other halogenated graphenes, electronic structures mainly consist of adatom- and carbon-dependent energy bands without the significant mixings; furthermore, they belong to the hole-doped metals. Under certain adatom adsorptions, the halogenated and hydrogenated graphenes, respectively, exhibit the metallic and semiconducting ferromagnetisms. In addition, the single- and multi-orbital hybridizations are, respectively, revealed in alkali- and Al-adsorbed graphenes, accounting for very high-density free electrons in the preserved Dirac cones and adatom-dominated conduction bands (Chaps. 7 \& 8).

The concentrations of halogen adatoms have been measured by various experimental methods. Both measurements from XPS and Raman spectroscopy reveal room-temperature fluorination saturates concentrations of 25\% for single-side exposure and 100\% for double-side exposure.\cite{robinson2010properties} Fluorinated graphenes, with double-side concentrations of 25\% and 50\% produced by exfoliating graphite fluoride with fluorinated ionic liquids, are examined by TEM and atomic force microscopy (AFM).\cite{chang2011facile} The precise control of the F/C ratio of fluorinated graphene is very important for opening the band gap, tuning the electrical conductivity and optical transparency, and resolving the structural transformation. This ratio can be controlled by changing the fluorination conditions, e.g., fluorination agents, temperature and time.\cite{yu2012increased,gong2012one} In additional to fluorine, XPS and EDS (energy-dispersive X-ray spectroscopy) measurements have identified the adatom concentrations in chlorinated and brominated graphenes, revealing the ranges of 18-27\% and 2-8\%, respectively.\cite{zheng2012production} Also, a 8.5\% coverage of Cl-adsorbed graphene is confirmed by XPS measurement.\cite{wu2011controlled} As to the in-plane lattice constant, the expansion by fluorination is verified from electron diffraction pattern.\cite{cheng2010reversible,nair2010fluorographene} The prominent geometric properties of halogenated graphenes, the optimal top-site positions, the adatom-dependent heights, and the fluorination-induced buckling structure, would require further experimental examinations, being useful in the identification of the single- or multi-orbital hybridizations in X-C bonds.

The metallic and semiconducting of halogenated graphenes have been examined by ARPES, optical spectroscopies and electrical transport measurements. The ARPES measurement on fluorinated graphenes shows a $\sim$0.79-eV redshift of the Fermi level below the Dirac point.\cite{walter2011highly} For chlorinated graphenes, the $p$-type doping is verified from the upshift of the graphitic G-band phonon in the Raman characterization.\cite{wu2011controlled} In addition, Br$_2$- and I$_2$-doped monolayer graphenes\cite{chu2012charge} present the $p$-type doping, since the Dirac voltage at the charge neutrality point where the 4-probe resistance is maximum is shifted to higher gate voltages in the increase of molecule concentration. A 3.8-eV energy gap of  high-concentration fluorographene is directly verified from photoluminescence spectrum and near edge X-ray absorption spectrum.\cite{jeon2011fluorographene} Also, the prominent features characteristic of the strong F-C bonds are confirmed by Fourier transform infrared spectroscopy and electron energy loss spectroscopy. A highly fluorinated graphene is verified to be transparent at visible frequencies and have the threshold absorption in the blue range, indicating a wide gap of ${E_g\ge\,3}$ eV.\cite{nair2010fluorographene} This is consistent with the high room-temperature resistance of $>$10G${\Omega}$ in the electrical measurements.\cite{nair2010fluorographene,cheng2010reversible} As for the theoretical predictions, electronic properties, being diversified by the X-X, X-C and C-C bonds, could be test by ARPES and STS measurements, such as the halogen-, carbon- and (X,C)-co-dominated energy bands, the destruction or distortion of Dirac cone, an enhanced DOS accompanied with a a bule-shift dip/plateau structure near the Fermi level, a pair of gap-related shoulder structures, and halogenation-induced many special structures. The spin-polarized spectroscopies are available in examining the spin-split energy bands and the highly asymmetric DOSs near $E_F$.

As a result of the remarkable properties and high stability, fluorinated graphenes are expected to exhibit potential applications in many areas. They have been utilized in lithium-related batteries.\cite{sun2014solvothermally} This material not only possesses abundant fluorine active sites for lithium storage but also facilitates the diffusion of Li$^+$ ions during charging and discharging, leading to high-performance lithium batteries. Edge-fluorinated graphene nanoplatelets could serve as high performance electrodes for lithium ion batteries and dye-sensitized solar cells.\cite{jeon2015edge} Fluorinated reduced GO is synthesized as an interlayer in lithium-sulfur batteries, which greatly improves the open circuit potential, cycling stability and capacitiy.\cite{vizintin2015fluorinated} Fluorinated graphenes could also be used as an electrode material for supercapacitors.\cite{zhao2014fluorinated} Moreover, this system presents many advantages in other fields, such as the biological scaffold for promoting neuro-induction of stem cells,\cite{wang2012fluorinated} the ink-jet printed technologies\cite{nebogatikova2016fluorinated} amonia detection,\cite{katkov2015backside} and adsoprtion of biomolecules.\cite{urbanova2016fluorinated} As for the other halogenated graphenes, they have some promising applications, e.g., Cl-graphene-based field-effect transistors,\cite{zhang2013impact} and brominatied few-layer graphenes for highly apparent conducting electrodes with low optical losses.\cite{mansour2015bromination}

{\bf 7. Alkali-adsorbed graphene-related systems}
\vskip 0.3 truecm

The chemical doings in graphitic systems have been extensively studied both experimentally and theoretically since 1984.\cite{schafhaeutl1840ueber} The alkali atoms, which include Li, Na, K, Rb \& Cs,\cite{dresselhaus1981intercalation,hannay1965superconductivity,magerl1985plane,belash1987superconductivity} are intercalated between graphitic layers. There exist various stage-N layered structures closely related to the alkali concentrations. Upon intercalation, electrons are effectively transferred from the intercalant layers to the carbon ones. Graphite interaction compounds can achieve a conductivity as good as copper, with the proper intercalations. Furthermore, they exhibit superconductivity, with transition temperature below 2 K, e.g., ${\sim\,0.15-1.9}$ K for alkali-doped compounds.\cite{hannay1965superconductivity,koike1980superconductivity,belash1989superconductivity} The similar chemical syntheses are successfully extended to the low-dimensional systems, such as, alkali-doped graphenes,\cite{ohta2006controlling,chen2008charged,yoo2008large,papagno2011large,bianchi2010electron,watcharinyanon2011rb} carbon nanotubes\cite{lee1997conductivity,rao1997evidence,ruzicka2000optical,suzuki2000work,sauvajol2003phonons} and fullerenes.\cite{hebard1991potassium,haddon1992electronic,guan2005direct} Specifically, the superconductivity temperature is largely enhanced to ${\sim\,19-38}$ K for alkali-doped C$_{60}$ fullerides.\cite{hebard1991potassium,ganin2008bulk,takeya2016superconductivity} From the ARPES measurements on alkali-doped graphenes,\cite{gruneis2008tunable,virojanadara2010epitaxial,sugawara2011fabrication,papagno2011large} the Dirac-cone structures have an obvious red shift, clearly indicating the high free carrier density. The creation of many conduction electrons is worthy of detailed examinations. The high doping efficiency could provide practical applications in nanoelectronics,\cite{obradovic2006analysis,wang2008room}  energy storages,\cite{yoo2008large,ataca2008high} and excellent performance batteries.\cite{yoo2008large,wang2009graphene}

From the point of view of geometric structure, each graphene nanoribbon is regarded as a finite-width graphene strip or an unzipped carbon nanotube. Up to now, graphene nanoribbons have been successfully produced by the various experimental techniques. Cutting graphene could be achieved by using metal catalysts,\cite{datta2008crystallographic,ci2008controlled} oxidation\cite{mcallister2007single,fujii2010cutting} lithographic patterning and etching,\cite{han2007energy,chen2007graphene} and sonochemical breaking.\cite{li2008chemically,wang2008room} Furthermore, the available routes in the unzipping of multi-walled carbon nanotubes cover the strong chemical reaction,\cite{kosynkin2009longitudinal,cataldo2010graphene} laser irradiation,\cite{kumar2011laser} metal-catalyzed cutting,\cite{eliias2009longitudinal,parashar2011single} plasma etching,\cite{jiao2009narrow,jiao2010aligned} STM tips,\cite{paiva2010unzipping} high-energy TEM,\cite{kim2010graphene} intercalation and exfoliation,\cite{cano2009ex} and electrochemical unzipping.\cite{shinde2011electrochemical} All the pristine graphene nanoribbons belong to semiconductors because of the finite-size confinement and the spin configuration on edge structure.\cite{son2006energy,chang2014configuration} A lot of theoretical and experimental studies show that the essential properties are dramatically changed by the chemical dopings. The semiconductor-metal transition is revealed in edge-decorated graphene nanoribbons, e.g., Li-, Be-, Na- and K-decorated systems.\cite{chang2014geometric,lin2015adatom,mao2013edge} The Li-adsorbed zigzag systems are expected to present the position-dependent spin configurations.\cite{uthaisar2009lithium} Hydrogen molecules\cite{johnson2010hydrogen} and tin oxide nanoparticles\cite{lin2013graphene} are successfully synthesized on layered systems. A thorough investigation on alkali-adsorbed systems is very useful in understanding how many free carriers created by adatoms.

When the alkali adatoms are adsorbed on graphene surfaces, the dramatic changes in band structure and free carrier density come to exist simultaneously. The critical orbital hybridizations in alkali-carbon (A-C) bonds are directly obtained from the detailed analyses on the alkali- and carbon-dependent energy bands, the slightly perturbed $\pi$ bonding, and the orbital-projected DOSs. The significant s-2$p_z$ hybridizations can account for the rich and unique essential properties. The first-principles calculations show that they dominate the optimal hollow-site position, the planar structure, the highest alkali concentration, and the feature-rich electronic properties. The unoccupied conduction bands become the partially occupied ones after the alkali-atom adsorptions. There are only few alkali-dominated conduction bands, and the other conduction and valence bands arise from carbon atoms. Energy bands, which are co-dominated by alkali and carbon atoms, are absent. Moreover, the free electron density in the occupied conduction bands are identified to be only proportional to the alkali concentration from the detailed investigations on the alkali-adsorbed graphene nanoribbons. These are quite different from the essential properties of the O-, H-, halogen- and Al-adsorbed graphenes, further illustrating the dominance of the orbital hybridizations in chemical bonds.

{\bf 7.1 Alkali-adsorbed graphenes}
\vskip 0.3 truecm
The essential properties of alkali-adsorbed graphenes are investigated for the distinct adatoms, concentrations and distributions. The optimal adatom position, as shown in Fig. 41, is located at the hollow site,\cite{chan2008first,jin2010crossover,praveen2015adsorption,lin2016feature} regardless of the above-mentioned factors. Specifically, two opposite alkali adatoms in the double-side adsorption are centered about the graphene plane. The hollow-site position has been verified by the low-energy electron microscopy.\cite{virojanadara2010epitaxial} The adatom height strongly depends on the atomic number, in which it varies from 1.74 {\AA} to 3.04 {\AA} as Li$\rightarrow$Cs (Table 6). The A-C bond length grows with the increasing adatom radius, ranging from 2.28 {\AA} to 3.36 {\AA}. Furthermore, the Li adatoms present the largest binding energy. In addition, the similar results could also be observed in graphene nanoribbons (not shown). These clearly illustrate that the Li adatoms exhibit the strongest bondings with carbon atoms among the alkali-adsorbed systems. The planar honeycomb structure keeps the same, so that the $\sigma$ bonding due to (2s,2$p_x$,2$p_y$) orbitals of carbon atoms is almost unchanged after alkali adsorption (also sees the vanishing $\Delta\rho$ between two carbons in Figs. 45(a)-45(f)). Apparently, these three orbitals do not take part in the orbital hybridizations in A-C bonds. The nearest C-C bond lengths are slightly lengthened, reflecting the minor modifications on the $\pi$ bondings.

The electronic structures are enriched by the alkali adsorption, especially for free electrons in partially occupied conduction bands. The single-side adsorption on a 4$\times\,$4 cell is chosen for a model study (Fig. 41(f)). All the alkali-adsorbed graphenes, as shown in Figs. 42(b)-42(f), exhibit the alkali- and carbon-dominated energy bands, accompanied with a modified Dirac-cone structure. The former appear as conduction bands, but not valence bands. They have weak energy dispersions at lower concentrations. Furthermore, among the alkali adatoms, the Li ones create the higher-energy conduction bands near $E_F$ (Fig. 42(b)). The latter present the quasi-rigid red shift, compared with those of pristine graphene (Fig. 42(a)). However, there are splitting  energy bands near the M point, as revealed in sliding bilayer graphene with the reduced rotation symmetry (Fig. 47). The higher-energy conduction Dirac cone becomes anisotropic and distorted in the presence of band mixing with the alkali-dominated bands. Apparently, the Fermi level is transferred from the Dirac point to the conduction-cone structure, indicating a lot of adsorption-induced free electrons. The alkali adatoms induce the similar n-type doping effects, as identified in alkali-doped graphites,\cite{dresselhaus1981intercalation,hannay1965superconductivity} carbon nanotubes,\cite{lee1997conductivity,rao1997evidence} and fullerenes.\cite{hebard1991potassium,haddon1992electronic} The carbon-dominated conduction Dirac cone nearest to $E_F$ is occupied by free electrons. They are almost the same for various alkali atoms at lower concentrations. Whether the alkali-related conduction bands are occupied are mainly determined by the free carrier density or the adatom concentration. Specifically, the codominant valence bands of carbon and alkali atoms are absent, while they are created by the O-, H-, F- and Al-absorbed systems. This means that the A-C bonds are relatively weak, compared with the other adatom-carbon bonds.

The alkali-adsorption modifications on electronic structures can be reached by tuning the concentration and distribution of adatom, as clearly indicated in Figs. 43(a)-43(f). For the stable Li-adsorbed systems, the highest concentrations, respectively, correspond to 50\% and 100\% armchair configurations in the single- and double-side adsorptions. The Dirac-cone structure appears near the K or $\Gamma$ point, in which it presents the gapless or separated Dirac points. The alkali-created energy spacing of Dirac points is about 0.1-0.4 eV (Figs. 43(b) and 43(d)).\cite{papagno2011large} This point has a red shift of $\sim$ 1 eV at lower concentrations (Figs. 43(b)-43(d)), while it becomes more than 2 eV at very high ones (Figs. 43(a) and 43(e)). In addition to anisotropy and distortion, the Dirac-cone structure near the $\Gamma$ point might exhibit an obvious splitting. Only the conduction Dirac cones are occupied by free electrons when the adatom concentration is sufficiently low (Figs. 42(b) and 43(d)). One and then two alkali-dominated parabolic conduction bands change into occupied ones with the increasing concentration. Furthermore, they have more strong energy dispersions or wider energy widths.

The alkali-induced free electrons appear between the Dirac point (the band-edge states) and the Fermi-momentum states, in which the carrier density ($\sigma_e$) is the enclosed area in the 2D wave-vector-space. They survive in carbon-related conduction cones and alkali-dominated parabolic conduction bands at sufficiently high adatom concentrations. $\sigma_e$ of the former (the latter) can be directly evaluated from ${k_F^2/\pi}$ (${k_F^2/2\pi}$), when such energy bands have the weakly anisotropic dispersions. The direction-dependent Fermi momenta in the anisotropic conduction cones need to be taken into account, and their average value is  used to estimate the total $\sigma_e$. Another estimation method is to calculate the ${\bf k}$-space area of the occupied conduction bands, while it is very cumbersome because of too much calculation time. Some alkali-adsorbed systems, with the well-behaved energy dispersions, are chosen to comprehend the specific relation between the 2D free electron density and the adtom concentration. Two sets of estimated values are close to each other, as indicated in Table 5, e.g., $\sigma_e\,\sim\, 1.10$-$1.13\times\,10^{14}/$cm$^2$ in the ${4\times\,4}$ case. The ratio of carrier density versus concentration is almost equal to one. That is, each alkali adatom contributes the outmost s-orbital state as free Fermions in the adsorbed system. This will be further examined and identified by introducing 1D alkali-adsorbed graphene nanoribbons, since their free carrier densities can be evaluated very accurately even in the presence of extra factors. The detailed results are discussed in Chap. 7.2. Finally, one free electron contributed by each alkali adatom is deduced to be independent of its kind and distribution, and dimension. $\sigma_e$ can reach ${10^{15}}$/cm$^2$ for alkali concentrations higher than 30\%. This very high 2D free electron density might be very useful in the great enhancement of electric currents, so that the alkali-adsorbed graphenes could be utilized as potential nanoelectronic devices\cite{obradovic2006analysis,wang2008room} and next-generation supercapacitors, such as the ultrafast rechargeable metal-ion battery,\cite{lin2015ultrafast,rani2013fluorinated} and the large reversible lithium storages.\cite{paek2008enhanced,wang2009graphene}

The significant orbital hybridization in A-C bond can create the variation of spatial charge distribution. Apparently, the $\sigma$ and $\pi$ bondings survive simultaneously in any alkali-adsorbed graphene-related systems, as clearly illustrated in Figs. 45(a)-45(f). The $\sigma$ orbital hybridizations, with very high charge density between two carbon atoms, are not affected by the alkali adsorptions, corresponding to the vanishing ${\Delta\rho}$ in Figs. 45(a)-45(f). This accounts for the rigid red-shift of $\sigma$ bands and the absence of  valence bands co-dominated by alkali and carbon atoms. On the other hand, the modifications on the $\pi$ bondings are observable through the charge variations between alkali and carbon atoms (the heavy dashed rectangles), and they are easily observed at high concentrations (Figs. 44(c) \& 44(d)). These indicate that only a single s-2$p_z$ orbital hybridization exists in the A-C bond. Moreover, the spatial charge distributions are almost identical in 2D and 1D systems. Specially, the observable charge variation on x-y plane is revealed between two neighboring alkali atoms under sufficiently high concentrations (inset of Fig. 44(c)), indicating the existence of A-A bonds. In short, the critical orbital hybridization and free electrons contributed by alkali are independent of distinct adatoms, distributions, concentrations, dimensions and edge structures.

The s-2$p_z$ orbital hybridization in the A-C bond is responsible for the unusual structures in DOSs. All the alkali-adsorbed graphenes, as clearly shown in Figs. 45(b)-45(h), present an obvious dip below the Fermi level. This Dirac-cone-induced structure has a large red shift at high concentrations, e.g., $-$2.2 eV and $-$3.4 eV for 50\% and 100\%, respectively (Figs. 45(g) \& 45(h)). There exist the splitting $\pi^\ast$ and $\pi$ peaks. The alkali-created structures are absent below the dip structure; that is, they do not mix with the $\pi$- and $\sigma$-dependent ones. At lower concentrations, a strong symmetric peak near $E_F$, which comes from a weak energy dispersion (Fig. 42), is merged with one of the $\pi^\ast$ peaks (red curves in Figs. 45(b)-45(e)). It changes into a largely extended structure at high concentrations, in which its width is more than 3 eV (Figs. 45(f)-45(h)). Furthermore, its initial shoulder is high and close to the dip structure. This further illustrates the significant s-s orbital interactions in A-A bonds under dense alkali distributions. The orbital-projected DOSs have demonstrated that the $\sigma$ orbitals of carbon atom do not take part in the A-C bond.

The main features of energy bands could be verified by ARPES and STS measurements. The high-resolution ARPES observations on chemically doped graphenes of Li,\cite{virojanadara2010epitaxial,sugawara2011fabrication} Na,\cite{papagno2011large} and K \cite{bostwick2007quasiparticle,ohta2006controlling} atoms have confirmed high density of free electrons in the linear conduction band. These systems reveal a red shift of ${\sim\,1-1.5}$ eV of the $\pi$ and $\pi^\ast$ bands at lower concentration, in good agreement with the present calculations. The further ARPES measurements are required to identify the adatom-dependent energy bands, the large red shift of Dirac point, the distortion and splitting of the Dirac-cone structure, the splitting middle-energy bands near the M point, and the occupied alkali-dominated conduction bands at sufficiently high concentration. These features could also be examined by the STS measurements on the special structures of DOS, including the splitting $\pi$ and $\pi^\ast$ peaks, the dip structure below $E_F$, and the alkali-induced extended structure.

{\bf 7.2 Alkali-adsorbed graphene nanoribbons}
\vskip 0.3 truecm

The detailed first-principles calculations on the alkali-adsorbed graphene nanoribbons are utilized to evaluate the free electron density exactly. They can provide the full information about how much conduction charge contributed by each alkali adatom. Even if the essential properties of the 1D systems are greatly diversified by the geometric structures, the complicated relations among the A-C chemical bonding, the finite-width confinement, the edge structure and the spin configuration will be demonstrated to hardly affect the carrier density. That is to say, this density will be identified to be only proportional to the adatom concentration by examining the various geometric structures and chemical adsorptions.

Two typical types of achiral graphene nanoribbons correspond to armchair and zigzag systems, with the hexagons normally arranged along the edge structure. Their widths are, respectively, characterized by the numbers of dimer and zigzag lines (N$_A$ and N$_z$) (Figs. 46(a) and 46(b)). The finite-size confinement directly induces a plenty of 1D energy bands, as shown in Fig. 47(a) for the N$_A$ = 12 armchair nanoribbon. This system has a direct energy gap of $E_g =$ 0.61 eV at the $\Gamma$ point. In general, $E_g$ declines quickly in the increment of ribbon width.\cite{son2006energy,chang2014configuration} The linearly intersecting bands at $E_F$=0 are absent, being quite different from those in armchair nanotubes (Fig. 11(a)). Energy bands present parabolic dispersions except for few of them with partially flat dispersions within a certain range of $k_x$. They possess the band-edge states at $k_x=$0 $\&$ 1, and the extra ones in between due to the subband anti-crossings. The semiconductor-metal transition occurs after the alkali-adatom adsorption on nanoribbon surface, as clearly revealed in Figs. 47(b)-47(d) for one adatom per unit cell. The Fermi momentum is proportional to the linear free electron density in each occupied conduction band by the relation ${\lambda=2k_F/\pi\,}$. The alkali-adsorbed systems  exhibit the metallic band structures, in which the conduction bands are easily modulated by adatoms (blue circles). The energy dispersions of valence bands are affected by the adatom adsorption, while their number and the dominance of carbon atoms keep unchanged. These arise from the complicated relations between the quantum confinement and the orbital hybridizations in A-C bonds.

The main features of band structures strongly depend on the concentration, relative position, single- or double-side adsorption, and edge structure. There are more occupied conduction bands in the increase of concentration. Five conduction bands are occupied under the single-side adsorption of two Li adatoms at the distinct edges, in which two low-lying ones mainly come from adatoms (Fig. 47(e)). They are further changed by tuning the relative position, e.g., four occupied conduction bands only with one Li-dominated band for two neighboring adatoms ($(3,7)_{single}$ in Fig. 47(f)). Concerning the variation from the single- to double-side absorptions, the conduction bands near $E_F$ present an obvious change, e.g., those of $(3,7)_{double}$ in Fig. 47(g). However, the total sum of the Fermi momenta is hardly affected by the relative position and the single- or double-side adsorption, i.e., the free carrier density is independent of the adatom distribution. The double-side adsorption is relatively easy to create more high carrier density (Fig. 47(h)), being closely related to the adatom concentration. Specifically, for the N$_A$=12 armchair nanoribbon, the highest $\lambda$ corresponds to twenty adatoms in the double-side adsorption (Table 7).

The edge structure can diversify band structures and spin configurations. A pristine zigzag system possesses the anti-ferromagnetic ordering across the nanoribbon and the ferromagnetic configuration at each edge,\cite{son2006energy,chang2014configuration} as clearly illustrated in Fig. 46(c) for N$_z$=8. Each energy band is doubly degenerate in the spin degree of freedom (Fig. 47(i)), independent of the spin-up- and spin-down-dominated configurations. A pair of valence and conduction bands near $E_F$ (blue triangles), which is partially flat at $k_x < $0.5, leads to a direct gap, e.g., $E_g$=0.46 eV at $k_x$=0.5. Energy gap is determined by the strong competition between spin arrangement and quantum confinement. The spatial wavefunctions are localized at edge boundaries, i.e., their energy bands are mostly contributed from the edge carbon atoms. The similar edge-localized states appear in alkali-adsorbed zigzag systems (Figs. 47(j)-47(l)). On the other hand, band structures and spin configurations are dramatically changed by the distribution and concentration of adatom. The spin-degenerate electronic states become split under one alkali adatom near zigzag edge (Fig. 47(j)). Two edge-localized energy bands crossing $E_F$ have an obvious splitting of $\sim$ 0.7 eV, while  another two ones below $E_F$ only exhibit a weak splitting. The former and the latter are, respectively, associated with edge carbon atoms away from and near adatoms. For the spin-split energy bands, there are more free carriers in the spin-down conduction band (the red curve crossing $E_F$), corresponding to the ferromagnetic ordering at one edge without adatoms (Fig. 46(d)). These clearly illustrate that spin configurations are seriously frustrated by the alkali-carbon interactions. The spin degeneracy and the anti-ferromagnetic configuration are further recovered, when adatoms are situated at the ribbon center. A pair of partial flat bands appears near $E_F$, in which only the conduction band has free electrons (Fig. 47(k)). Furthermore, the absence of spin distributions near adatoms, as shown in Fig. 46(e), presents the magnetic suppression. According to the specific relation between spin distribution and adatom position, the spin-dependent properties are deduced to be absent for two adatoms at distinct edges, e.g., the spin-degenerate energy bands without magnetism in Fig. 47(l). Specifically, two edge-localized energy bands are merged together at $E^v$ = $-$1.35 eV for $k_x < $0.5. In addition, the typical magnetic momenta of the edge carbon atoms are about 0.10-0.15 $\mu_B$ under the ferromagnetic ordering along the zigzag edge. In short, three kinds of magnetic configurations in alkali-adsorbed zigzag systems, the anti-ferromagnetic ordering across the ribbon, ferromagnetic ordering only along one edge and non-magnetism, are mainly determined by the adatom positions. All the alkali-adsorbed zigzag systems are 1D metals, accompanied with the diverse band structures. For the second kind of spin configuration, the spin-split energy bands, with the different free carrier densities, are expected to create the spin-polarized currents under the transport measurements, indicating potential applications in spintronic devices.[refs]

By the detailed calculations and analyses, the total free electron density in conduction bands below $E_F$ is linearly proportional to the adatom concentration. This is independent of the adatom kind, relative position, single- or double-side adsorption, edge structure and ribbon width, as clearly shown in Table 7. It directly reflects the fact that the orbital hybridizations in A-C bonds (or the $\pi$ bondings) almost keep the same under various adatom adsorptions. Specifically, the alkali adatoms contribute the outmost s-orbital electrons as free carriers in adsorbed systems. For N$_A$=12 alkali-adsorbed systems, the free electron density is $\lambda\sim$ 2.31$\times\,10^7$ e/cm for a single adatom in a unit cell, and it can reach  $\lambda\sim$ 4.64$\times\,10^8$ e/cm for the double-side adsorption of twenty adatoms. A simple linear relation is also suitable for the highest adatom concentration. Apparently, the almost same spatial charge distributions are revealed in alkali-adsorbed graphene and graphene nanoribbon (Figs. 44(a)-44(f) and 44(g)-44(h)). One s-orbital from each alkali adatom in forming the Fermi sea should be proper for alkali-adsorbed graphenes. That is, the 2D electron density is deduced to be dominated by the adatom concentration, but not the distribution configurations. On the other hand, the charge transfer between alkali and carbon atoms could also be obtained from the Bader analysis. However, it is very sensitive to the changes in the kind, position and concentration of adatom, clearly indicating that this analysis cannot be utilized to evaluate the free carrier density in adatom-adsorbed systems. The free electron density is greatly enhanced with the increasing adatom concentration, so that the electrical conductance is expected to behave so. The alkali-adsorbed graphene nanoribbons might be promising materials in nanoelectronic devices. \cite{obradovic2006analysis,wang2008room}

The main features of special structures in DOS are drastically changed by the dimensions, e.g., the important differences between 1D and 2D systems (Figs. 48 and 45). The alkali-adsorbed graphene nanoribbons exhibit many asymmetric peaks and few symmetric ones, as clearly shown in Figs. 48(a)-48(h). The former and the latter respectively, present the square-root and delta-function-like divergent forms as a result of the parabolic and partially flat energy dispersions. Their intensities are proportional to the inverse of curvature and the dispersionless k$_x$-range. For a pristine armchair system, an energy gap, with zero DOS, appears between one pair of opposite-side anti-symmetric peaks (Fig. 48(a)). Its value keeps the same after alkali adsorption, but becomes an energy spacing of valence and conduction bands. There are somewhat adsorption modifications on the carbon-dependent peak structures (Figs. 48(b) and 48(a)). A red shift of DOS could be roughly observed from the change of $E_F$. Specifically, adatoms induce the pronounced asymmetric peaks near $E_F$, owing to the alkali-dominated occupied or unoccupied conduction bands. The energy and number of special structures rely on the distribution and concentration (Figs. 48(c) and 48(d)). Such peaks might merge with the carbon-dependent ones. The low-lying peaks are further diversified by the distinct edge structures. The zigzag systems can present the delta-function-like symmetric peaks due to the edge-localized energy bands, being sensitive to the spin configurations. The anti-ferromagnetic, ferromagnetic and non-magnetic zigzag systems, respectively, exhibit a pair of symmetric peaks (blue triangles in Figs. 48(e) and 48(g)), three peaks (Fig. 48(f)), and a merged peak (Fig. 48(h)). The intensity is reduced in the spin-split energy bands (a pair of symmetric peaks across $E_F$ in Fig. 48(f)).

The experimental verifications on the unique electronic properties of the alkali-adsorbed graphene nanoribbons are absent up to now. The ARPES measurements, as done for pristine systems,\cite{ruffieux2012electronic,sugawara2006fermi} could be utilized to examine the occupied valence and conduction bands close to $E_F$, such as the alkali-dominated conduction bands, the carbon-related conduction and valence bands, the distribution-, concentration- and edge-dependent bands, and the spin degeneracy of partially flat edge-localized bands. They are very useful in identifying the single s-2$p_z$ orbital hybridization in the A-C bond and the specific spin configurations. These two critical pictures could also be identified from the STS measurements on the special peak structures,\cite{huang2012spatially,sode2015electronic,chen2013tuning} e.g., the alkali-induced asymmetric peaks near $E_F$, many carbon-created asymmetric peaks in a whole energy range, and the low-lying symmetric peaks of edge-localized bands with or without spin splitting.

{\bf 8. Metallic adatom-doped systems}
\vskip 0.3 truecm

Each Al atom has three outer-orbital electrons and thus creates the diverse chemical bondings. The Al-related graphene systems are one of the widely studied materials. The high sensitivity of Al atoms is quite suitable for the energy and environment engineering. Very importantly, Al atoms have been identified to play a critical role in the great enhancement of current density in an aluminium-ion battery,\cite{lin2015ultrafast} where the predominant AlCl${_4^-}$ anions are intercalated and de-intercalated between graphite layers during charge and discharge reactions, respectively. Also, an aluminum-ion battery composed of fluorinated natural graphite cathode is demonstrated to have very stable electrochemical behavior.\cite{rani2013fluorinated} As to the theoretical predictions, when a certain C atom on the hexagon lattice of graphene is replaced by Al, this system could serve as a toxic gas sensor.\cite{chi2009adsorption,ao2008enhancement} The formaldehyde (H$_2$CO) sensor is attributed to the ionic bonds associated with the charge transfer and the covalent bonds due to the orbital overlaps.\cite{chi2009adsorption} Carbon monoxide can be detected by the drastic change in the electrical conductivity before and after molecule adsorption.\cite{ao2008enhancement} Also, the Al-subtituted systems are expected to be a potential hydrogen storage material at room temperature, in which the functional capacity is largely enhanced by the Al atoms.\cite{ao2009doped} The similar functionality is revealed in Al-adsorbed graphenes.\cite{ao2010high} The rich physical and chemical phenomena are closely related to various orbital interactions. The multi-orbital hybridizations between Al adatoms and graphene deserves a thorough investigation.

The bismuth-related systems have been becoming one of the main-stream materials, especially for the topological insulators.\cite{hsieh2009tunable}  Bulk bismuth, with rhombohedral symmetry, is a well-known semimetal, being similar to ABC-stacked graphite.\cite{lin2012high,zhou2014interpretation,mcclure1969electron} Bismuth is widely studied in the fields of environment, biochemistry and energy engineering. For example, Bi-based nanoelectrode arrays can detect heavy metals,\cite{wanekaya2011applications} polycrystalline bismuth oxide film serves as a biosensor,\cite{shan2009polycrystalline} and bismuth oxide on nickel foam covered with thin carbon layers is an anode of lithium-ion battery.\cite{li2013bismuth} Recently, bismuth-atom adsorptions on monlayer graphene are clearly observed at room temperature.\cite{chen2015long,chen2015tailoring} The 4H-SiC(0001) substrate, corrugated buffer layer, slightly deformed monolayer graphene and temperature-dependent adatom configuration are identified from the STM measurements. An increase in temperature, a hexagonal array of Bi adatoms is transformed into triangular and rectangular nanoclusters. Furthermore, the STS measurements on the dI/dV spectrum confirm the red shift of Dirac point, free conduction electrons, and Bi-induced structures. These shed the light on controlling the nucleation of adatoms and subsequent growth of nanostructures on graphene surface. The critical roles played by the configuration of Bi adatoms, buffer layer, and substrate will be explored in detail. In addition, some theoretical studies are focused on the geometric and electronic structures of Bi-substituted\cite{akturk2010bismuth} and (Bi,Sb)-intercalated graphenes.\cite{hsu2013first}

Three active electrons of (3s,3p$_x$,3p$_y$) in each Al adatom are predicted  to present the multi-orbital hybridizations after adsorption on graphene surface. The Al-adsorbed graphenes possess the Al-C, C-C and Al-Al bonds. The first-principles calculations shows that the critical interactions of 3s/(3p$_x$,3p$_y$) and 2p$_z$ orbitals will determine the optimal adatom position, the n-type doping, the reformed $\pi$ and $\pi^\ast$ bands, the adatom-dependent valence and conduction bands, and the extra special structures in DOS. Furthermore, the important similarities and differences between the Al- and alkali-adsorbed graphenes cover the red shift of Dirac cone, free electron density, and adatom-induced energy bands. These mainly arise from the multi- or single-orbital hybridizations in adatom-C bonds. The theoretical predictions could be examined by various experimental methods. As to bismuth-adsorbed monolayer graphene, the strong effects of SiC substrate on the geometric and electronic properties are explored using the semi-empirical DFT-D2 correction of Grimme.\cite{grimme2006semiempirical} The six-layered substrate, corrugated buffer layer, and slightly deformed monolayer graphene are all simulated. Adatom arrangements are optimized through the detailed analyses on the ground-state energies, bismuth binding energies, and Bi-Bi interaction energies of different heights, inter-adatom distances, adsorption sites, and hexagonal positions. The Van der Waals interactions between buffer layer and monolayer graphene are shown to dominate the most stable and meta-stable structures. Moreover, the Bi-induced changes in DOS include a finite value at the Fermi level, a red shift of dip structure, and a adatom-dependent peak at lower energy. The calculated results can account for the STM and STS measurements.\cite{chen2015long,chen2015tailoring}

{\bf 8.1  Al-adsorbed graphenes}
\vskip 0.3 truecm

The optimal Al adsorption position is the hollow site,\cite{chan2008first,ao2010high} regardless of concentration and distribution. The highest Al-concentration in the single- and double-side adsorptions corresponds to a ${2\times\,2}$ unit cell without buckled structure (Fig. 49(a)). The C-C bond length nearest to the adatom is expanded only within 1.6\% for various concentrations and distributions, compared with that of pristine graphene (Table 8). The Al-C bond length and adatom height are, respectively, in the ranges of 2.54-2.57 \AA $\,$and 2.09-2.13 \AA. Both Al and Na adatoms have the almost same geometric structure (Table 6). These features clearly indicate the slight changes in the $\pi$ and $\sigma$ bondings of carbon atoms (Fig. 51), as observed in alkali-adsorbed graphenes. The Al- and alkali-related graphenes are expected to have certain similar properties.

The carbon-dominated energy bands are dramatically shifted by the Al-adatom absorption. The Dirac-cone structure, as indicated in Figs. 50(a)-50(f), is almost preserved in the Al-absorbed graphene. However, it might have the slightly separated Dirac points, or the splitting energy bands. This structure presents a quasi-rigid red shift; so that the Fermi level is located at the conduction Dirac cone. The energy difference between the Dirac point and the Fermi level grows with the increasing Al-concentration, indicating more electrons transferred from adatoms to carbons. Such free carriers only occupy the conduction bands arising from the C-2p$_z$ orbitals, being almost independent of the Al-dependent energy bands (blue circles). In addition, part of them can survive in the higher conduction bands at the highest concentration (Fig. 50(f)). The 3s and (3p$_x$,3p$_y$) orbitals of Al adatoms can built the valence and conduction bands in the ranges of $-$4.2 eV${\le\,E^v\le\,-}$2.0 eV and 0${\le\,E^c\le}$2 eV, respectively (details in DOS of Fig. 52(b)-52(f)). Their energy dispersions become weak at low concentrations ($\le$5.6\%; Figs. 50(c)-50(e)).

Both aluminum- and alkali-adsorbed systems exhibit the quasi-rigid red shift in the Dirac cone and $\sigma$ band. Specifically, the almost same conduction cone is revealed at low concentration, as shown in Fig. 50(e) and Figs. 42(b)-42(f) under 3.1\%. The free carrier density in conduction cone, the wave-vector space enclosed by this cone, is identical for these two systems. Al and alkali adatoms can create the same conduction electrons; that is, each Al adatom provides one electron as conduction carrier. However, there exist certain important differences between them, especially for high adatom concentration. The adatom-dependent valence bands are absent in alkali adsorption, since the outmost s-orbital only forms the conduction bands. The highest adatom concentration is about 25\% under the double-side Al adsorption, while it can reach 100\% in the alkali doping. The alkali-doped graphenes can generate more high electron density, in which free carriers occupy the carbon- and alkali-dominated conduction bands simultaneously.

The multi-orbital hybridizations of Al-adsorbed graphenes are directly revealed in the spatial charge distributions. The $\pi$ and $\sigma$ bondings of carbon atoms could survive under various concentrations and distributions, as indicated in $\rho$ of Figs. 51(a)-51(c). Their slight variations are responsible for the quasi-rigid shifts of the Dirac cone and the $\sigma$ band. According to the variation of charge density (${\Delta\rho}$ in Figs. 51(d)-51(f)), the 3s-orbital electrons present a redistribution between Al and the six nearest C atoms (dark blue; triangle), revealing a significant hybridization of 3s and 2p$_z$ orbitals (grey rectangle; red ring in the inset). The  charge distribution of (3p$_x$,3p$_y$) orbitals extends from the light blue ring near Al to the red ring between C and Al atoms. These orbitals make less contributions to the Al-C bonds. Specifically, at higher concentration, there also exist observable charge distributions on x-y plane between two neighboring Al atoms (blue region in the right-hand-side plot of Fig. 51(d)). This presents the evidence for the 3s-3s and (3p$_x$,3p$_y$) interactions in Al-Al bonds. The multi-orbital hybridizations of 3s and (3p$_x$,3p$_y$), respectively, lead to the Al-dependent valence and conduction bands.

The multi-orbital hybridizations in C-C, Al-C and Al-Al bonds can diversify the special structures in DOS. The 2p$_z$- and (2p$_x$,2p$_y$)-dependent DOSs of carbon atoms are dramatically changed by the Al adsorption, as clearly shown in Figs. 52(a)-52(f). The V-shape structure due to the Dirac cone (the red curve) and the $\sigma$-band shoulder structure (the blue curve) present an obvious red shift, as measured from the Fermi level. The finite DOS at $E_F$ principally comes from the $\pi^\ast$-electronic states, but not the Al-related conduction ones. The prominent $\pi$ peak becomes several sub-peaks because of the zone-folding effect, in which part of them are merged with the sharp peaks of the Al-3s orbitals (the green curve). This further illustrates the significant hybridization of 3s and 2p$_z$ orbitals in Al-C bonds, being associated with the Al-dependent valence bands. The $\pi^\ast$ peak exhibits the similar splitting, and the adsorption-induced subpeaks are distributed more close to $E_F$. Such structures might combine with two strong peaks arising from the Al-(3p$_x$,3p$_y$) orbitals (the cyan curve). Electrons in these two orbitals are transferred from Al to C, and they change into conduction carriers in Al-doped graphenes. The widths of the Al-dependent energy bands are widened at higher concentrations (Figs. 52(f) and 52(b)), an evidence of the Al-Al bonds.

The predicted geometric and electronic properties of Al-adsorbed graphenes are worthy of further experimental examinations. STM and TEM are powerful tools to determine the optimal adsorption position and the Al-C bond length. Whether the highest Al-concentration is about 25\% could be examined from the TEM, AFM, XPS, and EDS measurements, as done for  the halogenated graphenes (Chap. 6). The ARPES measurements are available in identifying the red shift of the Dirac cone and the $\sigma$ band, and the Al-dependent valence bands. The main features of DOSs, the slightly distorted V-shape structure, the red shift of the Fermi level, the splitting $\pi$ and $\pi^\ast$ peaks, the preserved $\sigma$ shoulder, and the adatom-created valence and conduction peaks, could be further verified by STS. All the measurements are useful in comprehending the specific orbital hybridizations and the doping effects. Specially, the Al-induced high free carrier density might have high potentials for future technological applications, e.g., high-capacity batteries,\cite{lin2015ultrafast,rani2013fluorinated} and energy storage.\cite{ao2009doped,ao2010high}

{\bf 8.2 Bi-adatom adsorption configurations on SiC-based graphene surface}
\vskip 0.3 truecm

The optimal Bi-adsorption configurations are strongly related to the substrate and buffer layer. Specifically, the atomic structure of buffer layer could be an intriguing subject. The previous work\cite{van1975leed} shows that the graphite-like layers
are bonded to the surface of SiC(0001). An important controversy lies in the STM image of $6\times6$ hexagonal reconstruction of the buffer layer,\cite{berger2006electronic,brar2007scanning,mallet2007electron} being different from the ($6\sqrt{3}\times6\sqrt{3}$)R$30^{\circ}$ reconstruction revealed in LEED patterns.\cite{berger2006electronic,chen2005atomic,de2007epitaxial} This difference might arise from the various substrates. In these reports, the geometric structure is determined from the evidences in STM/LEED measurements and numerical calculations. Recently, based on the comparison between experimental analyses
\cite{chen2015long,chen2015tailoring} and theoretical calculations,\cite{mattausch2007ab,varchon2008ripples} the existence of the long-range ripple structure related to the ($4\sqrt{3}\times4\sqrt{3}$)R$30^{\circ}$ reconstruction is obtained. The in-plane lattice constants in this work are $3.06$ {\AA}, $2.30$ {\AA}, and $2.65$ {\AA} for SiC, buffer layer, and graphene, respectively. They play a critical role in the optimal geometric structures. The buffer layer is identified to be a rippled shape. This indicates that the non-uniform Van der Walls interactions between buffer layer and monolayer graphene might dominate the Bi adsorption sites and the adatom nano-strucutres (discussed in detail later).

The six-layer Si-terminated 4H-SiC (0001) substrate is taken into account in order to accurately simulate the Bi-adsorbed monolayer graphene. The optimized results show that the four-layer substrate (region I in Fig. 53), being reduced from the six-layer one, presents the almost identical geometric properties. Second, a buffer layer is in a periodic ripple shape after relaxation (region II of Fig. 53), clearly revealing the significant bonding of carbon atoms at troughs with silicon ones. This periodic corrugation is consistent with the STM measurements.\cite{chen2015long} Third, the nearly flat monolayer graphene possesses a slightly extended C-C bond length of 1.50 {\AA}, as shown in region III. The interlayer distance between the monolayer graphene
and the buffer layer changes from 3.21 {\AA} to 5.45 {\AA} at the crests and troughs, respectively. This clearly illustrates that there exist the non-uniform Van der Waals interactions between them, and thus dominates the distribution of the Bi adatoms. Bismuth atoms can be adsorbed on monolayer graphene in self-consistent calculations (region IV in Fig. 53). Up to date, two kinds of adatom distributions have been observed based on the different experimental environments, namely a uniform hexagonal distribution and bismuthnanoclusters.\cite{chen2015long,chen2015tailoring} They are examined to be dependent on the adsorption energies.

The adsorption energy $\Delta E$ is very useful for understanding the optimized geometric structure, characterizing the reduced energy due to the bismuth adatomsn on graphene. It is defined as $\Delta E = E_{sys}-E_{SiC}-E_{buf}-E_{gra}$, where $E_{sys}$, $E_{SiC}$, $E_{buf}$, and $E_{gra}$ are the total energies of the composite system, silicon carbide substrate, buffer layer, and pristine monolayer graphene, respectively. The bismuth atom on the hollow site of the carbon hexagon has the highest adsorption energy (Table 9), leading to the less stable configuration. The bridge and top sites have comparable adsorption energies in which the former possesses the lower one. This suggests that bismuth atoms are most likely to be observed at the bridge sites. Moreover, the optimal distance $h$ between the adatoms and graphene surface depends on absorption sites. The shorter distance at the bridge sites indicates the stronger interactions, being responsible for the structural stability. Importantly, the adatom height of 2.32 {\AA} is consistent with that of STM measurements.\cite{chen2015long}

The distribution of bismuth atoms can be clearly explored by calculating the ground-state energy. The ground-state energies for the bridge sites in different hexagons along the periodic armchair direction are calculated, and they can provide the full information needed to determine the most stable position in monolayer graphene, as shown in Fig. 54(a). These positions are further divided into the three regions, red, yellow, and gray ones. The red hexagonal region is closest to the buffer layer with a distance of 3.21 {\AA}, corresponding to the lowest ground-state energy. This energy is set to be zero in order to compare it with the other bridge sites. The bridge sites between red and yellow sticks have higher ground-state energies of about $17-23$ meV (arrows in Fig. 54(b)). The other parts associated with the gray hexagons possess comparable ground-state energies to one another, exhibiting the highest energy difference in the range of $48$-$52$ meV. This illustrates that Bi adatoms are hardly transported from the red to the other regions at room temperature. The Bi atoms are most stable at the red hexagonal rings, since they have the strongest Van der Walls interactions with the crests of the buffer layer. The energy barrier of $\sim 50$ meV creates a potential well for the transportation of Bi adatoms; therefore, it will play an important role in the dramatic change of adatom distribution during the variation of temperature. The shortest, longest, and average distances between two Bi adatoms are, respectively, $14.2$ {\AA}, $18.1$ {\AA}, and $15.9$ {\AA}. The latter two are, respectively, indicated by brown and black arrows between neighboring red regions. This means that the most stable interatomic distance corresponds to the range of $14.2$-$18.1$ {\AA}. The simulated pattern deserves a closer examination with the STM measurements.\cite{chen2015long,chen2015tailoring} The stable uniform distribution in room temperature should be very useful in the future applications in energy engineering.\cite{su2015bismuth}

There are metastable nano-structures observed by STM after the annealing treatment,\cite{chen2015tailoring} including triangular and rectangular arrangements. These special configurations could be understood by a look at the ground-state energies and Bi-Bi interaction energies of optimized structures, as shown in Table 9. Various configurations of Bi nanoclusters are taken into account for different adatom numbers (from 1 to 6 in Fig. 55(a)), such as the internal structures and the adsorption positions associated with the rippled buffer layer. The Bi adatoms are set at the bridge sites of a hexagonal ring to have stronger interactions with graphene. The optimized distance between two isolated Bi atoms is about 2.88 {\AA}, which means unstable nano-structures for the nearest two bridge sites. On the other hand, the optimized adsorption positions for 2- and 3-adatom nanoclusters are the red hexagonal ring (Fig. 54(b)), possessing lower total energies of ~50 meV compared to those clusters at the grey hexagonal ring (close to the trough of ripple). For the 4-, 5-, and 6-adatom nanoclusters, some Bi adatoms are adsorbed at the second stable bridge site (between the red and yellow carbons) in order to avoid the repulsive force between the nearest two Bi adatoms. For various adatom numbers, the higher the number of adatoms is, the lower the $E_{sys}$. The stronger attractive Bi-Bi interactions, compared to the Bi-C bonds, are responsible for this result. The latter, $E_{Bi-C}=-1.01$ eV, is obtained by subtracting the $E_{gra}$ from that of single-Bi-adsorbed graphene. The reduced energy due to the Bi-Bi interactions is evaluated from $\Delta E_{Bi-Bi} = (E_{sys}-E_{gra}-E_{buf}-nE_{Bi}- \sum\limits_{i=1}^{n} E_{Bi-C})/n$.

Among the various Bi-adatom numbers, $\Delta E_{Bi-Bi}$'s in the 3-, 4-, 5-, and 6-adatom nanoclusters are much lower than that of the 2-adatom one. This indicates that the former four are metastable structures. The experimental measurements show that most of the nanoclusters are composed of 3 and 4 adatoms. If the bismuth coverage is insufficient, the 5- and 6-adatom nanoclusters are absent. A critical factor is that the bismuth atoms from the large-scale hexagonal array need to overcome the energy barrier ($\sim 50$ meV in Fig. 54(a)) in order to form these patterns. The temperature increase causes the transportation of Bi adatoms between two neighboring unit cells (red hexagons in Fig. 54(b)), where they have the same probability of moving toward or away from the hexagonal unit cell. The hexagonal symmetry results in the 4-adatom-dominated nanoclusters. However, the relatively few vacancies in the large-scale hexagonal array could create a non-uniform transport environment and thus the 3-adatom nanoclusters. It should also be noted that the nearest distance between two bismuth clusters is about $16$ {\AA}, revealing the energetic favorable adsorption sites of them situated at the red hexagonal rings in Fig. 54(b). This further illustrates that the buffer layer plays an important role at various temperatures.

DOS, as shown in Fig. 55(b), directly reflects the primary electronic properties. For a hexagonal array of Bi-adsorbed graphene, DOS is finite at $E=0$; it possesses a dip at low energy, and a peak structure at $E \sim -0.6$ eV. The first feature means that there exists a certain amount of free carriers. The second one at $-0.2$ eV is attributed to the existence of Dirac-cone structure. The third one originates from the contribution of bismuth adatoms. Such electronic properties are the critical characteristicsin identifying the Bi-adatom adsorption. The STS measurements can provide an accurate and efficient way to examine the theoretical predictions. The recent measurements\cite{chen2015long,chen2015tailoring} show the main features similar to those of the DFT calculations, covering a DOS at $E=0$, a small dip at low energy, and a peak at $\sim -0.7$ V originating from the bismuth atoms. The above-mentioned comparison of electronic properties might promote potential applications in electronic devices.\cite{cho2011insulating}

{\bf 9. Concluding Remarks}
\vskip 0.3 truecm

A systematic review has been made on the geometric and electronic properties of graphene-related systems. The first-principles calculations are used to explore the rich essential properties. Both structure-enriched and adatom-doped graphenes can exhibit the semiconducting, semi-metallic and metallic behaviors. The dramatic transitions of electronic spectra are achieved by sliding, rippling and chemical adsorption. The responsible mechanisms come from the $sp^2$ and $sp^3$ bondings in the angle-dependent C-C bonds, the specific orbital hybridizations in adatom-adatom bonds, and the simple and complex orbital hybridizations in various C-adatom bonds. They dominate the optimal ripple, the various adatom positions, and the stable adsorption distribution. The hexagonal symmetry, rotational symmetry, inversion/mirror symmetry, layer number, boundary condition, distribution and concentration of adatoms, Van der Waals interactions, and spin configuration are taken into consideration. The physical and chemical pictures are very useful in discussing materials science and applications. The calculated results are consistent with those from other theoretical calculations and validated by the experimental measurements, while most of predictions require further experimental verifications. The theoretical framework could promote the future studies on other 2D materials. Specifically, by the detailed analyses on the first-principles calculations, the critical orbital hybridizations in various chemical bonds are obtained from the atom-dominated energy bands, the spatial charge distributions, and the orbital-projected DOSs. The current analysis method could be further developed and generalized to the chemically doped materials.

The layered graphenes are formed by the intralayer $\sigma$ and $\pi$ bondings, and the interlayer Van der Waals interactions. The intralayer and interlayer atomic interactions of $2p_z$ orbitals are responsible for the rich electronic properties. There are five kinds of energy dispersions, depending on the stacking configuration and layer number. The trilayer AAA-, ABA-, ABC- and AAB-stacked graphene, respectively, present the unique low-energy spectra: (1) three vertical cones with separated Dirac points, the overlap of monolayer structure, (2) a slightly separated Dirac-cone structure and two pairs of parabolic bands, a superposition of monolayer- and bilayer-like energy bands, (3) the partially flat (surface-localized), sombrero-shaped and linear bands and (4) the oscillatory, sombrero-shaped and parabolic bands. Also, the middle-energy $\pi$ and $\pi^\ast$ bands have the specific saddle points. The critical points in the energy-wave-vector space induce many van Hove singularities, including the V-shaped form, the shoulder structures, the delta-function-like peaks, the 1D-like asymmetric peaks, and the logarithmically divergent peaks. Only the AAB stacking is a narrow-gap semiconductor, and others are semimetals. This lower-symmetry system has more complicated interlayer atomic interactions, as seen from the tight-binding model calculations.\cite{do2015configuration,do2015rich} In addition, the $\sigma$ and $\sigma^\ast$ bands are almost independent of stacking symmetry. The stacking-created differences further diversify other physical properties, e.g., magnetic properties,\cite{lin2015magneto,ho2016evolution,lin2014energy,lin2014stacking,koshino2011landau,do2015configuration} optical spectra,\cite{do2015rich,lin2015magneto,linyp2015magneto,lin2015electric,chiu2013critical,chiu2014layer,lu2006absorption,chiu2016influence} Coulomb excitations,\cite{ho2006coulomb,wu2014combined,lin2012electrically}
and transport properties.\cite{jhang2011stacking,zou2013transport,khodkov2015direct}

The dramatic transition of distinct band structures occurs during the transformation of stacking configuration or the change of geometric curvature. For the sliding bilayer graphene with $AA\rightarrow\,AB\rightarrow\,AA^\prime\,\rightarrow\,AA$, the electronic properties are mainly determined by the various interlayer atomic interactions of $2p_z$ orbitals. Two pairs of vertical Dirac cones, parabolic bands, and non-vertical Dirac cones are transformed into each other, accompanied by the creation of an arc-shaped stateless region, distorted energy dispersions, extra low-energy critical points, separation of cone structures, and splitting of M- and M$^\prime$-related states. The rotational symmetry is reduced, while the inversion symmetry keeps the same. There are more special structures in the low- and middle-energy DOS. All the stacking systems belong to gapless semimetals, and the K and K$^\prime$ valleys are doubly degenerate. Also, it is expected to induce the unusual magnetic quantization and optical selection rule.\cite{huang2014feature} On the other hand, the curvature effects in rippled graphenes account for the rich essential properties. The charge distributions, bond lengths, energy bands, and DOS are sensitive to the structure, curvature and period of ripple. The misorientation of $2p_z$ orbitals and hybridization of ($2s,2p_x,2p_y,2p_z$) orbitals on a curved surface grow with the increasing curvature. Both $2p_z$ and $2p_x$ orbitals could make comparable contributions to the low- and middle-energy states, and four orbitals dominate the deep-energy states. All the armchair ripples present the gapless Dirac cones with a curvature-dependent red shift in the Fermi-momentum state. Armchair ripples and nanotubes, respectively, belong to 2D semiconductors and 1D conductors. However, zigzag ripples exhibit the semiconductor-semimetal transition, in which there exist the distorted Dirac cones and the extra van Hove singularities. Furthermore, the strong curvature effects can create the metallic or narrow-gap zigzag nanotubes, depending on radii. For any graphene ripples, a plenty of middle-energy van Hove singularities mainly come from the effects of curvature and splitting.

For graphene oxides, the competition or cooperation among the critical chemical bondings in C-C, O-O and C-O bonds leads to the diversified properties. The various orbital hybridizations are identified from the atom-related energy bands, the spatial charge density, and the orbital-projected DOS. They create the feature-rich electronic structures, with the O-, (C,O)- and C-dominated energy bands at the different energy ranges, destruction or distortion of the Dirac cone, opening of a band gap, anisotropic energy dispersions, and many extra critical points. Moreover, five types of $\pi$ bondings, which are responsible for the magnitude of energy gap, are clearly evidenced in the band-decomposed charge density during the decrease of O-concentration. The bonding of carbon $2p_z$ orbitals presents the full termination, the partial suppression, the 1D extension, the distorted 2D form, or the well-behaved planar form. The first two types induce larger gaps at higher concentration, the last two types generate zero gaps at lower concentrations, and the third type relates to small or vanishing gaps at intermediate concentration. For example, monolayer graphene oxide possesses the finite and zero gaps, corresponding to the O-concentrations of $>25$\% and $<3$\% in the single-side adsorption, respectively. The above-mentioned electronic properties could be observed in any layered graphene oxides, even if the essential properties strongly depend on the concentration, distribution, single- or double-side adsorption, layer number and stacking configuration. For the single- and double-side adsorptions, the bilayer (trilayer) graphene oxides could be, respectively, regarded as the superposition of monolayer (bilayer) graphene and monolayer graphene oxide, and two monolayer graphene oxides (two monolayer graphene oxides and monolayer graphene). The layered graphene oxides are semiconductors (semimetals) in the absence (presence) of non-passivated graphenes. Apparently, the multi-orbital hybridizations lead to more atom- and orbital-dependent special structures in DOS. As to the optimal position, O adatoms are situated at the bridge site, being in sharp contrast with the top site for H- and halogen-doped systems, and the hollow site for alkali-, Al- and Bi-doped ones.

The strong bondings between hydrogen and carbon atoms can largely enrich the fundamental chemical and physical properties. The geometric, electronic and magnetic properties are closely related to the complicated hybridizations of 1s and (2$p_z$,2$p_x$,2$p_y$,2s) orbitals. The buckling structure due to the top-site H-adsorption is directly evidenced from the H-C-C functional groups. The (H-C,C-C) lengths, H-C-C angles and heights of passivated carbons are sensitive to concentration and distributions, as revealed in single- and double-side adsorptions. Graphane has the shortest H-C bonds, the longest C-C bonds and the lowest height. The dramatic change from the sp$^2$ to sp$^3$s bondings clearly indicates that the serious orbital hybridizations, the strong $\sigma$ bondings and the weak $\pi$ bondings greatly diversify band structures. The $\pi$ bands are absent under the full hydrogenation, so that energy gap of graphane is determined by the $\sigma$ and $\pi^\ast$ bands. Hydrogenated graphenes might exhibit the recovered Dirac-cone structure and $\pi$ bands, the pairs of weakly dispersive bands across $E_F$ dominated by the second- and fourth-nearest C atoms, and the (C,H)-co-dominated bands at deeper energies. There exist the middle-, narrow- and zero-gap systems with the non-monotonous dependence on adatom concentration. Specifically, for the passivation of the same sublattice, the spin-split energy bands across $E_F$ appear and become pairs of valence and conduction bands, leading to the ferromagnectic semiconductors. The spatial spin density can create the magnetic moment of ${\sim\,1-4 \mu_B}$, being proportional to the adatom number in a unit cell. Concerning DOS, graphane only exhibits two threshold shoulder structures related to energy gap, while the other hydrogenated systems possess two group of peaks structures centered at $E_F$. The curvature effects on the extra orbital hybridizations are revealed in DOS as a lot of merged structures at ${E\le\,-3}$ eV; that is, the special structures associated with the five orbitals come to exist together. The decrease of H-concentration results in  more low-lying special structures arising from the $\pi$- and $\pi^\ast$-electronic states.

Halogenated graphenes, with the top-side adatom adsorption, exhibit the diverse and unique chemical bondings, especially for the great differences between fluorination and other halogenations. There are X-X, X-C and C-C bonds, in which the former is presented at higher concentrations. (2p$_x$,2p$_y$,2p$_z$) orbitals of F and C have very strong hybridizations among one another. This leads to the buckled structure, the destruction or distortion of Dirac cone, and the absence of energy bands formed by F-2p$_z$ orbitals. (2p$_x$,2p$_y$) orbitals of F adatoms can built energy bands, combined with those of C atoms by the F-C bonds. As to low fluorination, they become narrow (F,C)-co-dominated bands at middle energy. Moreover, fluorinated graphenes are metals or semiconductors, sensitive to concentrations and distributions. On the other hand, only the significant interactions of $p_z$ orbitals exist in other X-C bonds, and the $\sigma$ bonding of graphene is unchanged under these halogenations. Electronic structures mainly consists of the carbon- and adatom-dependent energy bands; mixings between them are weak. The $\sigma$ bands almost keep the same, and the Dirac cone is slightly distorted. The other halogenated graphenes belong to metals. (p$_x$,p$_y$) and p$_z$ orbitals of halogen adatoms, respectively, create the distinct energy bands. They have the weakly  energy dispersions at low concentrations, and make much contribution to DOS near the Fermi level. The unusual hybridization-induced features are clearly evidenced in many special structures of DOSs. Both halogenated and hydrogenated graphenes might present the ferromagnetic spin configurations, while they are metals and semiconductors, respectively.  Moreover, the carbon- and adatom-dominated magnetisms are, respectively, revealed in the fluorinated and other halogenated systems.

The outermost s orbitals of alkali atoms only have a single s-2$p_z$ hybridization in bonding with carbon, so that the adatom-adsorbed graphenes exhibit the unique essential properties. The hollow-site optimal positions lead to the absence of buckled structure, being rather different from the top- and bridge-site ones. The Dirac-cone structure, with a slight distortion or separated Dirac points, has a red shift. This strongly depends on the adatom concentration, but not the kind and distribution of alkali adatom. Only few conduction bands belong to the alkali-dominated ones, while most of energy bands originate from carbon atoms. The former present the weak dispersions at low concentrations. However, part of them, with observable energy widths, become occupied under the sufficiently high alkali concentrations. This indicates the existence of s-s orbital hybridizations in  A-A bonds. The (alkali,C)-co-dominated energy bands are almost absent, while the (adatom,C)-related ones are revealed in O-, H-, F- and Al-adsorbed systems. Whether such energy bands could survive is determined by the hybridization strength of adatom-C bond. The alkali adsorption can create a lot of free electrons in conduction Dirac cone and even the occupied alkali-dominated bands. The metallic behavior is directly reflected in the high DOS near $E_F$, accompanied with the alkai-induced special structures. The above-mentioned A-C and A-A bonds, C- and alkali-dominated energy bands, and free carriers are also observed the alkali-adsorbed graphene nanoribbons. The free carrier density in such systems could be evaluated exactly. It is deduced that any alkali adatoms contribute one s-orbital electron to become free carriers in graphene-based systems, regardless of the kind, distribution, and concentration.

The aluminum-adsorbed graphenes present the multi-orbital interactions in chemical bonds and thus the feature-rich essential properties. The highest Al-concentration is about 25\% under the double-side adsorption, much lower than 100\% in O-, H-, halogen- and alkali-adsorbed graphenes. The $\pi$ and $\sigma$ bondings of carbon atoms are slightly changed after the hollow-site adsorption. The significant hybridizations of (3s,3p$_x$,3p$_y$) and 2p$_z$ orbitals in the Al-C bonds can create the red shift of carbon-dominated energy bands, the charge transfer from adatoms to carbons, the greatly enhanced DOS near $E_F$, and the Al-related energy bands. Furthermore, the multi-orbital hybridizations of (3s,3o$_x$,3p$_y$) exist in Al-Al bonds at higher concentrations. The Al-induced high free carrier density is comparable with that of the alkali dopings. Such electrons only occupy the conduction Dirac cone of C-2p$_z$ orbitals. Specifically, 3s and (3p$_x$,3p$_y$) orbitals of Al adatoms, respectively, built valence and conduction bands, combined with the $\pi$ and $\pi^\ast$ bands of C atoms. This results in many extra structures in DOS. The adatom-dependent valence bands are absent in alkali-adsorbed systems. As for Bi-adsorbed graphenes, the geometric and electronic structures are enriched by the substrate, rippled buffer layer, Van der Waals interactions, and Bi-Bi and Bi-C bonds. The optimized structure, with an inter-adatom distance of $\sim$14-17 \AA, presents a large-scale hexagonal array. The energy barrier for the formation of the meta-stable triangular and rectangular Bi-nanoclusters is about 50 meV. Furthermore, DOS exhibits a finite value at the Fermi level, a dip structure at ${E\sim\,-}$0.2 eV, and a prominent peak at ${E\sim\,-}$0.6 eV, corresponding to the free conduction electrons, the shifted Dirac cone, and the Bi-dependent electronic states, respectively.

The experimental measurements are consistent with part of theoretical predictions. Adatom concentrations are estimated from various experimental methods, especially for O,\cite{marcano2010improved,erickson2010determination} H,\cite{pumera2013graphane} F,\cite{robinson2010properties,chang2011facile} and Cl.\cite{wu2011controlled,zheng2012production} The STM measurements on geometric structures have confirmed the interlayer distance of AA and AB stackings,\cite{lee2008growth,rong1993electronic} the bridge, top and hollow sites, respectively, corresponding to the (oxygen,bismuth),\cite{erickson2010determination,chen2015long,chen2015tailoring} (hydrogen,fluorine)\cite{balog2009atomic} and alkali adsorptions,\cite{virojanadara2010epitaxial} and the hexagonal array and the metastable nanoclusters of Bi adatoms on monolayer graphene/buffer layer/SiC.\cite{chen2015long,chen2015tailoring} Further examinations are required for the C-C bond lengths, the O-, H-, alkali-, halogen- and Al-C bond lengths, the curvatures of rippled graphenes, and the buckled structures of hydrogenated and fluorinated graphenes. The verifications for band structures cover the linear and parabolic bands of tri-layer ABA stacking,\cite{coletti2013revealing,ohta2007interlayer} the partially flat, sombrero-shaped and linear bands of tri-layer ABC stacking,\cite{coletti2013revealing} the concentration-dependent energy gap and destruction of Dirac cone in graphene oxides,\cite{schulte2013bandgap} and the red-shift cone structure in fluorinated\cite{walter2011highly} and alkali-adsorbed systems.\cite{ohta2006controlling,bostwick2007quasiparticle,papagno2011large} The similar ARPES measurements could be done for AAA- and AAB-stacked graphenes, rippled graphenes, H-, halogen-, Al-, and Bi-adsorbed graphenes. Furthermore, the spin-resolved STS measurements are available in identifying whether hydrogenated and halogenated graphenes, respectively, exhibit the spin-split energy bands as the semiconducting and metallic characteristics. As to the main features of DOS, the STS identifications include a prominent peak at $E_F$ due to partial flat bands of tri-layer ABC stacking,\cite{que2015stacking} a dip structure at $E_F$ surrounded by a pair of asymmetric peaks in tri-layer AAB stacking,\cite{que2015stacking} a V-shaped spectrum and low-energy peaks, respectively, for the low- and high-curvature zigzag graphene ripples,\cite{meng2013strain} the oxidation- and hydrogenation-induced energy gaps,\cite{pandey2008scanning,guisinger2009exposure,castellanos2012reversible} the alkali- and Bi-enhanced finite value at $E_F$,[Rs]\cite{chen2015tailoring} and the Bi-dominated structure below $E_F$.\cite{chen2015tailoring} The other special structures, which are revealed in tri-layer AAA and ABA stackings, armchair graphene ripples, and O-, H-, halogen-, alkali-, and Al-adsorbed graphenes, are worthy of detailed examinations, especially for the adatom- and (adatom,carbon)-dominated structures. Moreover, the spin-polarized STS measurements on the low-lying prominent structures can directly verify the spin splitting in hydrogenated and halogenated graphenes. In addition, experimental measurements of optical spectra and electrical resistances are also utilized to investigate the adatom-induced semiconducting and metallic behaviors, e.g., those for graphene oxides,\cite{loh2010graphene,becerril2008evaluation} and fluorinated graphenes.\cite{cheng2010reversible,nair2010fluorographene,jeon2011fluorographene}

The above-mentioned theoretical and experimental results demonstrate that the essential properties of layered graphenes are dramatically changed by the various geometric structures and the chemical adatom adsorptions. The structure-enriched and adatom-adsorbed graphenes are expected to be promising materials in functionalized nanodevices. Rippled graphenes might be suitable for the applications of chemical gas sensor,\cite{chi2009adsorption,ao2008enhancement,johnson2010hydrogen} hydrogen storage\cite{tozzini2011reversible,ning2012high,tozzini2013prospects} and electronic device,\cite{yan2012high} owing to an active chemical environment on a curved surface. Graphene oxides exhibit the wide-range variation among the large-gap insulators, narrow-gap semiconductors and semimetals, so that they have high potentials in FETs,\cite{trung2014flexible,truong2014reduced,joung2010high} electrical sensors,\cite{eda2010chemically,basu2012recent,su2014electrical} transparent conductors,\cite{zheng2014graphene,nekahi2014transparent} and optoelectronic devices.\cite{murray2011graphene,yin2010organic,saha2014solution,lu2012novel} The tunable electronic properties, with magnetic configurations, are observed in hydrogenated and halogenated graphenes, clearly indicating the potential importance in transistors,\cite{gharekhanlou2010bipolar} superconductors,\cite{lu2016ferromagnetism,zhou2014graphene} lithium-related batteries,\cite{sun2014solvothermally} supercapacitors,\cite{zhao2014fluorinated} biological scaffold materials,\cite{wang2012fluorinated} printing technologies,\cite{nebogatikova2016fluorinated} molecule detectors,\cite{katkov2015backside} biosensors,\cite{urbanova2016fluorinated} and FETs.\cite{zhang2013impact} On the other hand, the alkali- and Al-adsorbed graphenes can create very large free electron density, being useful in developing the next-generation high-capacity batteries\cite{lin2015ultrafast,rani2013fluorinated} and energy storage.\cite{ao2009doped,ao2010high}

The current review work could be further generalized to the layered condensed-matter systems, with the nano-scaled thickness, the special lattice symmetries, and the distinct stacking configurations. In addition to graphene, the emergent 2D materials include silicene,\cite{tao2015silicene,meng2013buckled} germanene,\cite{li2014buckled,davila2014germanene} stanene,\cite{zhu2015epitaxial} phosphorene,\cite{liu2015semiconducting,li2014black} MoS$_2$\cite{mak2010atomically,coleman2011two} and so on. They are very suitable for studying the novel physical, chemical and material phenomena and have shown very high potentials in near-future technological applications.\cite{tao2015silicene,liu2015semiconducting,li2014black} Such systems possess the rich intrinsic properties in terms of lattice symmetries, planar or buckling structures, intra- and inter-layer atomic interactions, multi-orbital chemical bondings, different site energies, and spin configurations (or spin-orbital interactions). Apparently, the composite interactions can create the critical Hamiltonians and thus the unique phenomena.\cite{chen2016magnetic,ho2014magneto} This work shows that the first-principles calculations, combined with the important orbital hybridizations in atom-atom bonds and the specific spin arrangements, are reliable in comprehending the orbital- and spin-dominated essential properties. For example, energy gaps, electronic structures and DOSs are determined by which kinds of atomic orbitals and spin states. The fundamental properties of 2D materials could be significantly altered and controlled through the adatom/molecule adsorptions and dopings, such as the successful synthesis of hydrogenated silicene\cite{qiu2015ordered} and silicene oxide.\cite{du2014tuning,molle2013hindering} Also, there are some first-principles researches on adatom-adsorbed layered systems, e.g., electronic properties of hydrogenated silicene and germanene.\cite{houssa2011electronic,wei2013many,lin2015h,si2014functionalized} The rich and diverse phenomena will be revealed under the various chemical processes and the external electric and magnetic fields.\cite{lin2015magneto} The cooperative or competitive relations among the geometric symmetries, the multi-orbital atomic interactions, the spin-orbital couplings and the external fields deserve a thorough and systematic investigation.

\par\noindent {\bf Acknowledgments}

This work was supported by the National Center for Theoretical Sciences (South) and the Nation Science Council of Taiwan, under the grant No. NSC 102-2112-M-006-007-MY3.

\newpage
\renewcommand{\baselinestretch}{0.2}

\newpage
\begin{table}[htb]
\small
\caption{The optimized C-O bond lengths, C-C bond lengths, binding energy ($E_b$) and energy gap ($E_g$) for various concentrations of single-side adsorbed GOs. The electronic structures of concentrations labeled with * will be illustrated in detail.}
\centering
\begin{tabular*}{1\textwidth}{@{\extracolsep{\fill}}lllllll}
\hline
Fig.&Atoms&O:C (\%)&C-O bond&C-C bond&$E_b$&$E_g$ \\
& & &length (\AA)&length (\AA)&(eV)&(eV)\\
\hline
(a) &  & 1:2 (50)$^*$ & 1.43 & 1.53 &-4.4651 &3.54\\
(b) &  & 3:6 (50)$^*$& 1.41 & 1.55 &-4.0943 &2.83\\
(c) &  & 7:14 (50)$^*$& 1.42 & 1.54 &-4.2957&3.92\\
(b) & 1,2 & 2:6 (33.3)$^*$& 1.45 & 1.48 &-3.7433 &2.53\\
(b) & 1,3 & 2:6 (33.3)& 1.45 & 1.48 &-3.5751&1.28\\
(e) & 1,2 & 5:18 (27.8)& 1.45 & 1.47 &-4.0788 &1.26\\
(d) & 1,2 & 2:8 (25)$^*$& 1.46 & 1.46&-3.6509 & 0.63\\
(e) & 2 & 4:18 (22.2)$^*$ & 1.46 & 1.45 &-4.0989& 0\\
(b) & 1 & 1:6 (16.7)& 1.46 & 1.44 &-3.8586& 0.09\\
(d) & 1 & 1:8 (12.5) &1.46 & 1.44 &-3.9487& 0\\
(g) & 1,2 & 5:50 (10) & 1.46 & 1.44 &-3.9576& 0.21\\
(e) & 1 & 1:18 (5.6)$^*$& 1.46  & 1.44  & -3.9589 & 0.14\\
(f) & 1 & 1:24 (4.2)$^*$ & 1.47 & 1.43&-3.9636 & 0.07\\
(h) & 1,2 & 3:98 (3)$^*$ & 1.47 & 1.42 &-3.9245& 0\\
(g) & 1 & 1:50 (2) & 1.47 & 1.42 &-3.9187& 0\\
(h) & 1 & 1:98 (1)$^*$ & 1.47 & 1.42 &-3.8918& 0\\
\hline
 \end{tabular*}
 \end{table}

\begin{table}[htb]
\small
\caption{The calculated binding energy ($E_b$) and energy gap ($E_g$) of double-side adsorbed GOs with various O-concentrations.}
\label{t1}
\centering
\begin{tabular}{ c c c c c c }
\hline
O Concentration& $E_b$ (eV)& $E_g$ (eV)\\
\hline
50\% (both-side, zigzag)& -4.8622& 3.14\\
33.3\% (both-side, armchair) &-4.6682 &1.98\\
25\% (both-side, zigzag) & -4.5079 &0.17\\
11.1\% (both-side, zigzag) &-4.4813 &0.42\\
4\% (both-side, zigzag) &-4.4650 &0.66\\
\hline
\end{tabular}
\end{table}

\newpage
\begin{table}[htb]
\small
\caption{The calculated C-C and C-H bond lengths, H-C-C angle, height (shift on z-axis) of passivated C, binding energy, energy gap, and total magnetic moments per unit cell for various H-coverages of sigle-side adsorption. The electronic structures of concentrations labeled with * have spin polarizations.}
\label{t5}
\centering
\begin{tabular*}{1\textwidth}{@{\extracolsep{\fill}}llllllllllll}
\hline
Type&H:C&\%& 1$^{st}$ C-C&2$^{nd}$ C-C&C-H  &H-C-C & Height &$E_b$&$E_g$&M$_{tot}$\\
 & & & (\AA)& (\AA)&(\AA)& angle ($^o$)& (\AA)&(eV)&(eV)&($\mu_B$)\\
\hline
Zigzag & 4:8&50$^*$ & 1.5 & & 1.16 & 103.26 &0.34&-2.1 &0.5&2\\
 &2:8&25$^*$ &1.49&1.39&1.14&103.95&0.37&-2.27 &0.86&1\\
&2:8&25 &1.49&1.39&1.12&104.69&0.38&-3.14 &3.56&0\\
 &1:8&12.5$^*$ &1.48&1.4&1.13&103.55&0.35&-2.66 &0.42&0.5\\
 &2:18&11.1$^*$ &1.48&1.41&1.13&104.29&0.36 &-2.63&0.2&1\\
&2:18&11.1&1.48&1.39&1.13&103.42&0.34&-2.9 &0.24&0\\
&2:32&6.3&1.46&1.35&1.12&107.57& 0.37&-2.45&0.38&0\\
Armchair & 2:6&33.3$^*$ & 1.47 & 1.4 & 1.14 & 104.41 &  0.37&-2.17& 0.9&1\\
 &1:6&16.7$^*$&1.48&1.4&1.13&104& 0.36&-2.33&0.24&0.5\\
&2:24&8.3&1.48&1.4&1.12&104.4&0.34&-2.81&0.01&0\\
Chiral&1:14&7.1$^*$ &1.48&1.4&1.13&103.95&0.36&-2.6 &0.32&0.5\\
\hline
\end{tabular*}
\end{table}

\newpage
\begin{table}[htb]
\small
\caption{The calculated C-C and C-H bond lengths, H-C-C angles, heights (shift on z-axis) of passivated carbons, binding energy, energy gaps, and total magnetic moments per unit cell for various H-concentrations of double-side adsorption. $\ast$ represents the spin-polarized system.}
\label{t6}
\centering
\begin{tabular*}{1\textwidth}{@{\extracolsep{\fill}}lllllllllll}
\hline
Type& H:C&\% &1$^{st}$ C-C &2$^{nd}$ C-C& C-H&H-C-C &Height&$E_b$ &$E_g$&M$_{tot}$\\
 & & &(\AA)&(\AA)&(\AA)& angle ($^o$)& (\AA)&(eV)&(eV)&($\mu_B$)\\
\hline
Zigzag & 2:2&100& 1.53 & & 1.12 & 107.1 & 0.22&-3.24 & 3.57&0\\
 &2:8&25&1.5&1.39&1.12&101&0.3&-3.57&2.5&0\\
&2:18&11.1$^*$ &1.5&1.43&1.13&102.6&0.3&-2.95&0.11&1\\
 &2:32&6.3&1.49&1.39&1.13&104.2&0.57&-3.37&0.25&0\\
 &2:50&4&1.49&1.4&1.13&102.8&0.47&-3.35&0.75&0\\
&2:98&2&1.49&1.4&1.13&103.1&0.56&-3.49&0.2&0\\
Armchair & 2:6&33.3& 1.47 & 1.4 & 1.15 & 100.5 & 0.27&-2.32 & 0&0\\
 &2:24&8.3$^*$&1.49&1.4&1.13&103.5&0.28&-2.67&0.96&1\\
Chiral& 2:14&14.3 &1.5&1.43&1.12&104.6&0.46&-3.36&0.05&0\\
 &2:126&1.6&1.48&1.4&1.13&101.5&0.49&-3.46&0&0\\
\hline
\end{tabular*}
\end{table}

\newpage
\begin{table}[htb]
\small
\caption{The calculated C-C and C-X bond lengths, heights (shift on z-axis) of passivated carbons, binding energies, energy gaps, and total magnetic moments per unit cell of halogen-absorbed graphene systems. The double-side structures are labeled with *. $\triangle$ represents the very strong Cl-C bonding.}
\label{t6}
\centering
\begin{tabular*}{1\textwidth}{@{\extracolsep{\fill}}llllllllll}
\hline
Adatom& X:C&\% & &Bond length& &Height &$E_b$ &$E_g$&M$_{tot}$\\
 & & &C-X (\AA)&1st C-C (\AA)&2nd C-C (\AA)& (\AA)&(eV)& (eV)&($\mu_B$)\\
\hline
F& 8:8$^*$ & 100 &1.38 &1.58 & &0.48& -3.00 &3.12  & 0 \\
 & 4:8$^*$ & 50 &1.48 &1.51 &  & 0.38  & -2.71  &  2.77 &0\\
 & 4:8 & 50 &1.48 &1.5 &  & 0.28 & -1.16  & 0  &1.94\\
 & 2:8 & 25 &1.54&  1.48&1.43  &0.34 & -2.94  & 2.94 &0\\
 & 1:6 & 16.7 & 1.53 & 1.48& 1.42&  0.32& -2.45&  0&0.22\\
 & 1:18 & 5.6 & 1.56 & 1.48 & 1.42 &0.32  &-2.93  & 0 &0\\
 & 1:32 & 3.1 &1.57  &1.45  &1.42  &  0.32&-2.04 &  0&0\\
Cl & 8:8$^*$ ($\triangle$)& 100 &  1.74& 1.76 &  & 0.51 &  -0.62& 1.41 &0\\
& 8:8$^*$ & 100 & 3.56 &  1.48&   &0  & -0.84&  0&0\\
&4:8$^*$& 50 & 3.5& 1.46&  &0 &-1.11  & 0&0\\
&2:8& 25 &3.2&  1.44&1.44  &0 &  -1.65& 0&0\\
&1:6& 16.7 & 3.2&  1.44 &  1.44&0 & -1.16 & 0&0.56\\
&1:8& 12.5 &3&   1.44&  1.43&0 &  -1.64& 0&0.57\\
& 1:32 & 3.1 & 2.96 & 1.42 & 1.42 &0  &  -0.97 &  0&0.36\\
Br& 8:8$^*$& 100 &3.56  &  1.56 &  &0  & -0.48 & 0 &0\\
&2:8& 25 & 3.59&  1.43 & 1.43 &0 & -1.91 & 0&0\\
&1:8& 12.5 & 3.56&  1.43 &1.43  &0 & -1.43 & 0& 0.48\\
& 1:32 & 3.1 & 3.23& 1.42 & 1.42 &0  &  -0.71 & 0 &0.41\\
I& 1:32 & 3.1 &  3.59& 1.42 & 1.42 &  0&  -0.48 & 0 &0.44\\
At& 1:32 & 3.1 & 3.72 &1.42  &1.42  & 0 & -0.36 &0  &0.5\\
\hline
\end{tabular*}
\end{table}

\newpage

\begin{table}[htb]
  \caption{Akali-absorbed graphene systems}
  \label{tbl:2}
\centering
  \begin{tabular}{lcccccc}
     \hline
    \hline
    configuration & nearest C-C  & bond length & height  & $E_{b}$ (eV)&  $\sigma_{e}$ ($10^{14}$ $ e/cm^{2}$)\\
    &(\AA{}) & A-C (\AA{}) &  (\AA{}) &  &\\

    \hline
    3:6$_{single}$ (A); Li & 1.460 & 2.347 & 1.838 & -1.710 & 15.45 \\
    1:6$_{single}$ (A) & 1.430 & 2.350 & 1.865 & -1.733 & 5.52\\
    1:8$_{single}$ (Z)& 1.435 & 2.329 & 1.834 & -1.602 &3.78\\
    1:18$_{single}$ (Z)& 1.427 & 2.254 & 1.744 & -1.706 &2.24\\
    1:24$_{single}$ (A)& 1.439 & 2.270 & 1.756 & -1.890 &1.83\\
    1:32$_{single}$ (Z)& 1.439 & 2.279 & 1.767 & -1.936 &1.10\\
    1:32$_{single}$ (Z); K & 1.437 & 3.086 & 2.749 &-1.735   & 1.13\\
    1:32$_{single}$ (Z); Cs & 1.436 & 3.357 & 3.041 & -0.831  & 1.11\\
    \hline
    6:6$_{double}$ (A); Li& 1.480 & 2.373 & 1.855 & -1.745 &\\
    2:6$_{double}$ (A) & 1.442 & 2.341 & 1.844 & -1.723 &\\
    2:8$_{double}$ (Z) & 1.447 & 2.321 & 1.815 & -1.610 &\\

    \hline
    \hline

\end{tabular}
\end{table}

\newpage
\begin{table}[htb]
\small
  \caption{\ Free electron densities for various numbers and positions of adatoms in $N_A=12$ and $N_z=8$ alkali-adsorbed graphene nanoribbons}
  \label{tbl:3}
\centering
  \begin{tabular}{ccccccc}
\hline
\hline
    graphene & configurations &  $\lambda$ ($10^7$ $ e/cm$)& $\lambda$ /adatom & Bader charge \\
    nanoribbon & &  & concentration (e)  & transfer (e) \\

    \hline
                        &  (1)$_{single}$; Li & 2.31  & 0.995 & 0.75  \\
                        &  (1)$_{single}$; K & 2.31  & 0.995 & 0.45  \\
                        &  (1)$_{single}$; Cs & 2.31  & 0.995 & 0.54  \\
                        &  (3,7)$_{single}$; Li & 4.67  & 1.008 & 0.69  \\
                         & (3,{\color{red}{7}})$_{double}$ & 4.62  & 0.989 & 0.75   \\
    Armchair     & (1,10)$_{single}$  & 4.68  & 1.008 & 0.74   \\
    N$_A$= 12 & (1,{\color{red}{10}})$_{double}$  & 4.67 & 1.008 & 0.74   \\
                        & (1,{\color{red}{4}},7,{\color{red}{10}})$_{double}$  & 9.24 & 0.994 & 0.71   \\
                        & (1,{\color{red}{2}},5,{\color{red}{6}},9,{\color{red}{10}})$_{double}$ & 13.96 & 1.007 & 0.70  \\
                        & ({\color{red}{1,3,5,7,9}},2,4,6,8,10)$_{double}$ & 23.03 & 0.988 & 0.68  \\
                        & ({\color{red}{1$\rightarrow$10}},1$\rightarrow$10)$_{double}$ & 46.36 & 0.995 & 0.68  \\

    \hline
                    &  (1)$_{single}$                   & 2.02  & 1.001 & 0.88  \\
     Zigzag         & (7)$_{single}$                    & 2.03  & 1.009 & 0.89   \\
    N$_Z$= 8        & (1,13)$_{single}$                 & 4.10  & 1.015 & 0.88  \\
                    & (1,{\color{red}{13}})$_{double}$  & 4.02 & 1.001 & 0.88   \\
                    & (1,7,13)$_{single}$               & 6.13 & 1.011 & 0.87   \\
                    & (1,{\color{red}{7}},13)$_{double}$& 6.08 & 1.002 & 0.88   \\

    \hline
    \hline
  \end{tabular}
\end{table}

\newpage
\begin{table}[htb]
\small
\caption{Bond lengths and heights for various Al-adsorbed graphenes.}
\label{t8}
\centering
\begin{tabular}{cccccc}
\hline
& \multicolumn{3}{c}{ Bond length ({\AA})} & & \\
Unit cell&Nearest&Next nea-&Al-C&Lattice& Al \\
 &C-C&rest C-C&&expansion&height({\AA})\\
\hline
$2\times2$ & 1.441 & 1.431 &2.544& 1.63 \% & 2.097\\
$3\times3$ & 1.440 & 1.431 &2.538& 1.36 \% & 2.090\\
$2\sqrt{3}\times2\sqrt{3}$ & 1.428 & 1.440 &2.543& 1.17 \% & 2.096\\
$4\times4$ & 1.436 & 1.428 &2.544& 1.12 \% & 2.110\\
$5\times5$ & 1.419 & 1.413 &2.571& 1.06 \% & 2.128 \\
$3\sqrt{3}\times3\sqrt{3}$ & 1.428 & 1.438 &2.563& 1.02 \% & 2.110\\
\hline
\end{tabular}
\end{table}

\newpage
\begin{table}[htb]
\small
\caption{Adsorption energies and heights for various atomic sites. Total energies and Bi-Bi interaction energies for various Bi-nanoclusters.} \label{t1}
\centering
  \begin{tabular*}{0.5\textwidth}{@{\extracolsep{\fill}}lll}
\hline site & $\Delta$E (eV) & h (\AA)\\ \hline hollow & 1.6452 & 2.51 \\ bridge & 1.9663 & 2.32 \\ top & 1.9162 & 2.34 \\ \hline
\hline $\#$ Bi atoms & $E_{Total}$ (eV)& $\Delta E_{Bi-Bi}$ (eV) \\ \hline 1 & -2317.86 & x \\ 2 & -2322.14 & -1.13 \\ 3 & -2326.77 & -1.62 \\ 4 & -2330.60 &
-1.67 \\ 5 & -2334.19 & -1.65 \\ 6 & -2337.86 & -1.65 \\ \hline \end{tabular*} \end{table}

\newpage \centerline {\Large \textbf {FIGURE CAPTIONS}}

\vskip0.5 truecm 
\begin{itemize}
\item[Figure 1:] Geometric structures of tri-layer graphenes: (a) AAA stacking, (b) ABA stacking, (c) ABC stacking, and (d) AAB stacking.
\bigskip
\item[Figure 2:] 2D band structures for (a) AAA stacking, (b) ABA stacking, (c) ABC stacking, and (d) AAB stacking.
\bigskip
\item[Figure 3:] Low-lying 3D band structures around the K point for (a) AAA stacking, (b) ABA stacking, (c) ABC stacking, and (d) AAB stacking.
\bigskip
\item[Figure 4:] DOSs for various tri-layer graphenes: (a) AAA stacking, (b) ABA stacking, (c) ABC stacking, and (d) AAB stacking.
\bigskip
\item[Figure 5:] (a) Geometric structure of bilayer graphene shifting along the armchair and zigzag directions. (b) The shift-dependent total ground state energy and interlayer distance (the dashed blue and red dots), as measured from those of the AA stacking.
\bigskip
\item[Figure 6:] Low-energy band structures around the K point (heavy dashed circles) for various stacking configurations along armchair ($\delta_a$) and zigzag ($\delta_z$) directions: (a) $\delta_a=$0, (b) $\delta_a=$1/8, (c) $\delta_a=$4/8, (d) $\delta_a=$1, (e) $\delta_a=$11/8, (f) $\delta_a=$12/8, (g) $\delta_z=$1/8; (h) $\delta_z=$3/8.
\bigskip
\item[Figure 7:] 2D band structures along the high-symmetry points for various stacking configurations: (a) $\delta_a=$0, (b) $\delta_a=$1/8, (c) $\delta_a=$4/8, (d) $\delta_a=$1, (e) $\delta_a=$11/8, (f) $\delta_a=$12/8, (g) $\delta_z=$1/8; (h) $\delta_z=$3/8.
\bigskip
\item[Figure 8:] DOSs due to 2$p_z$ orbitals for various stacking configurations: (a) $\delta_a=$0, (b) $\delta_a=$1/8, (c) $\delta_a=$4/8, (d) $\delta_a=$1, (e) $\delta_a=$11/8, (f) $\delta_a=$12/8, (g) $\delta_z=$1/8; (h) $\delta_z=$3/8.
\bigskip
\item[Figure 9:] The shift dependence of van Hove singularities at (a) low and (b) middle energies.
\bigskip
\item[Figure 10:] Geometric structures of (a) armchair ripple and nanotube and (b) zigzag ripple and nanotube. (c) The rippling period $l$ and the amplitude $h$. (d) The rectangular Brillouin zone of graphene ripple.
\bigskip
\item[Figure 11:] (a) Energy bands of armchair ripples along the $\Gamma\,$Y direction at distinct curvatures and periods. Band structures in energy-wave-vector space for a $N_\lambda=6$ armchair ripple at (b) $C_r=0.09$ and (c) $C_r=0.53$. Also shown in (a) is the dashed curve for a ($3,3$) armchair nanotube.
\bigskip
\item[Figure 12:] Energy bands of a $N_\lambda=5$ zigzag ripple at various curvatures along the (a)-(d) $\Gamma\,$X and (e) $\Gamma\,$Y directions. (f) and (g) Energy bands for the critical curvature with periods of $N_\lambda=5$ and $9$. Band structures in energy-wave-vector space for a $N_\lambda=5$ zigzag ripple at (h) $C_r=0.05$ and (i) $C_r=0.27$. Also shown in (e) is the dashed curve for a ($5,0$) zigzag nanotube.
\bigskip
\item[Figure 13:] The spatial charge distributions: (a) a planar graphene, (c) a $N_\lambda=6$ armchair ripple at $C_r=0.53$ and (e) a $N_\lambda=5$ zigzag ripple at $C_r=0.27$. (b), (d) and (f) are those for the charge differences. The right-hand sides present four kinds of chemical bondings between two carbons, and the similar distributions for ($3,3$) and ($5,0$) nanotubes.
\bigskip
\item[Figure 14:] Density of states: (a) armchair and (b) zigzag ripples with various curvatures and periods. The orbital-projected DOS for (c) a $N_\lambda=6$ armchair ripple at $C_r=0.53$ and (d) a $N_\lambda=5$ zigzag ripple at $C_r=0.27$. DOSs of ($3,3$) and ($5,0$) nanotubes are shown in (a) and (b), respectively.
\bigskip
\item[Figure 15:] Geometric structures of GOs with various unit cells and concentrations: (a) O:C=1:2 (Z), (b) O:C=3:6 (A), (c) O:C=7:14 (C) (d) O:C=1:8 (Z), (e) O:C=1:18 (Z), (f) O:C=1:24 (A), (g) O:C=1:50 (Z), and (h) O:C=1:98 (Z). Various oxygen distributions are denoted by numbers, corresponding to different concentrations listed in Table 1.
\bigskip
\item[Figure 16:] Band structures of GOs with various concentrations: (a) 50\% (Z), (b) 50\% (A), (c) 50\% (C), (d) 33.3\% (A), (e) 25\% (Z), (f) 22.2\% (Z), (g) 5.6\% (Z), (h) 4.2\% (A), (i) 3\% (Z), and (j) 1\% (Z). Superscripts $c$ and $v$ correspond to the conduction and valence bands, respectively. The red and blue circles, respectively, correspond to the contributions of passivated C atoms and O atoms, in which the dominance is proportional to the radius of circle. Also shown in the inset of (a) is the first Brillouin zone.
\bigskip
\item[Figure 17:] The band-decomposed charge density of (a) pristine graphene, (b) 50\% (Z), (c) 33.3\% (A), (d) 4.2\% (A), and (e) 1\% (Z) O-concentrations, in which the side views of $\pi$ bondings are plotted along the dashed orange line of the corresponding top view one.
\bigskip
\item[Figure 18:] The charge density (plotted in xz-plane) of (a) pristine graphene, (b) 50\% (Z), (c) 50\% (A), (d) 33.3\% (A), (e) double-side 50\% (Z), and (f) double-side 25\% (Z). The top view of charge density difference and its projections on xz- and yz-planes for (g) pristine graphene (only xz-plane is shown), (h) 50\% (Z), (i) 50\% (A), (j) 33.3\% (A), (k) double-side 50\% (Z), and (l) double-side 25\% (Z). (Z). The xy-plane projections are also shown to clarify the orbital hybridizations in the O-O bonds.
\bigskip
\item[Figure 19:] Orbital-projected DOSs of GOs with various O-concentrations: (a) 50\% (Z), (b) 50\% (A), (c) 50\% (C), (d) 33.3\% (A), (e) 25\% (Z), (f) 22.2\% (Z), (g) 5.6\% (Z), (h) 4.2\% (A), (i) 3\% (Z), and (j) 1\% (Z).
\bigskip
\item[Figure 20:] Band structures of oxygen adsorbed few-layer graphenes: (a) AA stacking, (b) AB stacking, (c) AAA stacking, and (d) ABA stacking.
\bigskip
\item[Figure 21:] Band structures of double-side adsorbed GOs with the oxygen concentration of (a) 50\% (Z), (b) 33.3\% (A), (c) 25\% (Z), (d) 11.1\% (Z), (e) 4\% (Z), (f) AA stacking, (g) AB stacking, and (h) AAA stacking.
\bigskip
\item[Figure 22:] Geometric structures of single-side hydrogenated graphene with top view and side view: (a) H:C = 4:8 = 50\% (Z), (b) H:C = 2:6 = 33.3\% (A), (c) H:C = 2:8 = 25\% (Z), (d) H:C = 1:6 = 16.7\% (A), (e) H:C = 2:18 = 11.1\% (Z), (f) H:C = 1:14 = 7.1\% (C), (g) H:C = 2:8 = 25\% (Z), (h) H:C = 2:18 = 11.1\% (Z), (i) H:C = 2:24 = 8.3\% (A), and (j) H:C = 2:32 = 6.3\% (Z). (g)-(j) do not have the spin configurations.
\bigskip
\item[Figure 23:] Band structures of single-side hydrogenated graphenes with spin polarizations: (a) 50\% (Z), (b) 33.3\% (A), (c) 25\% (Z), (d) 16.7\% (A), (e) 11.1\% (Z), and (f) 7.1\% (C) H-concentrations.
\bigskip
\item[Figure 24:] Band structures of single-side hydrogenated graphenes without spin polarizations: (a) 25\% (Z), (b) 11.1\% (Z), (c) 8.3\% (A), and (d) 6.3\% (Z) H-concentrations.
\bigskip
\item[Figure 25:] The spin-density distribution of single-side hydrogenated graphenes with top- and side-views: (a) 50\% (Z), (b) 33.3\% (A), (c) 25\% (Z), (d) 16.7\% (A), (e) 11.1\% (Z), and (f) 7.1\% (C) H-concentrations. The red isosurfaces represent the charge density of spin-up configuration, whereas the green circles point out the position of H atoms.
\bigskip
\item[Figure 26:] The charge densities of (a) 50\% (Z), (b) 11.1\% (Z), (c) 33.3\% (A), (d) 7.1\% (C) , (e) double-side 100\% (Z), and (f) double-side 8.3\% (A) hydrogenated graphenes. The corresponding charge density differences are, respectively, shown in (g)-(l).
\bigskip
\item[Figure 27:] The orbital-projected DOSs of single-side hydrogenated graphenes for various H-concentrations: (a) 50\% (Z), (b) 33.3\% (A), (c) 25\% (Z), (d) 16.7\% (A), (e) 11.1\% (Z), (f) 7.1\% (C), (g) 25\% (Z), (h) 11.1\% (Z), (i) 8.3\% (A), and (j) 6.3\% (Z). The non-magnetic systems in (g)-(j) present the spin-degenerate DOSs.
\bigskip
\item[Figure 28:] Geometric structures of double-side hydrogenated graphene with top and side views for (a) 100\% (Z), (b) 33.3\% (A), (c) 25\% (Z), (d) 14.3\% (C), (e) 6.3\% (Z), (f) 4\% (Z), (g) 11.1\% (Z), and (h) 8.3\% (A) H-concentrations.
\bigskip
\item[Figure 29:] Band structures of double-side hydrogenated graphenes without spin polarizations: (a) 100\% (Z), (b) 25\% (Z), (c) 6.3\% (Z), (d) 4 \% (Z), (e) 2\% (Z), (f) 33.3\% (A), (g) 14.3\% (C), and (h) 1.6\% (C) H-concentrations.
\bigskip
\item[Figure 30:] Band structures of double-side hydrogenated graphenes with spin polarizations and their corresponding spin-density distributions: (a) 11.1\% (Z) and (b) 8.3\% (A) H-concentrations.
\bigskip
\item[Figure 31:] The orbital-projected DOSs of double-side hydrogenated graphenes: (a) 100\% (Z), (b) 25\% (Z), (c) 2\% (Z), (d) 33.3\% (A), (e) 14.3\% (C), (f) 1.6\% (C), (g) 11.1\% (Z), and (h) 8.3\% (A) H-concentrations. The red isosurfaces represent the charge density of spin-up configuration, whereas the green circles point out the position of H atoms.
\bigskip
\item[Figure 32:] Geometric structures of halogen-doped graphene for (a) double-side 100\% (Z), (b) double-side 50\% (Z), (c) 50\% (Z), (d) 25\% (Z), (e) 16.7\% (A), and (f) 3.1\% (Z) halogen-concentrations.
\bigskip
\item[Figure 33:] Band structures for (a) a pristine graphene, and the (b) F-, (c) Cl-, (d) Br-, (e) I-, and (f) At-doped graphenes with concentration X:C = 1:32 = 3.1\% (Z). The green circles correspond to the contributions of adatoms, in which the dominance is proportional to the radius of circle.
\bigskip
\item[Figure 34:] Band structures of F-doped graphenes: (a) double-side F:C = 8:8 = 100\% (Z), (b) double-side F:C = 4:8 = 50\% (Z), (c) F:C = 4:8 = 50\% (Z), (d) F:C = 2:8 = 25\% (Z), (e) F:C = 1:6 = 16.7\% (A), and (f) F:C = 1:18 = 5.6\% (Z) concentrations. The green and red circles correspond to the contributions of adatoms and passivated C atoms, respectively.
\bigskip
\item[Figure 35:] Band structures of Cl- and Br-doped graphenes: (a) double-side Cl:C = 8:8 = 100\% (Z), (b) double-side Cl:C = 4:8 = 50\% (Z), (c) Cl:C = 2:8 = 25\% (Z), (d) Cl:C = 1:6 = 16.7\% (A), (e) Cl:C = 1:8 = 12.5\% (Z), (f) double-side Br:C = 8:8 = 100\% (Z), (g) Br:C = 2:8 = 25\% (Z), and (h) Br:C = 1:8 = 12.5\% (Z) concentrations.
\bigskip
\item[Figure 36:] The spin-density distributions with top and side views, for various concentrations: (a) F:C = 50\% (Z), (b) F:C = 16.7\% (A), (c) Cl:C = 12.5\% (Z), (d) Br:C = 12.5\%, (e) Cl:C = 3.1\%, and (f) Br:C = 3.1\%. The red isosurfaces represent the charge density of spin-up configuration.
\bigskip
\item[Figure 37:] The charge densities of: (a) pristine graphene, (b) F:C = 8:8 = 100\% (Z), (c) F:C = 2:8 = 25\% (Z), and (d) F:C = 1:32 = 3.1\% (Z). The corresponding charge density differences of fluorinated graphenes are, respectively, shown in (e)-(g).
\bigskip
\item[Figure 38:] The charge densities of chlorinated graphenes: (a) Cl:C = 8:8 = 100\% (Z), (b) Cl:C = 2:8 = 25\% (Z), and (c) Cl:C = 1:32 = 3.1\% (Z). The corresponding charge density differences are, respectively, shown in (d)-(f).
\bigskip
\item[Figure 39:] Orbital-projected DOSs of fluorinated graphenes for various concentrations: (a) double-side F:C = 100\% (Z), (b) F:C = 25\% (Z), (c) F:C = 50\% (Z), (d) F:C = 16.7\% (A), (e) F:C = 5.6\% (Z), and (f) F:C = 3.1\% (Z).
\bigskip
\item[Figure 40:] Orbital-projected DOSs of Cl- and Br-doped graphenes for various concentrations: (a) double-side Cl:C = 100\% (Z), (b) double-side Cl:C = 50\% (Z), (c) Cl:C = 16.7\% (A), (d) Cl:C = 12.5\% (Z), (e) Cl:C = 3.1\% (Z), (f) double-side Br:C = 100\% (Z), (g) Br:C = 25\% (Z), and (h) Br:C = 3.1\% (Z).
\bigskip
\item[Figure 41:] Geometric structures of alkali-adsorbed graphene for (a) 50\% (A), (b) 16.7\% (A), (c) 12.5\% (Z), (d) 5.6\% (Z), (e) 4.2\% (A), and (f) 3.1\% (Z) alkali-concentrations.
\bigskip
\item[Figure 42:] Band structures for (a) a pristine graphene, and the (b) Li-, (c) Na-, (d) K-, (e) Rb-, and (f) Cs-adsorbed graphenes in $4\times4$ supercell. The blue circles correspond to the contributions of adatoms, in which the dominance is proportional to the radius of circle.
\bigskip
\item[Figure 43:] Band structures of Li-adsorbed graphenes with various concentrations: (a) 50\%, (b) 16.7\%, (c) 12.5\%, (d) 5.6\%, (e) 100\%, and (f) 25\%, in which the last two belong to double side adsorption. The blue circles correspond to the contributions of adatoms.
\bigskip
\item[Figure 44:] The spatial charge distributions of Li-adsorbed graphenes with (a) 3.1\%, (b) 5.6\%, (c) 50\%, and (d) 100\% concentrations, and (e) K-, (f) Cs-adsorbed graphenes with 3.1\% concentrations. Those of nanoribbon systems are shown for (g) N$_{Z} = 8$ \& (1)$_{single}$ and (h) N$_{A} = 12$ \& (1)$_{single}$.
\bigskip
\item[Figure 45:] Orbital-projected DOSs for (a) pristine, (b) Li-, (c) K-, and (d) Cs-adsorbed graphenes with 3.1\% concentration, and Li-adsorbed graphene with (e) 5.6\%, (f) 16.7\%, (g) 50\%, (h) 100\% concentrations.
\bigskip
\item[Figure 46:] Geometric structures of alkali-adsorbed graphene nanoribbons for (a) $N_{A} =12$ armchair and (b) $N_{Z} = 8$ zigzag systems. The dashed rectangles represent unit cells used in the calculations. The lattice constants I$_{x}$ are, respectively, $a = 3b$ and $a = 2\sqrt{3}b$ for armchair and zigzag systems. Numbers inside hexagons denote the adatom adsorption positions. The spin configurations of pristine, (1)$_{single}$ and (7)$_{single}$ are indicated in (c), (d) and (e), respectively. Blue and red circles, respectively, represent spin-up and spin-down arrangements.
\bigskip
\item[Figure 47:] Band structures of N$_{A} = 12$ armchair systems for (a) a pristine , (b) Li \& (1)$_{single}$, (c) K \& (1)$_{single}$, (d) Cs \& (1)$_{single}$, (e) Li \& (1,10)$_{single}$, (f) Li \& (3,7)$_{single}$, (g) Li \& (3,{\color{red}{7}})$_{double}$, (h) Li \& (1,{\color{red}{4}},7,{\color{red}{10}})$_{double}$, and those of N$_{Z} = 8$ zigzag systems for (i) a pristine, (j) Li \& (1)$_{single}$, (k) Li \& (7)$_{single}$, and (l) Li \&  (1,13)$_{single}$. Blue circles represent the contributions of alkali adatoms.
\bigskip
\item[Figure 48:] Orbital-projected DOSs in Li-adsorbed graphene nanoribbons of N$_{A} = 12$ armchair systems for (a) a pristine nanoribbon, (b) (1)$_{single}$, (c) (1,10)$_{single}$, and (d) (3,7)$_{single}$. Those of N$_{Z} = 8$ zigzag systems are shown for (e) a pristine nanoribbon, (f) (1)$_{single}$, (g) (7)$_{single}$, and (h) (1,13)$_{single}$.
\bigskip
\item[Figure 49:] Side view and top view geometric structures of aluminum-adsorbed graphene for (a) 12.5\% (A), (b) 5.6\% (A), (c) 4.2\% (Z), and (d) 3.1\% aluminum-concentrations.
\bigskip
\item[Figure 50:] Band structures of (a) pristine graphene and Al-adsorbed graphenes with various concentrations: (b) 12.5\%, (c) 5.6\%, (d) 4.2\%, (e) 3.1\%, and (f) 25\%, in which the last one belong to double side adsorption. The blue circles correspond to the contributions of adatoms.
\bigskip
\item[Figure 51:] The spatial charge distributions of Al-adsorbed graphenes with (a) 12.5\%, (b) 5.6\%, and (c) 3.1\% concentrations.
\bigskip
\item[Figure 52:] Orbital-projected DOSs for (a) pristine graphene and Al-adsorbed graphenes in (b) 12.5\%, (c) 5.6\%, (d) 4.2\%, (e) 3.1\%, and (f) 25\% concentrations.
\bigskip
\item[Figure 53:] Geometric structure of 4H-SiC (0001) substrate, buffer layer, monolayer rgaphene, and bismuth adatoms. C, Si and Bi are, respectively, indicated by the gray, yellow and purple solid circles.
\bigskip
\item[Figure 54:] (a) Ground-state energies of bismuth adsorption on different sites above monolayer graphene. (b) Geometric structures of hexagonal Bi array.
\bigskip
\item[Figure 55:] (a) Geometric structures of Bi nanoclusters. (b) Density of states for hexagonal Bi array.

\end{itemize}

\newpage
\begin{figure}
\centering
\includegraphics[scale=0.8]{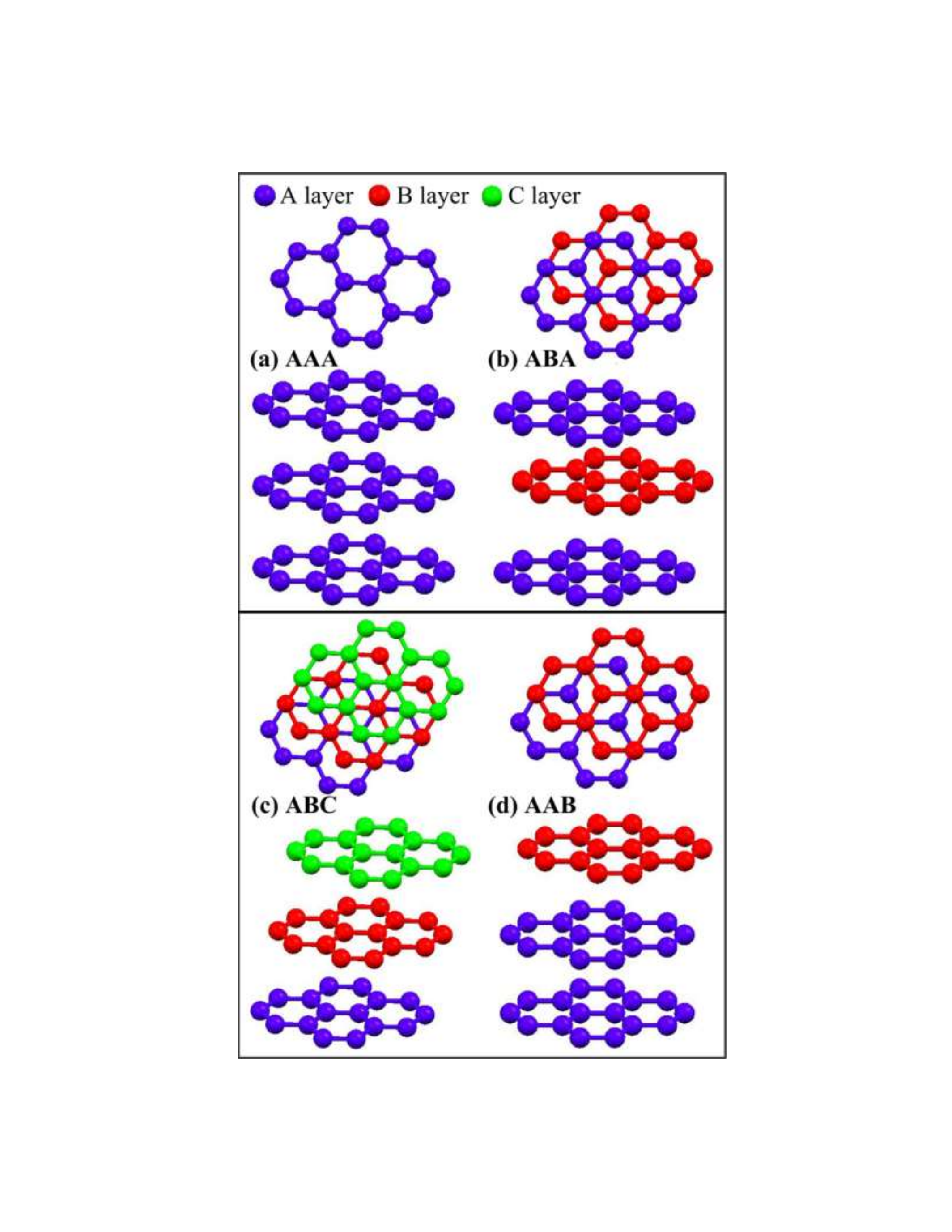}
\centering\caption{}
\end{figure}

\newpage
\begin{figure}
\centering
\includegraphics[scale=0.8]{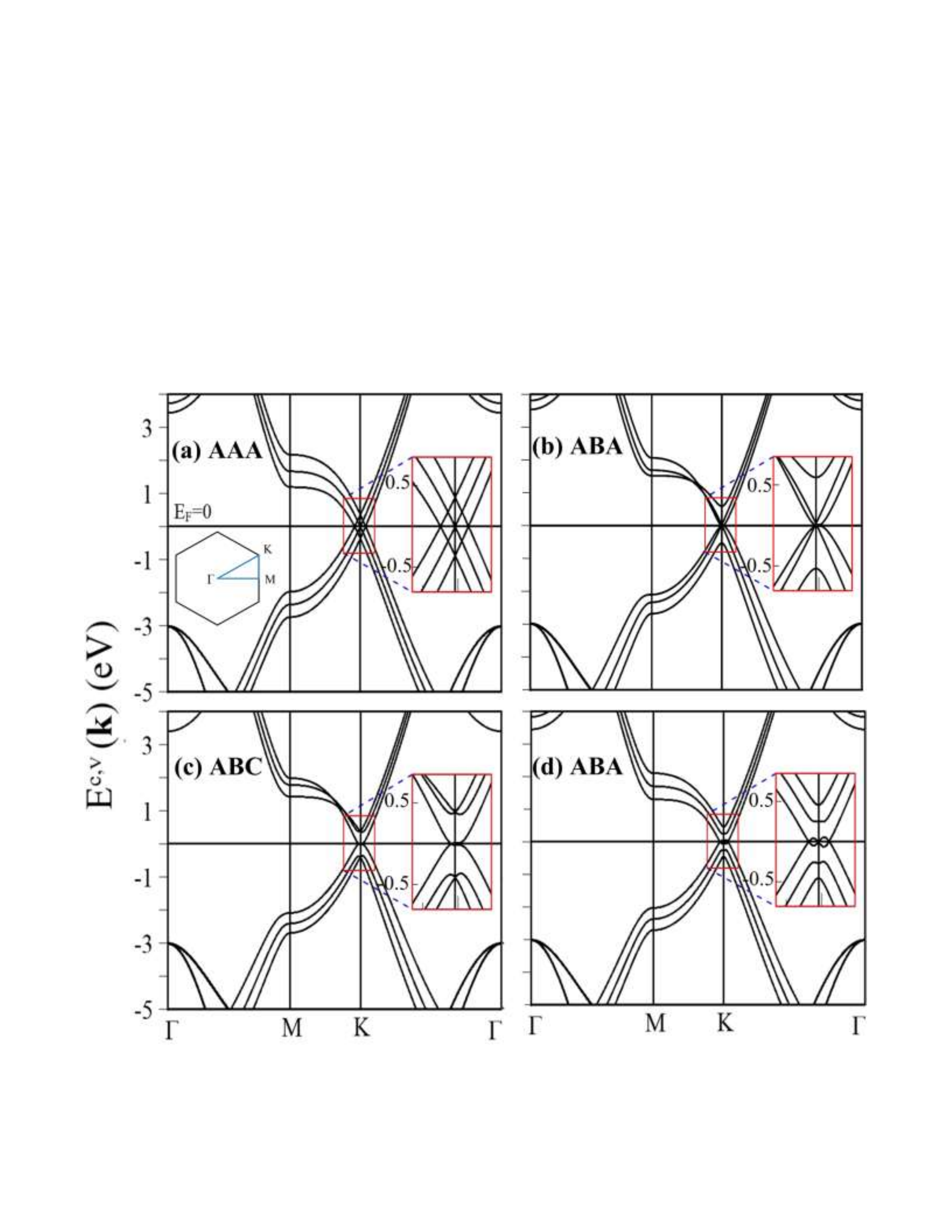}
\centering\caption{}
\end{figure}

\newpage
\begin{figure}
\centering
\includegraphics[scale=0.8]{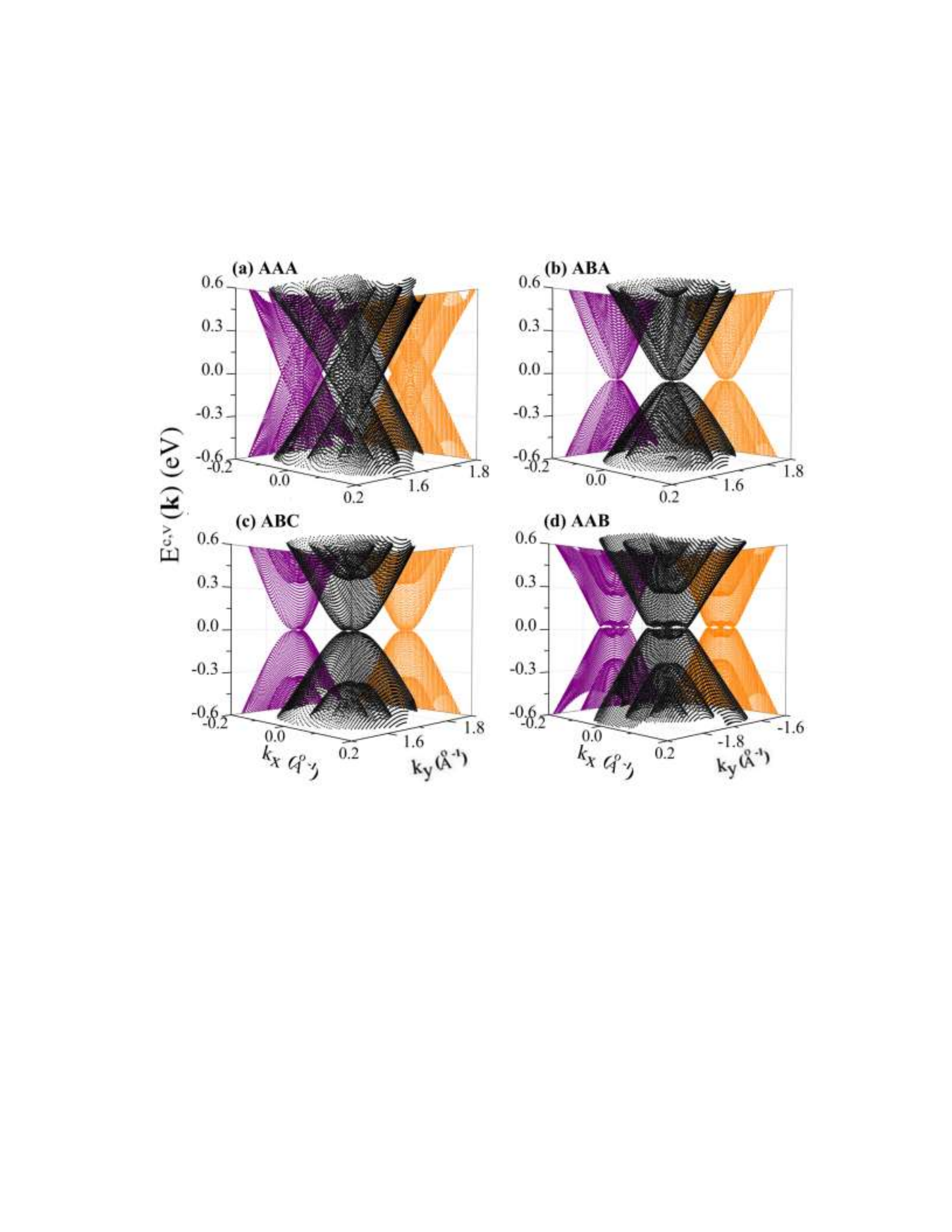}
\centering\caption{}
\end{figure}

\newpage
\begin{figure}
\centering
\includegraphics[scale=0.8]{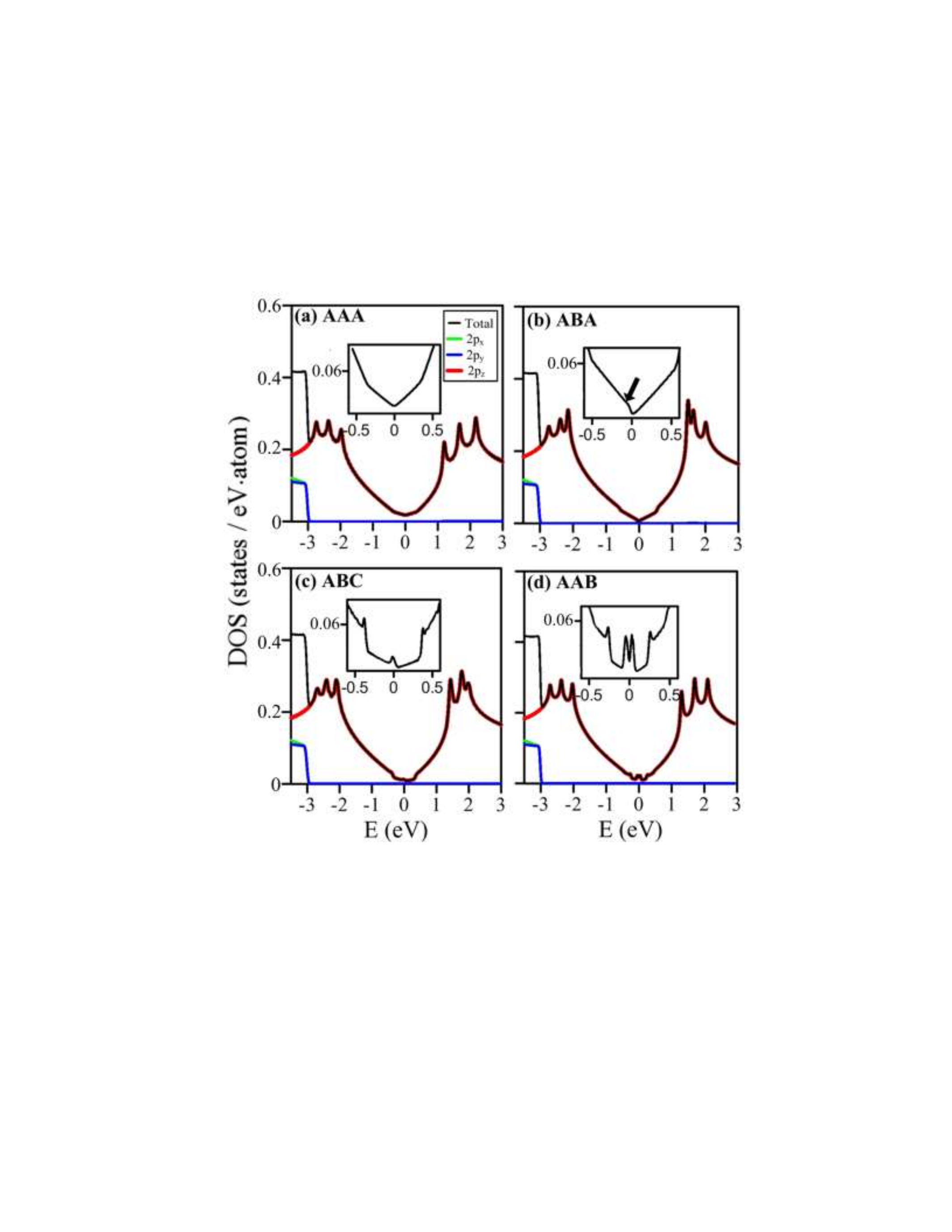}
\centering\caption{}
\end{figure}

\newpage
\begin{figure}
\centering
\includegraphics[scale=0.8]{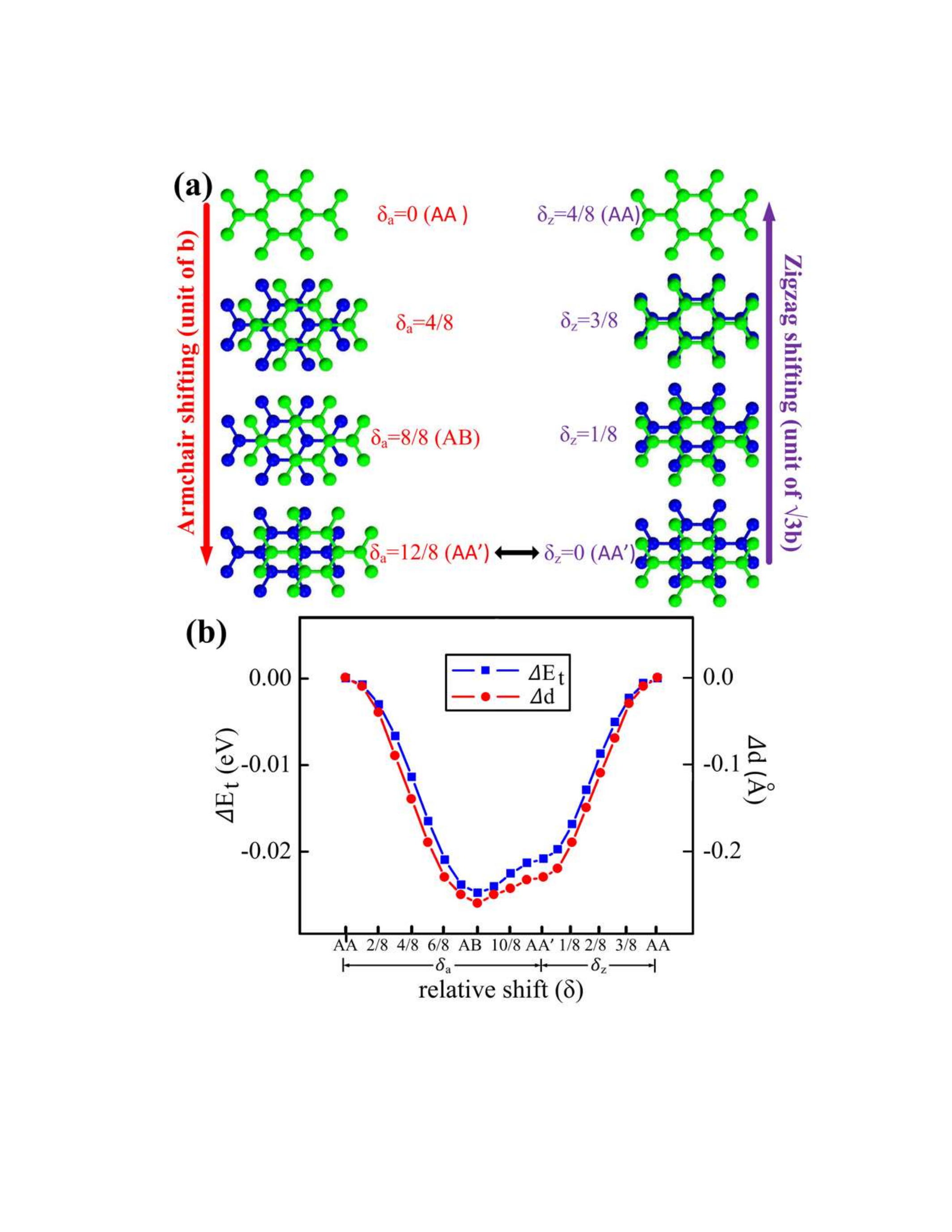}
\centering\caption{}
\end{figure}

\newpage
\begin{figure}
\centering
\includegraphics[scale=0.8]{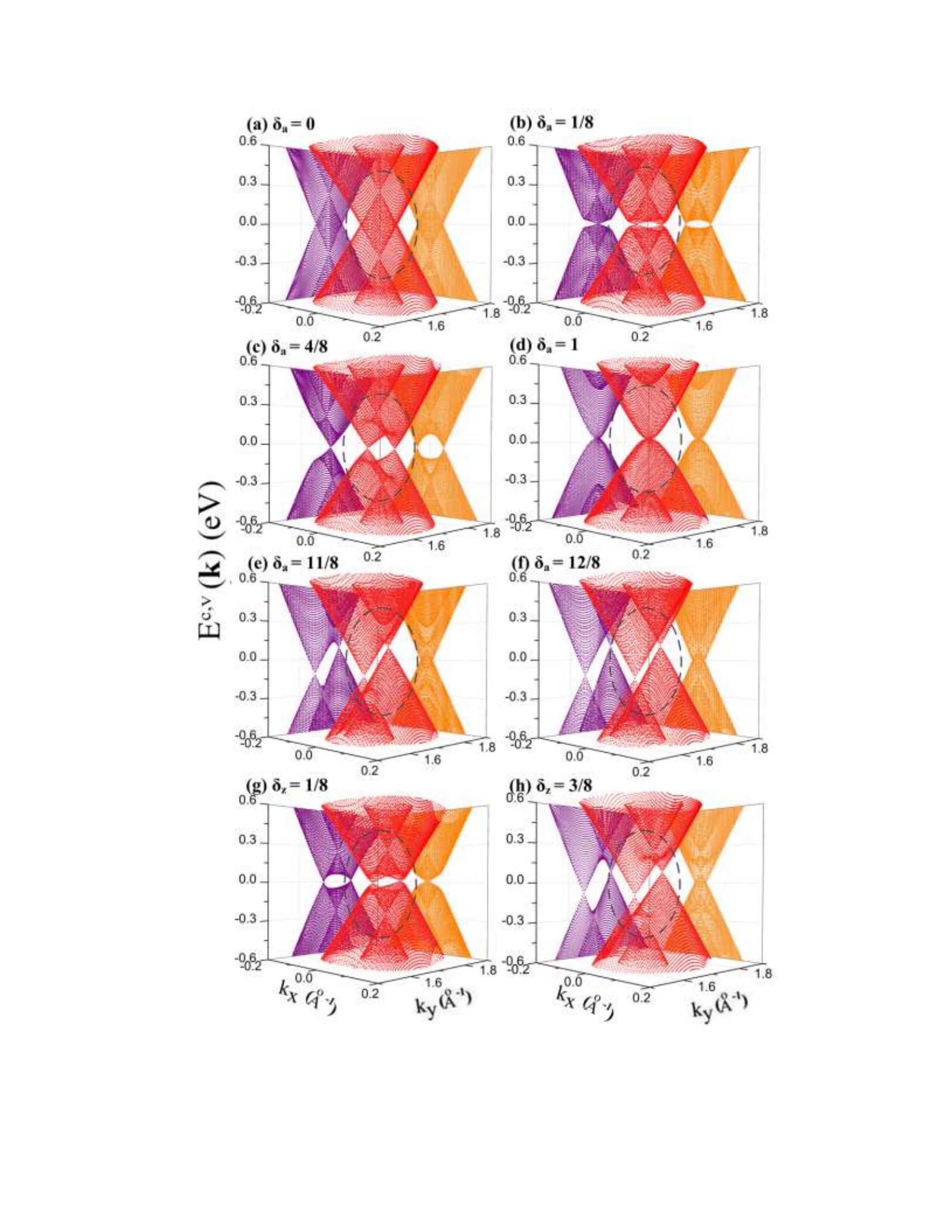}
\centering\caption{}
\end{figure}

\newpage
\begin{figure}
\centering
\includegraphics[scale=0.8]{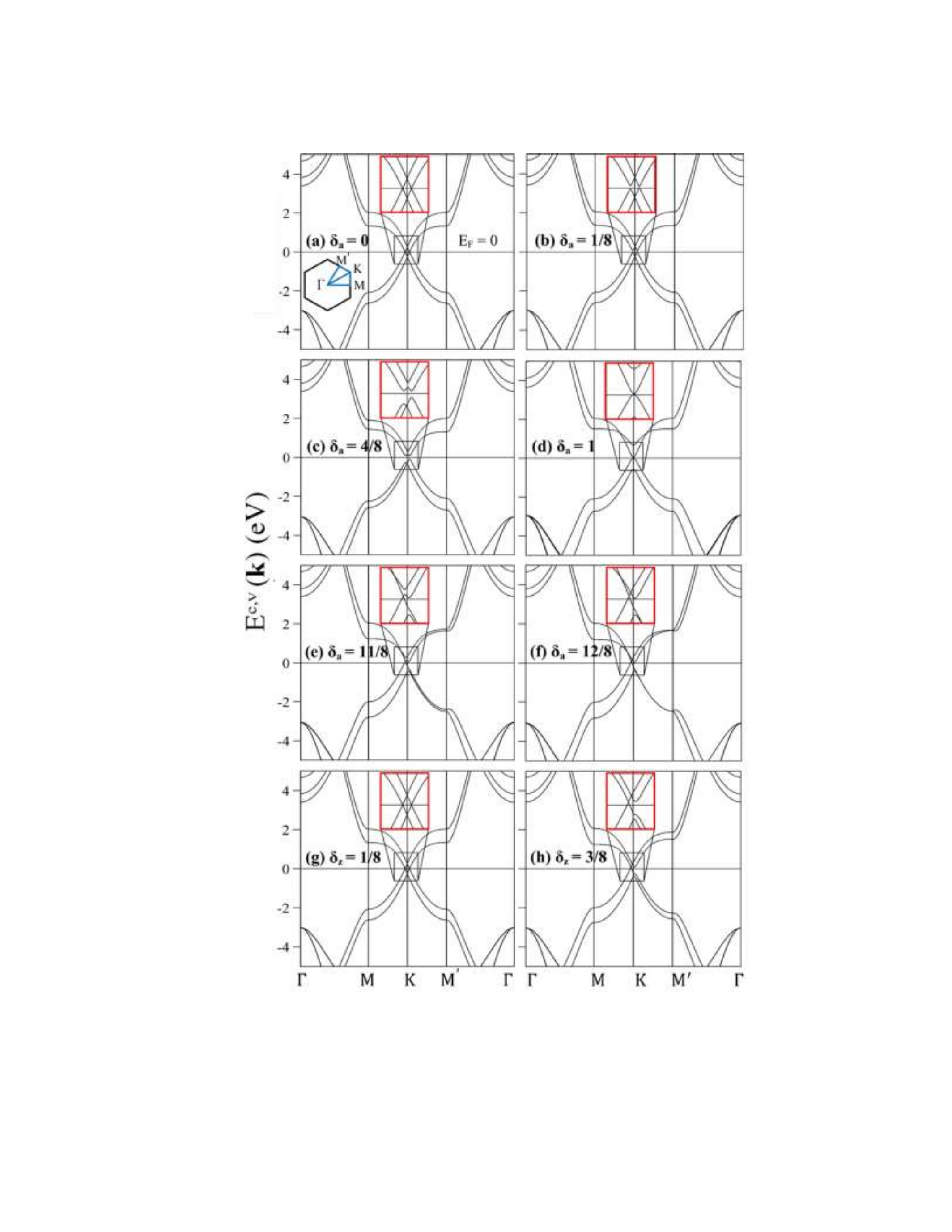}
\centering\caption{}
\end{figure}

\newpage
\begin{figure}
\centering
\includegraphics[scale=0.8]{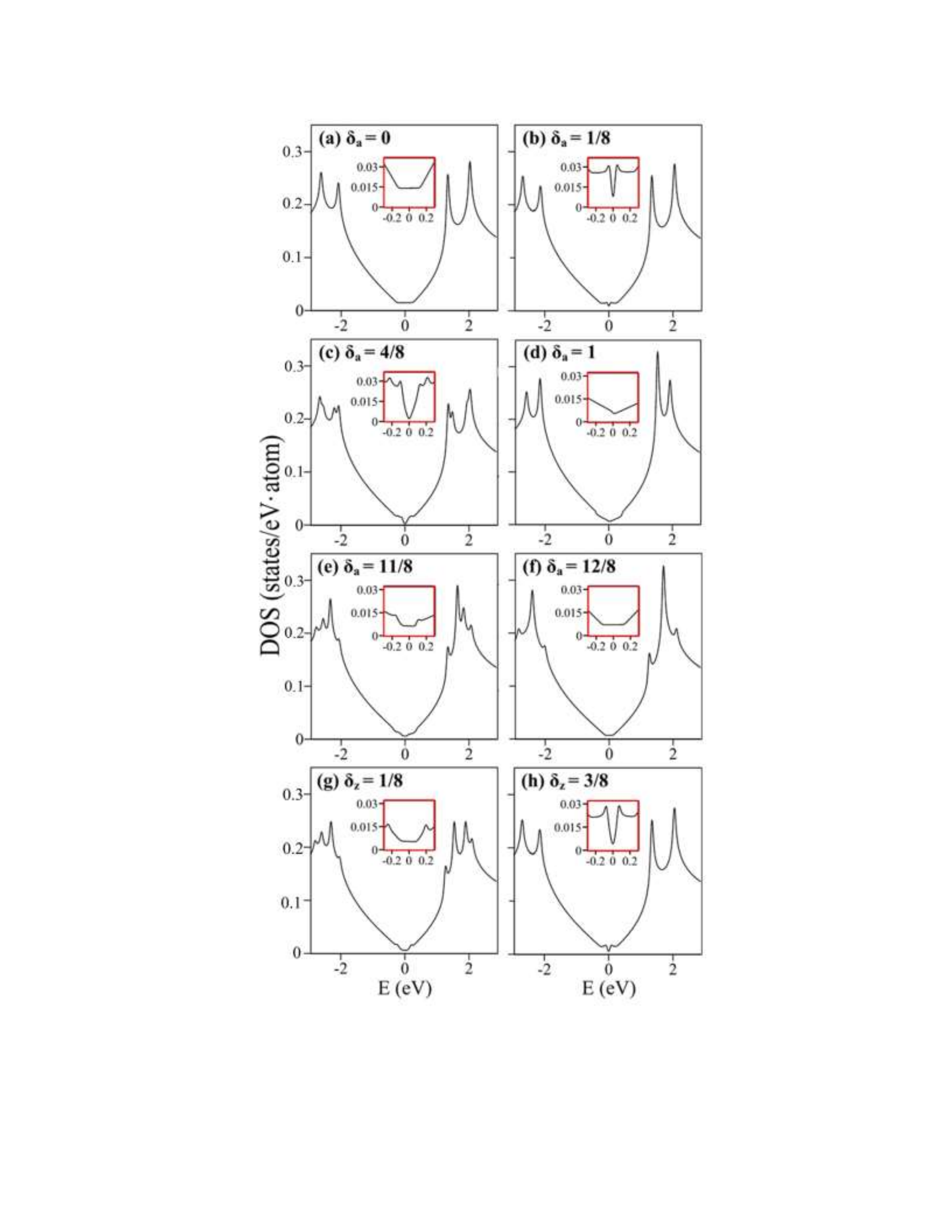}
\centering\caption{}
\end{figure}

\newpage
\begin{figure}
\centering
\includegraphics[scale=0.8]{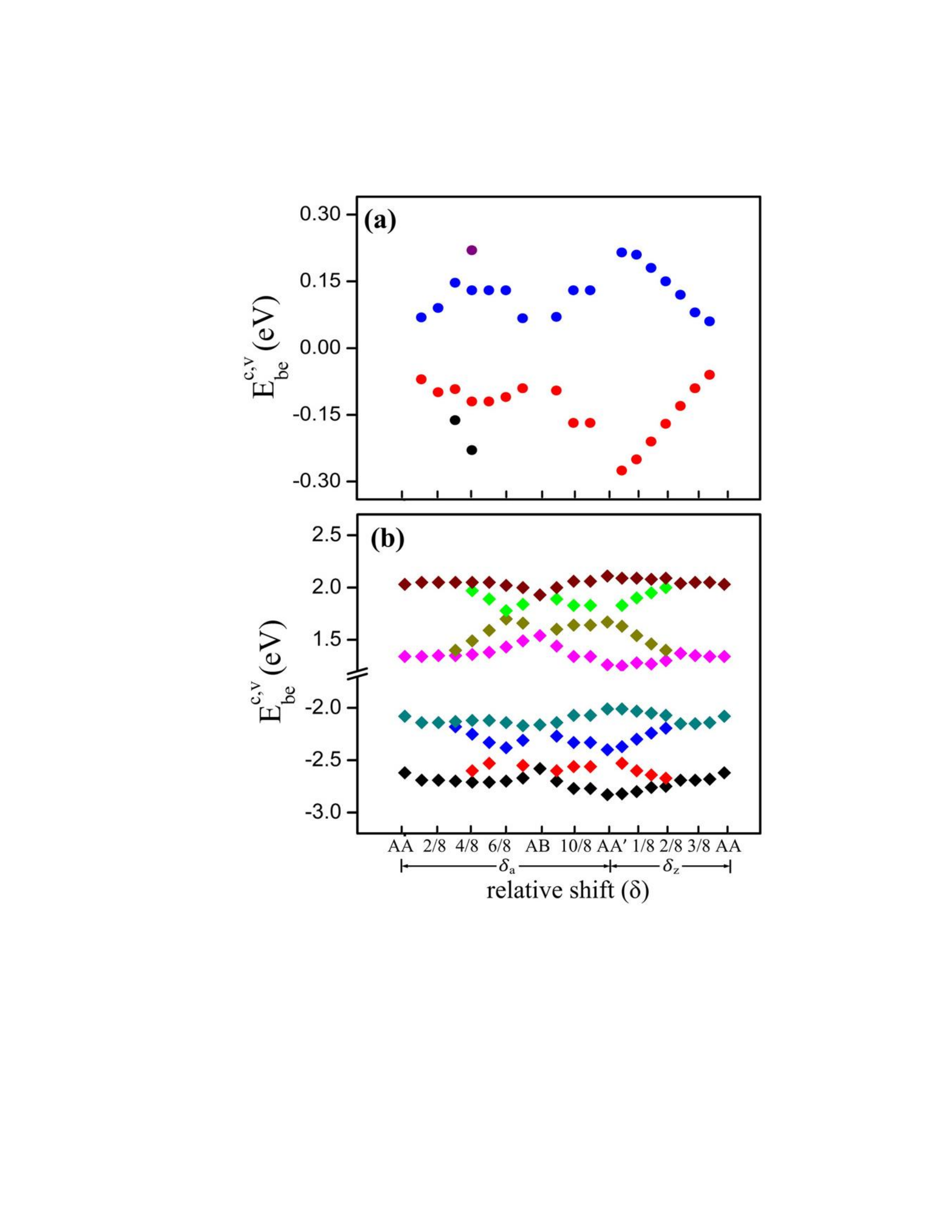}
\centering\caption{}
\end{figure}

\newpage
\begin{figure}
\centering
\includegraphics[scale=0.8]{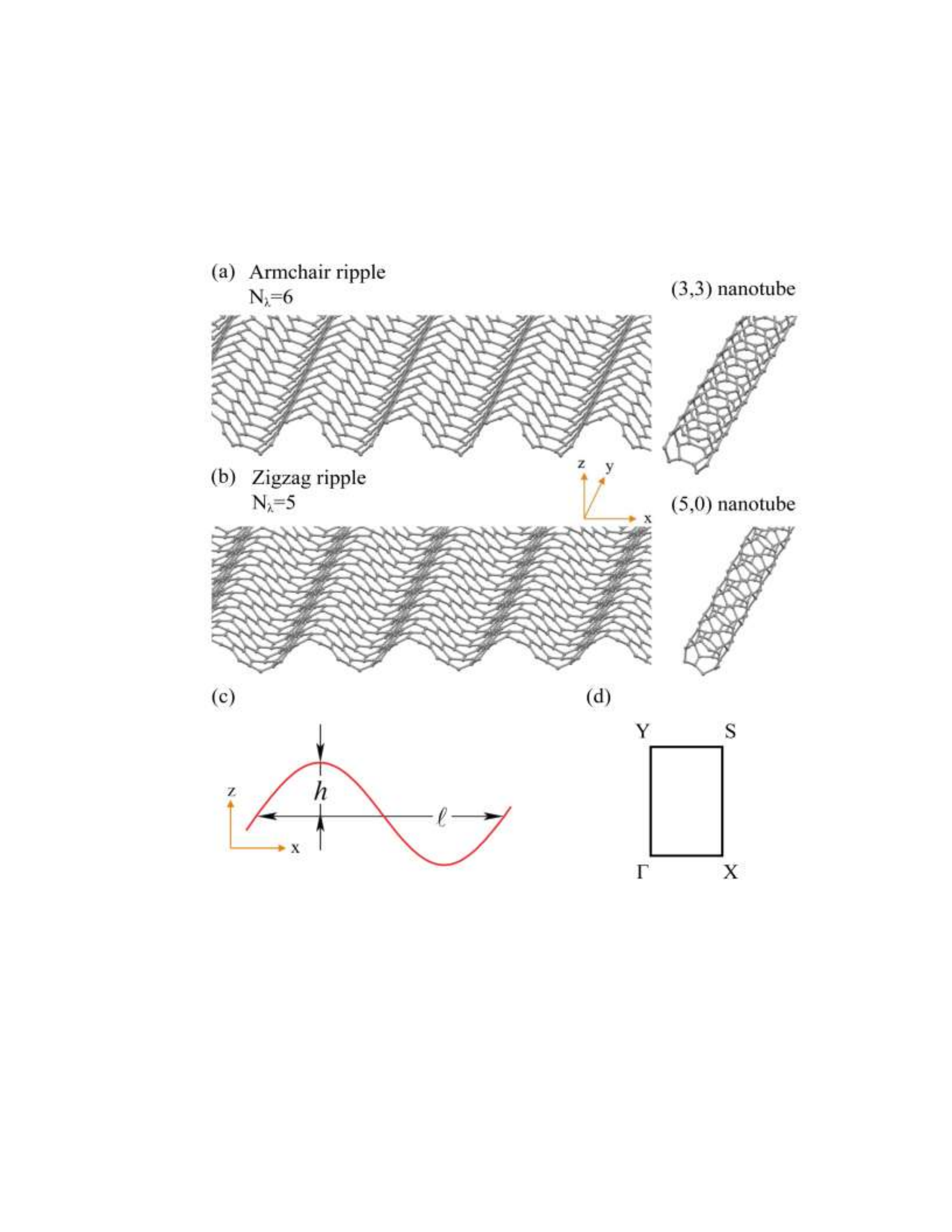}
\centering\caption{}
\end{figure}

\newpage
\begin{figure}
\centering
\includegraphics[scale=0.8]{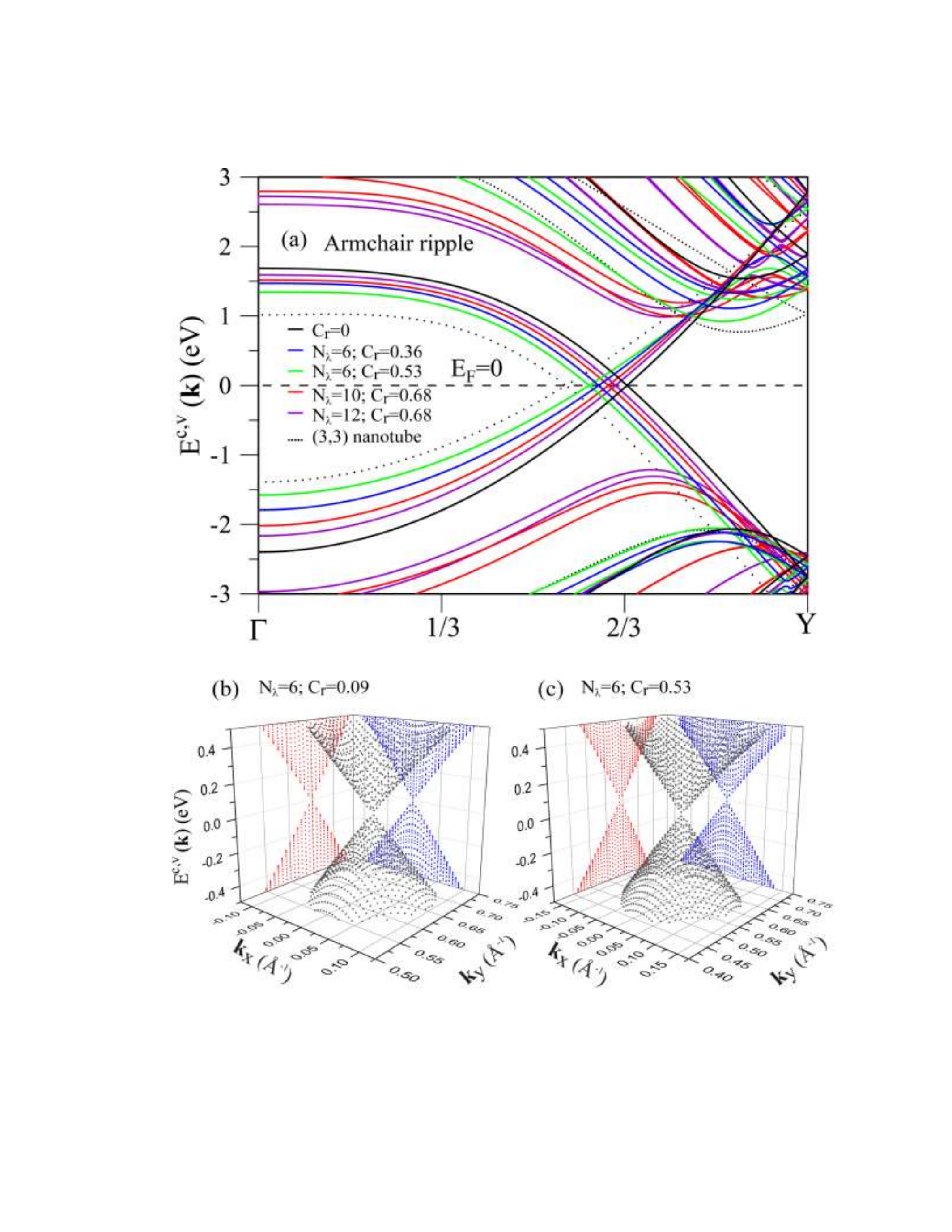}
\centering\caption{}
\end{figure}

\newpage
\begin{figure}
\centering
\includegraphics[scale=0.8]{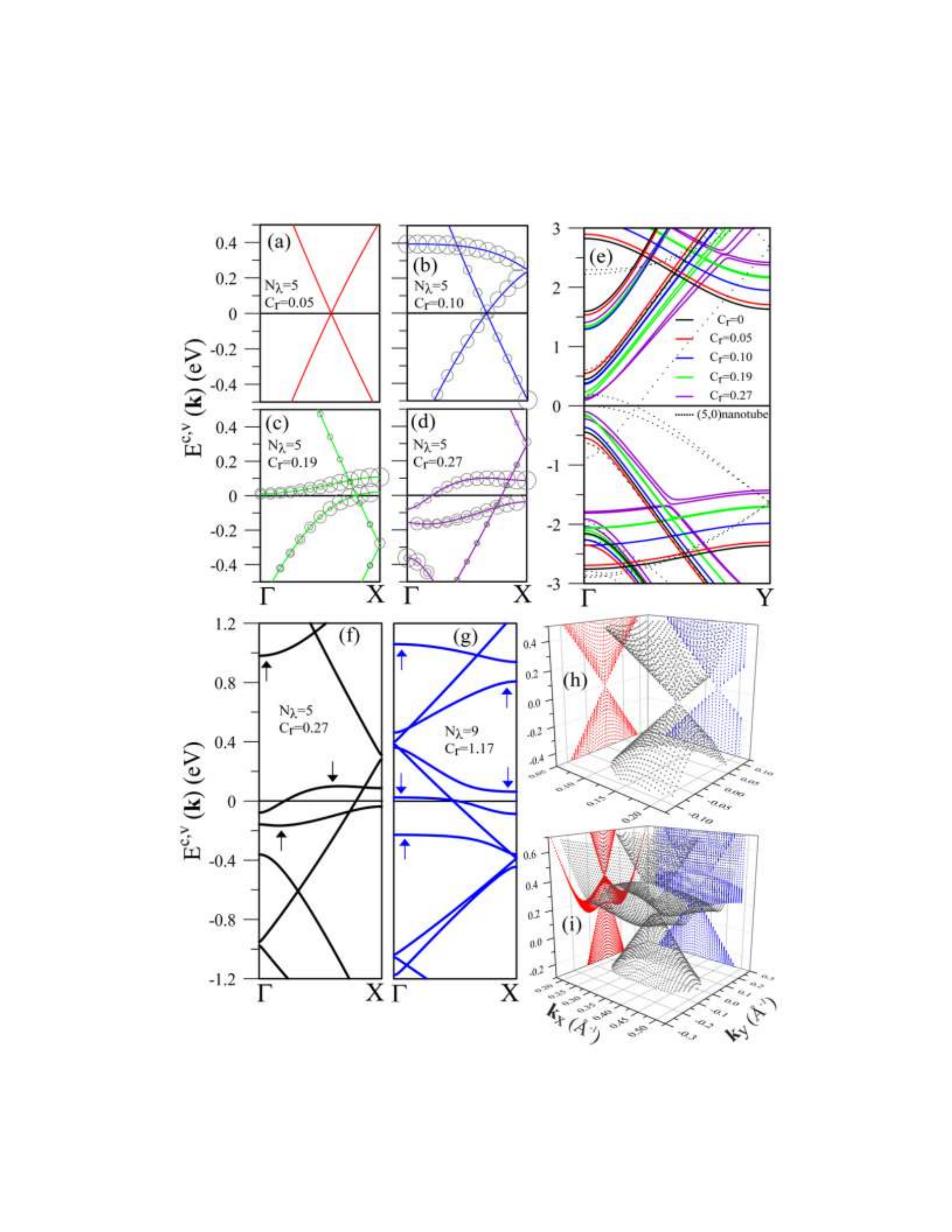}
\centering\caption{}
\end{figure}

\newpage
\begin{figure}
\centering
\includegraphics[scale=0.7]{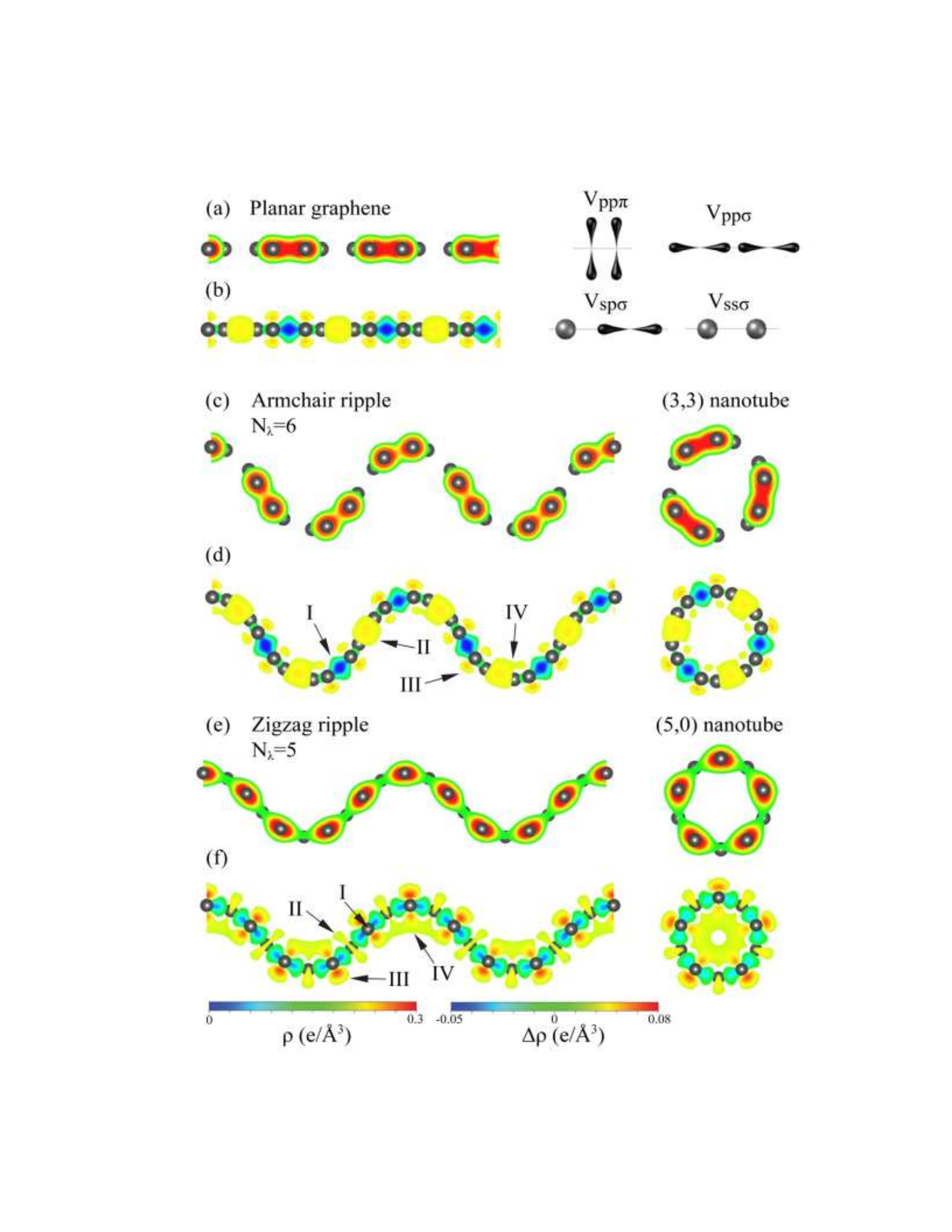}
\centering\caption{}
\end{figure}

\newpage
\begin{figure}
\centering
\includegraphics[scale=0.7]{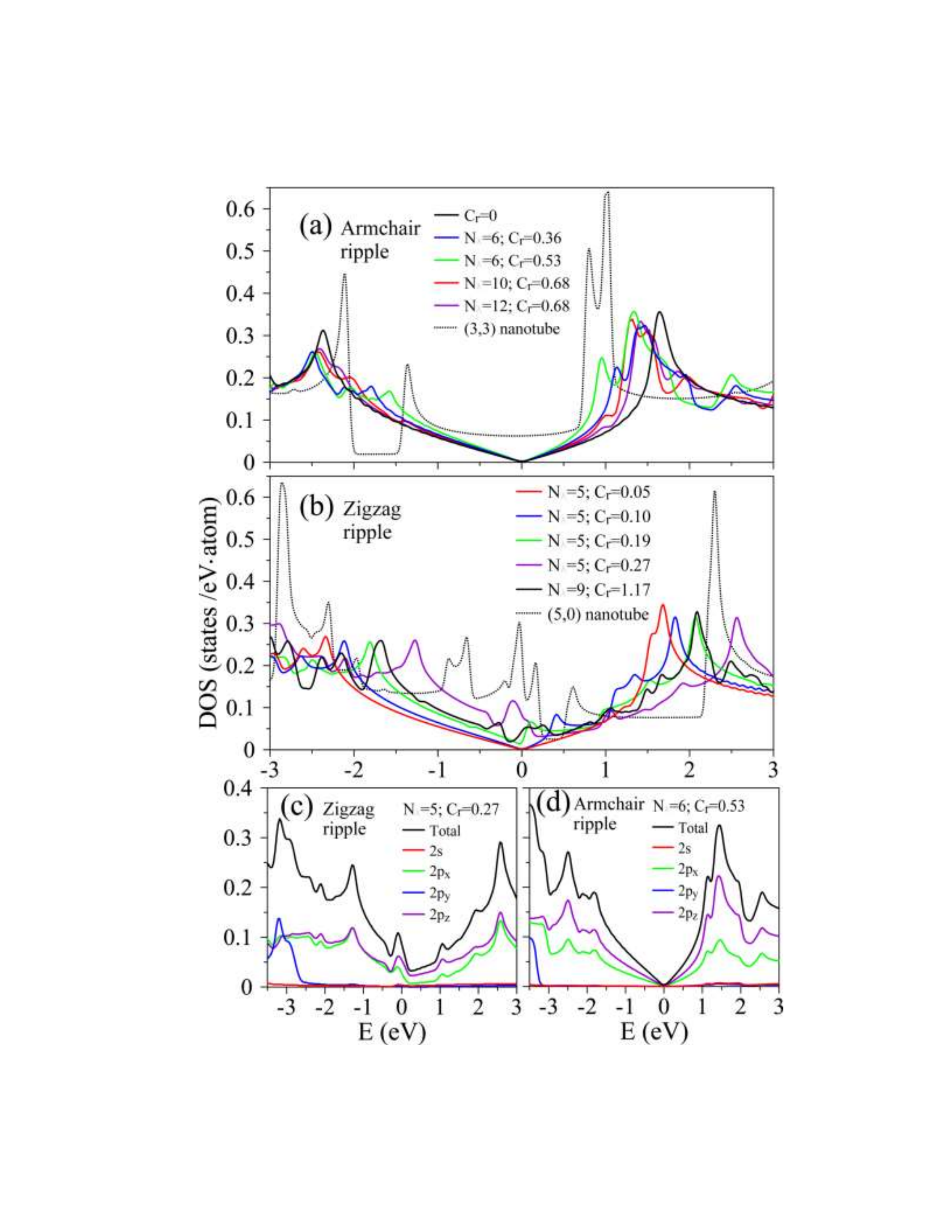}
\centering\caption{}
\end{figure}

\newpage
\begin{figure}
\centering
\includegraphics[scale=0.8]{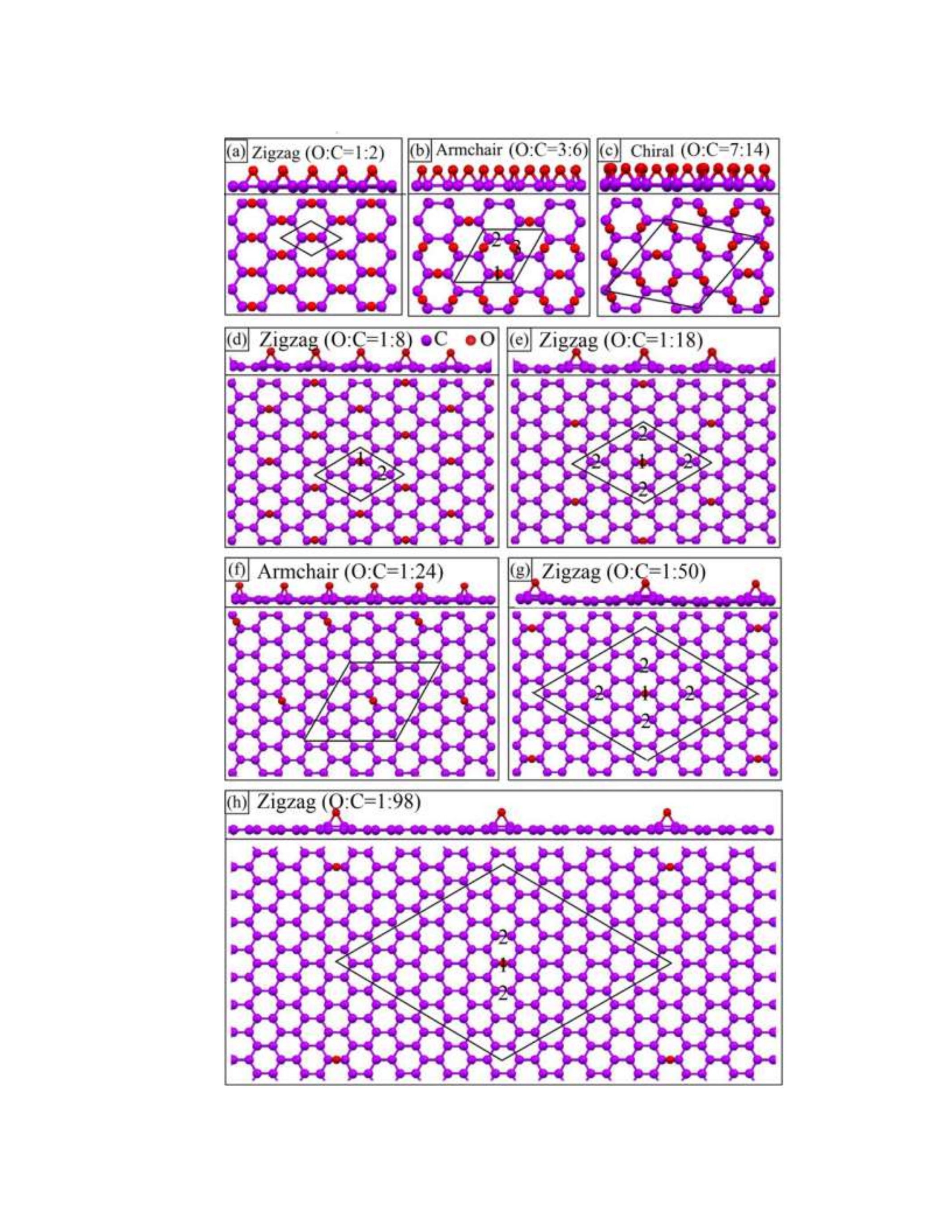}
\centering\caption{}
\end{figure}

\clearpage

\newpage
\begin{figure}
\centering
\includegraphics[scale=0.8]{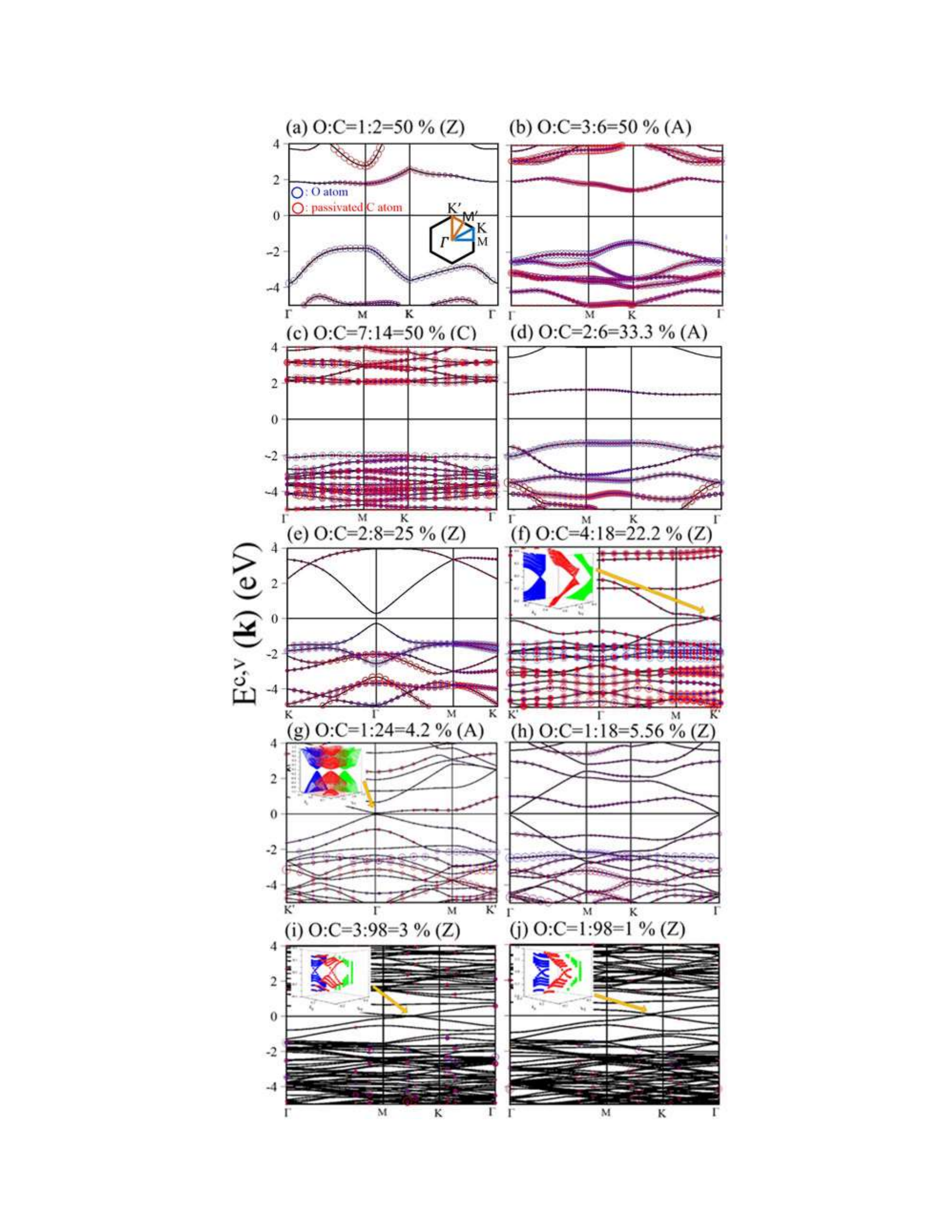}
\centering\caption{}
\end{figure}

\newpage
\begin{figure}
\centering
\includegraphics[scale=0.8]{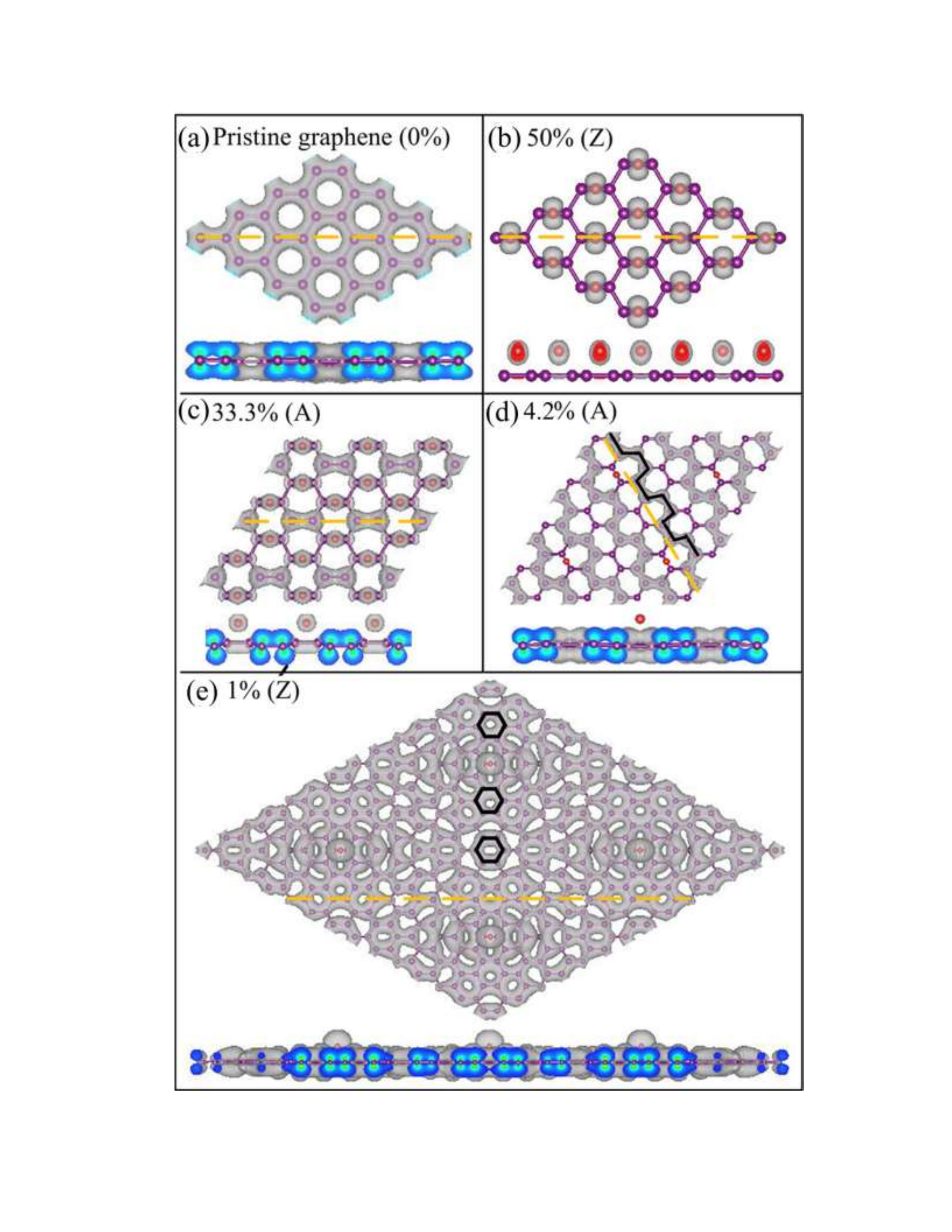}
\centering\caption{}
\end{figure}

\newpage
\begin{figure}
\centering
\includegraphics[scale=0.8]{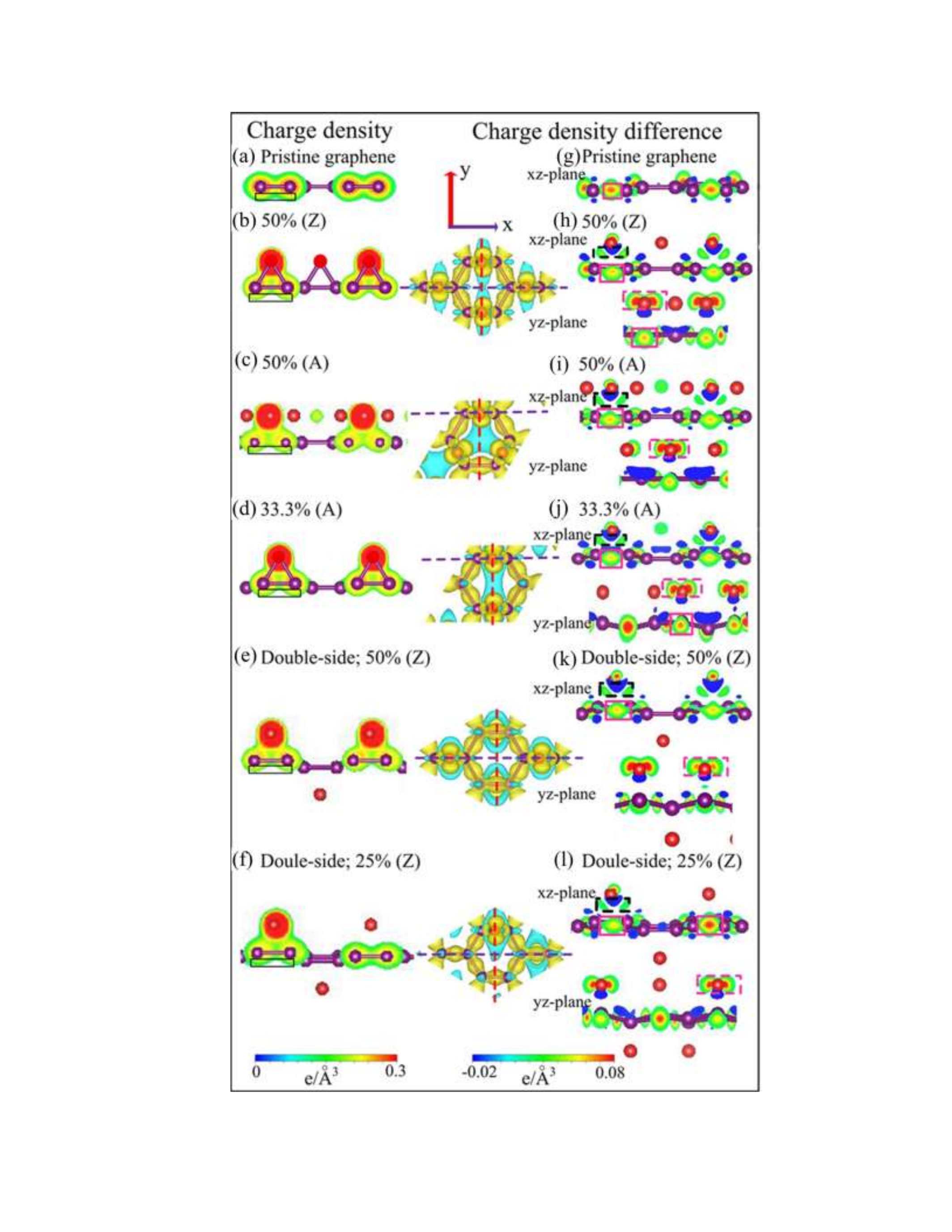}
\centering\caption{}
\end{figure}

\newpage
\begin{figure}
\centering
\includegraphics[scale=0.8]{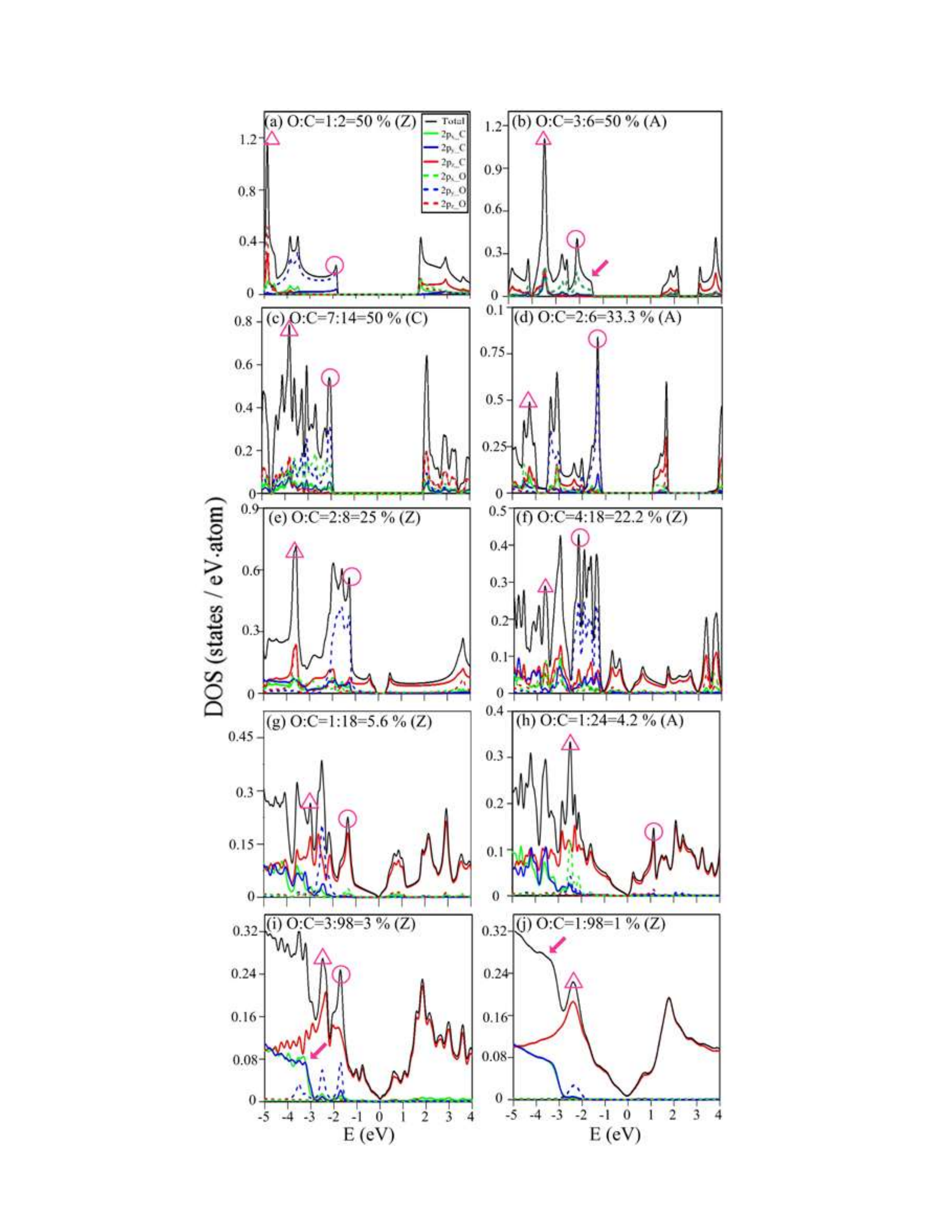}
\centering\caption{}
\end{figure}

\newpage
\begin{figure}
\centering
\includegraphics[scale=0.8]{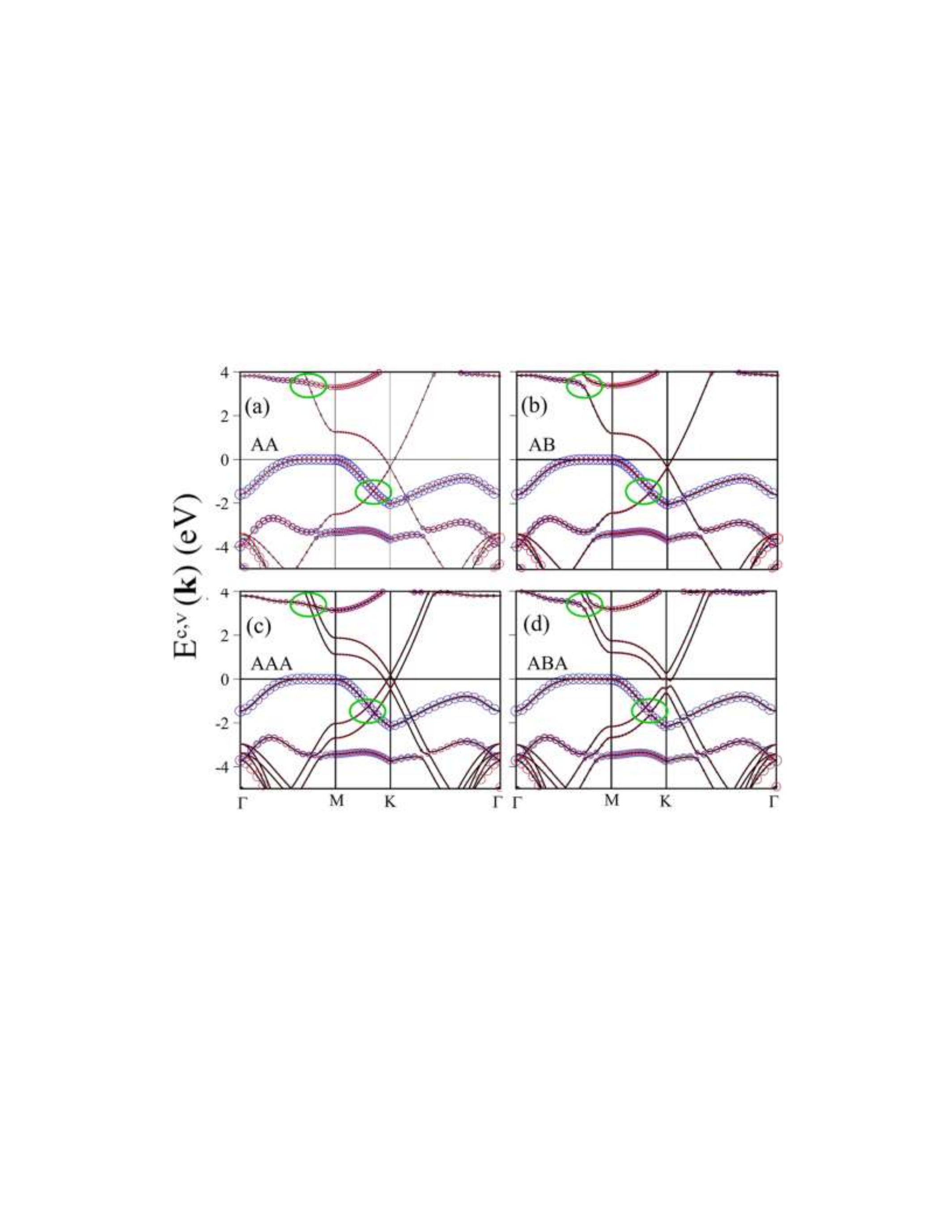}
\centering\caption{}
\end{figure}

\newpage
\begin{figure}
\centering
\includegraphics[scale=0.8]{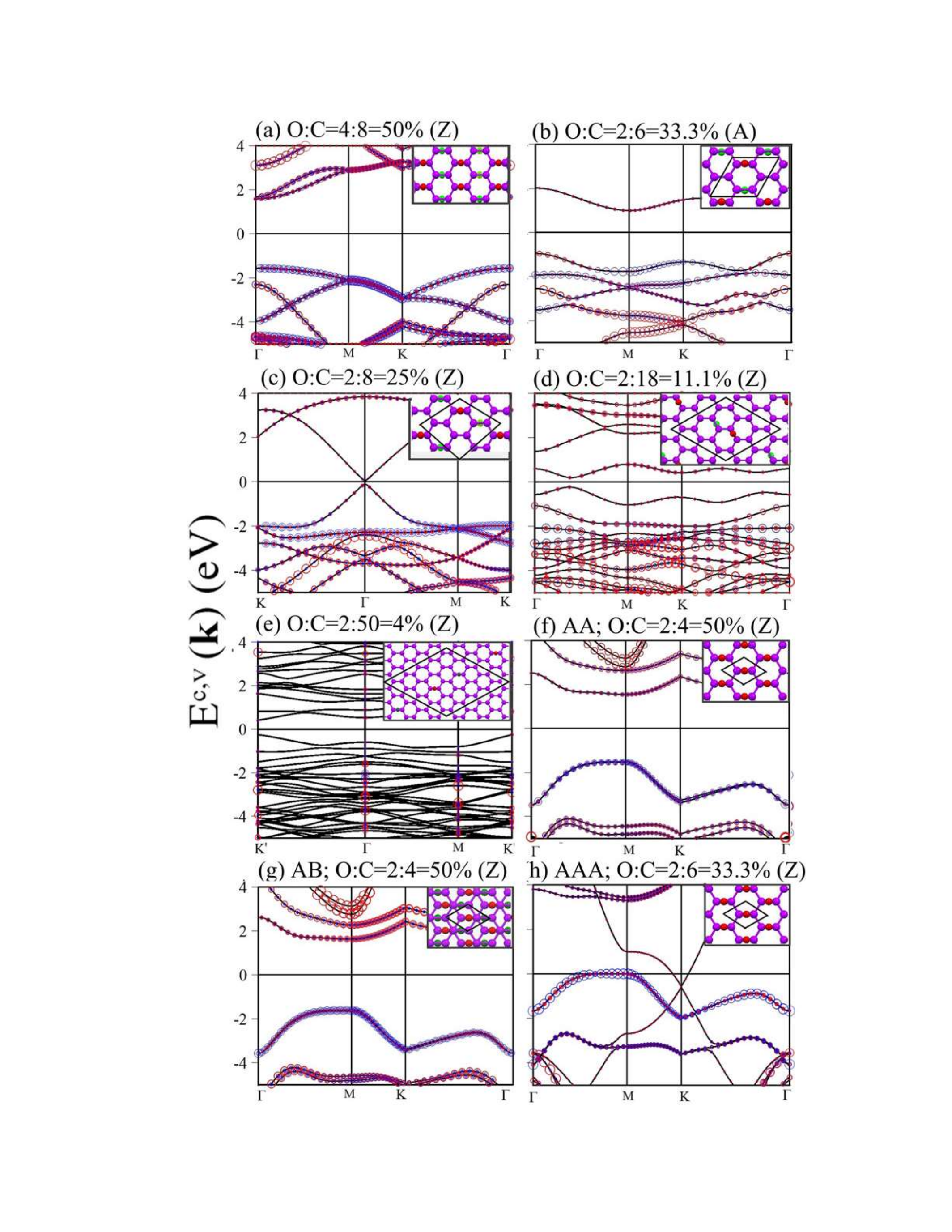}
\centering\caption{}
\end{figure}

\newpage
\begin{figure}
\centering
\includegraphics[scale=0.8]{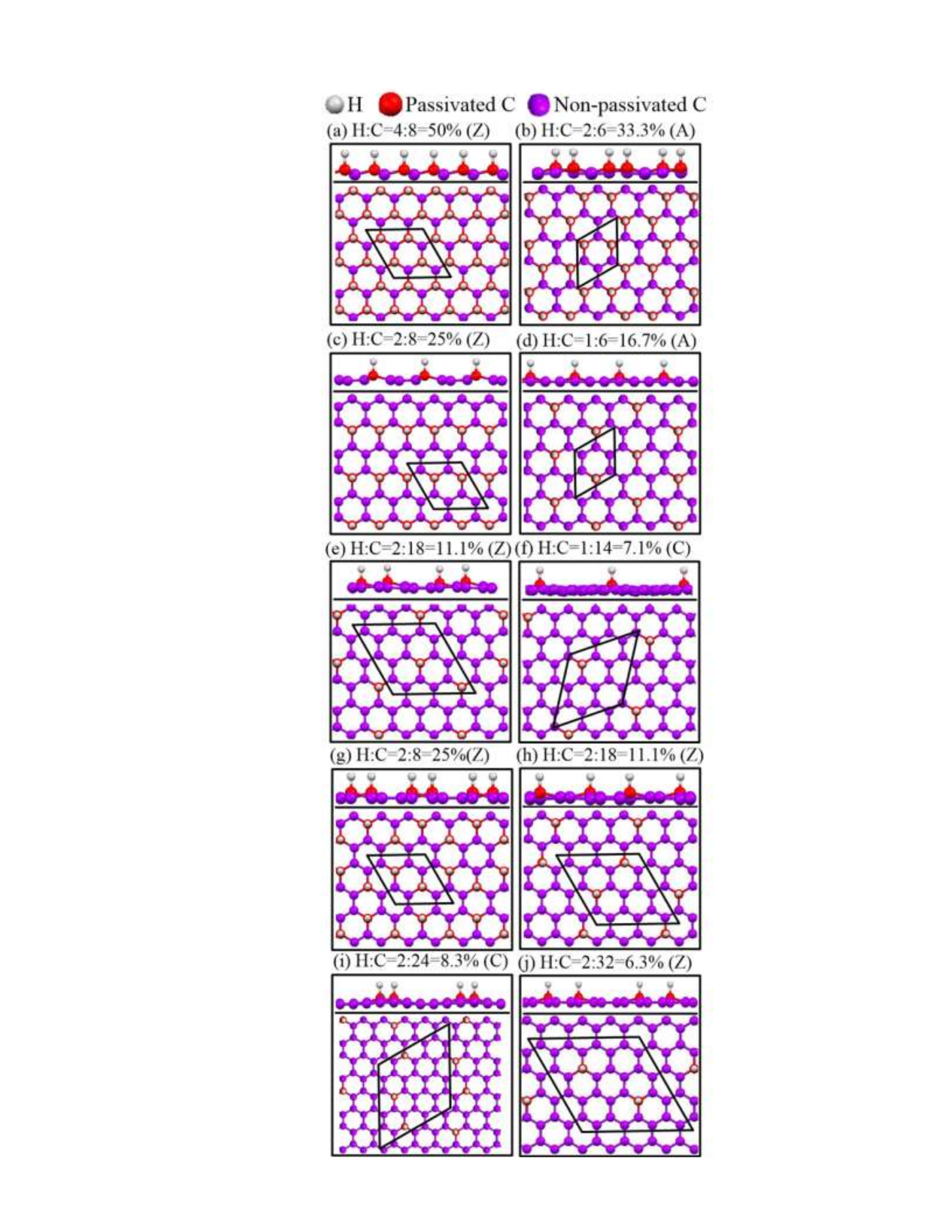}
\centering\caption{}
\end{figure}

\newpage
\begin{figure}
\centering
\includegraphics[scale=0.8]{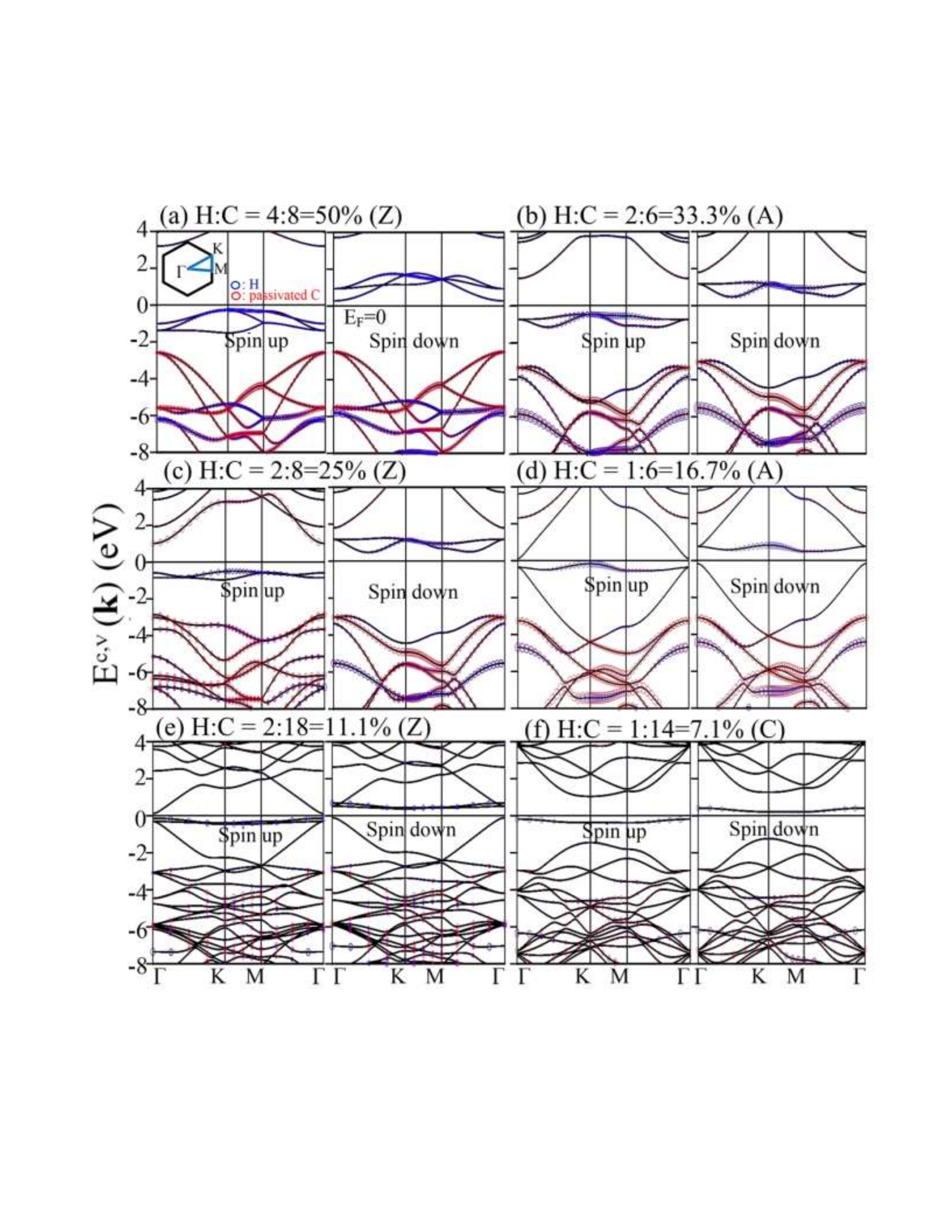}
\centering\caption{}
\end{figure}

\newpage
\begin{figure}
\centering
\includegraphics[scale=0.8]{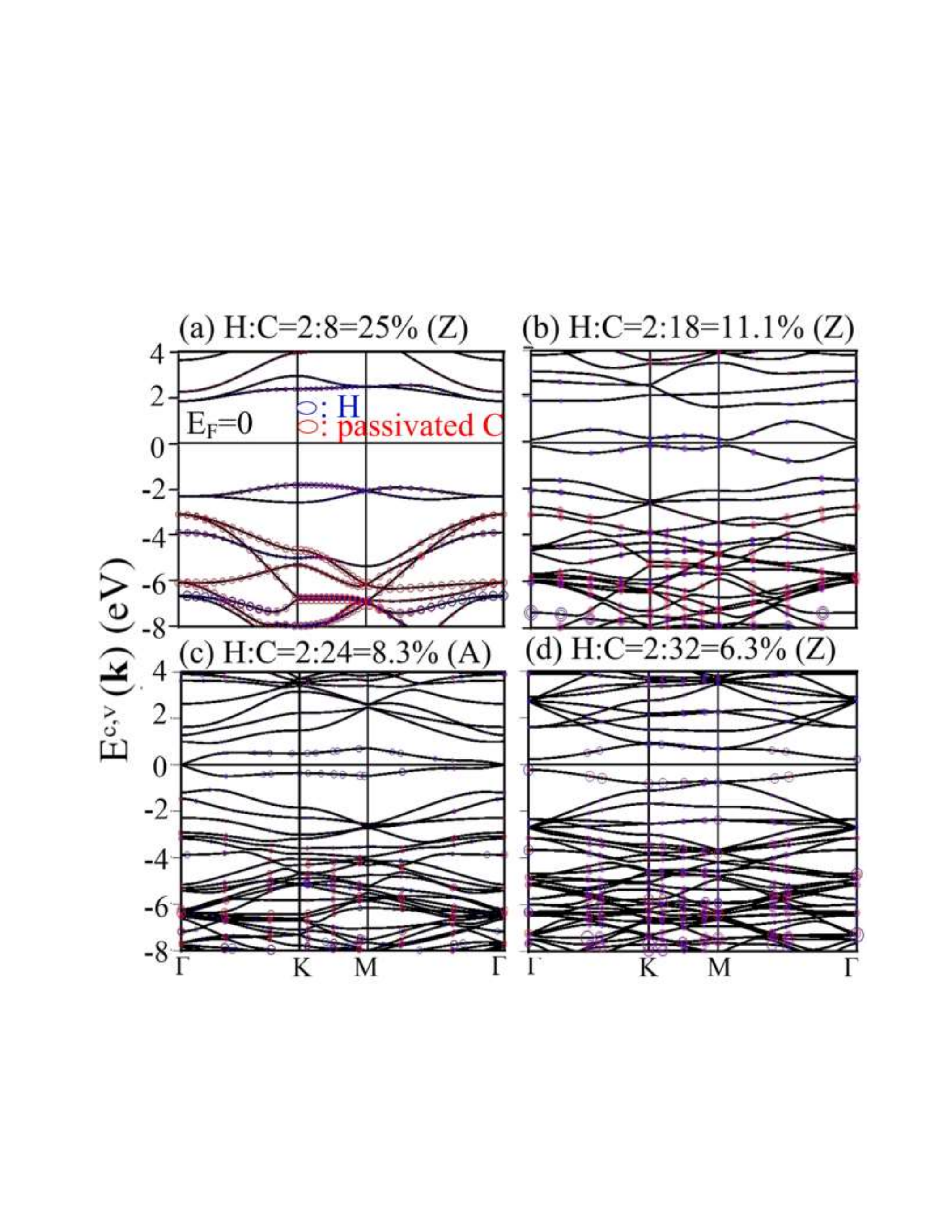}
\centering\caption{}
\end{figure}

\newpage
\begin{figure}
\centering
\includegraphics[scale=0.8]{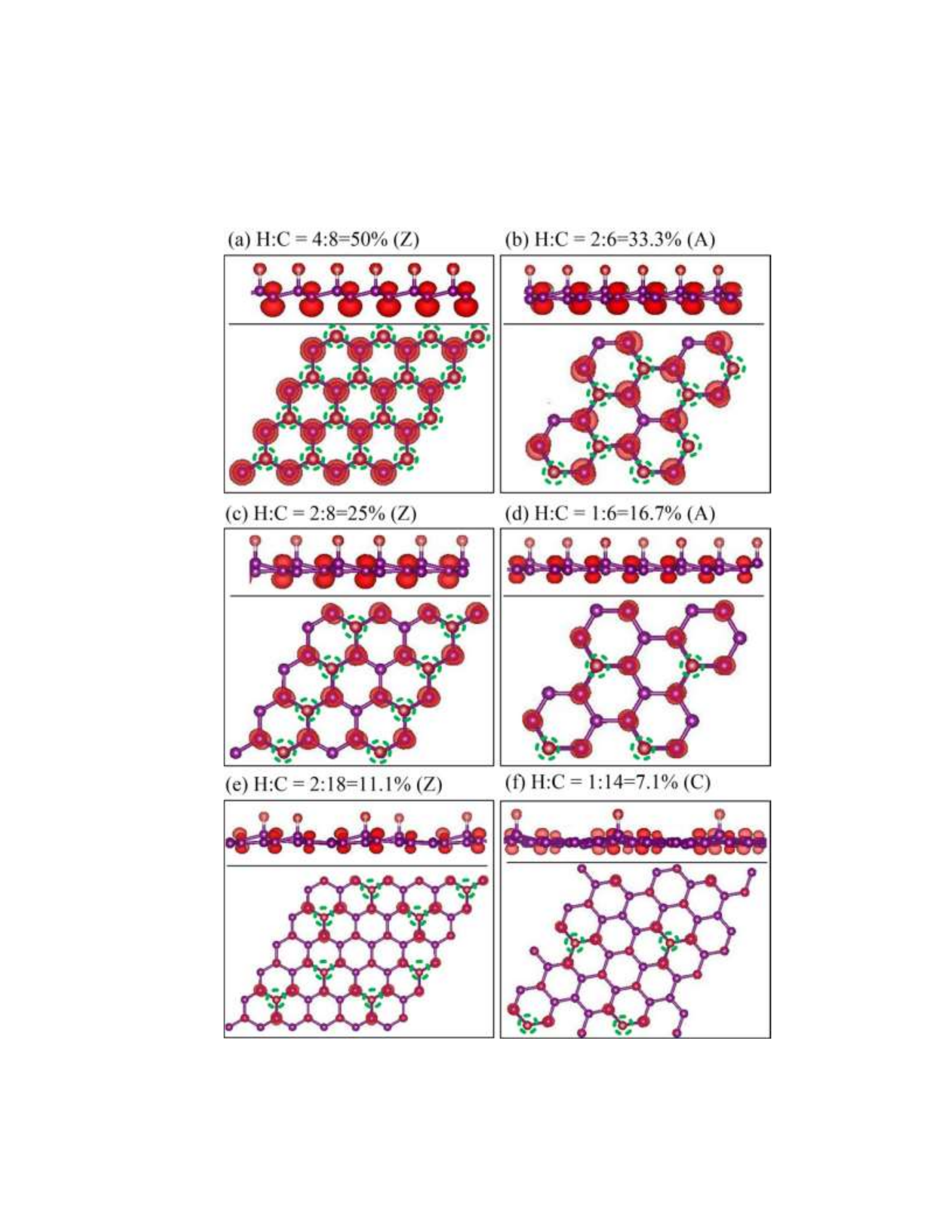}
\centering\caption{}
\end{figure}

\newpage
\begin{figure}
\centering
\includegraphics[scale=0.8]{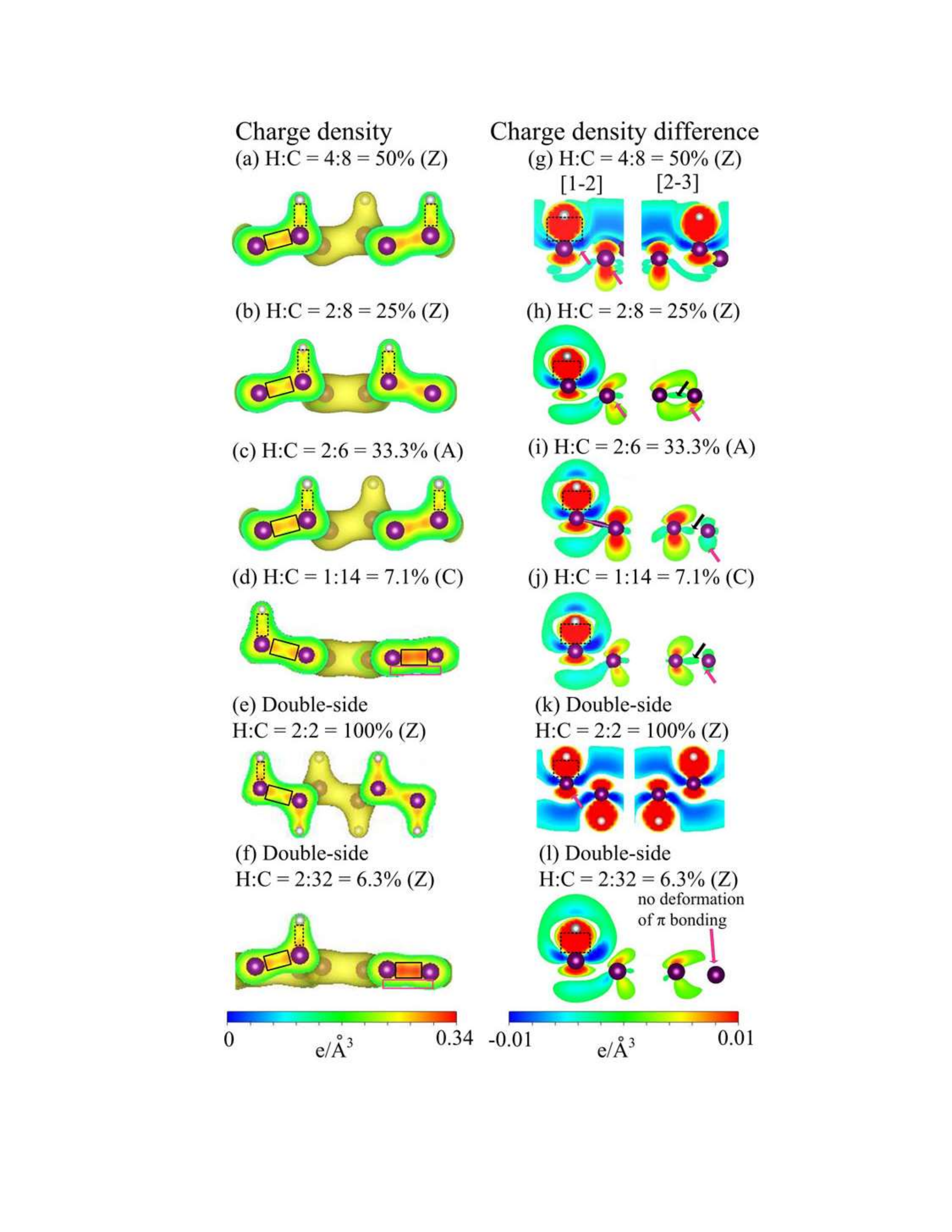}
\centering\caption{}
\end{figure}

\newpage
\begin{figure}
\centering
\includegraphics[scale=0.8]{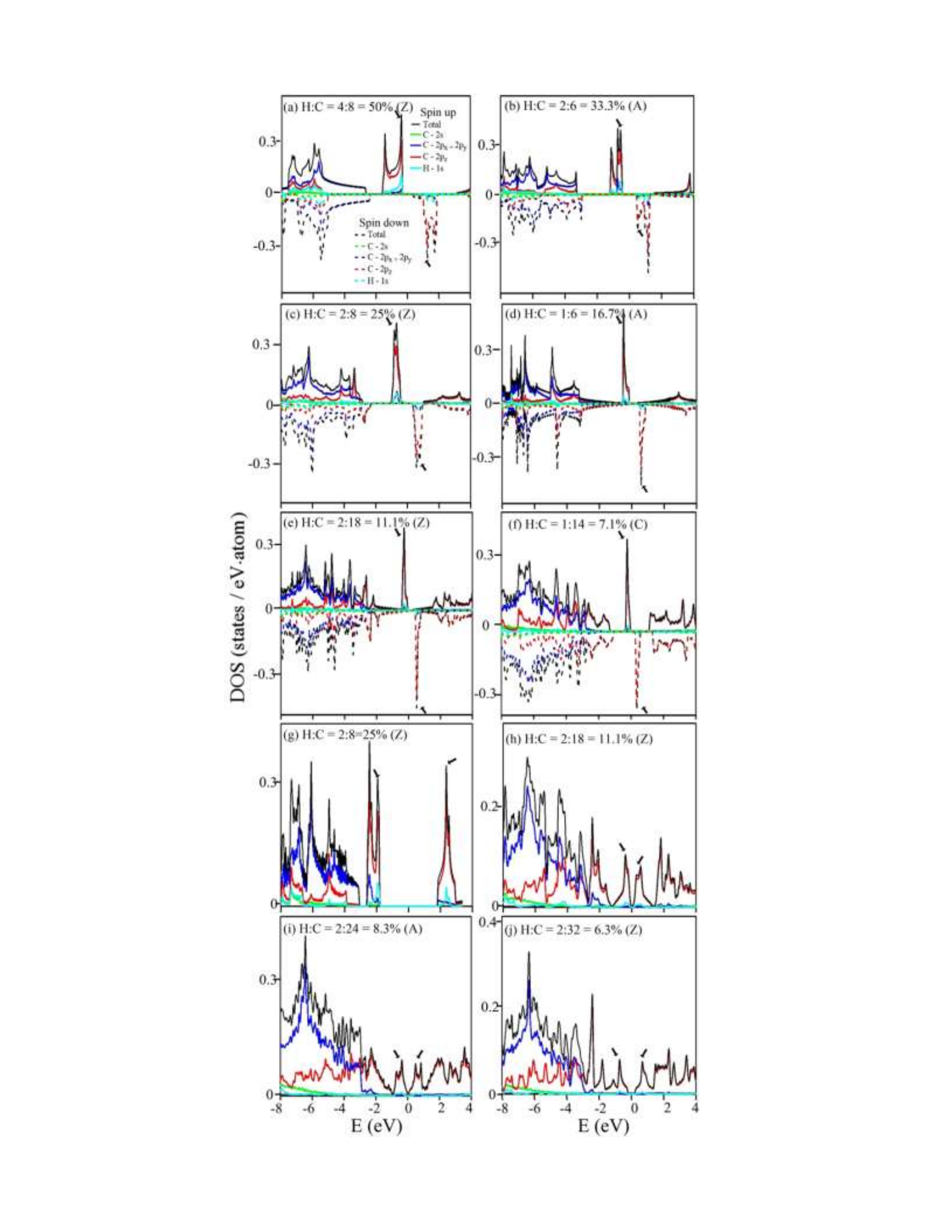}
\centering\caption{}
\end{figure}

\newpage
\begin{figure}
\centering
\includegraphics[scale=0.8]{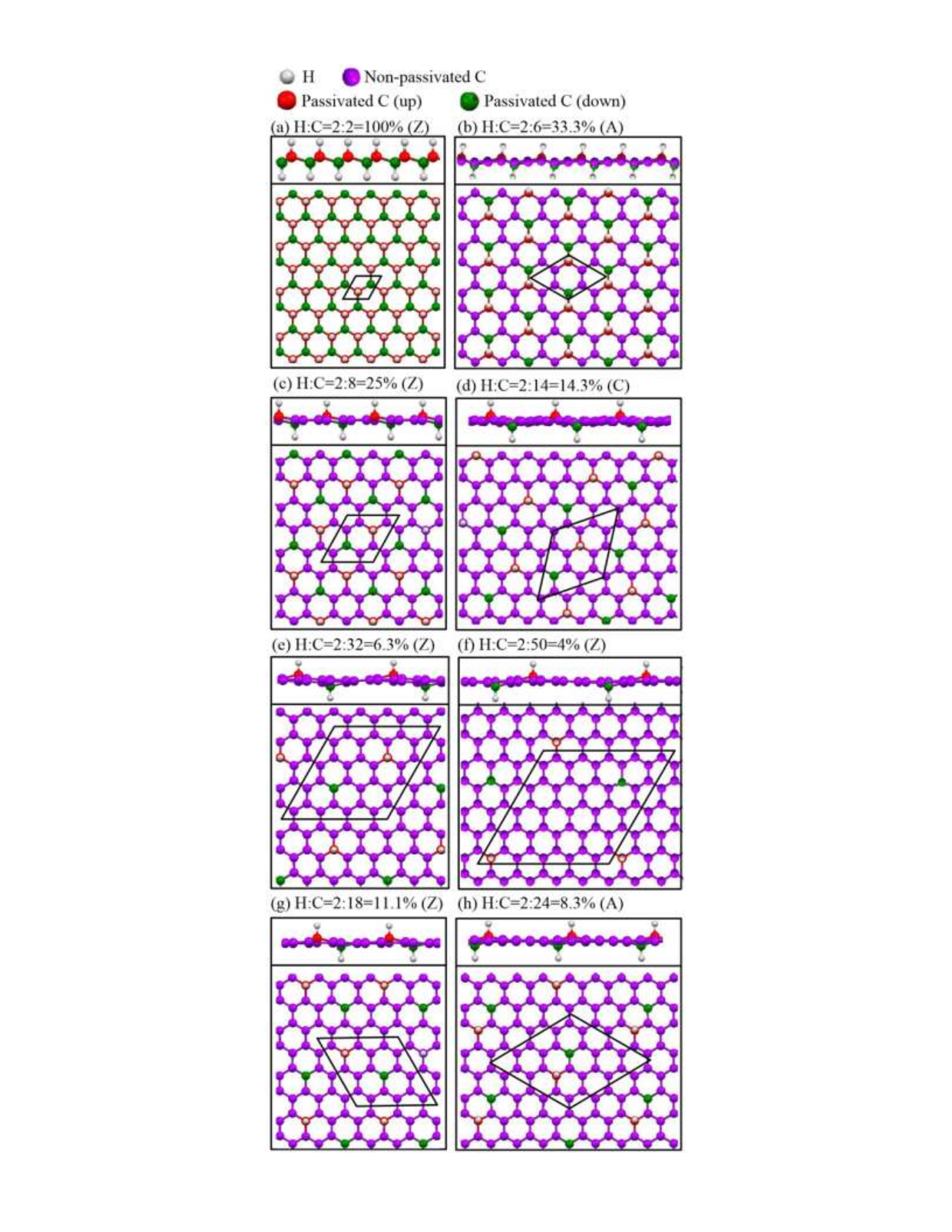}
\centering\caption{}
\end{figure}

\newpage
\begin{figure}
\centering
\includegraphics[scale=0.8]{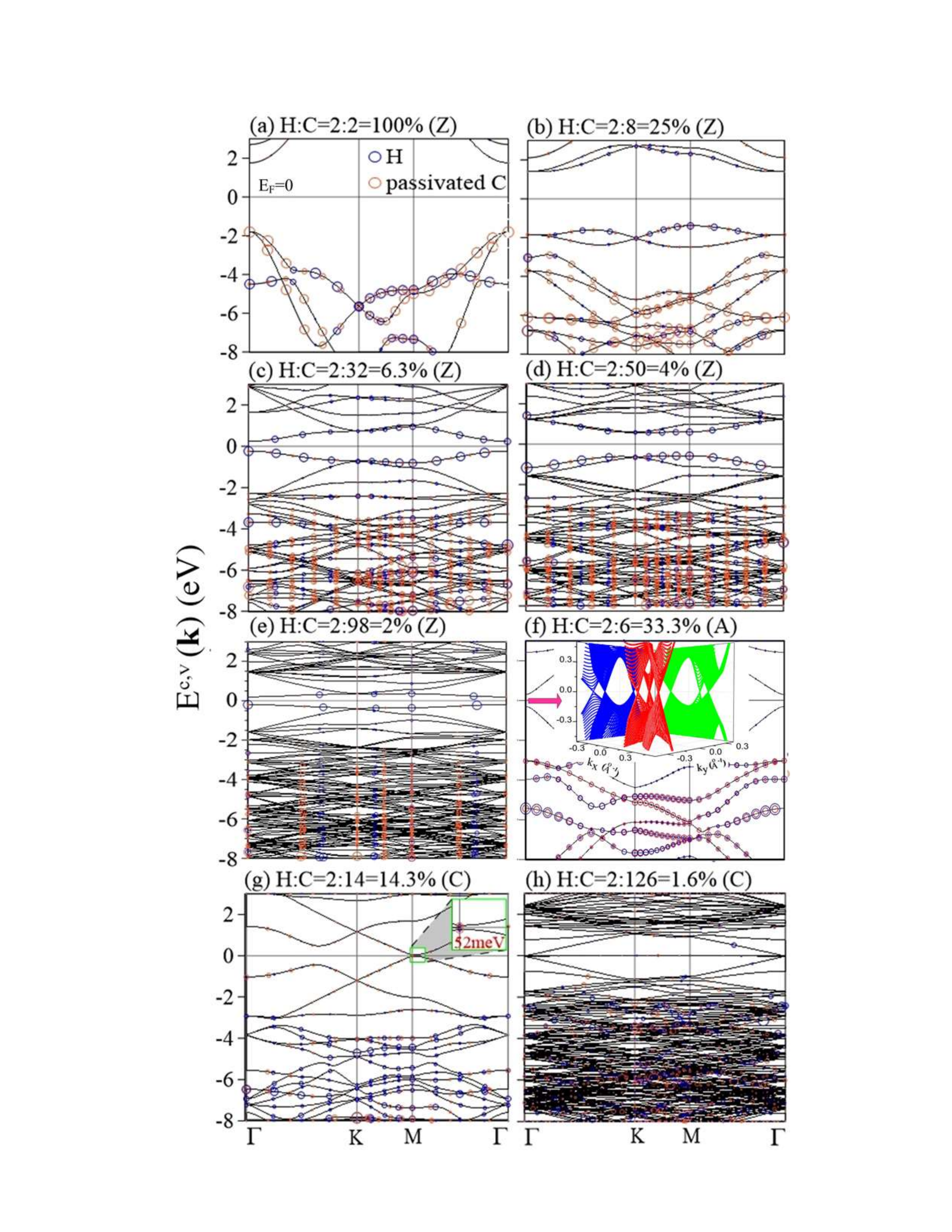}
\centering\caption{}
\end{figure}

\newpage
\begin{figure}
\centering
\includegraphics[scale=0.8]{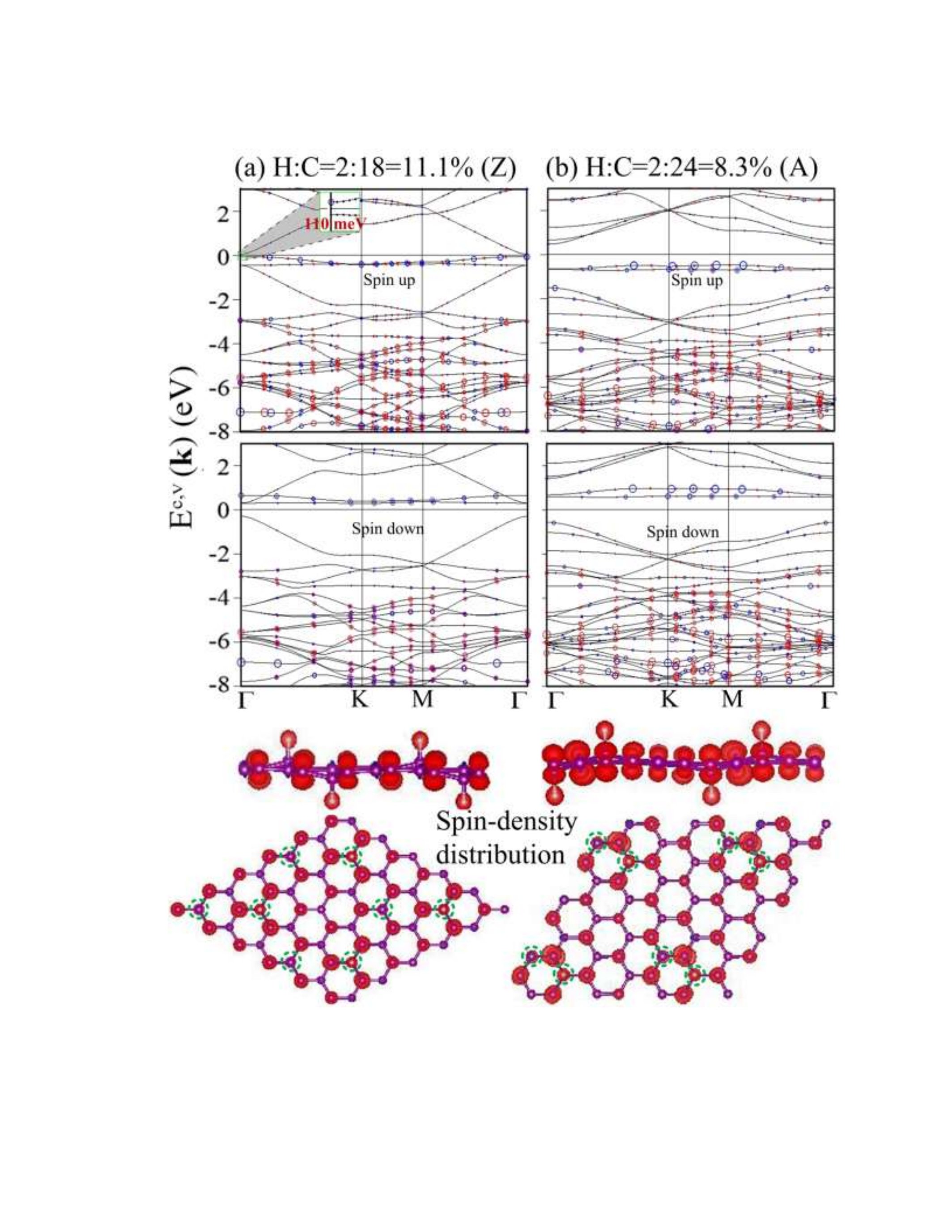}
\centering\caption{}
\end{figure}

\clearpage

\newpage
\begin{figure}
\centering
\includegraphics[scale=0.8]{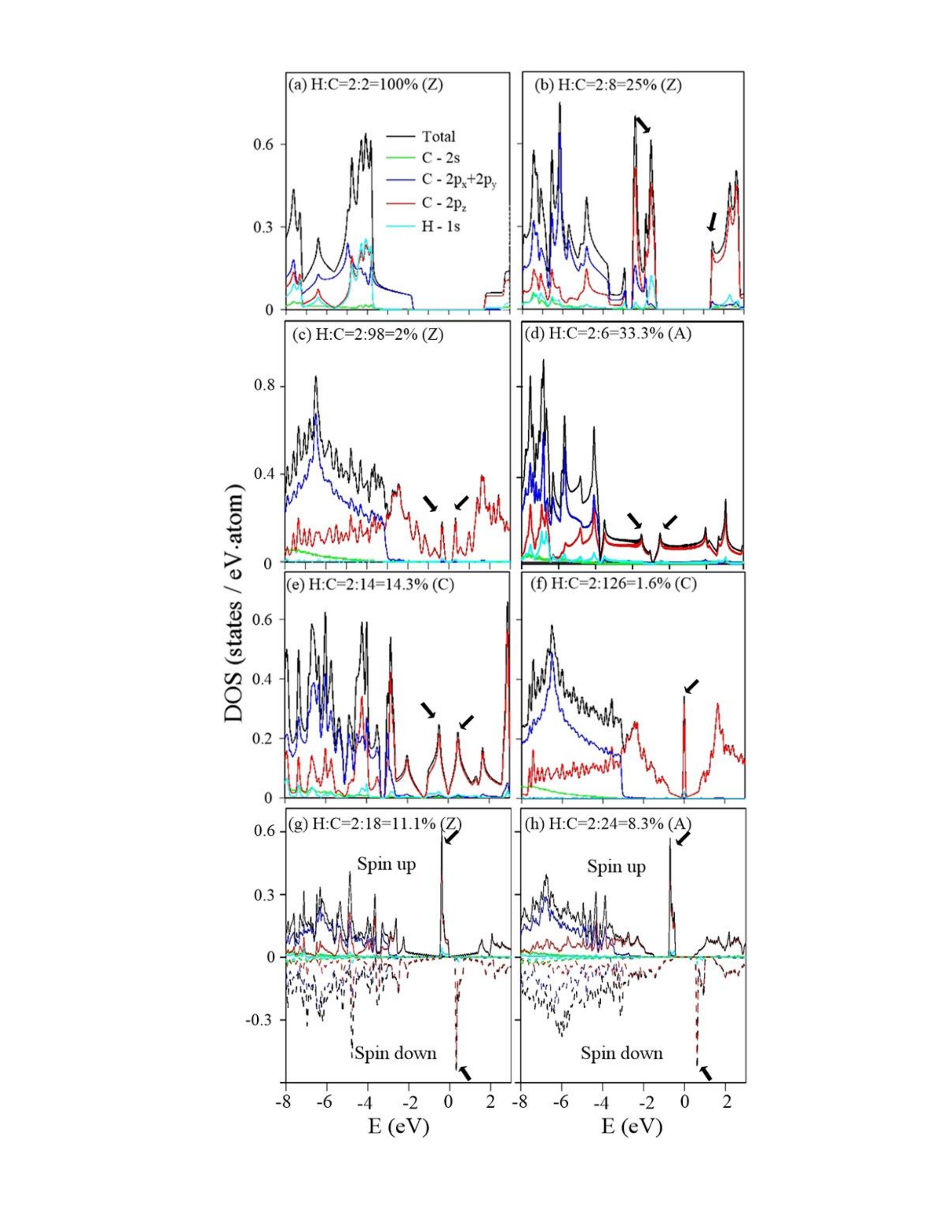}
\centering\caption{}
\end{figure}

\newpage
\begin{figure}
\centering
\includegraphics[scale=0.8]{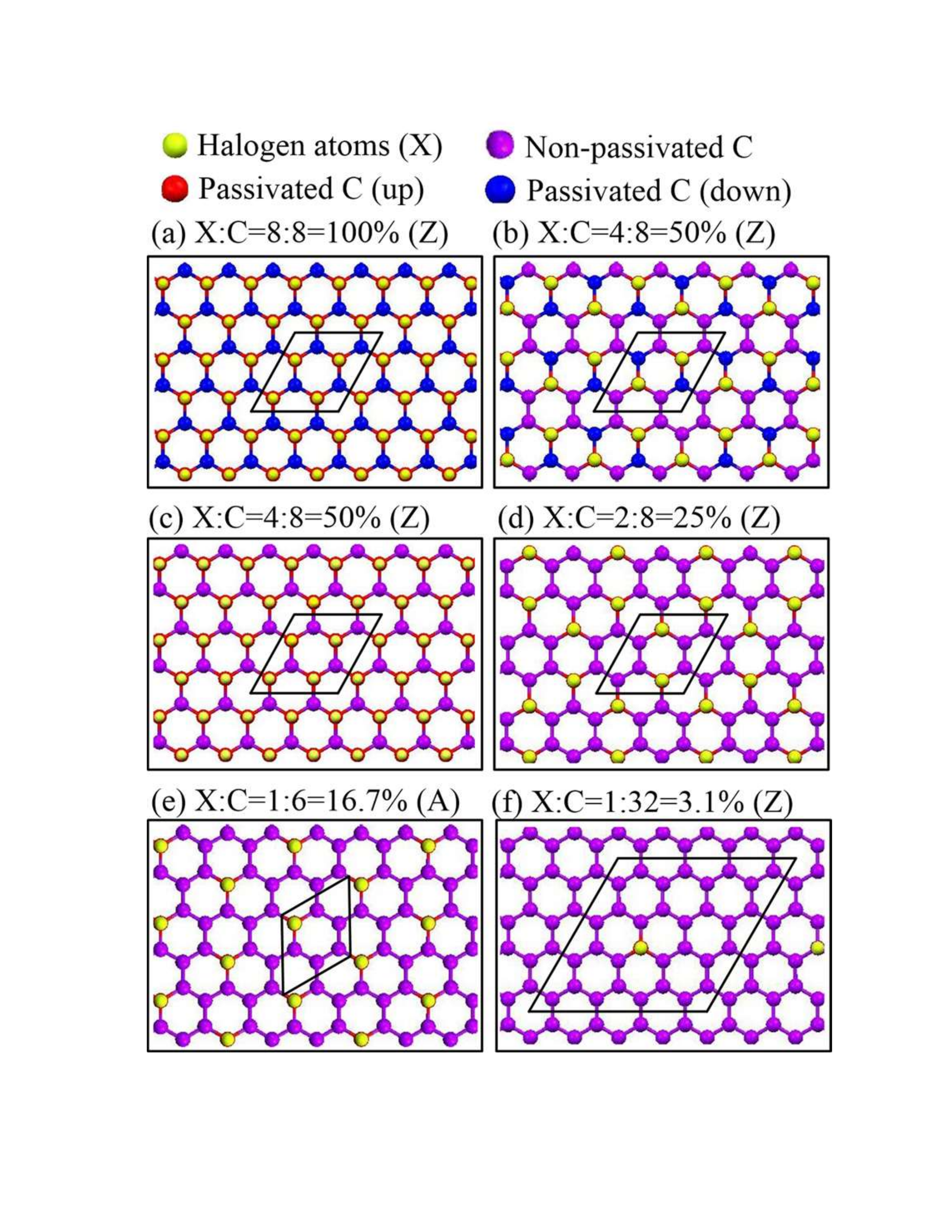}
\centering\caption{}
\end{figure}

\newpage
\begin{figure}
\centering
\includegraphics[scale=0.8]{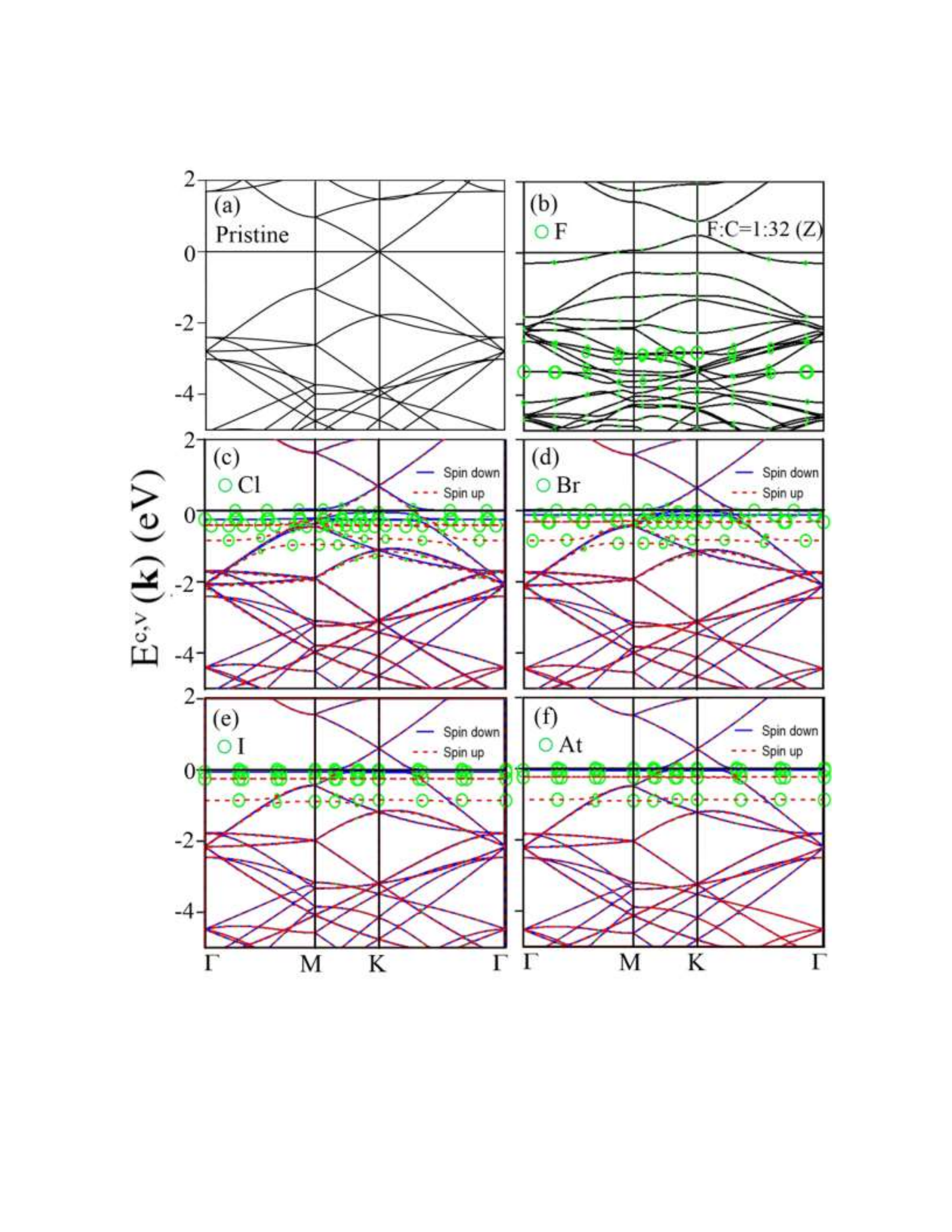}
\centering\caption{}
\end{figure}

\newpage
\begin{figure}
\centering
\includegraphics[scale=0.8]{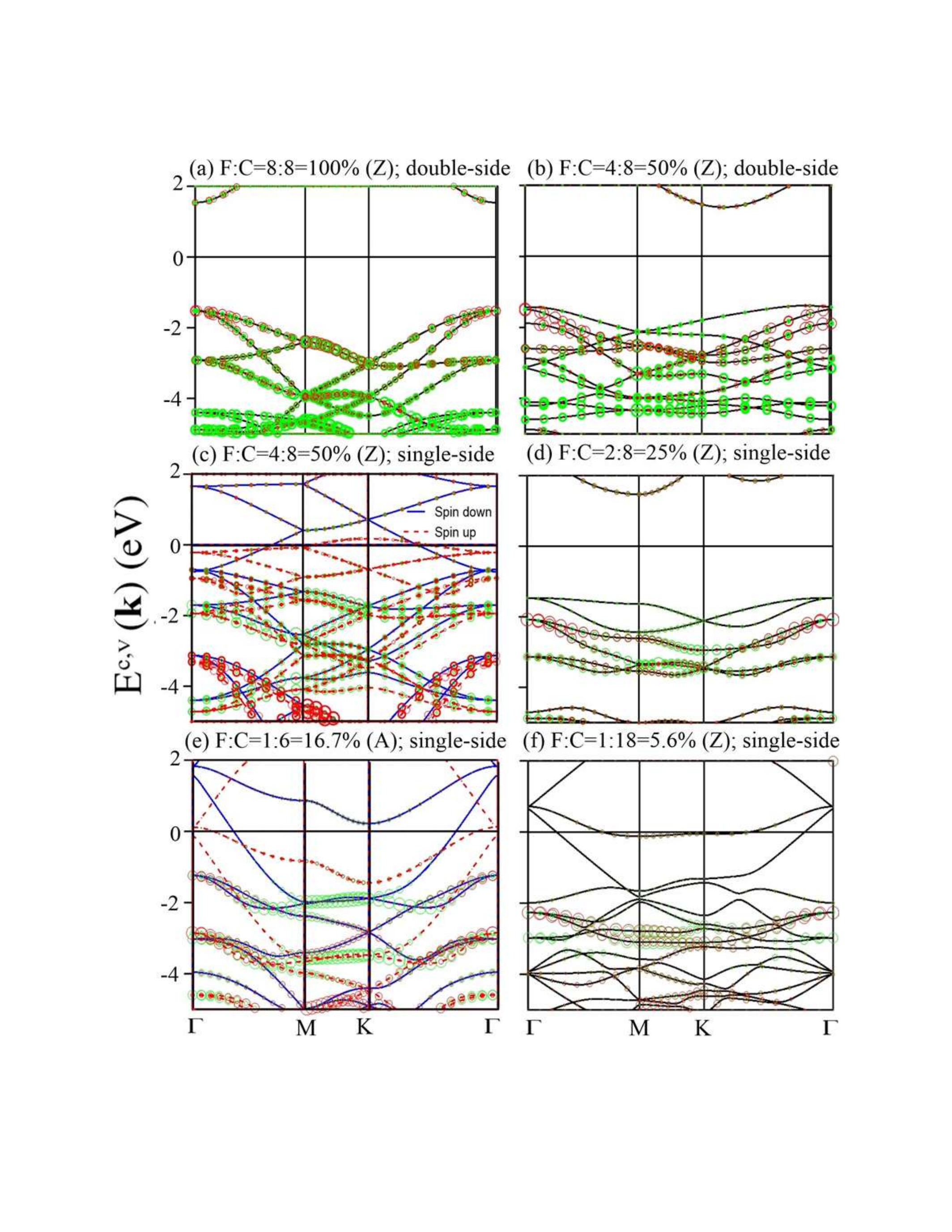}
\centering\caption{}
\end{figure}

\newpage
\begin{figure}
\centering
\includegraphics[scale=0.8]{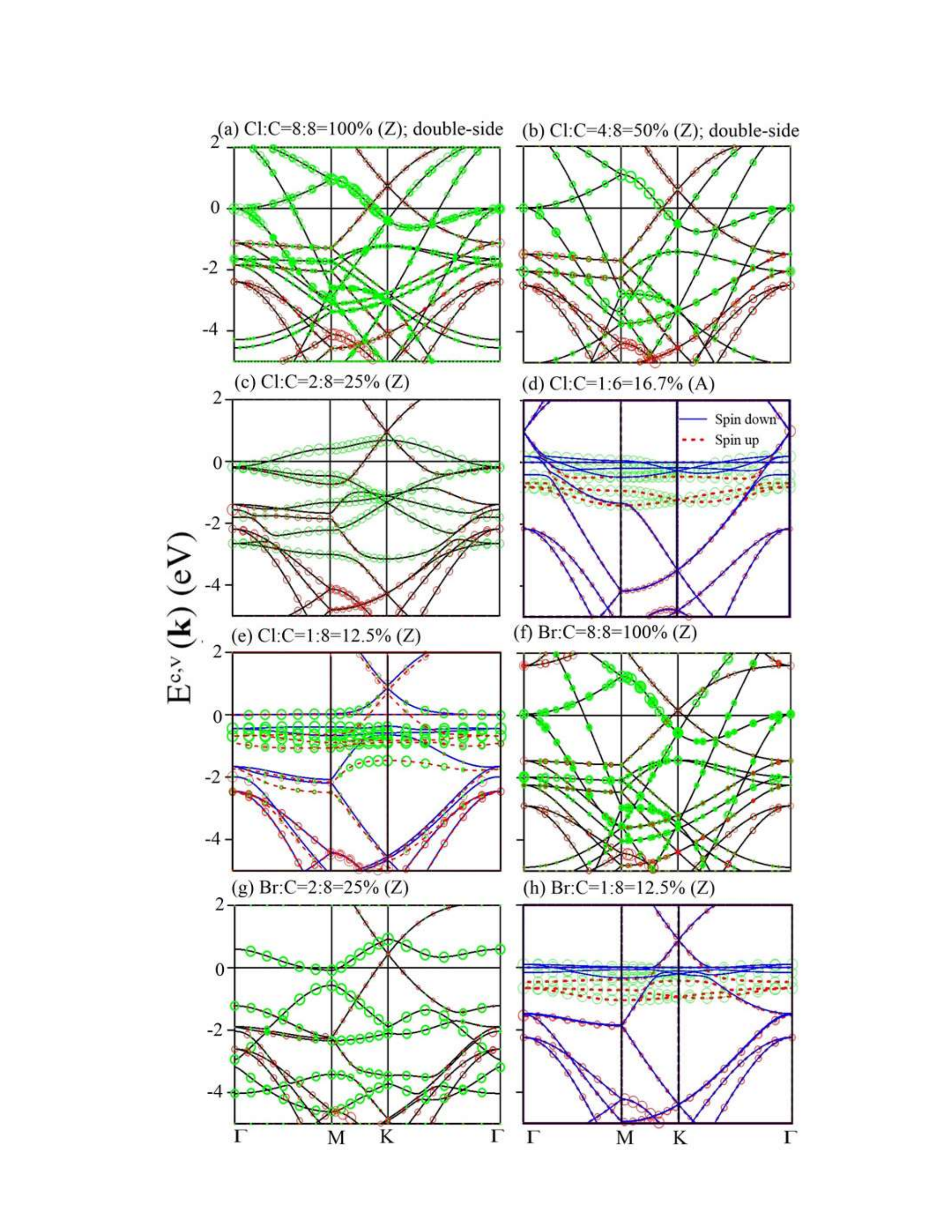}
\centering\caption{}
\end{figure}

\newpage
\begin{figure}
\centering
\includegraphics[scale=0.8]{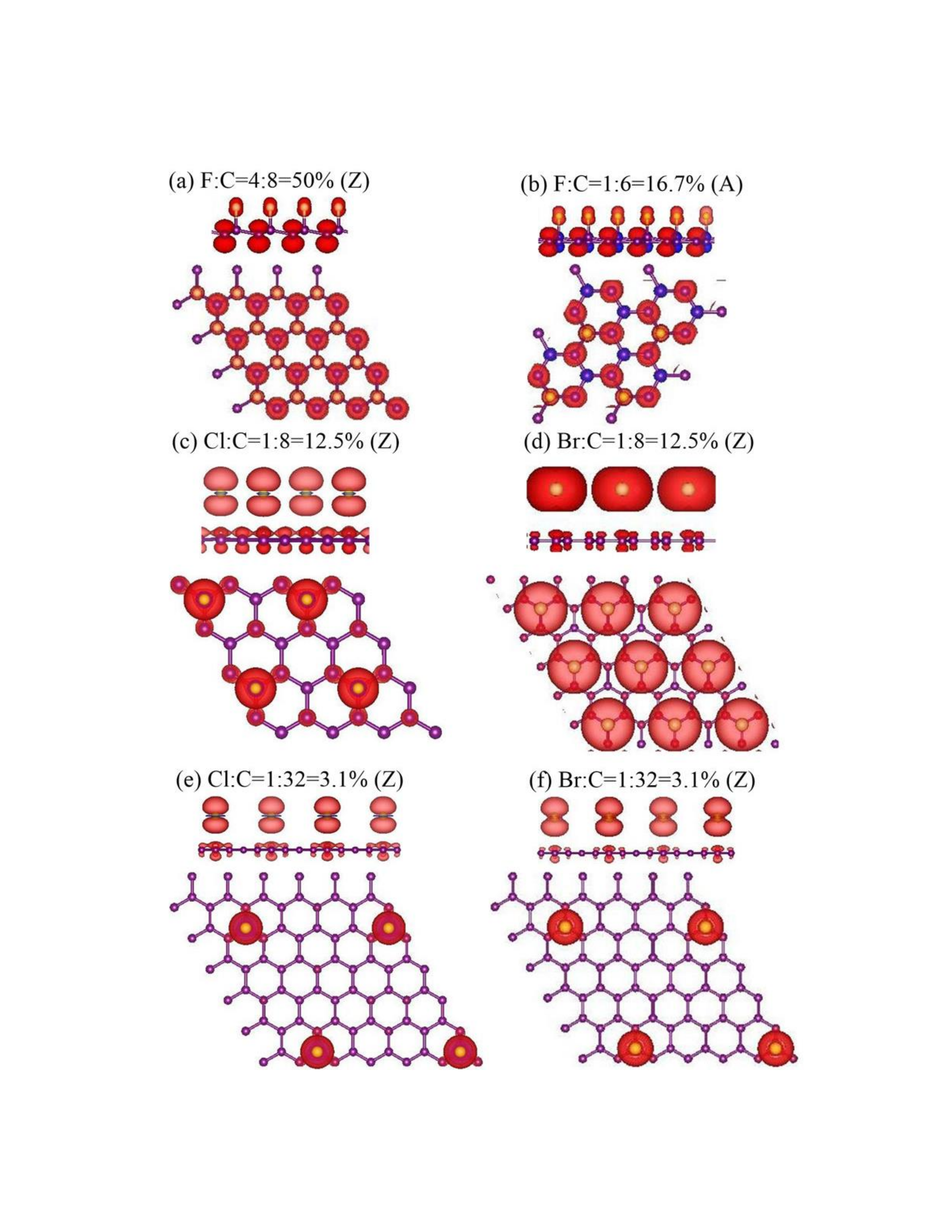}
\centering\caption{}
\end{figure}

\newpage
\begin{figure}
\centering
\includegraphics[scale=0.8]{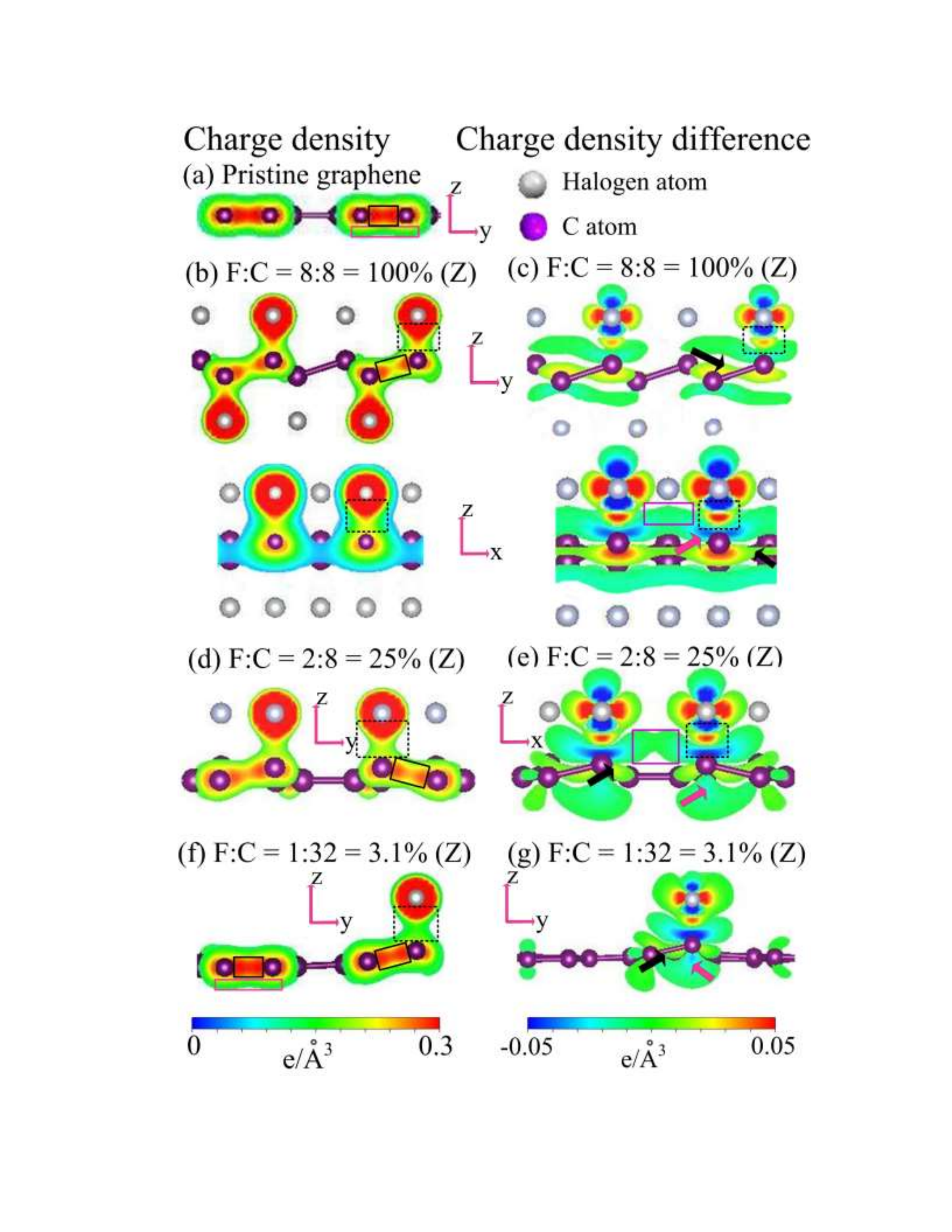}
\centering\caption{}
\end{figure}

\newpage
\begin{figure}
\centering
\includegraphics[scale=0.8]{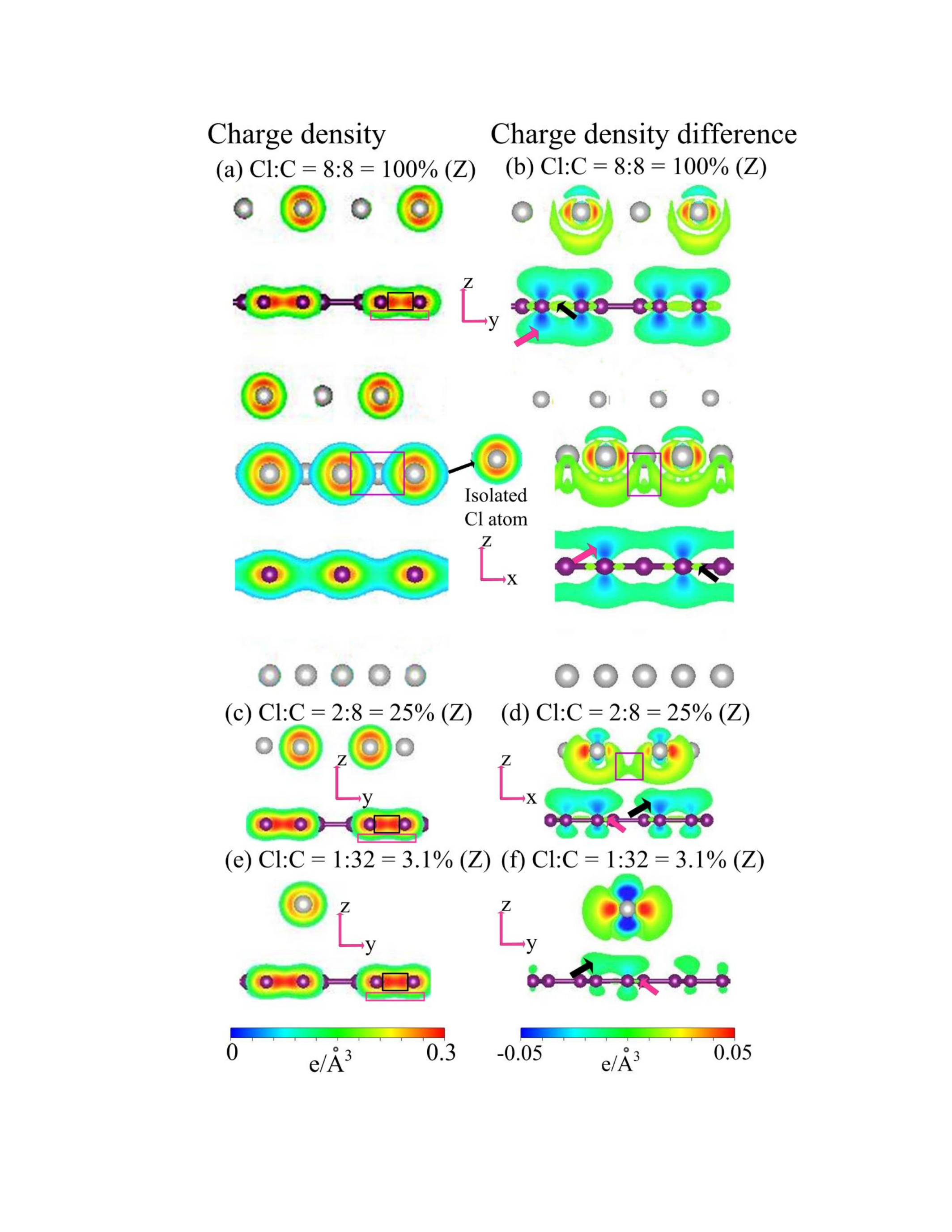}
\centering\caption{}
\end{figure}

\newpage
\begin{figure}
\centering
\includegraphics[scale=0.8]{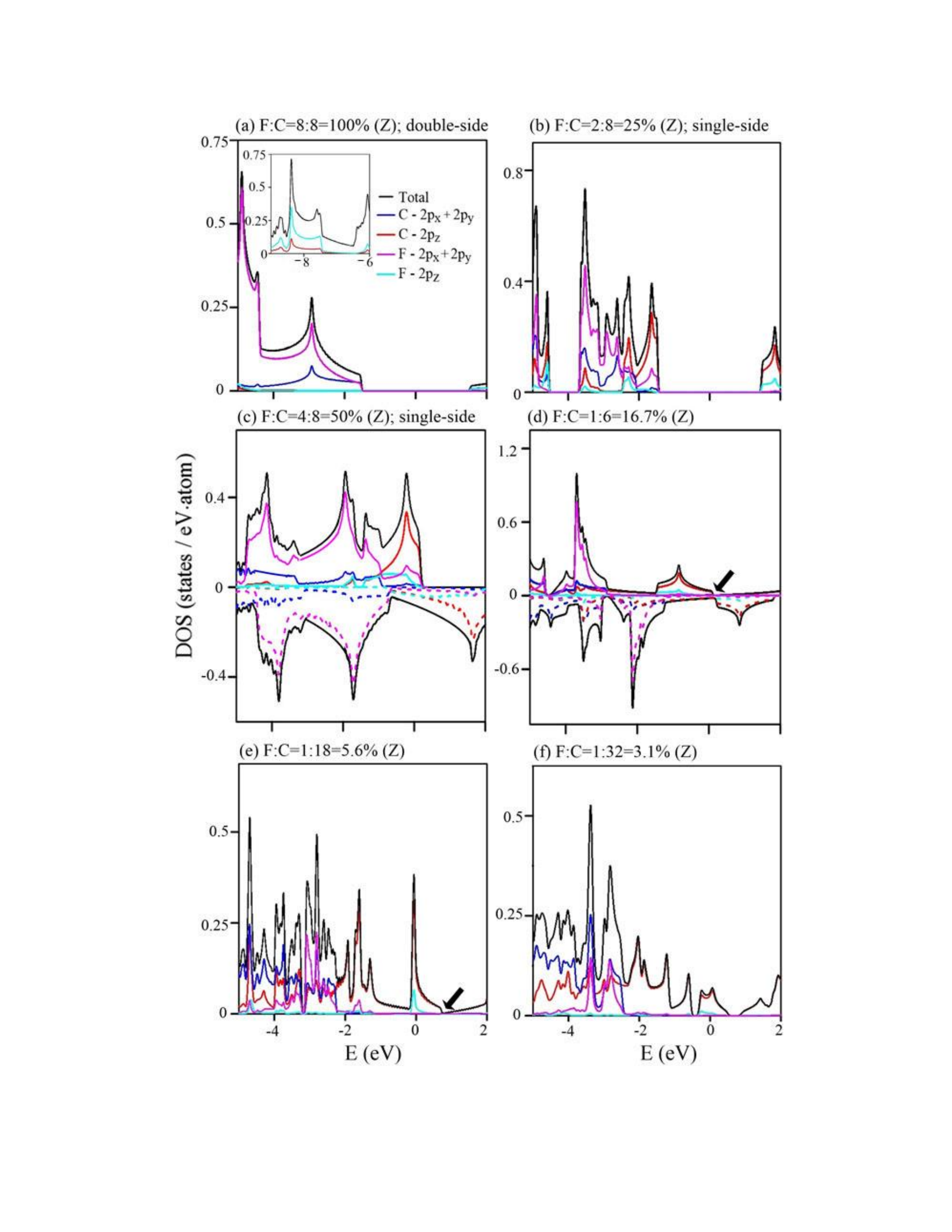}
\centering\caption{}
\end{figure}

\newpage
\begin{figure}
\centering
\includegraphics[scale=0.8]{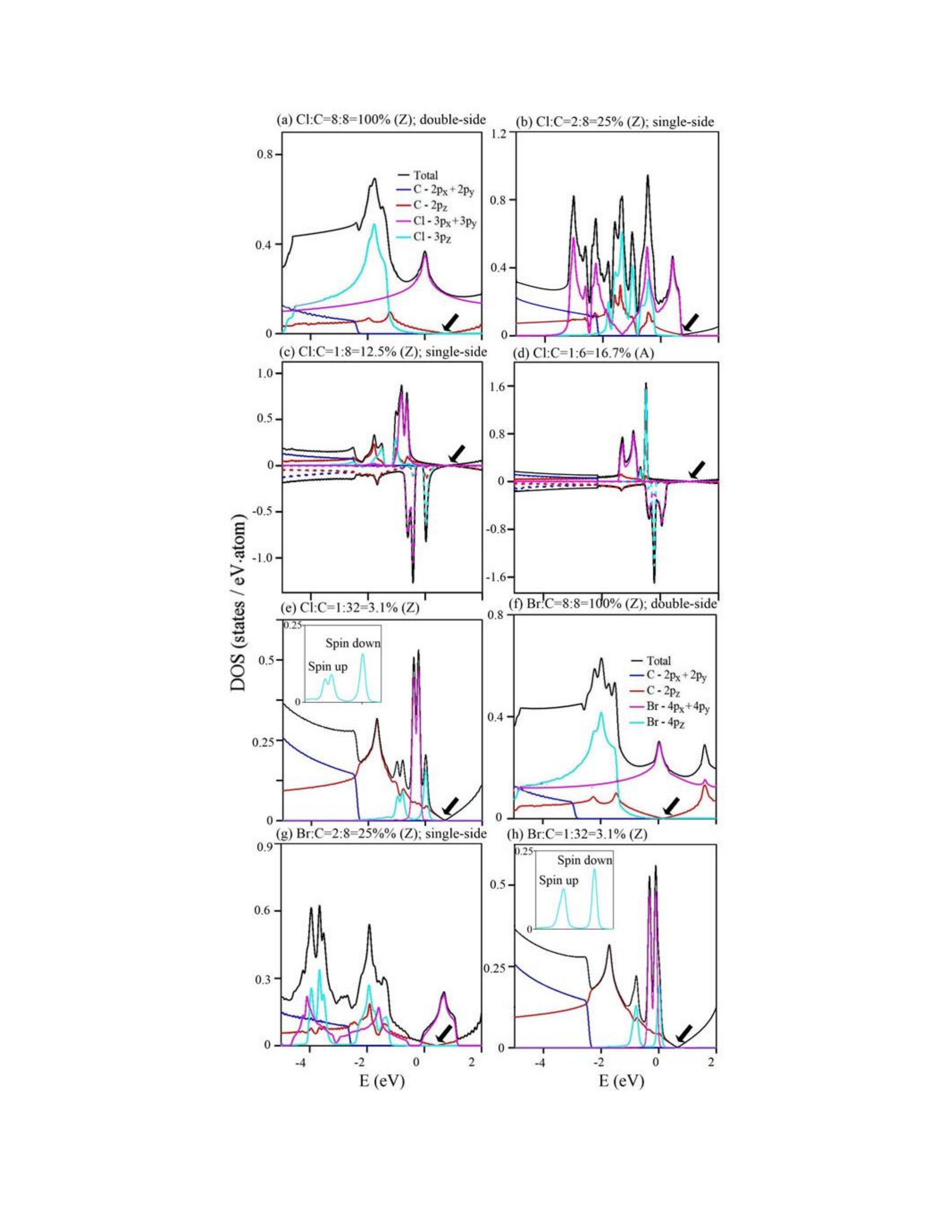}
\centering\caption{}
\end{figure}

\newpage
\begin{figure}
\centering
\includegraphics[scale=0.7]{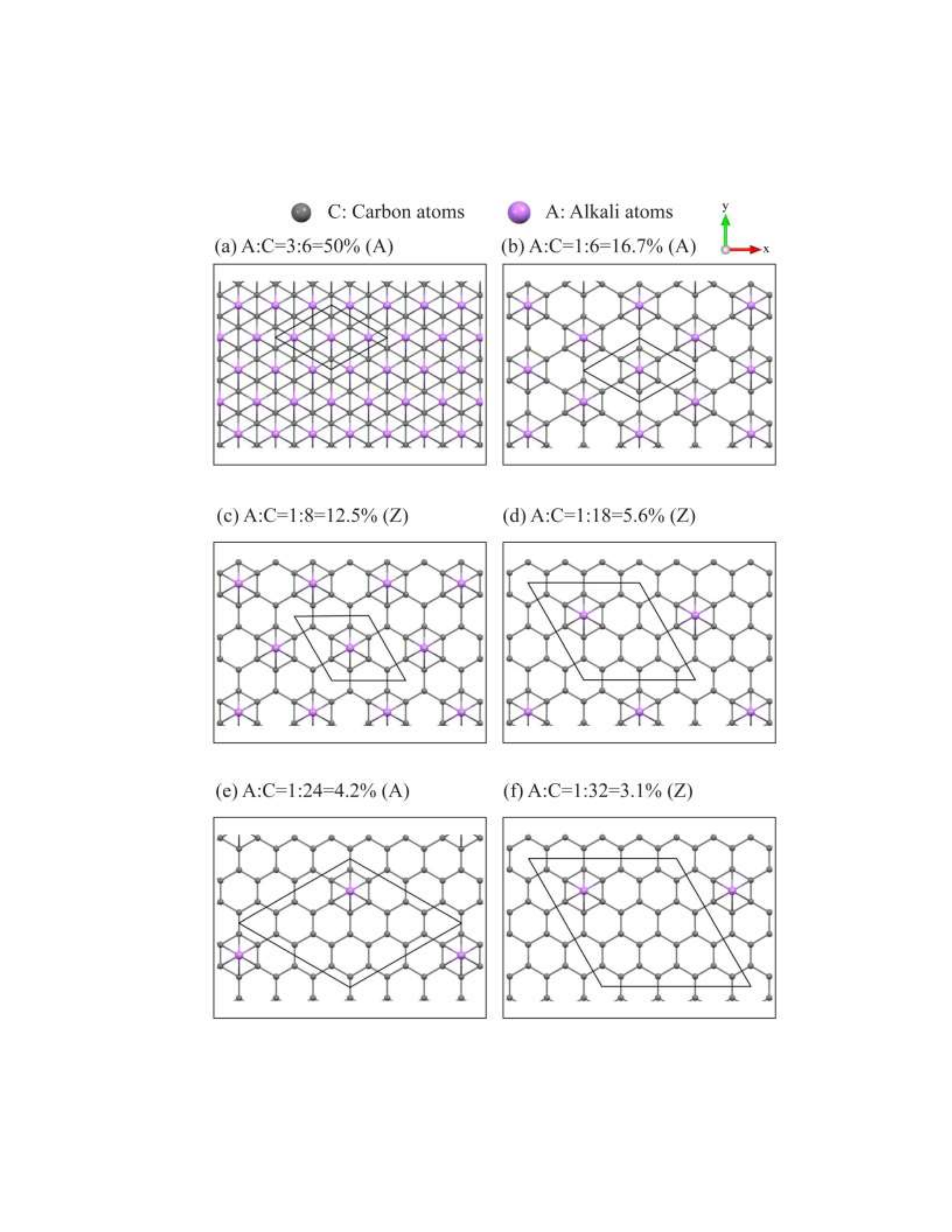}
\centering\caption{}
\end{figure}

\newpage
\begin{figure}
\centering
\includegraphics[scale=0.7]{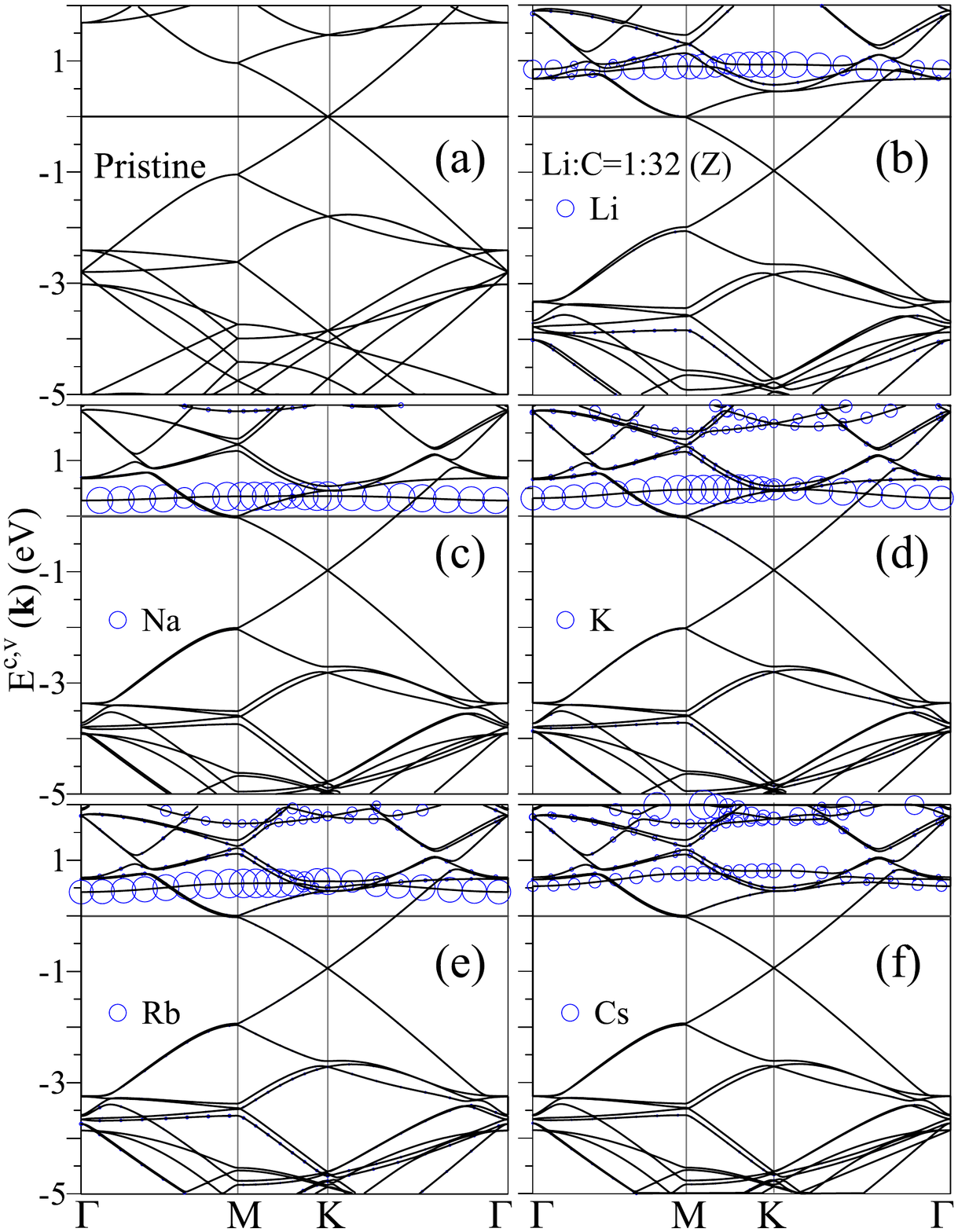}
\centering\caption{}
\end{figure}

\newpage
\begin{figure}
\centering
\includegraphics[scale=0.7]{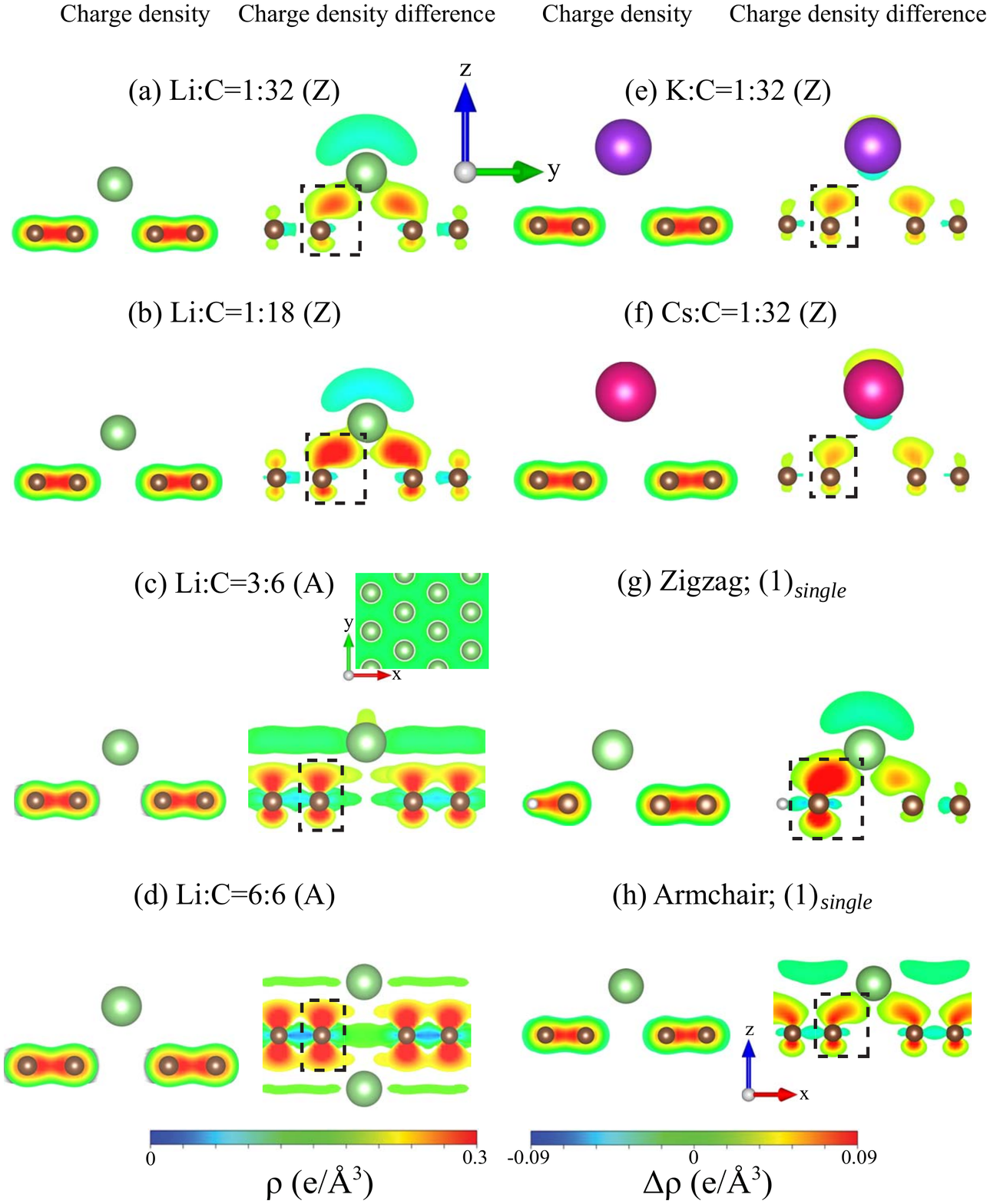}
\centering\caption{}
\end{figure}

\newpage
\begin{figure}
\centering
\includegraphics[scale=0.7]{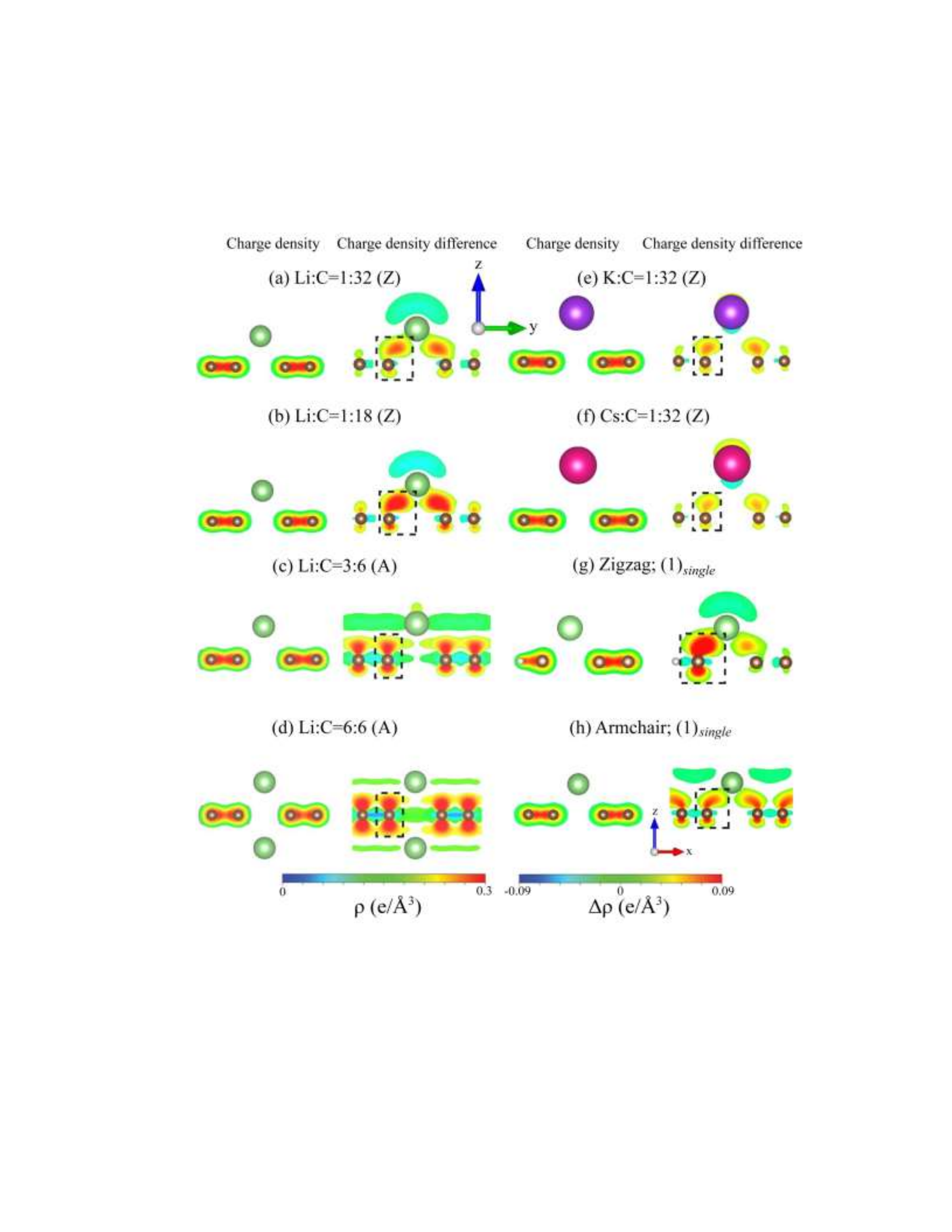}
\centering\caption{}
\end{figure}

\newpage
\begin{figure}
\centering
\includegraphics[scale=0.7]{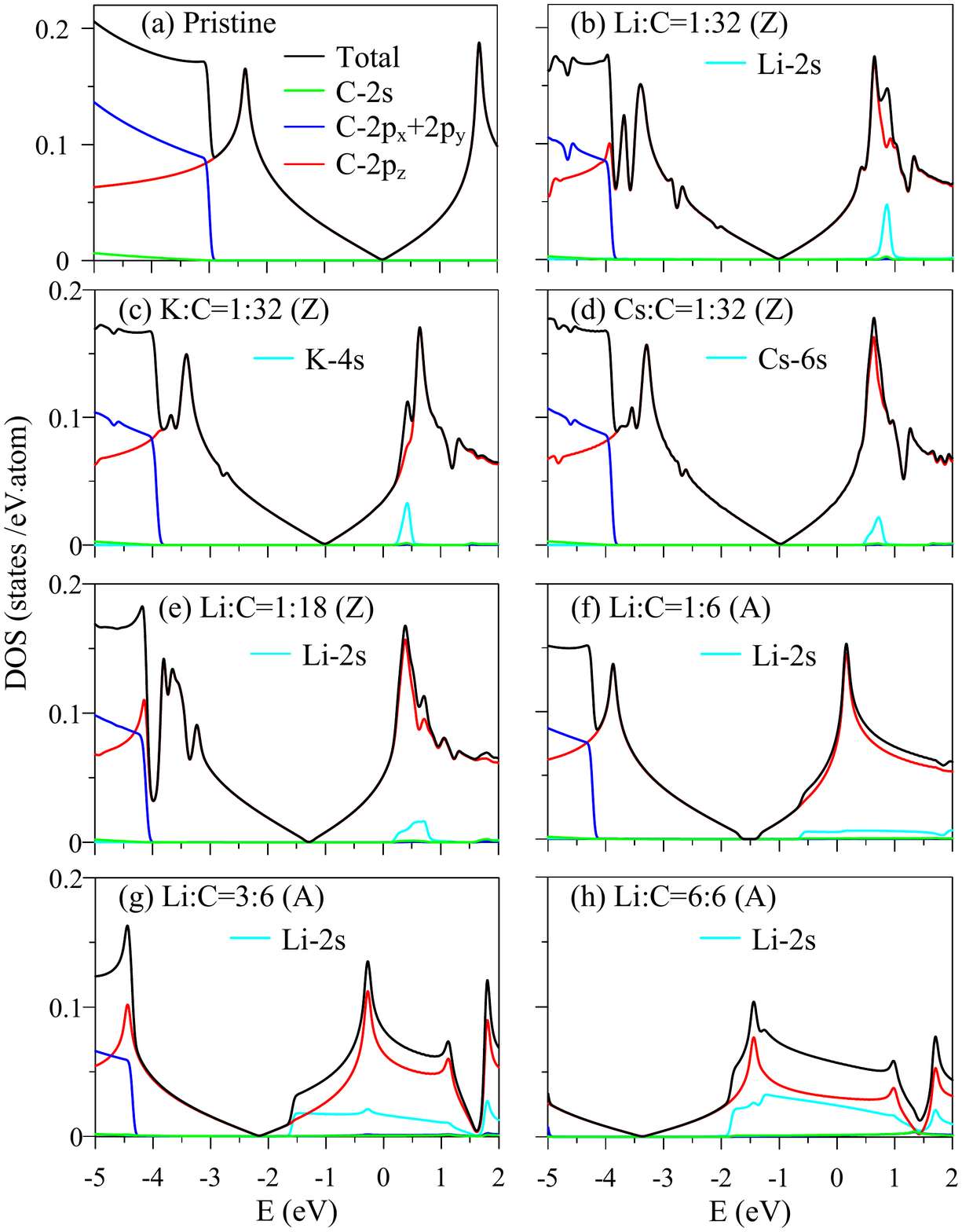}
\centering\caption{}
\end{figure}

\clearpage

\newpage
\begin{figure}
\centering
\includegraphics[scale=0.7]{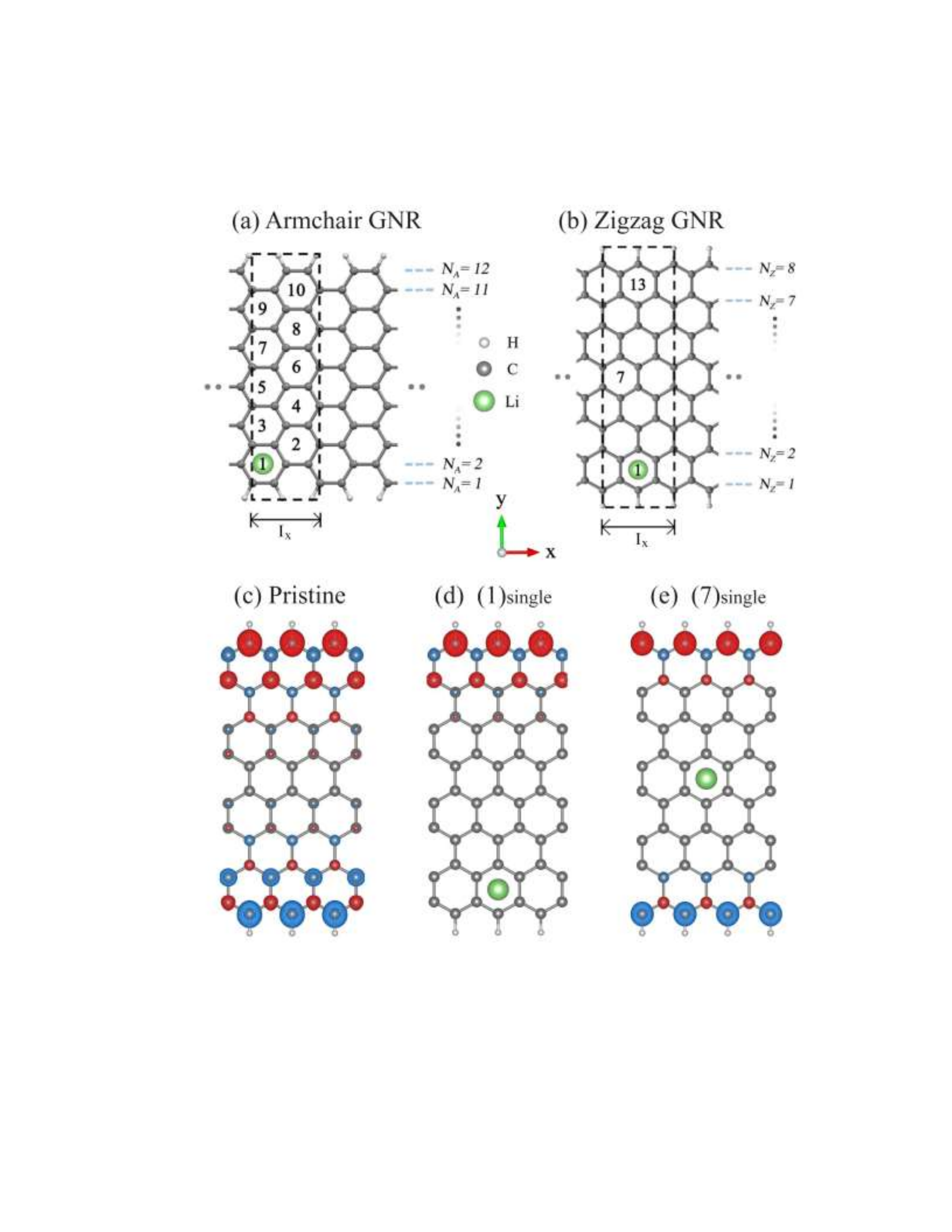}
\centering\caption{}
\end{figure}

\newpage
\begin{figure}
\centering
\includegraphics[scale=0.7]{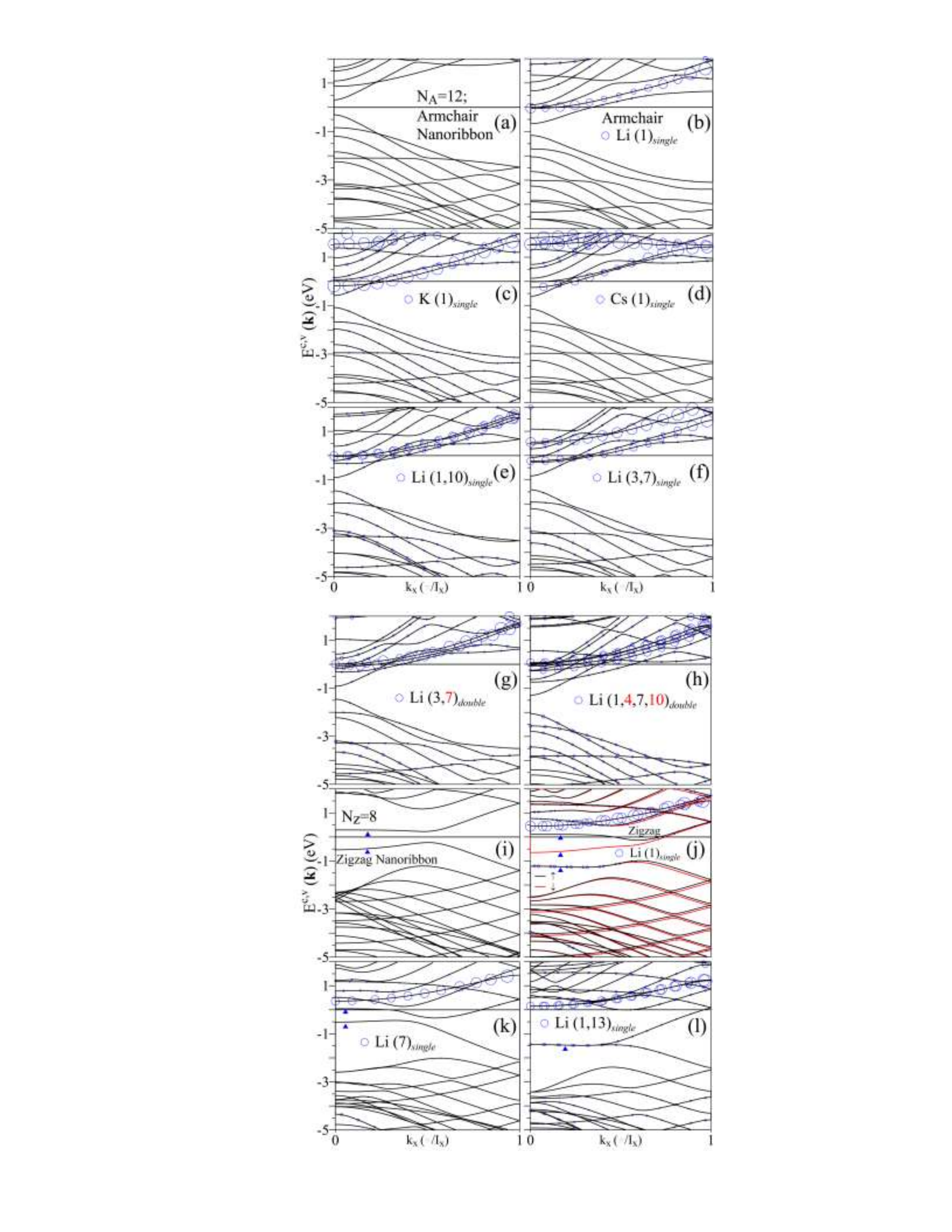}
\centering\caption{}
\end{figure}

\newpage
\begin{figure}
\centering
\includegraphics[scale=0.7]{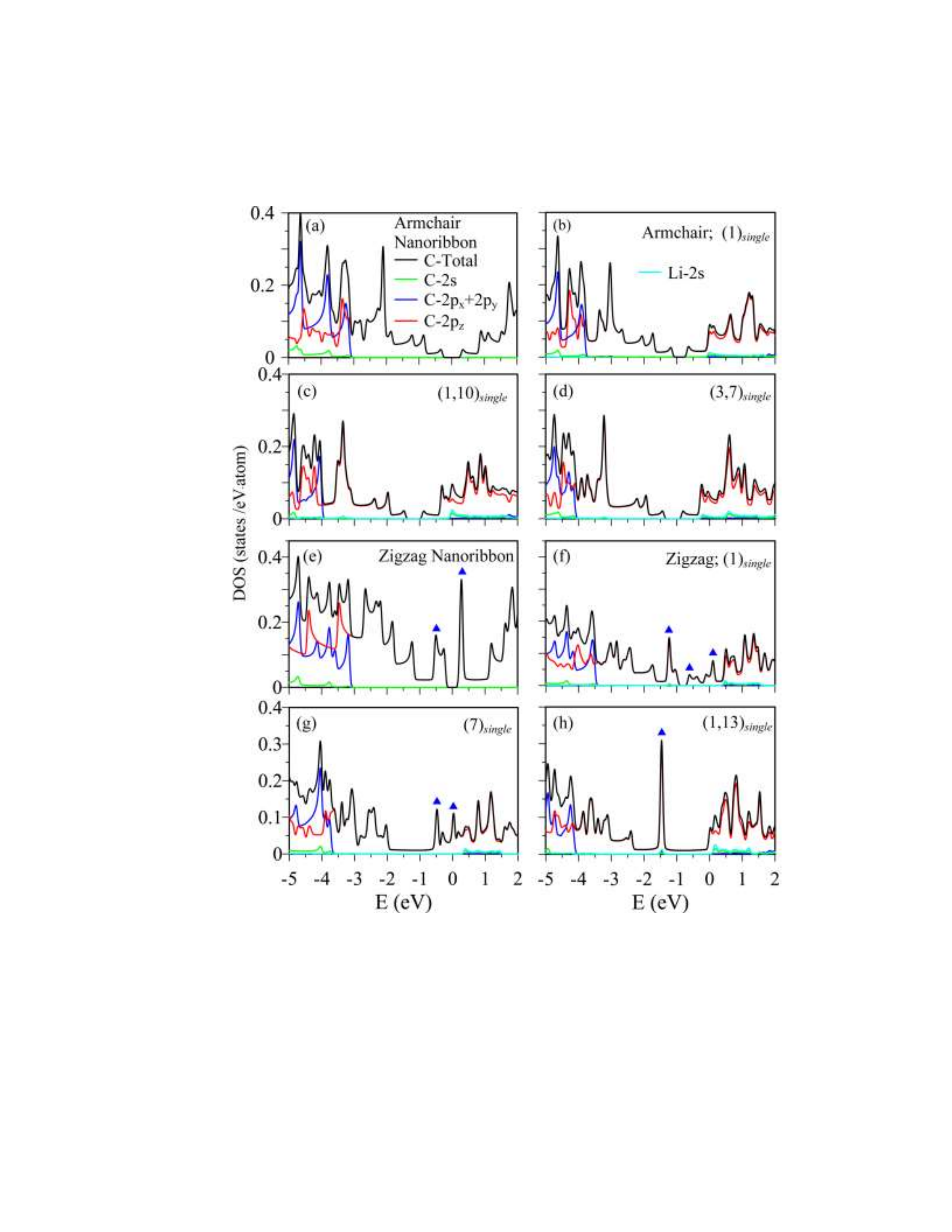}
\centering\caption{}
\end{figure}

\newpage
\begin{figure}
\centering
\includegraphics[scale=0.7]{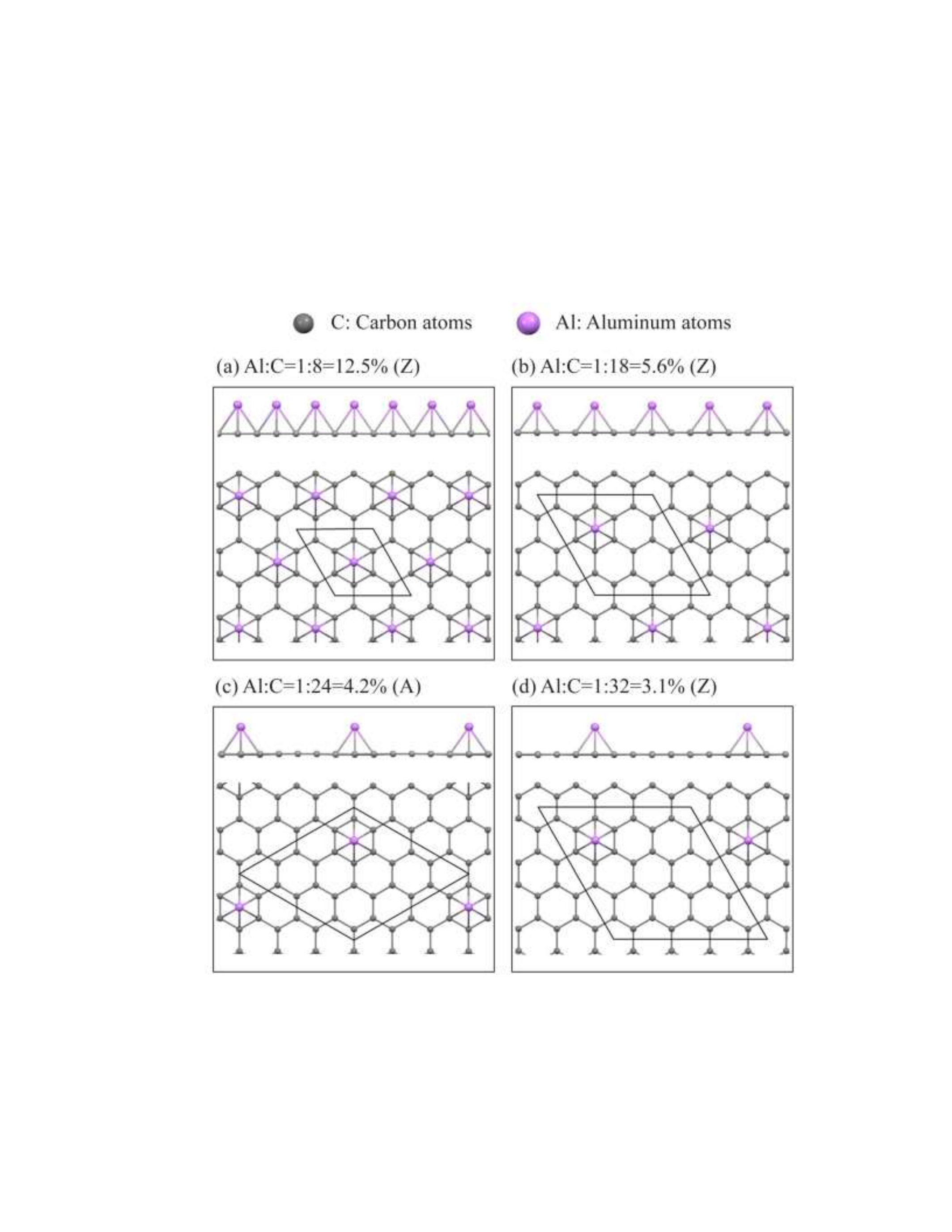}
\centering\caption{}
\end{figure}

\newpage
\begin{figure}
\centering
\includegraphics[scale=0.7]{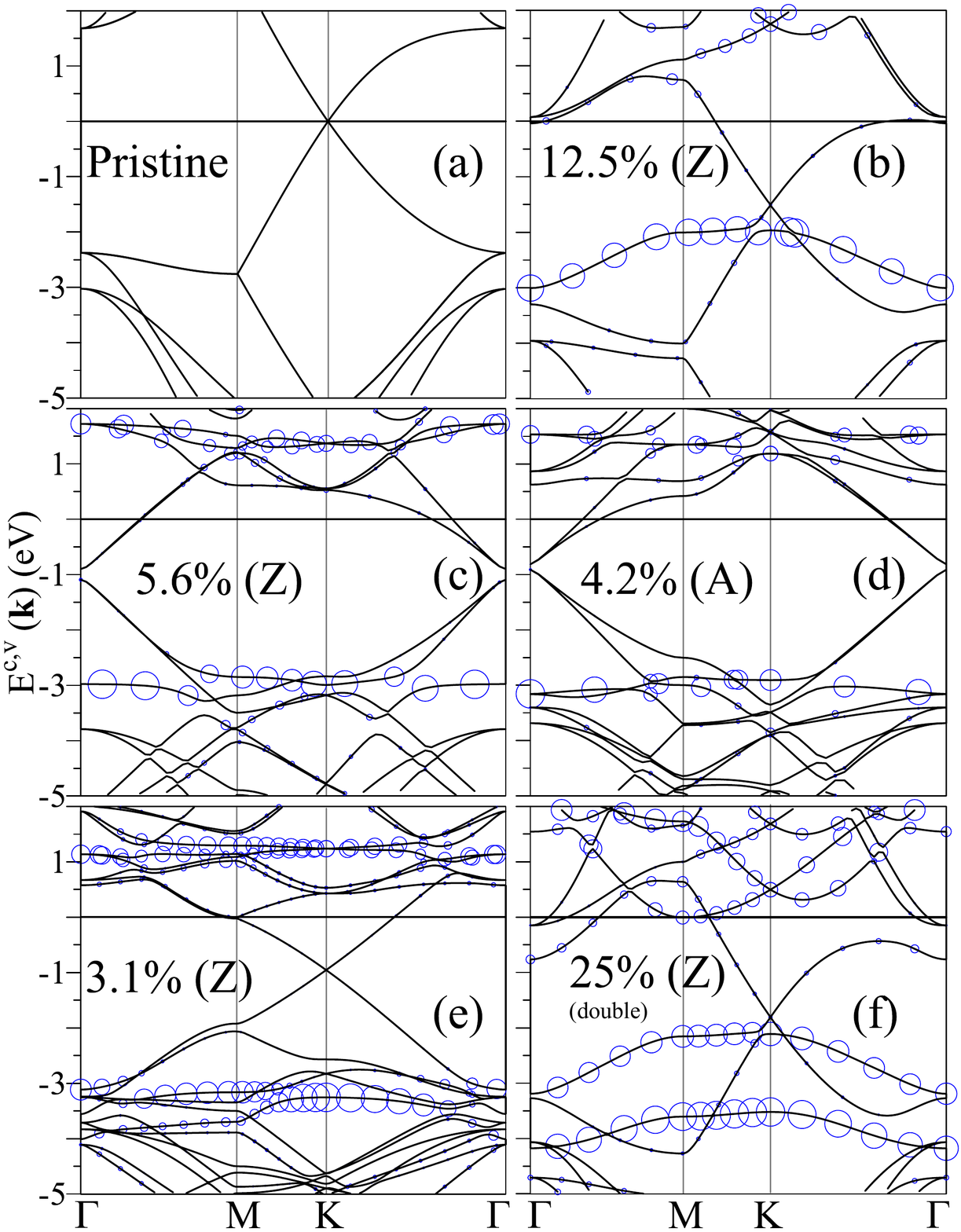}
\centering\caption{}
\end{figure}

\newpage
\begin{figure}
\centering
\includegraphics[scale=0.7]{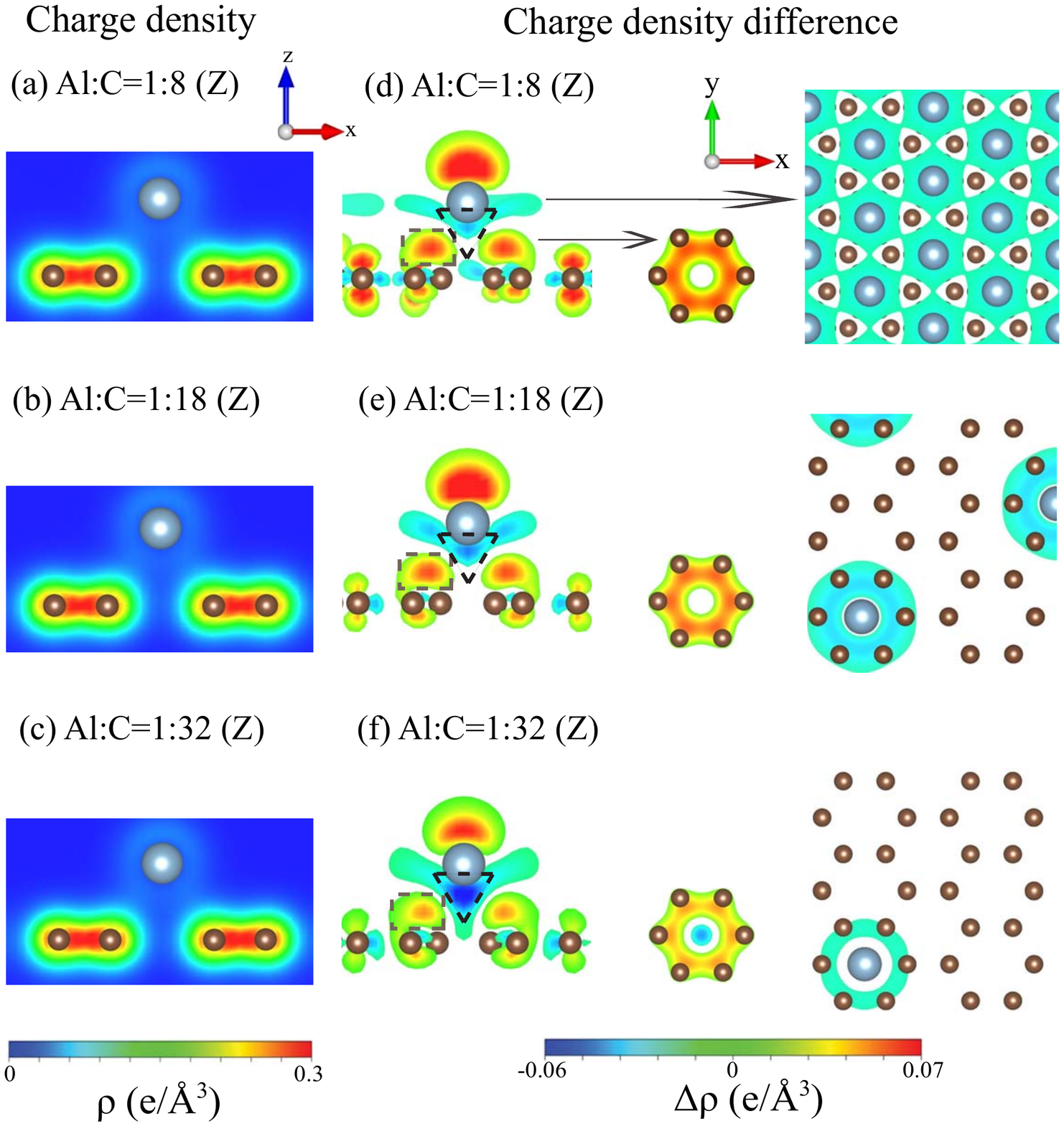}
\centering\caption{}
\end{figure}

\newpage
\begin{figure}
\centering
\includegraphics[scale=0.7]{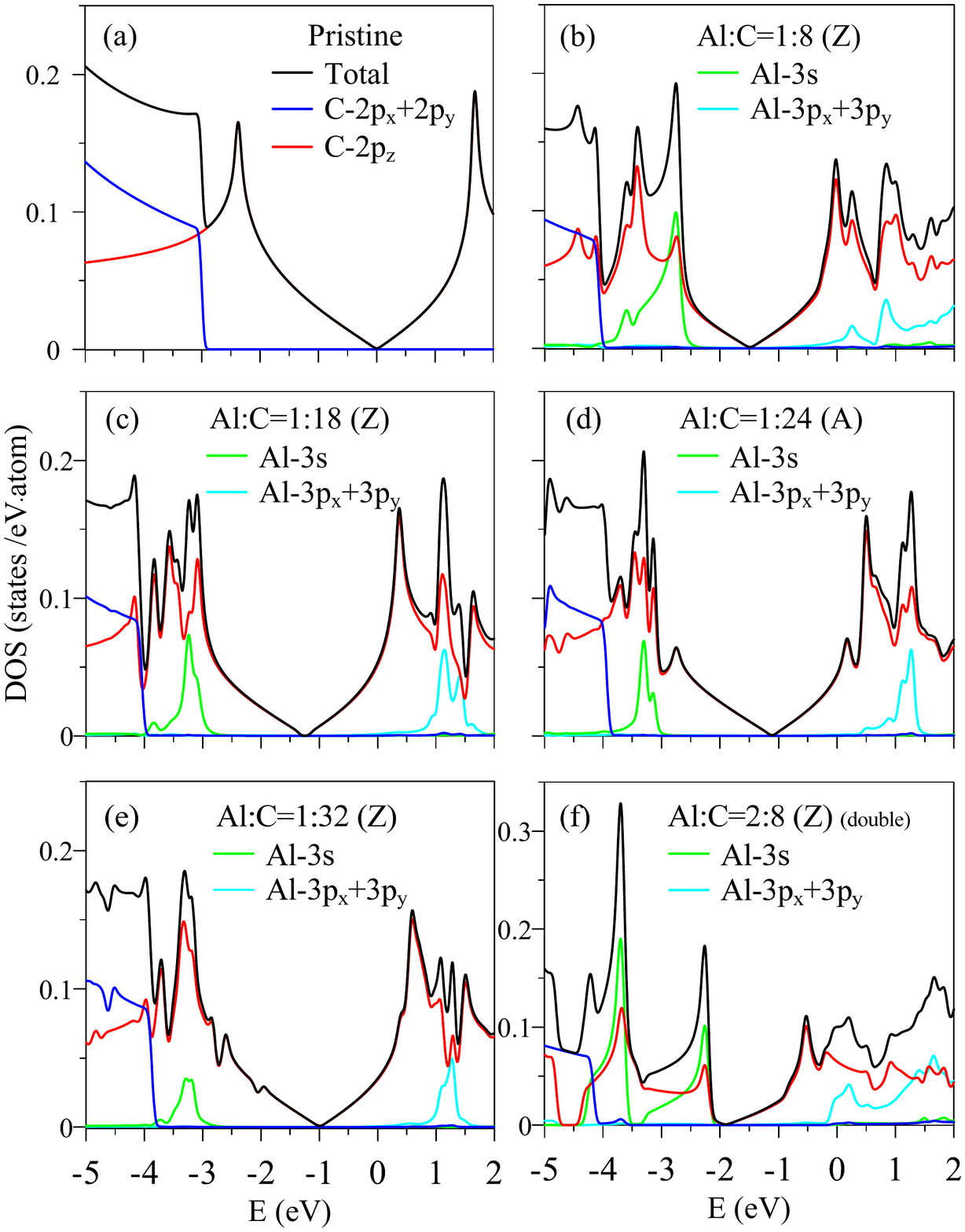}
\centering\caption{}
\end{figure}

\newpage
\begin{figure}
\centering
\includegraphics[scale=0.7]{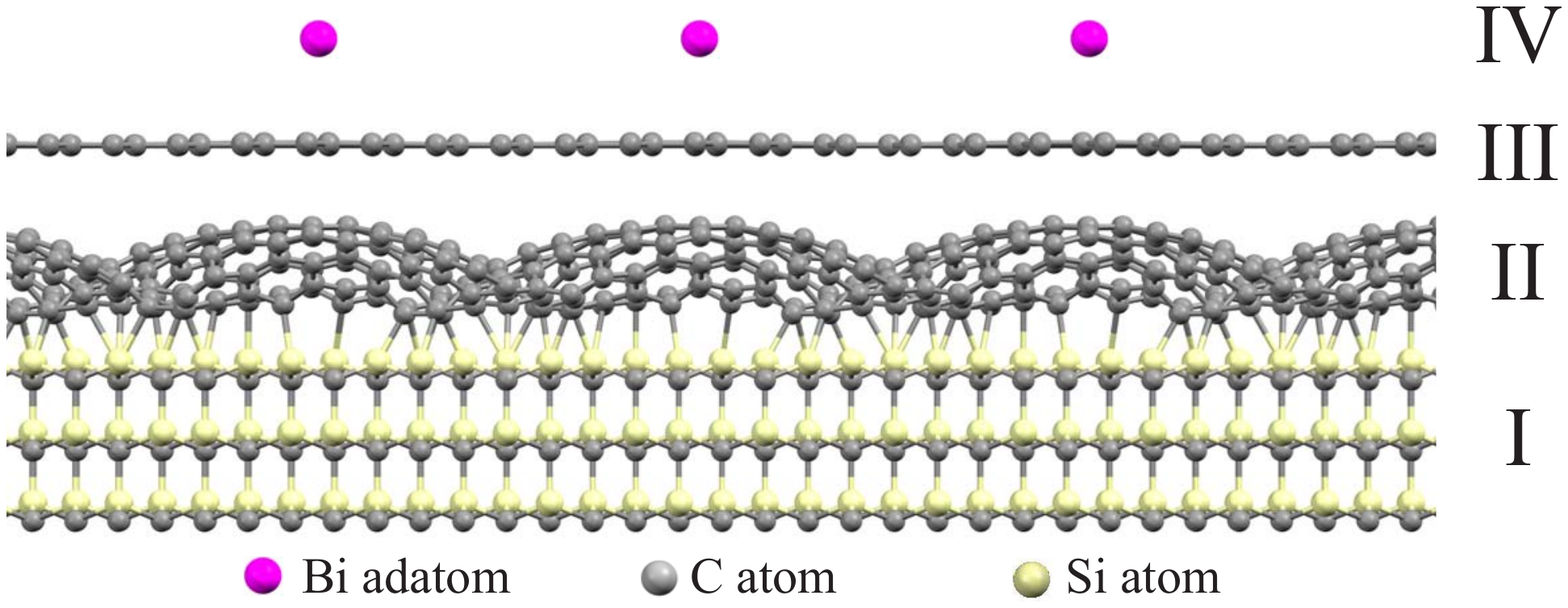}
\centering\caption{}
\end{figure}

\newpage
\begin{figure}
\centering
\includegraphics[scale=0.7]{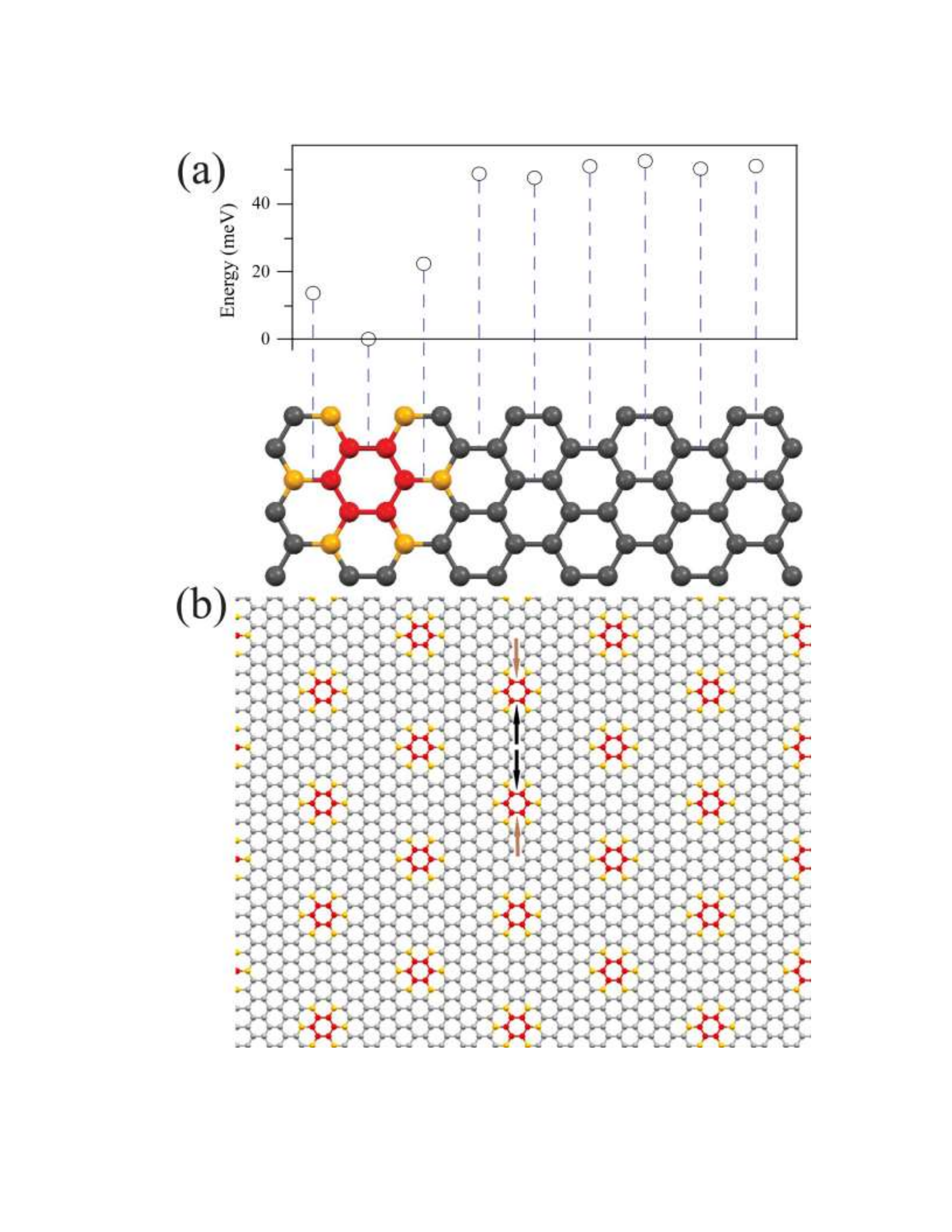}
\centering\caption{}
\end{figure}

\newpage
\begin{figure}
\centering
\includegraphics[scale=0.7]{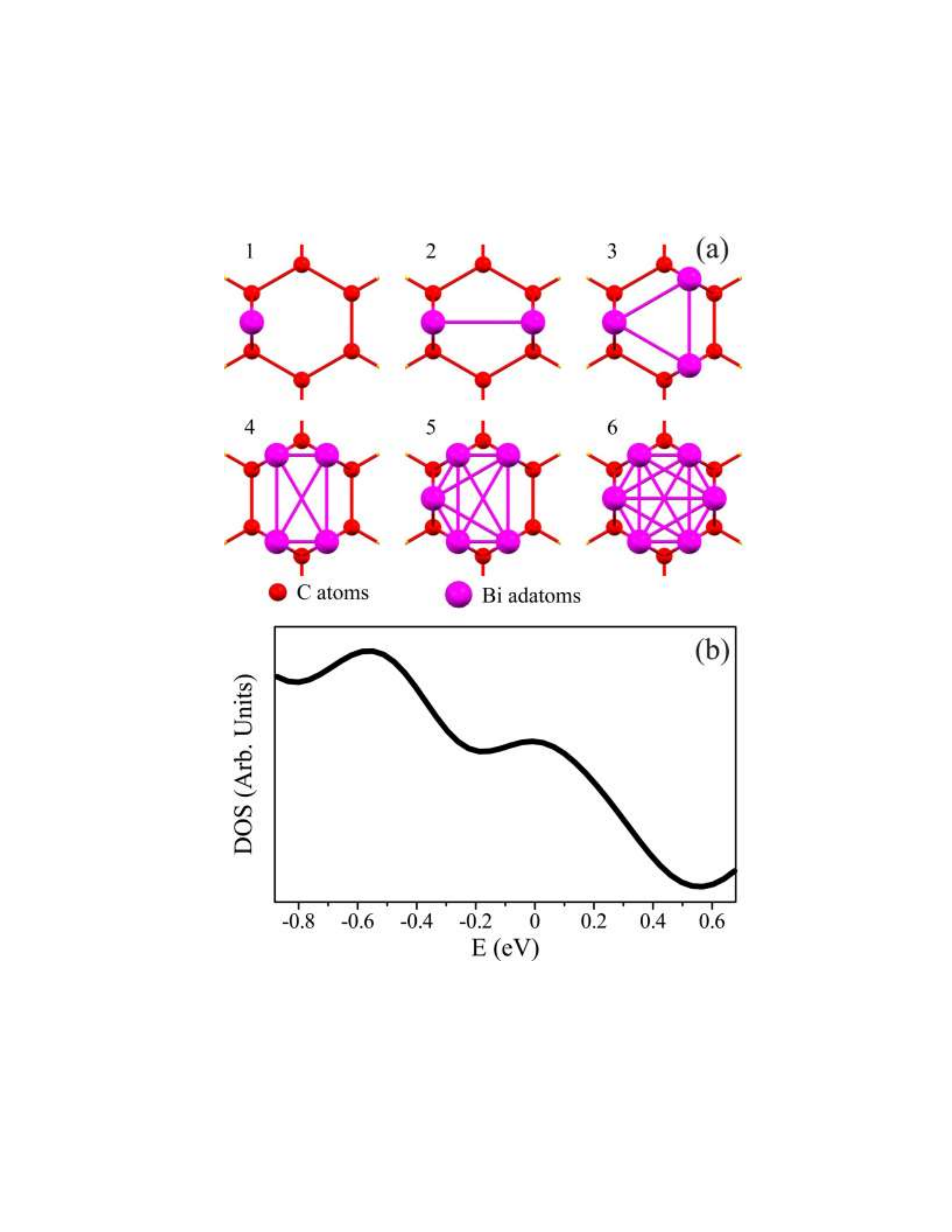}
\centering\caption{}
\end{figure}


\begin{thebibliography}{99}
\bibitem{kamo1983diamond} Kamo, M.; Sato, Y.; Matsumoto, S.; Setaka, N. Diamond synthesis from gas phase in microwave plasma. J. Crystal Growth 1983, 62, 642–644.
\bibitem{tuinstra1970raman} Tuinstra, F.; Koenig, J. L. Raman spectrum of graphite. J. Chem. Phys. 1970, 53, 1126–1130.
\bibitem{huang1997collective} Huang, C. S.; Lin, M. F.; Chuu, D. S. Collective excitations in graphite layers. Solid State commun. 1997, 103, 603–606.
\bibitem{novoselov2004electric} Novoselov, K. S.; Geim, A. K.; Morozov, S.; Jiang, D.; Zhang, Y.; Dubonos, S.; Grigorieva, I.; Firsov, A. Electric field effect in atomically thin carbon films. Science 2004, 306, 666–669.
\bibitem{novoselov2005two} Novoselov, K.; Geim, A. K.; Morozov, S.; Jiang, D.; Katsnelson, M.; Grigorieva, I.; Dubonos, S.; Firsov, A. Two-dimensional gas of massless Dirac fermions in graphene. Nature 2005, 438, 197–200.
\bibitem{son2006half} Son, Y. W.; Cohen, M. L.; Louie, S. G. Half-metallic graphene nanoribbons. Nature 2006, 444, 347–349.
\bibitem{li2008chemically} Li, X.; Wang, X.; Zhang, L.; Lee, S.; Dai, H. Chemically derived, ultrasmooth graphene nanoribbon semiconductors. Science 2008, 319, 1229–1232.
\bibitem{cai2010atomically} Cai, J. et al. Atomically precise bottom-up fabrication of graphene nanoribbons. Nature 2010, 466, 470–473.
\bibitem{iijima1991helical} Iijima, S. Helical microtubules of graphitic carbon. Nature 1991, 354, 56–58.
\bibitem{iijima1993single} Iijima, S.; Ichihashi, T. Single-shell carbon nanotubes of 1-nm diameter. Nature 1993, 363, 603–605.
\bibitem{de1995carbon} De Heer, W. A.; Chatelain, A.; Ugarte, D. A carbon nanotube field-emission electron source. Science 1995, 270, 1179–1180.
\bibitem{haddon1997electronic} Haddon, R. Electronic properties of carbon toroids. Nature 1997, 388, 31–32.
\bibitem{lin1998persistent} Lin, M.; Chuu, D. Persistent currents in toroidal carbon nanotubes. Phys. Rev. B 1998, 57, 6731.
\bibitem{mintmire1992fullerene} Mintmire, J.; Dunlap, B.; White, C. Are fullerene tubules metallic? Phys. Rev. Lett. 1992, 68, 631.
\bibitem{liu1998fullerene} Liu, J. et al. Fullerene pipes. Science 1998, 280, 1253–1256.
\bibitem{kratschmer1990solid} Kratschmer, W.; Lamb, L. D.; Fostiropoulos, K.; Huffman, D. Solid C60: a new form of carbon. Nature 1990, 347, 27.
\bibitem{chen2001new} Chen, X.; Deng, F.; Wang, J.; Yang, H.; Wu, G.; Zhang, X.; Peng, J.; Li, W. New method of carbon onion growth by radio-frequency plasma-enhanced chemical vapor deposition. Chem. Phys. Lett. 2001, 336, 201–204.
\bibitem{zhang2005experimental} Zhang, Y.; Tan, Y. W.; Stormer, H. L.; Kim, P. Experimental observation of the quantum Hall effect and Berry’s phase in graphene. Nature 2005, 438, 201–204.
\bibitem{zhang2011experimental} Zhang, L.; Zhang, Y.; Camacho, J.; Khodas, M.; Zaliznyak, I. The experimental observation of quantum Hall effect of l= 3 chiral quasiparticles in trilayer graphene. Nat. Phys. 2011, 7, 953–957.
\bibitem{bolotin2008ultrahigh} Bolotin, K. I.; Sikes, K.; Jiang, Z.; Klima, M.; Fudenberg, G.; Hone, J.; Kim, P.; Stormer, H. Ultrahigh electron mobility in suspended graphene. Solid State Commun. 2008, 146, 351–355.
\bibitem{orlita2008approaching} Orlita, M. et al. Approaching the Dirac point in high-mobility multilayer epitaxial graphene. Phys. Rev. Lett. 2008, 101, 267601.
\bibitem{nair2008fine} Nair, R. R.; Blake, P.; Grigorenko, A. N.; Novoselov, K. S.; Booth, T. J.; Stauber, T.; Peres, N. M.; Geim, A. K. Fine structure constant defines visual transparency of graphene. Science 2008, 320, 1308–1308.
\bibitem{chen2013transparent} Chen, T.; Xue, Y.; Roy, A. K.; Dai, L. Transparent and stretchable high-performance supercapacitors based on wrinkled graphene electrodes. ACS Nano 2013, 8, 1039–1046.
\bibitem{lee2008measurement} Lee, C.; Wei, X.; Kysar, J. W.; Hone, J. Measurement of the elastic properties and intrinsic strength of monolayer graphene. Science 2008, 321, 385–388.
\bibitem{lee2012estimation} Lee, J. U.; Yoon, D.; Cheong, H. Estimation of Young's modulus of graphene by Raman spectroscopy. Nano Lett. 2012, 12, 4444–4448.
\bibitem{ho2010magneto} Ho, Y. H.; Chiu, Y. H.; Lin, D. H.; Chang, C. P.; Lin, M. F. Magneto-optical selection rules in bilayer Bernal graphene. ACS Nano 2010, 4, 1465–1472.
\bibitem{chung2011exploration} Chung, H. C.; Lee, M. H.; Chang, C. P.; Lin, M. F. Exploration of edge-dependent optical selection rules for graphene nanoribbons. Optics Express 2011, 19, 23350–23363.
\bibitem{guinea2006electronic} Guinea, F.; Neto, A. C.; Peres, N. Electronic states and Landau levels in graphene stacks. Phys. Rev. B 2006, 73, 245426.
\bibitem{lin2015magneto} Lin, C. Y.; Wu, J. Y.; Ou, Y. J.; Chiu, Y. H.; Lin, M. F. Magneto-electronic properties of multilayer graphenes. Phys. Chem. Chem. Phys. 2015, 17, 26008–26035.
\bibitem{ho2006coulomb} Ho, J. H.; Lu, C. L.; Hwang, C. C.; Chang, C. P.; Lin, M. F. Coulomb excitations in AA-and AB-stacked bilayer graphites. Phys. Rev. B 2006, 74, 085406.
\bibitem{ju2011graphene} Ju, L. et al. Graphene plasmonics for tunable terahertz metamaterials. Nat. Nanotechnol. 2011, 6, 630–634.
\bibitem{castro2007biased} Castro, E. V.; Novoselov, K.; Morozov, S.; Peres, N.; Dos Santos, J. L.; Nilsson, J.; Guinea, F.; Geim, A.; Neto, A. C. Biased bilayer graphene: semiconductor with a gap tunable by the electric field effect. Phys. Rev. Lett. 2007, 99, 216802.
\bibitem{bao2011stacking} Bao, W. et al. Stacking-dependent band gap and quantum transport in trilayer graphene. Nat. Phys. 2011, 7, 948–952.
\bibitem{ho2009semimetallic} Ho, J.; Chiu, Y. H.; Tsai, S.; Lin, M. F. Semimetallic graphene in a modulated electric potential. Phys. Rev. B 2009, 79, 115427.
\bibitem{wu2009synthesis} Wu, Z. S.; Ren, W.; Gao, L.; Liu, B.; Jiang, C.; Cheng, H. M. Synthesis of high-quality graphene with a pre-determined number of layers. Carbon 2009, 47, 493–499.
\bibitem{hao2010probing} Hao, Y.; Wang, Y.; Wang, L.; Ni, Z.; Wang, Z.; Wang, R.; Koo, C. K.; Shen, Z.; Thong, J. T. Probing Layer Number and Stacking Order of Few-Layer Graphene by Raman Spectroscopy. Small 2010, 6, 195–200.
\bibitem{guinea2010energy} Guinea, F.; Katsnelson, M.; Geim, A. Energy gaps and a zero-field quantum Hall effect in graphene by strain engineering. Nat. Phys. 2010, 6, 30–33.
\bibitem{wong2012strain} Wong, J. H.; Wu, B. R.; Lin, M. F. Strain effect on the electronic properties of single layer and bilayer graphene. J. Phys. Chem. C 2012, 116, 8271–8277.
\bibitem{son2011electronic} Son, Y. W.; Choi, S. M.; Hong, Y. P.; Woo, S.; Jhi, S. H. Electronic topological transition in sliding bilayer graphene. Phys. Rev. B 2011, 84, 155410.
\bibitem{tran2015configuration} Tran, N. T. T.; Lin, S. Y.; Glukhova, O. E.; Lin, M. F. Configuration-induced rich electronic properties of bilayer graphene. J. Phys. Chem. C 2015, 119, 10623–10630.
\bibitem{giovannetti2008doping} Giovannetti, G.; Khomyakov, P.; Brocks, G.; Karpan, V. v.; Van den Brink, J.; Kelly, P. Doping graphene with metal contacts. Phys. Rev. Lett. 2008, 101, 026803.
\bibitem{dai2009gas} Dai, J.; Yuan, J.; Giannozzi, P. Gas adsorption on graphene doped with B, N, Al, and S: a theoretical study. Appl. Phys. Lett. 2009, 95, 232105.
\bibitem{lu2006influence} Lu, C. L.; Chang, C. P.; Huang, Y. C.; Chen, R. B.; Lin, M. L. Influence of an electric field on the optical properties of few-layer graphene with AB stacking. Phys. Rev. B 2006, 73, 144427.
\bibitem{l2007electronic} Lu, C. L.; Chang, C. P.; Huang, Y. C.; Ho, J. H.; Hwang, C. C.; Lin, M. F. Electronic properties of AA-and ABC-stacked few-layer graphites. J. Phys. Soc. Jpn. 2007, 76, 024701.
\bibitem{yan2007electric} Yan, J.; Zhang, Y.; Kim, P.; Pinczuk, A. Electric field effect tuning of electron-phonon coupling in graphene. Phys. Rev. Lett. 2007, 98, 166802.
\bibitem{lai2008magnetoelectronic} Lai, Y. H.; Ho, J. H.; Chang, C. P.; Lin, M. F. Magnetoelectronic properties of bilayer Bernal graphene. Phys. Rev. B 2008, 77, 085426.
\bibitem{goerbig2011electronic} Goerbig, M. Electronic properties of graphene in a strong magnetic field. Rev. Modern Phys. 2011, 83, 1193.
\bibitem{bolotin2008temperature} Bolotin, K.; Sikes, K.; Hone, J.; Stormer, H.; Kim, P. Temperature-dependent transport in suspended graphene. Phys. Rev. Lett. 2008, 101, 096802.
\bibitem{wu2011plasma} Wu, J. Y.; Chen, S. C.; Roslyak, O.; Gumbs, G.; Lin, M. F. Plasma excitations in graphene: Their spectral intensity and temperature dependence in magnetic field. ACS Nano 2011, 5, 1026–1032.
\bibitem{lee2008growth} Lee, J. K.; Lee, S. C.; Ahn, J. P.; Kim, S. C.; Wilson, J. I.; John, P. The growth of AA graphite on (111) diamond. J. Chem. Phys. 2008, 129, 234709–234709.
\bibitem{yankowitz2014electric} Yankowitz, M. et al. Electric field control of soliton motion and stacking in trilayer graphene. Nature Mater. 2014, 13, 786–789.
\bibitem{coletti2013revealing} Coletti, C. et al. Revealing the electronic band structure of trilayer graphene on SiC: An angle-resolved photoemission study. Phys. Rev. B 2013, 88, 155439.
\bibitem{ohta2007interlayer} Ohta, T.; Bostwick, A.; McChesney, J. L.; Seyller, T.; Horn, K.; Rotenberg, E. Interlayer interaction and electronic screening in multilayer graphene investigated with angle-resolved photoemission spectroscopy. Phys. Rev. Lett. 2007, 98, 206802.
\bibitem{norimatsu2010selective} Norimatsu, W.; Kusunoki, M. Selective formation of ABC-stacked graphene layers on SiC (0001). Phys. Rev. B 2010, 81, 161410.
\bibitem{rong1993electronic} Rong, Z. Y.; Kuiper, P. Electronic effects in scanning tunneling microscopy: Moiré pattern on a graphite surface. Phys. Review. B 1993, 48, 17427.
\bibitem{xu2012electronic} Xu, P.; Yang, Y.; Qi, D.; Barber, S.; Schoelz, J.; Ackerman, M.; Bellaiche, L.; Thibado, P. Electronic transition from graphite to graphene via controlled movement of the top layer with scanning tunneling microscopy. Phys. Rev. B 2012, 86, 085428.
\bibitem{pong2007observation} Pong, W. T.; Bendall, J.; Durkan, C. Observation and investigation of graphite superlattice boundaries by scanning tunneling microscopy. Surf. Sci. 2007, 601, 498–509.
\bibitem{kazemi2013stacking} Kazemi, A. S.; Crampin, S.; Ilie, A. Stacking-dependent superstructures at stepped armchair interfaces of bilayer/trilayer graphene. Appl. Phys. Lett. 2013, 102, 163111.
\bibitem{ferrari2007raman} Ferrari, A. C. Raman spectroscopy of graphene and graphite: disorder, electron–phonon coupling, doping and nonadiabatic effects. Solid state commun. 2007, 143, 47–57.
\bibitem{li2008carbon} Li, Q.; Ni, Z.; Gong, J.; Zhu, D.; Zhu, Z. Carbon nanotubes coated by carbon nanoparticles of turbostratic stacked graphenes. Carbon 2008, 46, 434–439.
\bibitem{lenski2011raman} Lenski, D. R.; Fuhrer, M. S. Raman and optical characterization of multilayer turbostratic graphene grown via chemical vapor deposition. J. Appl. Phys. 2011, 110, 013720.
\bibitem{lu2006low} Lu, C. L.; Chang, C. P.; Huang, Y. C.; Lu, J. M.; Hwang, C. C.; Lin, M. F. Low-energy electronic properties of the AB-stacked few-layer graphites. J. Phys. Condens. Matter 2006, 18, 5849.
\bibitem{koshino2010interlayer} Koshino, M. Interlayer screening effect in graphene multilayers with ABA and ABC stacking. Phys. Rev. B 2010, 81, 125304.
\bibitem{mak2010electronic} Mak, K. F.; Shan, J.; Heinz, T. F. Electronic structure of few-layer graphene: experimental demonstration of strong dependence on stacking sequence. Phys. Rev. Lett. 2010, 104, 176404.
\bibitem{lui2011observation} Lui, C. H.; Li, Z.; Mak, K. F.; Cappelluti, E.; Heinz, T. F. Observation of an electrically tunable band gap in trilayer graphene. Nat. Phys. 2011, 7, 944–947.
\bibitem{mohammadi2014effects} Mohammadi, Y.; Moradian, R.; Tabar, F. S. Effects of doping and bias voltage on the screening in AAA-stacked trilayer graphene. Solid State Commun. 2014, 193, 1–5.
\bibitem{do2015configuration} Do, T. N.; Lin, C. Y.; Lin, Y. P.; Shih, P. H.; Lin, M. F. Configuration-enriched magneto-electronic spectra of AAB-stacked trilayer graphene. Carbon 2015, 94, 619–632.
\bibitem{yuan2011landau} Yuan, S.; Roldán, R.; Katsnelson, M. I. Landau level spectrum of ABA-and ABC-stacked trilayer graphene. Phys. Rev. B 2011, 84, 125455.
\bibitem{zhang2012hund} Zhang, F.; Tilahun, D.; MacDonald, A. H. Hund’s rules for the N=0 Landau levels of trilayer graphene. Phys. Rev. B 2012, 85, 165139.
\bibitem{koshino2013stacking} Koshino, M. Stacking-dependent optical absorption in multilayer graphene. New J. Phys. 2013, 15, 015010.
\bibitem{do2015rich} Do, T. N.; Shih, P. H.; Chang, C. P.; Lin, C. Y.; Lin, M. F. Rich magneto-absorption spectra in AAB-stacked trilayer graphene. Phys. Chem. Chem. Phys. 2016, 18, 17597–605.
\bibitem{chiu2016influence} Chiu, C. W.; Chen, R. B. Influence of electric fields on absorption spectra of AABstacked trilayer graphene. Appl. Phys. Express 2016, 9, 065103.
\bibitem{yuan2010electronic} Yuan, S.; De Raedt, H.; Katsnelson, M. I. Electronic transport in disordered bilayer and trilayer graphene. Phys. Rev. B 2010, 82, 235409.
\bibitem{ghosh2012monte} Ghosh, B.; Misra, S. Monte Carlo simulation study of spin transport in trilayer graphene: A comparison between ABA and ABC stacking. J. Appl. Phys. 2012, 112, 073720.
\bibitem{ma2012stacking} Ma, R.; Sheng, L.; Liu, M.; Sheng, D. Stacking-order dependence in thermoelectric transport of biased trilayer graphene. Phys. Rev. B 2012, 86, 115414.
\bibitem{khodkov2012electrical} Khodkov, T.; Withers, F.; Hudson, D. C.; Craciun, M. F.; Russo, S. Electrical transport in suspended and double gated trilayer graphene. Appl. Phys. Lett. 2012, 100, 013114.
\bibitem{yan2008phonon} Yan, J. A.; Ruan, W.; Chou, M. Phonon dispersions and vibrational properties of monolayer, bilayer, and trilayer graphene: Density-functional perturbation theory. Phys. Rev. B 2008, 77, 125401.
\bibitem{yan2009electron} Yan, J. A.; Ruan, W.; Chou, M. Electron-phonon interactions for optical-phonon modes in few-layer graphene: First-principles calculations. Phys. Rev. B 2009, 79, 115443.
\bibitem{lui2013tunable} Lui, C. H. et al. Tunable infrared phonon anomalies in trilayer graphene. Phys. Rev. Lett. 2013, 110, 185504.
\bibitem{cong2015magnetic} Cong, C.; Jung, J.; Cao, B.; Qiu, C.; Shen, X.; Ferreira, A.; Adam, S.; Yu, T. Magnetic oscillation of optical phonon in ABA-and ABC-stacked trilayer graphene. Phys. Rev. B 2015, 91, 235403.
\bibitem{aoki2007dependence} Aoki, M.; Amawashi, H. Dependence of band structures on stacking and field in layered graphene. Solid State Commun. 2007, 142, 123–127.
\bibitem{zhang2010band} Zhang, F.; Sahu, B.; Min, H.; MacDonald, A. H. Band structure of ABC-stacked graphene trilayers. Phys. Rev. B 2010, 82, 035409.
\bibitem{avramov2011ab} Avramov, P. V.; Sakai, S.; Entani, S.; Matsumoto, Y.; Naramoto, H. Ab initio LCDFT study of graphene, multilayer graphenes and graphite. Chem. Phys. Lett. 2011, 508, 86–89.
\bibitem{menezes2014ab} Menezes, M. G.; Capaz, R. B.; Louie, S. G. Ab initio quasiparticle band structure of ABA and ABC-stacked graphene trilayers. Phys. Rev. B 2014, 89, 035431.
\bibitem{avetisyan2010stacking} Avetisyan, A.; Partoens, B.; Peeters, F. Stacking order dependent electric field tuning of the band gap in graphene multilayers. Phys. Rev. B 2010, 81, 115432.
\bibitem{siegel2013charge} Siegel, D. A.; Regan, W.; Fedorov, A. V.; Zettl, A.; Lanzara, A. Charge-carrier screening in single-layer graphene. Phys. Rev. Lett. 2013, 110, 146802.
\bibitem{bostwick2007quasiparticle} Bostwick, A.; Ohta, T.; Seyller, T.; Horn, K.; Rotenberg, E. Quasiparticle dynamics in graphene. Nat. Phys. 2007, 3, 36–40.
\bibitem{ohta2006controlling} Ohta, T.; Bostwick, A.; Seyller, T.; Horn, K.; Rotenberg, E. Controlling the electronic structure of bilayer graphene. Science 2006, 313, 951–954.
\bibitem{luican2011single} Luican, A.; Li, G.; Reina, A.; Kong, J.; Nair, R.; Novoselov, K. S.; Geim, A. K.; Andrei, E. Single-layer behavior and its breakdown in twisted graphene layers. Phys. Rev. Lett. 2011, 106, 126802.
\bibitem{li2010observation} Li, G.; Luican, A.; Dos Santos, J. L.; Neto, A. C.; Reina, A.; Kong, J.; Andrei, E. Observation of Van Hove singularities in twisted graphene layers. Nat. Phys. 2010, 6, 109–113.
\bibitem{cherkez2015van} Cherkez, V.; de Laissardiere, G. T.; Mallet, P.; Veuillen, J. Y. Van Hove singularities in doped twisted graphene bilayers studied by scanning tunneling spectroscopy. Phys. Rev. B 2015, 91, 155428.
\bibitem{lauffer2008atomic} Lauffer, P.; Emtsev, K.; Graupner, R.; Seyller, T.; Ley, L.; Reshanov, S.; Weber, H. Atomic and electronic structure of few-layer graphene on SiC (0001) studied with scanning tunneling microscopy and spectroscopy. Phys. Rev. B 2008, 77, 155426.
\bibitem{yankowitz2013local} Yankowitz, M.; Wang, F.; Lau, C. N.; LeRoy, B. J. Local spectroscopy of the electrically tunable band gap in trilayer graphene. Phys. Rev. B 2013, 87, 165102.
\bibitem{que2015stacking} Que, Y.; Xiao, W.; Chen, H.; Wang, D.; Du, S.; Gao, H.-J. Stacking-dependent electronic property of trilayer graphene epitaxially grown on Ru (0001). Appl. Phys. Lett. 2015, 107, 263101.
\bibitem{pierucci2015evidence} Pierucci, D. et al. Evidence for flat bands near the Fermi level in epitaxial rhombohedral multilayer graphene. ACS Nano 2015, 9, 5432–5439.
\bibitem{feng2013superlubric} Feng, X.; Kwon, S.; Park, J. Y.; Salmeron, M. Superlubric sliding of graphene nanoflakes on graphene. ACS Nano 2013, 7, 1718–1724.
\bibitem{liu2012observation} Liu, Z.; Yang, J.; Grey, F.; Liu, J. Z.; Liu, Y.; Wang, Y.; Yang, Y.; Cheng, Y.; Zheng, Q. Observation of microscale superlubricity in graphite. Phys. Rev. Lett. 2012, 108, 205503.
\bibitem{xu2013graphene} Xu, P. et al. Graphene Manipulation on 4H-SiC (0001) Using Scanning Tunneling Microscopy. Japanese Journal of Applied Physics 2013, 52, 035104.
\bibitem{meng2013strain} Meng, L. et al. Strain-induced one-dimensional Landau level quantization in corrugated graphene. Phys. Rev. B 2013, 87, 205405.
\bibitem{yan2013strain} Yan, W.; He, W. Y.; Chu, Z. D.; Liu, M.; Meng, L.; Dou, R. F.; Zhang, Y.; Liu, Z.; Nie, J. C.; He, L. Strain and curvature induced evolution of electronic band structures in twisted graphene bilayer. Nat. Commun. 2013, 4 .
\bibitem{bai2014creating} Bai, K. K.; Zhou, Y.; Zheng, H.; Meng, L.; Peng, H.; Liu, Z.; Nie, J. C.; He, L. Creating one-dimensional nanoscale periodic ripples in a continuous mosaic graphene monolayer. Phys. Rev. Lett. 2014, 113, 086102.
\bibitem{huang2014feature} Huang, Y. K.; Chen, S. C.; Ho, Y. H.; Lin, C. Y.; Lin, M. F. Feature-Rich Magnetic Quantization in Sliding Bilayer Graphenes. Sci. Rep. 2014, 4, 7509.
\bibitem{chen2014shift} Chen, S. C.; Chiu, C. W.; Wu, C. L.; Lin, M. F. Shift-enriched optical properties in bilayer graphene. RSC Adv. 2014, 4, 63779–63783.
\bibitem{koshino2013electronic} Koshino, M. Electronic transmission through AB-BA domain boundary in bilayer graphene. Phys. Rev. B 2013, 88, 115409.
\bibitem{choi2013anomalous} Choi, S. M.; Jhi, S. H.; Son, Y. W. Anomalous optical phonon splittings in sliding bilayer graphene. ACS Nano 2013, 7, 7151–7156.
\bibitem{lin2016optical} Lin, C. Y.; Do, T. N.; Huang, Y. K.; Lin, M. F. Optical Properties of Graphene in Magnetic and Electric fields. arXiv preprint arXiv:1603.02797 2016,
\bibitem{wehling2008midgap} Wehling, T.; Balatsky, A.; Tsvelik, A.; Katsnelson, M.; Lichtenstein, A. Midgap states in corrugated graphene: Ab initio calculations and effective field theory. EPL (Europhys. Lett.) 2008, 84, 17003.
\bibitem{lin2015feature} Lin, S. Y.; Chang, S. L.; Shyu, F. L.; Lu, J. M.; Lin, M. F. Feature-rich electronic properties in graphene ripples. Carbon 2015, 86, 207–216.
\bibitem{hamada1992new} Hamada, N.; Sawada, S.; Oshiyama, A. New one-dimensional conductors: graphitic microtubules. Phys. Rev. Lett. 1992, 68, 1579.
\bibitem{kane1997size} Kane, C. L.; Mele, E. Size, shape, and low energy electronic structure of carbon nanotubes. Physical Review Letters 1997, 78, 1932.
\bibitem{l2002electronic} Shyu, F. L.; Lin, M. F. Electronic and optical properties of narrow-gap carbon nanotubes. J. Phys. Soc.Jpn. 2002, 71, 1820–1823.
\bibitem{blase1994hybridization} Blase, X.; Benedict, L. X.; Shirley, E. L.; Louie, S. G. Hybridization effects and metallicity in small radius carbon nanotubes. Phys. Rev. Lett. 1994, 72, 1878.
\bibitem{gulseren2002systematic} Gülseren, O.; Yildirim, T.; Ciraci, S. Systematic ab initio study of curvature effects in carbon nanotubes. Phys. Rev. B 2002, 65, 153405.
\bibitem{brodie1859atomic} Brodie, B. C. On the atomic weight of graphite. Philos. Trans. R. Soc. London 1859, 149, 249–259.
\bibitem{staudenmaier1898verfahren} Staudenmaier, L. Verfahren zur darstellung der graphitsäure. Ber. Dtsch. Chem. Ges. 1898, 31, 1481–1487.
\bibitem{hummers1958preparation} Hummers Jr, W. S.; Offeman, R. E. Preparation of graphitic oxide. J. Am. Chem. Soc. 1958, 80, 1339–1339.
\bibitem{tang2012bottom} Tang, L.; Li, X.; Ji, R.; Teng, K. S.; Tai, G.; Ye, J.; Wei, C.; Lau, S. P. Bottomup synthesis of large-scale graphene oxide nanosheets. J. Mater. Chem. 2012, 22, 5676–5683.
\bibitem{tran2016chemical} Tran, N. T. T.; Lin, S. Y.; Lin, Y. T.; Lin, M. F. Chemical bonding-induced rich electronic properties of oxygen adsorbed few-layer graphenes. Phys. Chem. Chem. Phys. 2016, 18, 4000–4007.
\bibitem{tran2016pi} Tran, N. T. T.; Lin, S. Y.; Glukhova, O. E.; Lin, M. F. $\pi$-Bonding-dominated energy gaps in graphene oxide. RSC Adv. 2016, 6, 24458–24463.
\bibitem{lian2013big} Lian, K. Y.; Ji, Y. F.; Li, X. F.; Jin, M. X.; Ding, D. J.; Luo, Y. Big bandgap in highly reduced graphene oxides. J. Phys. Chem. C 2013, 117, 6049–6054.
\bibitem{ito2008semiconducting} Ito, J.; Nakamura, J.; Natori, A. Semiconducting nature of the oxygen-adsorbed graphene sheet. J. Appl. Phys. 2008, 103, 113712.
\bibitem{mkhoyan2009atomic} Mkhoyan, K. A.; Contryman, A. W.; Silcox, J.; Stewart, D. A.; Eda, G.; Mattevi, C.; Miller, S.; Chhowalla, M. Atomic and electronic structure of graphene-oxide. Nano Lett. 2009, 9, 1058–1063.
\bibitem{schulte2013bandgap} Schulte, K.; Vinogradov, N.; Ng, M. L.; Mårtensson, N.; Preobrajenski, A. Bandgap formation in graphene on Ir (111) through oxidation. Applied Surface Science 2013, 267, 74–76.
\bibitem{pandey2008scanning} Pandey, D.; Reifenberger, R.; Piner, R. Scanning probe microscopy study of exfoliated oxidized graphene sheets. Surf. Sci. 2008, 602, 1607–1613.
\bibitem{sofo2007graphane} Sofo, J. O.; Chaudhari, A. S.; Barber, G. D. Graphane: a two-dimensional hydrocarbon. Phys. Rev. B 2007, 75, 153401.
\bibitem{chandrachud2010systematic} Chandrachud, P.; Pujari, B. S.; Haldar, S.; Sanyal, B.; Kanhere, D. A systematic study of electronic structure from graphene to graphane. J. Phys. Condens. Matter 2010, 22, 465502.
\bibitem{yang2010two} Yang M. et al. Two-dimensional graphene superlattice made with partial hydrogenation. Appl. Phys. Lett. 2010, 96, 193115.
\bibitem{gao2011band} Gao, H.; Wang, L.; Zhao, J.; Ding, F.; Lu, J. Band gap tuning of hydrogenated graphene: H coverage and configuration dependence. J. Phys. Chem. C 2011, 115, 3236–3242.
\bibitem{huang2016configuration} Huang, H. C.; Lin, S. Y.; Wu, C. L.; Lin, M. F. Configuration-and concentration-dependent electronic properties of hydrogenated graphene. Carbon 2016, 103, 84–93.
\bibitem{elias2009control} Elias, D. C. et al. Control of graphene’s properties by reversible hydrogenation: evidence for graphane. Science 2009, 323, 610–613.
\bibitem{jaiswal2011controlled} Jaiswal, M.; Yi Xuan Lim, C. H.; Bao, Q.; Toh, C. T.; Loh, K. P.; Ozyilmaz, B. Controlled hydrogenation of graphene sheets and nanoribbons. ACS Nano 2011, 5, 888–896.
\bibitem{jones2010formation} Jones, J.; Mahajan, K.; Williams, W.; Ecton, P.; Mo, Y.; Perez, J. Formation of graphane and partially hydrogenated graphene by electron irradiation of adsorbates on graphene. Carbon 2010, 48, 2335–2340.
\bibitem{wang2010toward} Wang, Y.; Xu, X.; Lu, J.; Lin, M.; Bao, Q.; Ozyilmaz, B.; Loh, K. P. Toward high throughput interconvertible graphane-to-graphene growth and patterning. ACS Nano 2010, 4, 6146–6152.
\bibitem{balog2010bandgap} Balog, R. et al. Bandgap opening in graphene induced by patterned hydrogen adsorption. Nat. Mater. 2010, 9, 315–319.
\bibitem{grassi2011scaling} Grassi, R.; Low, T.; Lundstrom, M. Scaling of the energy gap in pattern-hydrogenated graphene. Nano Lett. 2011, 11, 4574–4578.
\bibitem{guisinger2009exposure} Guisinger, N. P.; Rutter, G. M.; Crain, J. N.; First, P. N.; Stroscio, J. A. Exposure of epitaxial graphene on SiC (0001) to atomic hydrogen. Nano Lett. 2009, 9, 1462–1466.
\bibitem{castellanos2012reversible} Castellanos-Gomez, A.; Wojtaszek, M.; Tombros, N.; vanWees, B. J. Reversible hydrogenation and bandgap opening of graphene and graphite surfaces probed by scanning tunneling spectroscopy. Small 2012, 8, 1607–1613.
\bibitem{balog2009atomic} Balog, R.; Jørgensen, B.; Wells, J.; Lægsgaard, E.; Hofmann, P.; Besenbacher, F.; Hornekær, L. Atomic hydrogen adsorbate structures on graphene. J. Am. Chem. Soc. 2009, 131, 8744–8745.
\bibitem{boukhvalov2008hydrogen} Boukhvalov, D.; Katsnelson, M.; Lichtenstein, A. Hydrogen on graphene: Electronic structure, total energy, structural distortions and magnetism from first-principles calculations. Phys. Rev. B 2008, 77, 035427
\bibitem{hussain2012polylithiated} Hussain, T.; Maark, T. A.; De Sarkar, A.; Ahuja, R. Polylithiated (OLi$_2$) functionalized graphane as a potential hydrogen storage material. Appl. Phys. Lett. 2012, 101, 243902.
\bibitem{hussain2012strain} Hussain, T.; De Sarkar, A.; Ahuja, R. Strain induced lithium functionalized graphane as a high capacity hydrogen storage material. Appl. Phys. Lett. 2012, 101, 103907.
\bibitem{antipina2012high} Antipina, L. Y.; Avramov, P. V.; Sakai, S.; Naramoto, H.; Ohtomo, M.; Entani, S.; Matsumoto, Y.; Sorokin, P. B. High hydrogen-adsorption-rate material based on graphane decorated with alkali metals. Phys. Rev. B 2012, 86, 085435.
\bibitem{nechaev2011high} Nechaev, Y. S. The high-density hydrogen carrier intercalation in graphane-like nanostructures, relevance to its on-board storage in fuel-cell-powered vehicles. The Open Fuel Cells . 2011, 4, 10–2174.
\bibitem{cheng2010reversible} Cheng, S.-H.; Zou, K.; Okino, F.; Gutierrez, H. R.; Gupta, A.; Shen, N.; Eklund, P.; Sofo, J.; Zhu, J. Reversible fluorination of graphene: evidence of a two-dimensional wide bandgap semiconductor. Phys. Rev. B 2010, 81, 205435.
\bibitem{mazanek2015tuning} Mazanek, V.; Jankovsky, O.; Luxa, J.; Sedmidubsky, D.; Janousek, Z.; Sembera, F.; Mikulics, M.; Sofer, Z. Tuning of fluorine content in graphene: towards large-scale production of stoichiometric fluorographene. Nanoscale 2015, 7, 13646–13655.
\bibitem{wang2014fluorination} Wang, B.; Wang, J.; Zhu, J. Fluorination of graphene: A spectroscopic and microscopic study. ACS Nano 2014, 8, 1862–1870.
\bibitem{sherpa2014local} Sherpa, S. D.; Kunc, J.; Hu, Y.; Levitin, G.; De Heer, W. A.; Berger, C.; Hess, D. W. Local work function measurements of plasma-fluorinated epitaxial graphene. Appl. Phys. Lett. 2014, 104, 081607.
\bibitem{gong2013photochemical} Gong, P.; Wang, Z.; Li, Z.; Mi, Y.; Sun, J.; Niu, L.; Wang, H.; Wang, J.; Yang, S. Photochemical synthesis of fluorinated graphene via a simultaneous fluorination and reduction route. RSC Adv. 2013, 3, 6327–6330.
\bibitem{lee2012selective} Lee, W. H.; Suk, J. W.; Chou, H.; Lee, J.; Hao, Y.; Wu, Y.; Piner, R.; Akinwande, D.; Kim, K. S.; Ruoff, R. S. Selective-area fluorination of graphene with fluoropolymer and laser irradiation. Nano Lett. 2012, 12, 2374–2378.
\bibitem{samanta2013highly} Samanta, K.; Some, S.; Kim, Y.; Yoon, Y.; Min, M.; Lee, S. M.; Park, Y.; Lee, H. Highly hydrophilic and insulating fluorinated reduced graphene oxide. Chem. Commun. 2013, 49, 8991–8993.
\bibitem{wang2012synthesis} Wang, Z.; Wang, J.; Li, Z.; Gong, P.; Liu, X.; Zhang, L.; Ren, J.; Wang, H.; Yang, S. Synthesis of fluorinated graphene with tunable degree of fluorination. Carbon 2012, 50, 5403–5410.
\bibitem{nair2010fluorographene} Nair, R. R. et al. Fluorographene: A Two-Dimensional Counterpart of Teflon. Small 2010, 6, 2877–2884.
\bibitem{jeon2011fluorographene} Jeon, K. J. et al. Fluorographene: a wide bandgap semiconductor with ultraviolet luminescence. ACS Nano 2011, 5, 1042–1046.
\bibitem{medeiros2010dft} Medeiros, P. V.; Mascarenhas, A. J.; de Brito Mota, F.; de Castilho, C. M. C. A DFT study of halogen atoms adsorbed on graphene layers. Nanotech. 2010, 21, 485701.
\bibitem{sahin2012chlorine} Sahin, H.; Ciraci, S. Chlorine adsorption on graphene: Chlorographene. J. Phys. Chem. C 2012, 116, 24075–24083.
\bibitem{walter2011highly} Walter, A. L. et al. Highly p-doped epitaxial graphene obtained by fluorine intercalation. Appl. Phys. Lett. 2011, 98, 184102.
\bibitem{sun2014solvothermally} Sun, C.; Feng, Y.; Li, Y.; Qin, C.; Zhang, Q.; Feng, W. Solvothermally exfoliated fluorographene for high-performance lithium primary batteries. Nanoscale 2014, 6, 2634–2641.
\bibitem{zhao2014fluorinated} Zhao, F. G.; Zhao, G.; Liu, X. H.; Ge, C. W.; Wang, J. T.; Li, B. L.; Wang, Q. G.; Li, W. S.; Chen, Q. Y. Fluorinated graphene: facile solution preparation and tailorable properties by fluorine-content tuning. J. Mater. Chem. A 2014, 2, 8782–8789.
\bibitem{wang2012fluorinated} Wang, Y.; Lee, W. C.; Manga, K. K.; Ang, P. K.; Lu, J.; Liu, Y. P.; Lim, C. T.; Loh, K. P. Fluorinated Graphene for Promoting Neuro-Induction of Stem Cells. Adv. Mater. 2012, 24, 4285–4290.
\bibitem{nebogatikova2016fluorinated} Nebogatikova, N.; Antonova, I.; Kurkina, I.; Soots, R.; Vdovin, V.; Timofeev, V.; Smagulova, S.; Prinz, V. Y. Fluorinated graphene suspension for inkjet printed technologies. Nanotechnolgy 2016, 27, 205601.
\bibitem{katkov2015backside} Katkov, M.; Sysoev, V.; Gusel’Nikov, A.; Asanov, I.; Bulusheva, L.; Okotrub, A. A backside fluorine-functionalized graphene layer for ammonia detection. Phys. Chem. Chem. Phys. 2015, 17, 444–450.
\bibitem{urbanova2016fluorinated} Urbanová, V. et al. Fluorinated graphenes as advanced biosensors–effect of fluorine coverage on electron transfer properties and adsorption of biomolecules. Nanoscale 2016.
\bibitem{liu2012electronic} Liu, H.; Hou, Z.; Hu, C.; Yang, Y.; Zhu, Z. Electronic and magnetic properties of fluorinated graphene with different coverage of fluorine. J. Phys. Chem. C 2012, 116, 18193–18201.
\bibitem{nair2012spin} Nair, R.; Sepioni, M.; Tsai, I. L.; Lehtinen, O.; Keinonen, J.; Krasheninnikov, A.; Thomson, T.; Geim, A.; Grigorieva, I. Spin-half paramagnetism in graphene induced by point defects. Nat. Phys. 2012, 8, 199–202.
\bibitem{li2011photochemical} Li, B.; Zhou, L.; Wu, D.; Peng, H.; Yan, K.; Zhou, Y.; Liu, Z. Photochemical chlorination of graphene. ACS Nano 2011, 5, 5957–5961.
\bibitem{wu2011controlled} Wu, J.; Xie, L.; Li, Y.; Wang, H.; Ouyang, Y.; Guo, J.; Dai, H. Controlled chlorine plasma reaction for noninvasive graphene doping. J. Am. Chem. Soc. 2011, 133, 19668–19671.
\bibitem{jankovsky2014towards} Jankovsky, O.; Simek, P.; Klimova, K.; Sedmidubsky, D.; Matejkova, S.; Pumera, M.; Sofer, Z. Towards graphene bromide: bromination of graphite oxide. Nanoscale 2014, 6, 6065–6074.
\bibitem{yao2012catalyst} Yao, Z.; Nie, H.; Yang, Z.; Zhou, X.; Liu, Z.; Huang, S. Catalyst-free synthesis of iodine-doped graphene via a facile thermal annealing process and its use for electrocatalytic oxygen reduction in an alkaline medium. Chem. Commun. 2012, 48, 1027–1029.
\bibitem{dresselhaus1981intercalation} Dresselhaus, M.; Dresselhaus, G. Intercalation compounds of graphite. Adv. Phys. 1981, 30, 139–326.
\bibitem{lee1997conductivity} Lee, R.; Kim, H.; Fischer, J.; Thess, A.; Smalley, R. E. Conductivity enhancement in single-walled carbon nanotube bundles doped with K and Br. Nature 1997, 388, 255–257.
\bibitem{rao1997evidence} Rao, A. M.; Eklund, P.; Bandow, S.; Thess, A.; Smalley, R. E. Evidence for charge transfer in doped carbon nanotube bundles from Raman scattering. Nature 1997, 388, 257–259.
\bibitem{chen2008charged} Chen, J. H.; Jang, C.; Adam, S.; Fuhrer, M.; Williams, E.; Ishigami, M. Charged impurity scattering in graphene. Nat. Phys. 2008, 4, 377–381.
\bibitem{hebard1991potassium} Hebard, A.; Rosseinky, M.; Haddon, R.; Murphy, D.; Glarum, S.; Palstra, T.; Ramirez, A.; Karton, A. Potassium-doped C60. Nature 1991, 350, 600–601.
\bibitem{haddon1992electronic} Haddon, R. Electronic structure, conductivity and superconductivity of alkali metal doped (C60). Acc. Chem. Res. 1992, 25, 127–133.
\bibitem{bianchi2010electron} Bianchi, M.; Rienks, E.; Lizzit, S.; Baraldi, A.; Balog, R.; Hornekær, L.; Hofmann, P. Electron-phonon coupling in potassium-doped graphene: Angle-resolved photoemission spectroscopy. Phys. Rev. B 2010, 81, 041403.
\bibitem{yoo2008large} Yoo, E.; Kim, J.; Hosono, E.; Zhou, H.-s.; Kudo, T.; Honma, I. Large reversible Li storage of graphene nanosheet families for use in rechargeable lithium ion batteries. Nano Lett. 2008, 8, 2277–2282.
\bibitem{papagno2011large} Papagno, M.; Rusponi, S.; Sheverdyaeva, P. M.; Vlaic, S.; Etzkorn, M.; Pacilé, D.; Moras, P.; Carbone, C.; Brune, H. Large band gap opening between graphene Dirac cones induced by Na adsorption onto an Ir superlattice. ACS Nano 2011, 6, 199–204.
\bibitem{watcharinyanon2011rb} Watcharinyanon, S.; Virojanadara, C.; Johansson, L. I. Rb and Cs deposition on epitaxial graphene grown on 6H-SiC (0001). Surf. Sci. 2011, 605, 1918–1922.
\bibitem{virojanadara2010epitaxial} Virojanadara, C.; Watcharinyanon, S.; Zakharov, A.; Johansson, L. I. Epitaxial graphene on 6 H-SiC and Li intercalation. Phys. Rev. B 2010, 82, 205402.
\bibitem{sugawara2011fabrication} Sugawara, K.; Kanetani, K.; Sato, T.; Takahashi, T. Fabrication of Li-intercalated bilayer graphene. AIP Adv. 2011, 1, 022103.
\bibitem{chan2008first} Chan, K. T.; Neaton, J.; Cohen, M. L. First-principles study of metal adatom adsorption on graphene. Phys. Rev. B 2008, 77, 235430.
\bibitem{jin2010crossover} Jin, K.-H.; Choi, S.-M.; Jhi, S.-H. Crossover in the adsorption properties of alkali metals on graphene. Phys. Rev. B 2010, 82, 033414.
\bibitem{praveen2015adsorption} Praveen, C.; Piccinin, S.; Fabris, S. Adsorption of alkali adatoms on graphene supported by the Au/Ni (111) surface. Phys. Rev. B 2015, 92, 075403.
\bibitem{lin2016feature} Lin, S. Y.; Lin, Y. T.; Tran, N. T. T.; Su, W. P.; Lin, M. F. Feature-rich electronic properties of aluminum-doped graphenes. arXiv preprint arXiv:1606.01624 2016.
\bibitem{obradovic2006analysis} Obradovic, B.; Kotlyar, R.; Heinz, F.; Matagne, P.; Rakshit, T.; Giles, M.; Stettler, M.; Nikonov, D. Analysis of graphene nanoribbons as a channel material for field-effect transistors. Appl. Phys. Lett. 2006, 88, 142102.
\bibitem{wang2008room} Wang, X.; Ouyang, Y.; Li, X.; Wang, H.; Guo, J.; Dai, H. Room-temperature allsemiconducting sub-10-nm graphene nanoribbon field-effect transistors. Phys. Rev. Lett. 2008, 100, 206803.
\bibitem{ao2009doped} Ao, Z.; Jiang, Q.; Zhang, R.; Tan, T.; Li, S. Al doped graphene: A promising material for hydrogen storage at room temperature. J. Appl. Phys. 2009, 105, 4307.
\bibitem{ao2008enhancement} Ao, Z.; Yang, J.; Li, S.; Jiang, Q. Enhancement of CO detection in Al doped graphene. Chemical Physics Letters 2008, 461, 276–279.
\bibitem{lin2015ultrafast} Lin, M. C. et al. An ultrafast rechargeable aluminium-ion battery. Nature 2015, 520, 324–328.
\bibitem{chen2015long} Chen, H. H.; Su, S.; Chang, S. L.; Cheng, B. Y.; Chong, C. W.; Huang, J.; Lin, M. F. Long-range interactions of bismuth growth on monolayer epitaxial graphene at room temperature. Carbon 2015, 93, 180–186.
\bibitem{chen2015tailoring} Chen, H. H.; Su, S.; Chang, S. L.; Cheng, B. Y.; Chen, S.; Chen, H. Y.; Lin, M. F.; Huang, J. Tailoring low-dimensional structures of bismuth on monolayer epitaxial graphene. Scientific reports 2015, 5, 11623.
\bibitem{kresse1996efficient} Kresse, G.; Furthmüller, J. Efficient iterative schemes for ab initio total-energy calculations using a plane-wave basis set. Phys. Rev. B 1996, 54, 11169.
\bibitem{perdew1996generalized} Perdew, J. P.; Burke, K.; Ernzerhof, M. Generalized gradient approximation made simple. Phys. Rev. Lett. 1996, 77, 3865.
\bibitem{blochl1994projector} Blöchl, P. E. Projector augmented-wave method. Phys. Rev. B 1994, 50, 17953.
\bibitem{grimme2006semiempirical} Grimme, S. Semiempirical GGA-type density functional constructed with a long-range dispersion correction. J. Comput. Chem. 2006, 27, 1787–1799.
\bibitem{hass2008growth} Hass, J.; De Heer, W. A.; Conrad, E. H. The growth and morphology of epitaxial multilayer graphene. J. Phys.: Condens. Matter 2008, 20, 323202.
\bibitem{coraux2008structural} Coraux, J.; N’Diaye, A. T.; Busse, C.; Michely, T. Structural coherency of graphene on Ir (111). Nano Lett. 2008, 8, 565–570.
\bibitem{ismach2010direct} Ismach, A.; Druzgalski, C.; Penwell, S.; Schwartzberg, A.; Zheng, M.; Javey, A.; Bokor, J.; Zhang, Y. Direct chemical vapor deposition of graphene on dielectric surfaces. Nano Lett. 2010, 10, 1542–1548.
\bibitem{park2010growth} Park, H. J.; Meyer, J.; Roth, S.; Skákalová, V. Growth and properties of few-layer graphene prepared by chemical vapor deposition. Carbon 2010, 48, 1088–1094.
\bibitem{reina2008large} Reina, A.; Jia, X.; Ho, J.; Nezich, D.; Son, H.; Bulovic, V.; Dresselhaus, M. S.; Kong, J. Large area, few-layer graphene films on arbitrary substrates by chemical vapor deposition. Nano Lett. 2008, 9, 30–35.
\bibitem{chae2009} Chae, S. J. et al. Synthesis of large-area graphene layers on poly-Nickel substrate by chemical vapor deposition: wrinkle formation. Adv. Mater. 2009, 21, 2328–2333.
\bibitem{zhao2011few} Zhao, G.; Li, J.; Ren, X.; Chen, C.; Wang, X. Few-layered graphene oxide nanosheets as superior sorbents for heavy metal ion pollution management. Environ. Sci. Technol. 2011, 45, 10454–10462.
\bibitem{stankovich2007synthesis} Stankovich, S.; Dikin, D. A.; Piner, R. D.; Kohlhaas, K. A.; Kleinhammes, A.; Jia, Y.; Wu, Y.; Nguyen, S. T.; Ruoff, R. S. Synthesis of graphene-based nanosheets via chemical reduction of exfoliated graphite oxide. Carbon 2007, 45, 1558–1565.
\bibitem{rao2016study} Rao, C.; Subrahmanyam, K.; Matte, H. R.; Abdulhakeem, B.; Govindaraj, A.; Das, B.; Kumar, P.; Ghosh, A.; Late, D. J. A study of the synthetic methods and properties of graphenes. Sci. Technol. Adv. Mater. 2016,
\bibitem{qin2016growth} Qin, B.; Zhang, T.; Chen, H.; Ma, Y. The growth mechanism of few-layer graphene in the arc discharge process. Carbon 2016, 102, 494–498.
\bibitem{wu2010efficient} Wu, Y.; Wang, B.; Ma, Y.; Huang, Y.; Li, N.; Zhang, F.; Chen, Y. Efficient and largescale synthesis of few-layered graphene using an arc-discharge method and conductivity studies of the resulting films. Nano Research 2010, 3, 661–669.
\bibitem{mahanandia2014electrochemical} Mahanandia, P.; Simon, F.; Heinrich, G.; Nanda, K. K. An electrochemical method for the synthesis of few layer graphene sheets for high temperature applications. Chem. Commun. 2014, 50, 4613–4615.
\bibitem{li2011flame} Li, Z.; Zhu, H.; Xie, D.; Wang, K.; Cao, A.; Wei, J.; Li, X.; Fan, L.; Wu, D. Flame synthesis of few-layered graphene/graphite films. Chem. Commun. 2011, 47, 3520–3522.
\bibitem{lobato2011multiple} Lobato, I.; Partoens, B. Multiple Dirac particles in AA-stacked graphite and multilayers of graphene. Phys. Rev. B 2011, 83, 165429.
\bibitem{chiu2010absorption} Chiu, C. W.; Lee, S. H.; Chen, S. C.; Shyu, F. L.; Lin, M. F. Absorption spectra of AA-stacked graphite. New J. Phys. 2010, 12, 083060.
\bibitem{chiu2011excitation} Chiu, C. W.; Huang, Y. C.; Shyu, F. L.; Lin, M. F. Excitation spectra of ABC-stacked graphene superlattice. Appl. Phys. Lett. 2011, 98, 261920.
\bibitem{shyu2000loss} Shyu, F. L.; Lin, M. F. Loss spectra of graphite-related systems: a multiwall carbon nanotube, a single-wall carbon nanotube bundle, and graphite layers. Phys. Rev. B 2000, 62, 8508.
\bibitem{ooi2006density} Ooi, N.; Rairkar, A.; Adams, J. B. Density functional study of graphite bulk and surface properties. Carbon 2006, 44, 231–242.
\bibitem{lin2000opticalJ} Lin, M. F.; Shyu, F. L. Optical properties of nanographite ribbons. J. Phys. Soc. Jpn 2000, 69, 3529–3532.
\bibitem{lin2000optical} Lin, M. F.; Shyu, F. L.; Chen, R. B. Optical properties of well-aligned multiwalled carbon nanotube bundles. Phys. Rev. B 2000, 61, 14114.
\bibitem{lin1994plasmons} Lin, M. F.; Shung, K. W. K. Plasmons and optical properties of carbon nanotubes. Phys. Rev. B 1994, 50, 17744.
\bibitem{lin1996collective} Lin, M. F.; Chuu, D. S.; Huang, C. S.; Lin, Y. K.; Shung, K. W. K. Collective excitations in a single-layer carbon nanotube. Phys. Rev. B 1996, 53, 15493.
\bibitem{gensterblum1991high} Gensterblum, G.; Pireaux, J.; Thiry, P.; Caudano, R.; Vigneron, J.; Lambin, P.; Lucas, A.; Krätschmer, W. High-resolution electron-energy-loss spectroscopy of thin films of C 60 on Si (100). Phys. Rev. Lett. 1991, 67, 2171.
\bibitem{bolognesi2012collective} Bolognesi, P.; Avaldi, L.; Ruocco, A.; Verkhovtsev, A.; Korol, A.; SolovâAZyov, A. Collective excitations in the electron energy loss spectra of C60. Eur. Phys. J. D 2012, 66, 1–9.
\bibitem{goto2013parity} Goto, H.; Uesugi, E.; Eguchi, R.; Kubozono, Y. Parity effects in few-layer graphene. Nano Lett. 2013, 13, 5153–5158.
\bibitem{linyp2015magneto} Lin, Y. P.; Lin, C. Y.; Ho, Y. H.; Do, T. N.; Lin, M. F. Magneto-optical properties of ABC-stacked trilayer graphene. Phys. Chem. Chem. Phys. 2015, 17, 15921–15927.
\bibitem{ho2016evolution} Ho, C. H.; Chang, C. P.; Lin, M. F. Evolution and dimensional crossover from the bulk subbands in ABC-stacked graphene to a three-dimensional Dirac cone structure in rhombohedral graphite. Phys. Rev. B 2016, 93, 075437.
\bibitem{lin2014energy} Lin, Y. P.; Wang, J.; Lu, J. M.; Lin, C. Y.; Lin, M. F. Energy spectra of ABC-stacked trilayer graphene in magnetic and electric fields. RSC Adv. 2014, 4, 56552–56560.
\bibitem{lin2014stacking} Lin, C. Y.; Wu, J. Y.; Chiu, Y. H.; Chang, C. P.; Lin, M. F. Stacking-dependent magnetoelectronic properties in multilayer graphene. Phys. Rev. B 2014, 90, 205434.
\bibitem{koshino2011landau} Koshino, M.; McCann, E. Landau level spectra and the quantum Hall effect of multilayer graphene. Phys. Rev. B 2011, 83, 165443.
\bibitem{ruffieux2012electronic} Ruffieux, P. et al. Electronic structure of atomically precise graphene nanoribbons. ACS Nano 2012, 6, 6930–6935.
\bibitem{sugawara2006fermi} Sugawara, K.; Sato, T.; Souma, S.; Takahashi, T.; Suematsu, H. Fermi surface and edge-localized states in graphite studied by high-resolution angle-resolved photoemission spectroscopy. Phys. Rev. B 2006, 73, 045124.
\bibitem{gruneis2008electron} Grüneis, A. et al. Electron-electron correlation in graphite: a combined angle-resolved photoemission and first-principles study. Phys. Rev. Lett. 2008, 100, 037601.
\bibitem{kim2013coexisting} Kim, K. S.; Walter, A. L.; Moreschini, L.; Seyller, T.; Horn, K.; Rotenberg, E.; Bostwick, A. Coexisting massive and massless Dirac fermions in symmetry-broken bilayer graphene. Nat. Mater. 2013, 12, 887–892.
\bibitem{sutter2009electronic} Sutter, P.; Hybertsen, M. S.; Sadowski, J. T.; Sutter, E. Electronic structure of fewlayer epitaxial graphene on Ru (0001). Nano Lett. 2009, 9, 2654–2660.
\bibitem{zhou2008metal} Zhou, S. Y.; Siegel, D. A.; Fedorov, A. V.; Lanzara, A. Metal to insulator transition in epitaxial graphene induced by molecular doping. Phys. Rev. Lett. 2008, 101, 086402.
\bibitem{gruneis2008tunable} Grüneis, A.; Vyalikh, D. V. Tunable hybridization between electronic states of graphene and a metal surface. Phys. Rev. B 2008, 77, 193401.
\bibitem{mazzola2013kinks} Mazzola, F. et al. Kinks in the $\sigma$ band of graphene induced by electron-phonon coupling. Phys. Rev. Lett. 2013, 111, 216806.
\bibitem{jung2016sublattice} Jung, S. W.; Shin, W. J.; Kim, J.; Moreschini, L.; Yeom, H. W.; Rotenberg, E.; Bostwick, A.; Kim, K. S. Sublattice Interference as the Origin of $\sigma$ Band Kinks in Graphene. Phys. Rev. Lett. 2016, 116, 186802.
\bibitem{huang2012spatially} Huang, H.; Wei, D.; Sun, J.; Wong, S. L.; Feng, Y. P.; Neto, A. H. C.; Wee, A. T. S. Spatially resolved electronic structures of atomically precise armchair graphene nanoribbons. Sci. Rep. 2012, 2 .
\bibitem{sode2015electronic} Söde, H.; Talirz, L.; Gröning, O.; Pignedoli, C. A.; Berger, R.; Feng, X.; Müllen, K.; Fasel, R.; Ruffieux, P. Electronic band dispersion of graphene nanoribbons via Fourier-transformed scanning tunneling spectroscopy. Phys. Rev. B 2015, 91, 045429.
\bibitem{chen2013tuning} Chen, Y. C.; De Oteyza, D. G.; Pedramrazi, Z.; Chen, C.; Fischer, F. R.; Crommie, M. F. Tuning the band gap of graphene nanoribbons synthesized from molecular precursors. ACS Nano 2013, 7, 6123–6128.
\bibitem{wilder1998electronic} Wilder, J. W. G.; Venema, L. C.; Rinzler, A. G.; Smalley, R. E.; Dekker, C. Electronic structure of atomically resolved carbon nanotubes. Nature 1998, 391, 59–62.
\bibitem{odom1998atomic} Odom, T. W.; Huang, J. L.; Kim, P.; Lieber, C. M. Atomic structure and electronic properties of single-walled carbon nanotubes. Nature 1998, 391, 62–64.
\bibitem{klusek1999investigations} Klusek, Z. Investigations of splitting of the $\pi$ bands in graphite by scanning tunneling spectroscopy. Appl. Surf. Sci. 1999, 151, 251–261.
\bibitem{li2009scanning} Li, G.; Luican, A.; Andrei, E. Y. Scanning tunneling spectroscopy of graphene on graphite. Phys. Rev. Lett. 2009, 102, 176804.
\bibitem{gyamfi2011fe} Gyamfi, M.; Eelbo, T.; Wasniowska, M.; Wiesendanger, R. Fe adatoms on graphene/Ru (0001): Adsorption site and local electronic properties. Phys. Rev. B 2011, 84, 113403.
\bibitem{alden2013strain} Alden, J. S.; Tsen, A. W.; Huang, P. Y.; Hovden, R.; Brown, L.; Park, J.; Muller, D. A.; McEuen, P. L. Strain solitons and topological defects in bilayer graphene. Proc. Natl. Acad. Sci. 2013, 110, 11256–11260.
\bibitem{guinea2008midgap} Guinea, F.; Katsnelson, M.; Vozmediano, M. Midgap states and charge inhomogeneities in corrugated graphene. Phys. Rev. B 2008, 77, 075422.
\bibitem{guinea2008gauge} Guinea, F.; Horovitz, B.; Le Doussal, P. Gauge field induced by ripples in graphene. Phys. Rev. B 2008, 77, 205421.
\bibitem{guinea2009gauge} Guinea, F.; Horovitz, B.; Le Doussal, P. Gauge fields, ripples and wrinkles in graphene layers. Solid State Commun. 2009, 149, 1140–1143.
\bibitem{isacsson2008electronic} Isacsson, A.; Jonsson, L. M.; Kinaret, J. M.; Jonson, M. Electronic superlattices in corrugated graphene. Physical Review B 2008, 77, 035423.
\bibitem{boukhvalov2009enhancement} Boukhvalov, D. W.; Katsnelson, M. I. Enhancement of chemical activity in corrugated graphene. J. Phys. Chem. C 2009, 113, 14176–14178.
\bibitem{atanasov2010tuning} Atanasov, V.; Saxena, A. Tuning the electronic properties of corrugated graphene: Confinement, curvature, and band-gap opening. Phys. Rev. B 2010, 81, 205409.
\bibitem{vozmediano2010gauge} Vozmediano, M. A.; Katsnelson, M.; Guinea, F. Gauge fields in graphene. Phys. Rep. 2010, 496, 109–148.
\bibitem{iannuzzi2011comparative} Iannuzzi, M.; Hutter, J. Comparative study of the nature of chemical bonding of corrugated graphene on Ru (0001) and Rh (111) by electronic structure calculations. Surf. Sci. 2011, 605, 1360–1368.
\bibitem{santos2012magnetism} Santos, E.; Riikonen, S.; Sánchez-Portal, D.; Ayuela, A. Magnetism of single vacancies in rippled graphene. J. Phys. Chem. C 2012, 116, 7602–7606.
\bibitem{zhang2012electron} Zhang, P.; Zhou, Y.; Wu, M. Electron spin relaxation in rippled graphene with low mobilities. J. Appl. Phys. 2012, 112, 073709.
\bibitem{mohammed2012tunnel} Mohammed, F.; Bhattacharyya, S. Tunnel transport in nitrogen-incorporated rippled graphene. Europhys. Lett. 2012, 100, 26009.
\bibitem{seyed2013computational} Seyed-Talebi, S. M.; Beheshtian, J. Computational study of ammonia adsorption on the perfect and rippled graphene sheet. Physica B: Conden. Matter 2013, 429, 52–56.
\bibitem{imam2014first} Imam, M.; Stojic, N.; Binggeli, N. First-Principles Investigation of a Rippled Graphene Phase on Ir (001): The Close Link between Periodicity, Stability, and Binding. J. Phys. Chem. C 2014, 118, 9514–9523.
\bibitem{zwierzycki2014transport} Zwierzycki, M. Transport properties of rippled graphene. J. Phys.: Conden. Matter 2014, 26, 135303.
\bibitem{koskinen2014graphene} Koskinen, P. Graphene cardboard: From ripples to tunable metamaterial. Appl. Phys. Lett. 2014, 104, 101902.
\bibitem{monteverde2015under} Monteverde, U.; Pal, J.; Migliorato, M.; Missous, M.; Bangert, U.; Zan, R.; Kashtiban, R.; Powell, D. Under pressure: Control of strain, phonons and bandgap opening in rippled graphene. Carbon 2015, 91, 266–274.
\bibitem{pudlak2015cooperative} Pudlak, M.; Pichugin, K.; Nazmitdinov, R. Cooperative phenomenon in a rippled graphene: Chiral spin guide. Phys. Rev. B 2015, 92, 205432.
\bibitem{wu2016tuning} Wu, H.; Liu, X. Tuning electromechanics of dynamic ripple pattern in graphene monolayer. Carbon 2016, 98, 510–518.
\bibitem{pereira2009tight} Pereira, V. M.; Neto, A. C.; Peres, N. Tight-binding approach to uniaxial strain in graphene. Phys. Rev. B 2009, 80, 045401.
\bibitem{choi2010effects} Choi, S. M.; Jhi, S. H.; Son, Y. W. Effects of strain on electronic properties of graphene. Phys. Rev. B 2010, 81, 081407.
\bibitem{yoosefian2015pd} Yoosefian, M.; Etminan, N. Pd-doped single-walled carbon nanotube as a nanobiosensor for histidine amino acid, a DFT study. RSC Adv. 2015, 5, 31172–31178.
\bibitem{josa2015tailoring} Josa, D.; González-Veloso, I.; Rodríguez-Otero, J.; Cabaleiro-Lago, E. M. Tailoring buckybowls for fullerene recognition. A dispersion-corrected DFT study. Phys. Chem. Chem. Phys. 2015, 17, 6233–6241.
\bibitem{capasso2014graphene} Capasso, A.; Placidi, E.; Zhan, H.; Perfetto, E.; Bell, J. M.; Gu, Y.; Motta, N. Graphene ripples generated by grain boundaries in highly ordered pyrolytic graphite. Carbon 2014, 68, 330–336.
\bibitem{deng2016wrinkled} Deng, S.; Berry, V. Wrinkled, rippled and crumpled graphene: an overview of formation mechanism, electronic properties, and applications. Mater. Today 2016, 19, 197–212.
\bibitem{lee2016multilayer} Lee, K. J.; Kim, D.; Jang, B. C.; Kim, D. J.; Park, H.; Jung, D. Y.; Hong, W.; Kim, T. K.; Choi, Y. K.; Choi, S. Y. Multilayer Graphene with a Rippled Structure as a Spacer for Improving Plasmonic Coupling. Adv. Funct. Mater. 2016, 26, 5093–5101.
\bibitem{rakic2016large} Rakic, I. S.; Capeta, D.; Plodinec, M.; Kralj, M. Large-scale transfer and characterization of macroscopic periodically nano-rippled graphene. Carbon 2016, 96, 243–249.
\bibitem{de2008periodically} De Parga, A. V.; Calleja, F.; Borca, B.; Passeggi Jr, M.; Hinarejos, J.; Guinea, F.; Miranda, R. Periodically rippled graphene: growth and spatially resolved electronic structure. Phys. Rev. Lett. 2008, 100, 056807.
\bibitem{levy2010strain} Levy, N.; Burke, S.; Meaker, K.; Panlasigui, M.; Zettl, A.; Guinea, F.; Neto, A. C.; Crommie, M. Strain-induced pseudo–magnetic fields greater than 300 tesla in graphene nanobubbles. Science 2010, 329, 544–547.
\bibitem{ramirez2016interference} Ramírez-Jíménez, R.; Álvarez-Fraga, L.; Jimenez-Villacorta, F.; Climent-Pascual, E.; Prieto, C.; de Andrés, A. Interference enhanced Raman effect in graphene bubbles. Carbon 2016, 105, 556–565.
\bibitem{savoskin2007carbon} Savoskin, M. V.; Mochalin, V. N.; Yaroshenko, A. P.; Lazareva, N. I.; Konstantinova, T. E.; Barsukov, I. V.; Prokofiev, I. G. Carbon nanoscrolls produced from acceptor-type graphite intercalation compounds. Carbon 2007, 45, 2797–2800.
\bibitem{xie2009controlled} Xie, X.; Ju, L.; Feng, X.; Sun, Y.; Zhou, R.; Liu, K.; Fan, S.; Li, Q.; Jiang, K. Controlled fabrication of high-quality carbon nanoscrolls from monolayer graphene. Nano Lett. 2009, 9, 2565–2570.
\bibitem{berman2015macroscale} Berman, D.; Deshmukh, S. A.; Sankaranarayanan, S. K.; Erdemir, A.; Sumant, A. V. Macroscale superlubricity enabled by graphene nanoscroll formation. Science 2015, 348, 1118–1122.
\bibitem{kosynkin2009longitudinal} Kosynkin, D. V.; Higginbotham, A. L.; Sinitskii, A.; Lomeda, J. R.; Dimiev, A.; Price, B. K.; Tour, J. M. Longitudinal unzipping of carbon nanotubes to form graphene nanoribbons. Nature 2009, 458, 872–876.
\bibitem{liu2012folded} Liu, F.; Song, S.; Xue, D.; Zhang, H. Folded structured graphene paper for high performance electrode materials. Adv. Mater. 2012, 24, 1089–1094.
\bibitem{yan2012high} Yan, J.; Liu, J.; Fan, Z.; Wei, T.; Zhang, L. High-performance supercapacitor electrodes based on highly corrugated graphene sheets. Carbon 2012, 50, 2179–2188.
\bibitem{chi2009adsorption} Chi, M.; Zhao, Y. P. Adsorption of formaldehyde molecule on the intrinsic and Al-doped graphene: a first principle study. Comput. Mater. Sci. 2009, 46, 1085–1090.
\bibitem{tozzini2011reversible} Tozzini, V.; Pellegrini, V. Reversible hydrogen storage by controlled buckling of graphene layers. J. Phys. Chem. C 2011, 115, 25523–25528.
\bibitem{ning2012high} Ning, G.; Xu, C.; Mu, L.; Chen, G.; Wang, G.; Gao, J.; Fan, Z.; Qian, W.; Wei, F. High capacity gas storage in corrugated porous graphene with a specific surface area-lossless tightly stacking manner. Chem. Commun. 2012, 48, 6815–6817.
\bibitem{tozzini2013prospects} Tozzini, V.; Pellegrini, V. Prospects for hydrogen storage in graphene. Phys. Chem. Chem. Phys. 2013, 15, 80–89.
\bibitem{saito1992electronic} Saito, R.; Fujita, M.; Dresselhaus, G.; Dresselhaus, u. M. Electronic structure of chiral graphene tubules. Appl. Phys. Lett. 1992, 60, 2204–2206.
\bibitem{saito1992electronicprb} Saito, R.; Fujita, M.; Dresselhaus, G.; Dresselhaus, M. S. Electronic structure of graphene tubules based on C60. Phys. Rev. B 1992, 46, 1804.
\bibitem{bhattacharyya2016lifshitz} Bhattacharyya, S.; Singh, A. K. Lifshitz transition and modulation of electronic and transport properties of bilayer graphene by sliding and applied normal compressive strain. Carbon 2016, 99, 432–438.
\bibitem{lee2015extreme} Lee, K. W.; Lee, C. E. Extreme sensitivity of the electric-field-induced band gap to the electronic topological transition in sliding bilayer graphene. Sci. Rep. 2015, 5, 17490.
\bibitem{chang2014geometric} Chang, S. L.; Lin, S. Y.; Lin, S. K.; Lee, C. H.; Lin, M. F. Geometric and electronic properties of edge-decorated graphene nanoribbons. Sci. Rep. 2014, 4 .
\bibitem{liu2002properties} Liu, H.; Chan, C. Properties of 4 Å carbon nanotubes from first-principles calculations. Phys. Rev. B 2002, 66, 115416.
\bibitem{vcervenka2009room} Cervenka, J.; Katsnelson, M.; Flipse, C. Room-temperature ferromagnetism in graphite driven by two-dimensional networks of point defects. Nat. Phys. 2009, 5, 840–844.
\bibitem{kondo2012atomic} Kondo, T. e. a. Atomic-scale characterization of nitrogen-doped graphite: Effects of dopant nitrogen on the local electronic structure of the surrounding carbon atoms. Phys. Rev. B 2012, 86, 035436.
\bibitem{tao2011spatially} Tao, C. et al. Spatially resolving edge states of chiral graphene nanoribbons. Nat. Phys. 2011, 7, 616–620.
\bibitem{nakada2011migration} Nakada, K.; Ishii, A. Migration of adatom adsorption on graphene using DFT calculation. Solid State Commun. 2011, 151, 13–16.
\bibitem{brar2011gate} Brar, V. W. et al. Gate-controlled ionization and screening of cobalt adatoms on a graphene surface. Nat. Phys. 2011, 7, 43–47.
\bibitem{zanella2008electronic} Zanella, I.; Fagan, S. B.; Mota, R.; Fazzio, A. Electronic and magnetic properties of Ti and Fe on graphene. J. Phys. Chem. C 2008, 112, 9163–9167.
\bibitem{gao2010first} Gao, H.; Zhou, J.; Lu, M.; Fa, W.; Chen, Y. First-principles study of the IVA group atoms adsorption on graphene. J. Appl. Phys. 2010, 107, 114311.
\bibitem{dai2010adsorption} Dai, J.; Yuan, J. Adsorption of molecular oxygen on doped graphene: Atomic, electronic, and magnetic properties. Phys. Rev. B 2010, 81, 165414.
\bibitem{fan2013nitrogen} Fan, W.; Xia, Y.-Y.; Tjiu, W. W.; Pallathadka, P. K.; He, C.; Liu, T. Nitrogendoped graphene hollow nanospheres as novel electrode materials for supercapacitor applications. J. Power Sources 2013, 243, 973–981.
\bibitem{karthika2013phosphorus} Karthika, P.; Rajalakshmi, N.; Dhathathreyan, K. Phosphorus-doped exfoliated graphene for supercapacitor electrodes. J. Nanosci. Nanotechnol 2013, 13, 1746–1751.
\bibitem{xue2015multiscale} Xue Y H, C. H. Q. J. D. L. M., Zhu L Multiscale patterning of graphene oxide and reduced graphene oxide for flexible supercapacitors. Carbon 2015, 92, 305–310.
\bibitem{chen2011high} Chen, Y.; Zhang, X.; Zhang, D.; Yu, P.; Ma, Y. High performance supercapacitors based on reduced graphene oxide in aqueous and ionic liquid electrolytes. Carbon 2011, 49, 573–580.
\bibitem{gao2011direct} Gao, W.; Singh, N.; Song, L.; Liu, Z.; Reddy, A. L. M.; Ci, L.; Vajtai, R.; Zhang, Q.; Wei, B.; Ajayan, P. M. Direct laser writing of micro-supercapacitors on hydrated graphite oxide films. Nat. Nanotechnol. 2011, 6, 496–500.
\bibitem{loh2010graphene} Loh, K. P.; Bao, Q.; Eda, G.; Chhowalla, M. Graphene oxide as a chemically tunable platform for optical applications. Nat. Chem. 2010, 2, 1015–1024.
\bibitem{wu2009organic} Wu, J.; Agrawal, M.; Becerril, H. A.; Bao, Z.; Liu, Z.; Chen, Y.; Peumans, P. Organic light-emitting diodes on solution-processed graphene transparent electrodes. ACS Nano 2009, 4, 43–48.
\bibitem{su2009composites} Su, Q.; Pang, S.; Alijani, V.; Li, C.; Feng, X.; Müllen, K. Composites of graphene with large aromatic molecules. Adv. Mater. 2009, 21, 3191–3195.
\bibitem{porro2015memristive} Porro, S.; Accornero, E.; Pirri, C. F.; Ricciardi, C. Memristive devices based on graphene oxide. Carbon 2015, 85, 383–396.
\bibitem{zhang2013synthesis} Zhang, C.; Mahmood, N.; Yin, H.; Liu, F.; Hou, Y. Synthesis of phosphorus-doped graphene and its multifunctional applications for oxygen reduction reaction and lithium ion batteries. Adv. Mater. 2013, 25, 4932–4937.
\bibitem{wang2013situ} Wang, Z. L.; Xu, D.; Wang, H. G.; Wu, Z.; Zhang, X. B. In situ fabrication of porous graphene electrodes for high-performance energy storage. ACS Nano 2013, 7, 2422–2430.
\bibitem{wang2010nitrogen} Wang, Y.; Shao, Y.; Matson, D. W.; Li, J.; Lin, Y. Nitrogen-doped graphene and its application in electrochemical biosensing. ACS Nano 2010, 4, 1790–1798.
\bibitem{sheng2012electrochemical} Sheng, Z. H.; Zheng, X. Q.; Xu, J. Y.; Bao, W. J.; Wang, F. B.; Xia, X. H. Electrochemical sensor based on nitrogen doped graphene: simultaneous determination of ascorbic acid, dopamine and uric acid. Biosens. Bioelectron 2012, 34, 125–131.
\bibitem{veerapandian2012synthesis} Veerapandian, M.; Lee, M. H.; Krishnamoorthy, K.; Yun, K. Synthesis, characterization and electrochemical properties of functionalized graphene oxide. Carbon 2012, 50, 4228–4238.
\bibitem{robinson2008reduced} Robinson, J. T.; Perkins, F. K.; Snow, E. S.; Wei, Z.; Sheehan, P. E. Reduced graphene oxide molecular sensors. Nano Lett. 2008, 8, 3137–3140.
\bibitem{nourbakhsh2010bandgap} Nourbakhsh, A.; Cantoro, M.; Vosch, T.; Pourtois, G.; Clemente, F.; van der Veen, M. H.; Hofkens, J.; Heyns, M. M.; De Gendt, S.; Sels, B. F. Bandgap opening in oxygen plasma-treated graphene. Nanotechnology 2010, 21, 435203.
\bibitem{azadeh2011tunable} Azadeh, M. S.; Kokabi, A.; Hosseini, M.; Fardmanesh, M. Tunable bandgap opening in the proposed structure of silicon-doped graphene. Micro Nano Lett. 2011, 6, 582–585.
\bibitem{zhang2016opening} Zhang, S.; Lin, S.; Li, X.; Liu, X.; Wu, H.; Xu, W.; Wang, P.; Wu, Z.; Zhong, H.; Xu, Z. Opening the band gap of graphene through silicon doping for the improved performance of graphene/GaAs heterojunction solar cells. Nanoscale 2016, 8, 226–232.
\bibitem{denis2010band} Denis, P. A. Band gap opening of monolayer and bilayer graphene doped with aluminium, silicon, phosphorus, and sulfur. Chem.l Phys. Lett. 2010, 492, 251–257.
\bibitem{denis2013concentration} Denis, P. A. Concentration dependence of the band gaps of phosphorus and sulfur doped graphene. Comp. Mater. Sci. 2013, 67, 203–206.
\bibitem{shinde2011direct} Shinde, P. P.; Kumar, V. Direct band gap opening in graphene by BN doping: Ab-initio calculations. Phys. Rev. B 2011, 84, 125401.
\bibitem{denis2011chemical} Denis, P. A. Chemical reactivity of lithium doped monolayer and bilayer graphene. J. Phys. Chem. C 2011, 115, 13392–13398.
\bibitem{xu2015electronic} Xu, C.; Brown, P. A.; Lu, J.; Shuford, K. L. Electronic Properties of Halogen-Adsorbed Graphene. J. Phys. Chem. C 2015, 119, 17271–17277.
\bibitem{chu2012charge} Chu, S. W.; Baek, S. J.; Kim, D. C.; Seo, S.; Kim, J. S.; Park, Y. W. Charge transport in graphene doped with diatomic halogen molecules ($I_2$, $Br_2$) near Dirac point. Synthetic Met. 2012, 162, 1689–1693.
\bibitem{jin2014high} Jin, E.; Kornblum, L.; Kumah, D.; Zou, K.; Broadbridge, C.; Ngai, J.; Ahn, C.; Walker, F. A high density two-dimensional electron gas in an oxide heterostructure on Si (001). APL Mater. 2014, 2, 116109.
\bibitem{xu2016quasi} Xu, P.; Droubay, T. C.; Jeong, J. S.; Mkhoyan, K. A.; Sushko, P. V.; Chambers, S. A.; Jalan, B. Quasi 2D Ultrahigh Carrier Density in a Complex Oxide Broken-Gap Heterojunction. Adv. Mater. Inter. 2016, 3 .
\bibitem{gomez2007electronic} Gómez-Navarro, C.; Weitz, R. T.; Bittner, A. M.; Scolari, M.; Mews, A.; Burghard, M.; Kern, K. Electronic transport properties of individual chemically reduced graphene oxide sheets. Nano Lett. 2007, 7, 3499–3503.
\bibitem{tung2009high} Tung, V. C.; Allen, M. J.; Yang, Y.; Kaner, R. B. High-throughput solution processing of large-scale graphene. Nat. Nanotechnol. 2009, 4, 25–29.
\bibitem{eda2008large} Eda, G.; Fanchini, G.; Chhowalla, M. Large-area ultrathin films of reduced graphene oxide as a transparent and flexible electronic material. Nat. Nanotechnol. 2008, 3, 270–274.
\bibitem{schafhaeutl1840lxxxvi} Schafhaeutl, C. LXXXVI. On the combinations of carbon with silicon and iron, and other metals, forming the different species of cast iron, steel, and malleable iron. Phil. Mag. 1840, 16, 570–590.
\bibitem{li2015graphene} Li, F.; Jiang, X.; Zhao, J.; Zhang, S. Graphene oxide: A promising nanomaterial for energy and environmental applications. Nano Energy 2015, 16, 488–515.
\bibitem{poh2012graphenes} Poh, H. L.; Šanek, F.; Ambrosi, A.; Zhao, G.; Sofer, Z.; Pumera, M. Graphenes prepared by Staudenmaier, Hofmann and Hummers methods with consequent thermal exfoliation exhibit very different electrochemical properties. Nanoscale 2012, 4, 3515–3522.
\bibitem{bai2011functional} Bai, H.; Li, C.; Shi, G. Functional composite materials based on chemically converted graphene. Adv. Mater. 2011, 23, 1089–1115.
\bibitem{marcano2010improved} Marcano, D. C.; Kosynkin, D. V.; Berlin, J. M.; Sinitskii, A.; Sun, Z.; Slesarev, A.; Alemany, L. B.; Lu, W.; Tour, J. M. Improved synthesis of graphene oxide. ACS Nano 2010, 4, 4806–4814.
\bibitem{C5RA02099A} Wu, R.; Wang, Y.; Chen, L.; Huang, L.; Chen, Y. Control of the oxidation level of graphene oxide for high efficiency polymer solar cells. RSC Adv. 2015, 5, 49182–49187.
\bibitem{saxena2011investigation} Saxena, S.; Tyson, T. A.; Shukla, S.; Negusse, E.; Chen, H.; Bai, J. Investigation of structural and electronic properties of graphene oxide. Appl. Phys. Lett. 2011, 99, 013104.
\bibitem{erickson2010determination} Erickson, K.; Erni, R.; Lee, Z.; Alem, N.; Gannett, W.; Zettl, A. Determination of the local chemical structure of graphene oxide and reduced graphene oxide. Adv. Mater. 2010, 22, 4467–4472.
\bibitem{huang2012oxygen} Huang, H.; Li, Z.; She, J.; Wang, W. Oxygen density dependent band gap of reduced graphene oxide. J. Appl. Phys. 2012, 111, 054317.
\bibitem{cai2008synthesis} Cai, W. et al. Synthesis and solid-state NMR structural characterization of 13C-labeled graphite oxide. Science 2008, 321, 1815–1817.
\bibitem{hontoria1995study} Hontoria-Lucas, C.; Lopez-Peinado, A.; López-González, J. d. D.; Rojas-Cervantes, M.; Martin-Aranda, R. Study of oxygen-containing groups in a series of graphite oxides: physical and chemical characterization. Carbon 1995, 33, 1585–1592.
\bibitem{gao2009new} Gao, W.; Alemany, L. B.; Ci, L.; Ajayan, P. M. New insights into the structure and reduction of graphite oxide. Nat. Chem. 2009, 1, 403–408.
\bibitem{lerf1998structure} Lerf, A.; He, H.; Forster, M.; Klinowski, J. Structure of graphite oxide revisited. J. Phys. Chem. B 1998, 102, 4477–4482.
\bibitem{casabianca2010nmr} Casabianca, L. B.; Shaibat, M. A.; Cai, W. W.; Park, S.; Piner, R.; Ruoff, R. S.; Ishii, Y. NMR-based structural modeling of graphite oxide using multi-dimensional 13C solid-state NMR and ab initio chemical shift calculations. J. Am. Chem. Soc. 2010, 132, 5672–5676.
\bibitem{du2004novel} Du, X.; Xiao, M.; Meng, Y.; Hay, A. Novel synthesis of conductive poly (arylene disulfide)/graphite nanocomposite. Synth. Metals 2004, 143, 129–132.
\bibitem{szabo2005composite} Szabó, T.; Szeri, A.; Dékány, I. Composite graphitic nanolayers prepared by self-assembly between finely dispersed graphite oxide and a cationic polymer. Carbon 2005, 43, 87–94.
\bibitem{stankovich2006graphene} Stankovich, S.; Dikin, D. A.; Dommett, G. H.; Kohlhaas, K. M.; Zimney, E. J.; Stach, E. A.; Piner, R. D.; Nguyen, S. T.; Ruoff, R. S. Graphene-based composite materials. Nature 2006, 442, 282–286.
\bibitem{neto2009electronic} Neto, A. C.; Guinea, F.; Peres, N.; Novoselov, K. S.; Geim, A. K. The electronic properties of graphene. Rev. Mod. Phys. 2009, 81, 109.
\bibitem{boukhvalov2008modeling} Boukhvalov, D. W.; Katsnelson, M. I. Modeling of graphite oxide. J. Am. Chem. Soc. 2008, 130, 10697–10701.
\bibitem{ohta2008morphology} Ohta, T.; El Gabaly, F.; Bostwick, A.; McChesney, J. L.; Emtsev, K. V.; Schmid, A. K.; Seyller, T.; Horn, K.; Rotenberg, E. Morphology of graphene thin film growth on SiC (0001). New J. Phys. 2008, 10, 023034.
\bibitem{liang2015band} Liang, H.; Smith, C.; Mills, C.; Silva, S. The band structure of graphene oxide examined using photoluminescence spectroscopy. J. Mater. Chem. C 2015, 3, 12484–12491.
\bibitem{xu2012direct} Xu, K. et al. Direct measurement of Dirac point energy at the graphene/oxide interface. Nano Lett. 2012, 13, 131–136.
\bibitem{becerril2008evaluation} Becerril, H. A.; Mao, J.; Liu, Z.; Stoltenberg, R. M.;Bao, Z.; Chen, Y. Evaluation of solution-processed reduced graphene oxide films as transparent conductors. ACS Nano 2008, 2, 463–470.
\bibitem{mohan2015characterisation} Mohan, V. B.; Brown, R.; Jayaraman, K.; Bhattacharyya, D. Characterisation of reduced graphene oxide: Effects of reduction variables on electrical conductivity. Mater. Sci. Eng. B 2015, 193, 49–60.
\bibitem{mattevi2009evolution} Mattevi, C.; Eda, G.; Agnoli, S.; Miller, S.; Mkhoyan, K. A.; Celik, O.; Mastrogiovanni, D.; Granozzi, G.; Garfunkel, E.; Chhowalla, M. Evolution of electrical, chemical, and structural properties of transparent and conducting chemically derived graphene thin films. Adv. Funct. Mater. 2009, 19, 2577–2583.
\bibitem{jung2008tunable} Jung, I.; Dikin, D. A.; Piner, R. D.; Ruoff, R. S. Tunable electrical conductivity of individual graphene oxide sheets reduced at low temperatures. Nano Lett. 2008, 8, 4283–4287.
\bibitem{trung2014flexible} Trung, T. Q.; Tien, N. T.; Kim, D.; Jang, M.; Yoon, O. J.; Lee, N.-E. A Flexible Reduced Graphene Oxide Field-Effect Transistor for Ultrasensitive Strain Sensing. Adv. Funct. Mater. 2014, 24, 117–124.
\bibitem{truong2014reduced} Truong, T. K.; Nguyen, T.; Trung, T. Q.; Sohn, I. Y.; Kim, D. J.; Jung, J. H.; Lee, N. E. Reduced graphene oxide field-effect transistor with indium tin oxide extended gate for proton sensing. Curr. Appl. Phys. 2014, 14, 738–743.
\bibitem{joung2010high} Joung, D.; Chunder, A.; Zhai, L.; Khondaker, S. I. High yield fabrication of chemically reduced graphene oxide field effect transistors by dielectrophoresis. Nanotech. 2010, 21, 165202.
\bibitem{he2010centimeter} He, Q.; Sudibya, H. G.; Yin, Z.; Wu, S.; Li, H.; Boey, F.; Huang, W.; Chen, P.; Zhang, H. Centimeter-long and large-scale micropatterns of reduced graphene oxide films: fabrication and sensing applications. ACS Nano 2010, 4, 3201–3208.
\bibitem{eda2010chemically} Eda, G.; Chhowalla, M. Chemically derived graphene oxide: towards large-area thinfilm electronics and optoelectronics. Adv. Mater. 2010, 22, 2392–2415.
\bibitem{basu2012recent} Basu, S.; Bhattacharyya, P. Recent developments on graphene and graphene oxide based solid state gas sensors. Sensor. Actuat. B Chem. 2012, 173, 1–21.
\bibitem{su2014electrical} Su, P. G.; Chiou, C. F. Electrical and humidity-sensing properties of reduced graphene oxide thin film fabricated by layer-by-layer with covalent anchoring on flexible substrate. Sensor. Actuat. B Chem. 2014, 200, 9–18.
\bibitem{zheng2014graphene} Zheng, Q.; Li, Z.; Yang, J.; Kim, J. K. Graphene oxide-based transparent conductive films. Prog. in Mater. Sci. 2014, 64, 200–247.
\bibitem{nekahi2014transparent} Nekahi, A.; Marashi, P.; Haghshenas, D. Transparent conductive thin film of ultra large reduced graphene oxide monolayers. Appl. Surf. Sci. 2014, 295, 59–65.
\bibitem{murray2011graphene} Murray, I. P. et al. Graphene oxide interlayers for robust, high-efficiency organic photovoltaics. J. Phys. Chem. Lett. 2011, 2, 3006–3012.
\bibitem{yin2010organic} Yin, Z.; Sun, S.; Salim, T.; Wu, S.; Huang, X.; He, Q.; Lam, Y. M.; Zhang, H. Organic photovoltaic devices using highly flexible reduced graphene oxide films as transparent electrodes. ACS Nano 2010, 4, 5263–5268.
\bibitem{saha2014solution} Saha, S. K.; Bhaumik, S.; Maji, T.; Mandal, T. K.; Pal, A. J. Solution-processed reduced graphene oxide in light-emitting diodes and photovoltaic devices with the same pair of active materials. RSC Adv. 2014, 4, 35493–35499.
\bibitem{lu2012novel} Lu, Y.; Jiang, Y.; Wei, W.; Wu, H.; Liu, M.; Niu, L.; Chen, W. Novel blue light emitting graphene oxide nanosheets fabricated by surface functionalization. J. Mater. Chem. 2012, 22, 2929–2934.
\bibitem{gharekhanlou2010bipolar} Gharekhanlou, B.; Tousaki, S.; Khorasani, S. Bipolar transistor based on graphane. 2010; p 012061.
\bibitem{lu2016ferromagnetism} Lu, H. Y.; Hao, L.; Wang, R.; Ting, C. Ferromagnetism and superconductivity with possible p + ip pairing symmetry in partially hydrogenated graphene. Phys. Rev. B 2016, 93, 241410 .
\bibitem{zhou2014graphene} Zhou, C.; Chen, S.; Lou, J.; Wang, J.; Yang, Q.; Liu, C.; Huang, D.; Zhu, T. Graphene’s cousin: the present and future of graphane. Nanoscale Res. Lett. 2014, 9, 1–9.
\bibitem{hussain2014enriching} Hussain, T.; Panigrahi, P.; Ahuja, R. Enriching physisorption of H$_2$S and NH$_3$ gases on a graphane sheet by doping with Li adatoms. Phys. Chem. Chem. Phys. 2014, 16, 8100–8105.
\bibitem{nechaev2011solid} Nechaev, Y. S. On the solid hydrogen carrier intercalation in graphane-like regions in carbon-based nanostructures. Int. J. Hydrog. Energy 2011, 36, 9023–9031.
\bibitem{zhou2009ferromagnetism} Zhou, J.; Wang, Q.; Sun, Q.; Chen, X.; Kawazoe, Y.; Jena, P. Ferromagnetism in semihydrogenated graphene sheet. Nano Lett. 2009, 9, 3867–3870.
\bibitem{boukhvalov2010stable} Boukhvalov, D. Stable antiferromagnetic graphone. Physica E Low Dimens Syst Nanostruct. 2010, 43, 199–201.
\bibitem{usachov2015observation} Usachov, D. et al. Observation of Single-Spin Dirac Fermions at the Graphene/Ferromagnet Interface. Nano Lett. 2015, 15, 2396–2401.
\bibitem{gierz2010giant} Gierz, I.; Dil, J. H.; Meier, F.; Slomski, B.; Osterwalder, J.; Henk, J.; Winkler, R.; Ast, C. R.; Kern, K. Giant anisotropic spin splitting in epitaxial graphene. arXiv preprint arXiv:1004.1573 2010,
\bibitem{ellis2013magneto} Ellis, C. T.; Stier, A. V.; Kim, M. H.; Tischler, J. G.; Glaser, E. R.; MyersWard, R. L.; Tedesco, J. L.; Eddy, C. R.; Gaskill, D. K.; Cerne, J. Magneto-optical fingerprints of distinct graphene multilayers using the giant infrared Kerr effect. Sci. Rep. 2013, 3, 3143 .
\bibitem{chen2016time} Chen, J. Y.; Zhu, J.; Zhang, D.; Lattery, D. M.; Li, M.; Wang, J. P.; Wang, X. Time-Resolved Magneto-Optical Kerr Effect of Magnetic Thin Films for Ultrafast Thermal Characterization. J. Phys. Chem. Lett. 2016, 7, 2328âAS2332.
\bibitem{friedman2015hydrogenated} Friedman, A. L.; vanâAZt Erve, O. M.; Robinson, J. T.; Whitener Jr, K. E.; Jonker, B. T. Hydrogenated graphene as a homoepitaxial tunnel barrier for spin and charge transport in graphene. ACS Nano 2015, 9, 6747–6755.
\bibitem{han2014graphene} Han, W.; Kawakami, R. K.; Gmitra, M.; Fabian, J. Graphene spintronics. Nat. Nanotechnol. 2014, 9, 794–807.
\bibitem{son2006energy} Son, Y. W.; Cohen, M. L.; Louie, S. G. Energy gaps in graphene nanoribbons. Phys. Rev. Lett. 2006, 97, 216803.
\bibitem{chang2014configuration} Chang, S. L.; Wu, B. R.; Wong, J. H.; Lin, M. F. Configuration-dependent geometric and electronic properties of bilayer graphene nanoribbons. Carbon 2014, 77, 1031–1039.
\bibitem{serrate2010imaging} Serrate, D.; Ferriani, P.; Yoshida, Y.; Hla, S.-W.; Menzel, M.; Von Bergmann, K.; Heinze, S.; Kubetzka, A.; Wiesendanger, R. Imaging and manipulating the spin direction of individual atoms. Nat. Nanotechnol. 2010, 5, 350–353.
\bibitem{wulfhekel2007spin} Wulfhekel, W.; Kirschner, J. Spin-polarized scanning tunneling microscopy of magnetic structures and antiferromagnetic thin films. Mater. Res. 2007, 37, 69.
\bibitem{corbetta2012magnetic} Corbetta, M.; Ouazi, S.; Borme, J.; Nahas, Y.; Donati, F.; Oka, H.; Wedekind, S.; Sander, D.; Kirschner, J. Magnetic response and spin polarization of bulk Cr tips for in-field spin-polarized scanning tunneling microscopy. Jpn. J. Appl. Phys. 2012, 51, 030208.
\bibitem{berbil2007spin} Berbil-Bautista, L.; Krause, S.; Bode, M.; Wiesendanger, R. Spin-polarized scanning tunneling microscopy and spectroscopy of ferromagnetic Dy (0001)/W (110) films. Phys. Rev. B 2007, 76, 064411.
\bibitem{wen2011graphane} Wen, X. D.; Hand, L.; Labet, V.; Yang, T.; Hoffmann, R.; Ashcroft, N.; Oganov, A. R.; Lyakhov, A. O. Graphane sheets and crystals under pressure. Proc. Natl. Acad. Sci. 2011, 108, 6833–6837.
\bibitem{samarakoon2009chair} Samarakoon, D. K.; Wang, X. Q. Chair and twist-boat membranes in hydrogenated graphene. ACS Nano 2009, 3, 4017–4022.
\bibitem{yi2015stability} Yi, D.; Yang, L.; Xie, S.; Saxena, A. Stability of hydrogenated graphene: a first-principles study. RSC Adv. 2015, 5, 20617–20622.
\bibitem{eng2013highly} Eng, A. Y. S.; Sofer, Z.; Simek, P.; Kosina, J.; Pumera, M. Highly hydrogenated graphene through microwave exfoliation of graphite oxide in hydrogen plasma: towards electrochemical applications. Chem. Eur. J. 2013, 19, 15583–15592.
\bibitem{luo2009thickness} Luo, Z.; Yu, T.; Kim, K.-j.; Ni, Z.; You, Y.; Lim, S.; Shen, Z.; Wang, S.; Lin, J. Thickness-dependent reversible hydrogenation of graphene layers. ACS Nano 2009, 3, 1781–1788.
\bibitem{wojtaszek2011road} Wojtaszek, M.; Tombros, N.; Caretta, A.; Van Loosdrecht, P.; Van Wees, B. A road to hydrogenating graphene by a reactive ion etching plasma. J. Appl. Phys. 2011, 110, 063715.
\bibitem{byun2011nanoscale} Byun, I. S. et al. Nanoscale lithography on monolayer graphene using hydrogenation and oxidation. ACS Nano 2011, 5, 6417–6424.
\bibitem{pumera2013graphane} Pumera, M.; Wong, C. H. A. Graphane and hydrogenated graphene. Chem. Soc. Rev. 2013, 42, 5987–5995.
\bibitem{zhou2012hydrogenated} Zhou, Y.; Wang, Z.; Yang, P.; Sun, X.; Zu, X.; Gao, F. Hydrogenated graphene nanoflakes: semiconductor to half-metal transition and remarkable large magnetism. J. Phys. Chem. C 2012, 116, 5531–5537.
\bibitem{seah2014towards} Seah, T. H.; Poh, H. L.; Chua, C. K.; Sofer, Z.; Pumera, M. Towards graphane applications in security: the electrochemical detection of trinitrotoluene in seawater on hydrogenated graphene. Electroanalysis 2014, 26, 62–68.
\bibitem{robinson2010properties} Robinson, J. T. et al. Properties of fluorinated graphene films. Nano Lett. 2010, 10, 3001–3005.
\bibitem{zhang2013two} Zhang, M.; Ma, Y.; Zhu, Y.; Che, J.; Xiao, Y. Two-dimensional transparent hydrophobic coating based on liquid-phase exfoliated graphene fluoride. Carbon 2013, 63, 149–156.
\bibitem{romero2013fluorinated} Romero-Aburto, R. e. a. Fluorinated graphene oxide: a new multimodal material for biological applications. Adv. Mater. 2013, 25, 5632–5637.
\bibitem{dubois2014thermal} Dubois, M. et al. Thermal exfoliation of fluorinated graphite. Carbon 2014, 77, 688–704.
\bibitem{zbovril2010graphene} Zboril, R.; Karlicky, F.; Bourlinos, A. B.; Steriotis, T. A.; Stubos, A. K.; Georgakilas, V.; Safárová, K.; Jancík, D.; Trapalis, C.; Otyepka, M. Graphene fluoride: a stable stoichiometric graphene derivative and its chemical conversion to graphene. Small 2010, 6, 2885–2891.
\bibitem{klintenberg2010theoretical} Klintenberg, M.; Lebegue, S.; Katsnelson, M.; Eriksson, O. Theoretical analysis of the chemical bonding and electronic structure of graphene interacting with Group IA and Group VIIA elements. Phys. Rev. B 2010, 81, 085433.
\bibitem{subrahmanyam2009comparative} Subrahmanyam, K.; Voggu, R.; Govindaraj, A.; Rao, C. A comparative Raman study of the interaction of electron donor and acceptor molecules with graphene prepared by different methods. Chem. Phys. Lett. 2009, 472, 96–98.
\bibitem{leenaerts2010first} Leenaerts, O.; Peelaers, H.; Hernández-Nieves, A.; Partoens, B.; Peeters, F. First-principles investigation of graphene fluoride and graphane. Phys. Rev. B 2010, 82, 195436.
\bibitem{chang2011facile} Chang, H.; Cheng, J.; Liu, X.; Gao, J.; Li, M.; Li, J.; Tao, X.; Ding, F.; Zheng, Z. Facile Synthesis of Wide-Bandgap Fluorinated Graphene Semiconductors. Chem. Eur. J. 2011, 17, 8896–8903.
\bibitem{yu2012increased} Yu, X.; Lin, K.; Qiu, K.; Cai, H.; Li, X.; Liu, J.; Pan, N.; Fu, S.; Luo, Y.; Wang, X. Increased chemical enhancement of Raman spectra for molecules adsorbed on fluorinated reduced graphene oxide. Carbon 2012, 50, 4512–4517.
\bibitem{gong2012one} Gong, P.; Wang, Z.; Wang, J.; Wang, H.; Li, Z.; Fan, Z.; Xu, Y.; Han, X.; Yang, S. One-pot sonochemical preparation of fluorographene and selective tuning of its fluorine coverage. J. Mater. Chem. 2012, 22, 16950–16956.
\bibitem{zheng2012production} Zheng, J.; Liu, H. T.; Wu, B.; Di, C.-A.; Guo, Y. L.; Wu, T.; Yu, G.; Liu, Y.-Q.; Zhu, D. B. Production of graphite chloride and bromide using microwave sparks. Sci. Rep. 2012, 2 .
\bibitem{jeon2015edge} Jeon, I. Y. et al. Edge-Fluorinated Graphene Nanoplatelets as High Performance Electrodes for Dye-Sensitized Solar Cells and Lithium Ion Batteries. Adv. Funct. Mater. 2015, 25, 1170–1179.
\bibitem{vizintin2015fluorinated} Vizintin, A.; LozinsÌNek, M.; Chellappan, R. K.; Foix, D.; Krajnc, A.; Mali, G.; Drazic, G.; Genorio, B.; Dedryvère, R.; Dominko, R. Fluorinated Reduced Graphene Oxide as an Interlayer in Li–S Batteries. Chem. Mater. 2015, 27, 7070–7081.
\bibitem{zhang2013impact} Zhang, X.; Hsu, A.; Wang, H.; Song, Y.; Kong, J.; Dresselhaus, M. S.; Palacios, T. Impact of chlorine functionalization on high-mobility chemical vapor deposition grown graphene. ACS Nano 2013, 7, 7262–7270.
\bibitem{mansour2015bromination} Mansour, A. E.; Dey, S.; Amassian, A.; Tanielian, M. H. Bromination of Graphene: A New Route to Making High Performance Transparent Conducting Electrodes with Low Optical Losses. ACS Appl. Mater. Inter. 2015, 7, 17692–17699.
\bibitem{schafhaeutl1840ueber} Schafhaeutl, C. Ueber die Verbindungen des Kohlenstoffes mit Silicium, Eisen und anderen Metallen, welche die verschiedenen Gallungen von Roheisen, Stahl und Schmiedeeisen bilden. Journal für Praktische Chemie 1840, 21, 129–157.
\bibitem{hannay1965superconductivity} Hannay, N.; Geballe, T.; Matthias, B.; Andres, K.; Schmidt, P.; MacNair, D. Superconductivity in graphitic compounds. Phys. Rev. Lett. 1965, 14, 225.
\bibitem{magerl1985plane} Magerl, A.; Zabel, H.; Anderson, I. In-plane jump diffusion of Li in LiC6. Phys. Rev. Lett. 1985, 55, 222.
\bibitem{belash1987superconductivity} Belash, I.; Bronnikov, A.; Zharikov, O.; Palnichenko, A. On the superconductivity of graphite intercalation compounds with sodium. Solid State Commun. 1987, 64, 1445–1447.
\bibitem{koike1980superconductivity} Koike, Y.; Suematsu, H.; Higuchi, K.; Tanuma, S.-i. Superconductivity in graphite-alkali metal intercalation compounds. Physica B+ C 1980, 99, 503–508.
\bibitem{belash1989superconductivity} Belash, I.; Bronnikov, A.; Zharikov, O.; Pal’nichenko, A. Superconductivity of graphite intercalation compound with lithium C2Li. Solid State Commun. 1989, 69, 921–923.
\bibitem{ruzicka2000optical} Ruzicka, B.; Degiorgi, L.; Gaal, R.; Thien-Nga, L.; Bacsa, R.; Salvetat, J.-P.; Forro, L. Optical and dc conductivity study of potassium-doped single-walled carbon nanotube films. Phys. Rev. B 2000, 61, R2468.
\bibitem{suzuki2000work} Suzuki, S.; Bower, C.; Watanabe, Y.; Zhou, O. Work functions and valence band states of pristine and Cs-intercalated single-walled carbon nanotube bundles. Appl. Phys. Lett. 2000, 76, 4007–4009.
\bibitem{sauvajol2003phonons} Sauvajol, J. L.; Bendiab, N.; Anglaret, E.; Petit, P. Phonons in alkali-doped single-wall carbon nanotube bundles. C R Phys 2003, 4, 1035–1045.
\bibitem{guan2005direct} Guan, L.; Suenaga, K.; Shi, Z.; Gu, Z.; Iijima, S. Direct imaging of the alkali metal site in K-doped fullerene peapods. Phys. Rev. Lett. 2005, 94, 045502.
\bibitem{ganin2008bulk} Ganin, A. Y.; Takabayashi, Y.; Khimyak, Y. Z.; Margadonna, S.; Tamai, A.; Rosseinsky, M. J.; Prassides, K. Bulk superconductivity at 38 K in a molecular system. Nat. Mater. 2008, 7, 367–371.
\bibitem{takeya2016superconductivity} Takeya, H.; Konno, T.; Hirata, C.; Wakahara, T.; Miyazawa, K.; Yamaguchi, T.; Tanaka, M.; Takano, Y. Superconductivity in alkali-doped fullerene nanowhiskers. J. Phys. Condens. Matter 2016, 28, 354003.
\bibitem{ataca2008high} Ataca, C.; Aktürk, E.; Ciraci, S.; Ustunel, H. High-capacity hydrogen storage by metallized graphene. Appl. Phys. Lett. 2008, 93, 043123.
\bibitem{wang2009graphene} Wang, G.; Shen, X.; Yao, J.; Park, J. Graphene nanosheets for enhanced lithium storage in lithium ion batteries. Carbon 2009, 47, 2049–2053.
\bibitem{datta2008crystallographic} Datta, S. S.; Strachan, D. R.; Khamis, S. M.; Johnson, A. C. Crystallographic etching of few-layer graphene. Nano Lett. 2008, 8, 1912–1915.
\bibitem{ci2008controlled} Ci, L.; Xu, Z.; Wang, L.; Gao,W.; Ding, F.; Kelly, K. F.; Yakobson, B. I.; Ajayan, P. M. Controlled nanocutting of graphene. Nano Res. 2008, 1, 116–122.
\bibitem{mcallister2007single} McAllister, M. J. et al. Single sheet functionalized graphene by oxidation and thermal expansion of graphite. Chem. Mater. 2007, 19, 4396–4404.
\bibitem{fujii2010cutting} Fujii, S.; Enoki, T. Cutting of oxidized graphene into nanosized pieces. J. Am. Chem. Soc. 2010, 132, 10034–10041.
\bibitem{han2007energy} Han, M. Y.; Özyilmaz, B.; Zhang, Y.; Kim, P. Energy band-gap engineering of graphene nanoribbons. Phys. Rev. Lett. 2007, 98, 206805.
\bibitem{chen2007graphene} Chen, Z.; Lin, Y.-M.; Rooks, M. J.; Avouris, P. Graphene nano-ribbon electronics. Physica E Low Dimens Syst Nanostruct. 2007, 40, 228–232.
\bibitem{cataldo2010graphene} Cataldo, F.; Compagnini, G.; Patané, G.; Ursini, O.; Angelini, G.; Ribic, P. R.; Margaritondo, G.; Cricenti, A.; Palleschi, G.; Valentini, F. Graphene nanoribbons produced by the oxidative unzipping of single-wall carbon nanotubes. Carbon 2010, 48, 2596–2602.
\bibitem{kumar2011laser} Kumar, P.; Panchakarla, L.; Rao, C. Laser-induced unzipping of carbon nanotubes to yield graphene nanoribbons. Nanoscale 2011, 3, 2127–2129.
\bibitem{eliias2009longitudinal} Elias, A. L.; Botello-Mendez, A. R.; Meneses-Rodriguez, D.; Jehova-Gonzalez, V.; Ramirez-Gonzalez, D.; Ci, L.; Munoz-Sandoval, E.; Ajayan, P. M.; Terrones, H.; Terrones, M. Longitudinal cutting of pure and doped carbon nanotubes to form graphitic nanoribbons using metal clusters as nanoscalpels. Nano Lett. 2009, 10, 366–372.
\bibitem{parashar2011single} Parashar, U. K.; Bhandari, S.; Srivastava, R. K.; Jariwala, D.; Srivastava, A. Single step synthesis of graphene nanoribbons by catalyst particle size dependent cutting of multiwalled carbon nanotubes. Nanoscale 2011, 3, 3876–3882.
\bibitem{jiao2009narrow} Jiao, L.; Zhang, L.; Wang, X.; Diankov, G.; Dai, H. Narrow graphene nanoribbons from carbon nanotubes. Nature 2009, 458, 877–880.
\bibitem{jiao2010aligned} Jiao, L.; Zhang, L.; Ding, L.; Liu, J.; Dai, H. Aligned graphene nanoribbons and crossbars from unzipped carbon nanotubes. Nano Res. 2010, 3, 387–394.
\bibitem{paiva2010unzipping} Paiva, M. C.; Xu, W.; Fernanda Proença, M.; Novais, R. M.; Lægsgaard, E.; Besenbacher, F. Unzipping of functionalized multiwall carbon nanotubes induced by STM. Nano Lett. 2010, 10, 1764–1768.
\bibitem{kim2010graphene} Kim, K.; Sussman, A.; Zettl, A. Graphene nanoribbons obtained by electrically unwrapping carbon nanotubes. ACS Nano 2010, 4, 1362–1366.
\bibitem{cano2009ex} Cano-Márquez, A. et al. Ex-MWNTs: graphene sheets and ribbons produced by lithium intercalation and exfoliation of carbon nanotubes. Nano Lett. 2009, 9, 1527–1533.
\bibitem{shinde2011electrochemical} Shinde, D. B.; Debgupta, J.; Kushwaha, A.; Aslam, M.; Pillai, V. K. Electrochemical unzipping of multi-walled carbon nanotubes for facile synthesis of high-quality graphene nanoribbons. J. Am. Chem. Soc. 2011, 133, 4168–4171.
\bibitem{lin2015adatom} Lin, Y. T.; Chung, H. C.; Yang, P. H.; Lin, S. Y.; Lin, M. F. Adatom bond-induced geometric and electronic properties of passivated armchair graphene nanoribbons. Phys. Chem. Chem. Phys. 2015, 17, 16545–16552.
\bibitem{mao2013edge} Mao, Y.; Hao, W.; Wei, X.; Yuan, J.; Zhong, J. Edge-adsorption of potassium adatoms on graphene nanoribbon: A first principle study. Appl. Surf. Sci. 2013, 280, 698–704.
\bibitem{uthaisar2009lithium} Uthaisar, C.; Barone, V.; Peralta, J. E. Lithium adsorption on zigzag graphene nanoribbons. J. Appl. Phys. 2009, 106, 113715.
\bibitem{johnson2010hydrogen} Johnson, J. L.; Behnam, A.; Pearton, S.; Ural, A. Hydrogen Sensing Using Pd- Functionalized Multi-Layer Graphene Nanoribbon Networks. Adv. Mater. 2010, 22, 4877–4880.
\bibitem{lin2013graphene} Lin, J.; Peng, Z.; Xiang, C.; Ruan, G.; Yan, Z.; Natelson, D.; Tour, J. M. Graphene nanoribbon and nanostructured SnO2 composite anodes for lithium ion batteries. ACS Nano 2013, 7, 6001–6006.
\bibitem{rani2013fluorinated} Rani, J. V.; Kanakaiah, V.; Dadmal, T.; Rao, M. S.; Bhavanarushi, S. Fluorinated natural graphite cathode for rechargeable ionic liquid based aluminum–ion battery. J. Electrochem. Soc. 2013, 160, A1781–A1784.
\bibitem{paek2008enhanced} Paek, S. M.; Yoo, E.; Honma, I. Enhanced cyclic performance and lithium storage capacity of SnO2/graphene nanoporous electrodes with three-dimensionally delaminated flexible structure. Nano Lett. 2008, 9, 72–75.
\bibitem{ao2010high} Ao, Z.; Peeters, F. High-capacity hydrogen storage in Al-adsorbed graphene. Phys. Rev. B 2010, 81, 205406.
\bibitem{hsieh2009tunable} Hsieh, D. et al. A tunable topological insulator in the spin helical Dirac transport regime. Nature 2009, 460, 1101–1105.
\bibitem{lin2012high} Lin, Q.; Li, T.; Liu, Z.; Song, Y.; He, L.; Hu, Z.; Guo, Q.; Ye, H. High-resolution TEM observations of isolated rhombohedral crystallites in graphite blocks. Carbon 2012, 50, 2369–2371.
\bibitem{zhou2014interpretation} Zhou, Z.; Bouwman, W.; Schut, H.; Pappas, C. Interpretation of X-ray diffraction patterns of (nuclear) graphite. Carbon 2014, 69, 17–24.
\bibitem{mcclure1969electron} McClure, J. Electron energy band structure and electronic properties of rhombohedral graphite. Carbon 1969, 7, 425–432.
\bibitem{wanekaya2011applications} Wanekaya, A. K. Applications of nanoscale carbon-based materials in heavy metal sensing and detection. Analyst 2011, 136, 4383–4391.
\bibitem{shan2009polycrystalline} Shan, D.; Zhang, J.; Xue, H.-G.; Zhang, Y.-C.; Cosnier, S.; Ding, S.-N. Polycrystalline bismuth oxide films for development of amperometric biosensor for phenolic compounds. Biosens. Bioelectron. 2009, 24, 3671–3676.
\bibitem{li2013bismuth} Li, Y.; Trujillo, M. A.; Fu, E.; Patterson, B.; Fei, L.; Xu, Y.; Deng, S.; Smirnov, S.; Luo, H. Bismuth oxide: a new lithium-ion battery anode. J. Mater. Chem. A 2013, 1, 12123–12127.
\bibitem{akturk2010bismuth} Aktürk, O. Ü.; Tomak, M. Bismuth doping of graphene. Appl. Phys. Lett. 2010, 96, 081914.
\bibitem{hsu2013first} Hsu, C. H.; Ozolins, V.; Chuang, F. C. First-principles study of Bi and Sb intercalated graphene on SiC (0001) substrate. Surf. Sci. 2013, 616, 149–154.
\bibitem{van1975leed} Van Bommel, A.; Crombeen, J.; Van Tooren, A. LEED and Auger electron observations of the SiC (0001) surface. Surf. Sci. 1975, 48, 463–472.
\bibitem{berger2006electronic} Berger, C. et al. Electronic confinement and coherence in patterned epitaxial graphene. Science 2006, 312, 1191–1196.
\bibitem{brar2007scanning} Brar, V. W.; Zhang, Y.; Yayon, Y.; Ohta, T.; McChesney, J. L.; Bostwick, A.; Rotenberg, E.; Horn, K.; Crommie, M. F. Scanning tunneling spectroscopy of inhomogeneous electronic structure in monolayer and bilayer graphene on SiC. Appl. Phys. Lett. 2007, 91, 122102.
\bibitem{mallet2007electron} Mallet, P.; Varchon, F.; Naud, C.; Magaud, L.; Berger, C.; Veuillen, J. Y. Electron states of mono-and bilayer graphene on SiC probed by scanning-tunneling microscopy. Phys. Rev. B 2007, 76, 041403.
\bibitem{chen2005atomic} Chen, W.; Xu, H.; Liu, L.; Gao, X.; Qi, D.; Peng, G.; Tan, S. C.; Feng, Y.; Loh, K. P.; Wee, A. T. S. Atomic structure of the 6H–SiC (0001) nanomesh. Surf. Sci. 2005, 596, 176–186.
\bibitem{de2007epitaxial} De Heer, W. A. et al. Epitaxial graphene. Solid State Commun. 2007, 143, 92–100.
\bibitem{mattausch2007ab} Mattausch, A.; Pankratov, O. Ab initio study of graphene on SiC. Phys. Rev. Lett. 2007, 99, 076802.
\bibitem{varchon2008ripples} Varchon, F.; Mallet, P.; Veuillen, J. Y.; Magaud, L. Ripples in epitaxial graphene on the Si-terminated SiC (0001) surface. Phys. Rev. B 2008, 77, 235412.
\bibitem{su2015bismuth} Su, D.; Shixue, D.; Guoxiu, W. Bismuth: A new anode for the Na-ion battery. Nano Energy 2015, 12, 88-95.
\bibitem{cho2011insulating} Cho, S.; Butch, N. P.; Paglione, J.; Fuhrer, M. S. Insulating behavior in ultrathin bismuth selenide field effect transistors. Nano Lett. 2011, 11, 1925-1927.
\bibitem{lin2015electric} Lin, Y. P.; Lin, C. Y.; Chang, C. P.; Lin, M. F. Electric-field-induced rich magneto-absorption spectra of ABC-stacked trilayer graphene. RSC Adv. 2015, 5, 80410–80414.
\bibitem{chiu2013critical} Chiu, C. W.; Chen, S. C.; Huang, Y. C.; Shyu, F. L.; Lin, M. F. Critical optical properties of AA-stacked multilayer graphenes. Appl. Phys. Lett. 2013, 103, 041907.
\bibitem{chiu2014layer} Chiu, C. W.; Ho, Y. H.; Shyu, F. L.; Lin, M. F. Layer-enriched optical spectra of AB-stacked multilayer graphene. Appl. Phys. Express 2014, 7, 115102.
\bibitem{lu2006absorption} Lu, C. L.; Lin, H. C.; Hwang, C. C.; Wang, J.; Lin, M. F.; Chang, C. P. Absorption spectra of trilayer rhombohedral graphite. Appl. Phys. Lett. 2006, 89, 221910.
\bibitem{wu2014combined} Wu, J. Y.; Gumbs, G.; Lin, M. F. Combined effect of stacking and magnetic field on plasmon excitations in bilayer graphene. Phys. Rev. B 2014, 89, 165407.
\bibitem{lin2012electrically} Lin, M. F.; Chuang, Y. C.; Wu, J. Y. Electrically tunable plasma excitations in AA-stacked multilayer graphene. Phys. Rev. B 2012, 86, 125434.
\bibitem{jhang2011stacking} Jhang, S. H. et al. Stacking-order dependent transport properties of trilayer graphene. Phys. Rev. B 2011, 84, 161408.
\bibitem{zou2013transport} Zou, K.; Zhang, F.; Clapp, C.; MacDonald, A.; Zhu, J. Transport studies of dualgated ABC and ABA trilayer graphene: band gap opening and band structure tuning in very large perpendicular electric fields. Nano Lett. 2013, 13, 369–373.
\bibitem{khodkov2015direct} Khodkov, T.; Khrapach, I.; Craciun, M. F.; Russo, S. Direct Observation of a Gate Tunable Band Gap in Electrical Transport in ABC-Trilayer Graphene. Nano Lett. 2015, 15, 4429–4433.
\bibitem{tao2015silicene} Tao, L.; Cinquanta, E.; Chiappe, D.; Grazianetti, C.; Fanciulli, M.; Dubey, M.; Molle, A.; Akinwande, D. Silicene field-effect transistors operating at room temperature. Nat. Nanotechnol. 2015, 10, 227–231.
\bibitem{meng2013buckled} Meng, L. et al. Buckled silicene formation on Ir (111). Nano Lett. 2013, 13, 685–690.
\bibitem{li2014buckled} Li, L.; Lu, S. Z.; Pan, J.; Qin, Z.; Wang, Y.-q.; Wang, Y.; Cao, G. Y.; Du, S.; Gao, H. J. Buckled germanene formation on Pt (111). Adv. Mater. 2014, 26, 4820–4824.
\bibitem{davila2014germanene} Dávila, M.; Xian, L.; Cahangirov, S.; Rubio, A.; Le Lay, G. Germanene: a novel twodimensional germanium allotrope akin to graphene and silicene. New J. Phys. 2014, 16, 095002.
\bibitem{zhu2015epitaxial} Zhu, F. F.; Chen, W. J.; Xu, Y.; Gao, C. L.; Guan, D. D.; Liu, C. H.; Qian, D.; Zhang, S. C.; Jia, J. F. Epitaxial growth of two-dimensional stanene. Nat. Mater. 2015, 14, 1020–1025.
\bibitem{liu2015semiconducting} Liu, H.; Du, Y.; Deng, Y.; Peide, D. Y. Semiconducting black phosphorus: synthesis, transport properties and electronic applications. Chem. Soc. Rev. 2015, 44, 2732–2743.
\bibitem{li2014black} Li, L.; Yu, Y.; Ye, G. J.; Ge, Q.; Ou, X.; Wu, H.; Feng, D.; Chen, X. H.; Zhang, Y. Black phosphorus field-effect transistors. Nat. Nanotechnol. 2014, 9, 372–377.
\bibitem{mak2010atomically} Mak, K. F.; Lee, C.; Hone, J.; Shan, J.; Heinz, T. F. Atomically thin MoS2: a new direct-gap semiconductor. Phys. Re. Lett. 2010, 105, 136805.
\bibitem{coleman2011two} Coleman, J. N. et al. Two-dimensional nanosheets produced by liquid exfoliation of layered materials. Science 2011, 331, 568–571.
\bibitem{chen2016magnetic} Chen, S. C.; Wu, C. L.; Wu, J. Y.; Lin, M. F. Magnetic quantization of sp3 bonding in monolayer gray tin. Phys. Rev. B 2016, 94, 045410.
\bibitem{ho2014magneto} Ho, Y. H.; Chiu, C. W.; Su, W. P.; Lin, M. F. Magneto-optical spectra of transition metal dichalcogenides: A comparative study. Appl. Phys. Lett. 2014, 105, 222411.
\bibitem{qiu2015ordered} Qiu, J.; Fu, H.; Xu, Y.; Oreshkin, A.; Shao, T.; Li, H.; Meng, S.; Chen, L.; Wu, K. Ordered and reversible hydrogenation of silicene. Phys. Rev. Lett. 2015, 114, 126101.
\bibitem{du2014tuning} Du, Y. et al. Tuning the band gap in silicene by oxidation. ACS Nano 2014, 8, 10019–10025.
\bibitem{molle2013hindering} Molle, A.; Grazianetti, C.; Chiappe, D.; Cinquanta, E.; Cianci, E.; Tallarida, G.; Fanciulli, M. Hindering the oxidation of silicene with non-reactive encapsulation. Adv. Funct. Mater. 2013, 23, 4340–4344.
\bibitem{houssa2011electronic} Houssa, M.; Scalise, E.; Sankaran, K.; Pourtois, G.; AfanasâAZEv, V.; Stesmans, A. Electronic properties of hydrogenated silicene and germanene. Appl. Phys. Lett. 2011, 98, 223107.
\bibitem{wei2013many} Wei, W.; Dai, Y.; Huang, B.; Jacob, T. Many-body effects in silicene, silicane, germanene and germanane. Phys. Chem. Chem. Phys. 2013, 15, 8789–8794.
\bibitem{lin2015h} Lin, S. Y.; Chang, S. L.; Tran, N. T. T.; Yang, P. H.; Lin, M. F. H–Si bonding-induced unusual electronic properties of silicene: a method to identify hydrogen concentration. Phys. Chem. Chem. Phys. 2015, 17, 26443–26450.
\bibitem{si2014functionalized} Si, C.; Liu, J.; Xu, Y.; Wu, J.; Gu, B. L.; Duan, W. Functionalized germanene as a prototype of large-gap two-dimensional topological insulators. Phys. Rev. B 2014, 89, 115429.

\end{thebibliography}
\end{document}